\documentclass[english]{article}
\usepackage[T1]{fontenc}
\usepackage[latin9]{inputenc}
\usepackage{geometry}
\usepackage{array}
\usepackage{float}
\usepackage{multirow}
\usepackage{amsmath}
\usepackage{amssymb}
\usepackage{setspace}
\usepackage{esint}
\usepackage[authoryear]{natbib}
\PassOptionsToPackage{normalem}{ulem}
\usepackage{ulem}
\usepackage{url}
\usepackage{graphicx}
\makeatletter
\newcommand{\lyxaddress}[1]{
\par {\raggedright #1
\noindent\par}
}
\makeatother

\usepackage{babel}
\begin{document}
\date{}

\title{LSMM: A statistical approach to integrating functional annotations with genome-wide association studies}

\author{Jingsi Ming$^{1}$, Mingwei Dai$^{2,5}$, Mingxuan Cai$^{1}$,\\ Xiang
Wan$^{3}$, Jin Liu$^{4}$\thanks{Correspondence should be addressed to Can Yang (macyang@ust.hk) and
Jin Liu (jin.liu@duke-nus.edu.sg)} and Can Yang$^{5*}$}

\maketitle

\lyxaddress{\begin{center}
$^{1}$Department of Mathematics, Hong Kong Baptist University, Hong
Kong\\
$^{2}$School of Mathematics and Statistics, Xi'an Jiaotong University,
Xi'an, China\\
$^{3}$Department of Computer Science, Hong Kong Baptist University,
Hong Kong\\
$^{4}$Centre for Quantitative Medicine, Duke-NUS Medical School,
Singapore\\
$^{5}$Department of Mathematics, The Hong Kong University of Science
and Technology, Hong Kong
\par\end{center}}

\begin{abstract}
Thousands of risk variants underlying complex phenotypes (quantitative traits and diseases) have been identified in genome-wide association studies (GWAS). However, there are still two major challenges towards deepening our understanding of the genetic architectures of complex phenotypes. First, the majority of GWAS hits are in the non-coding region and their biological interpretation is still unclear. Second, accumulating evidence from GWAS suggests the polygenicity of complex traits, i.e., a complex trait is often affected by many variants with small or moderate effects, whereas a large proportion of risk variants with small effects remains unknown. The availability of functional annotation data enables us to address the above challenges. In this study, we propose a latent sparse mixed model (LSMM) to integrate functional annotations with GWAS data. Not only does it increase statistical power of the identification of risk variants, but also offers more biological insights by detecting relevant functional annotations. To allow LSMM scalable to millions of variants and hundreds of functional annotations, we developed an efficient variational expectation-maximization (EM) algorithm for model parameter estimation and statistical inference. We first conducted comprehensive simulation studies to evaluate the performance of LSMM. Then we applied it to analyze 30 GWAS of complex phenotypes integrated with 9 genic category annotations and 127 tissue-specific functional annotations from the Roadmap project. The results demonstrate that our method possesses more statistical power over conventional methods, and can help researchers achieve deeper understanding of genetic architecture of these complex phenotypes. The LSMM software is available at https://github.com/mingjingsi/LSMM.
\end{abstract}
\pagebreak{}

\section{Introduction}
Since the success of the first GWAS on age-related macular degeneration \citep{klein2015complement}, more than 40,000 single-nucleotide polymorphisms (SNPs) have been reported in about 3,100 GWAS at the genome-wide significance level (see GWAS Catalog http://www.ebi.ac.uk/gwas/) \citep{welter2014nhgri}. Despite these fruitful discoveries, the emerging evidence from GWAS presents great challenges towards deeper understanding of the genetic architectures of complex phenotypes. First, more than 85$\%$ genome-wide significant hits are located in the non-coding region \citep{welter2014nhgri} and thus their functional roles are still largely elusive. Second, complex phenotypes are often highly polygenic, i.e., they are affected by a vast number of risk variants with individually small effects. For example, 70$\%$-80$\%$ of the variation in human height can be attributed to genetics \citep{visscher2008heritability}. However, \citet{wood2014defining} collected more than 250,000 samples and identified 697 variants at genome-wide significance level, and all these variants together can only explain 20$\%$ of heritability. A recent estimate \citep{boyle2017expanded} suggests that about 100,000 variants may be associated with human height. Given current sample sizes, a large proportion of risk variants underlying complex phenotypes remain unknown yet.

Fortunately, an increasing number of reports suggest that the functional importance of SNPs may not be equal \citep{schork2013all}, which provides a direction to address the above challenges. On one hand, SNPs in or near genic regions can explain more heritability of complex phenotypes \citep{yang2011genome, smith2011genome}. For example, the partition of genic category annotations for SNPs have revealed that SNPs in 5' UTR, exon and 3' UTR are significantly enriched across diverse complex traits \citep{schork2013all}. On the other hand, tissue-specific functional annotations can provide information that is complementary to genic category annotations, for dissecting genetic contribution to complex diseases in a tissue-specific manner. To name a few, genetic variants related to functions of immune cells are significantly enriched for immune diseases, such as rheumatoid arthritis, coeliac disease and type 1 diabetes; variants with liver functions are enriched for metabolic traits, such as LDL, HDL and total cholesterol; variants with pancreatic islet functions are enriched for fasting glucose \citep{kundaje2015integrative}. Additionally, SNPs in genes that are preferentially expressed in the central nervous system are significantly enriched in psychiatric disorders (e.g., schizophrenia and bipolar disorder) \citep{chung2014gpa}.

A large amount of functional annotation data has become publicly available and the volume is still expanding. The Encyclopedia of DNA Elements (ENCODE) project \citep{encode2012integrated} have conducted more than 1,650 experiments on 147 cell lines to access functional elements across the human genome, such as DNase I hypersensitive sites and transcription factor binding. The NIH Roadmap Epigenomics Mapping Consortium \citep{kundaje2015integrative} is generating high-quality genome-wide human epigenomic maps of histone modifications, chromatin accessibility, DNA methylation and mRNA expression across more than one hundred of human cell types and tissues.

With the availability of rich functional annotations, we aim to (1) integrate genic category annotations and tissue-specific functional annotations with GWAS to increase the statistical power of the identification of risk SNPs, and (2) detect relevant tissue-specific functional annotations among a large amount of available annotation data to have a more biologically insightful interpretation of GWAS results. Statistical methods to incorporate genic category annotations have been proposed, e.g., stratified FDR methods \citep{schork2013all}, cmfdr \citep{zablocki2014covariate}, GPA \citep{chung2014gpa} and EPS \citep{liu2016eps}. However, these methods were designed to handle a few number of functional annotations and can not be scalable to a large-scale integrative analysis.

In this study, we propose a \underline{L}atent \underline{S}parse \underline{M}ixed \underline{M}odel (LSMM) to integrate genic category annotations and tissue-specific functional annotations with GWAS data. The ``latent" statuses are used to connect the observed summary statistics from GWAS with functional annotations. ``Mixed" models are designed to simultaneously consider both genic category and tissue-specific annotations, where genic category annotations are put into the design matrix of fixed effects, and tissue-specific annotations are encoded in the design matrix of random effects. We further impose a ``sparse" structure on the random effects to adaptively select relevant tissue-specific annotations. We conducted comprehensive simulations to investigate the properties of LSMM and then applied LSMM to real data. We integrated summary statistics from 30 GWAS with 9 genic category annotations and 127 tissue-specific functional annotations from the Roadmap project. Compared with conventional methods, our method is able to increase the statistical power in the identification of risk variants and detection of tissue-specific functional annotations and providing a deeper understanding of genetic architecture of complex phenotypes.

\section{Latent Sparse Mixed Model (LSMM)}

\subsection{Model}
Suppose we have the summary statistics ($p$-values) of $M$ SNPs from GWAS. Consider the two-groups model \citep{efron2008microarrays}, i.e., SNPs either belong to null or non-null group. Let $\gamma_{j}$ be the latent variable indicating the membership of the $j$-th SNP, i.e., $\gamma_j = 0$ or $\gamma_j = 1$ indicates the $j$-th SNP from null or non-null group, respectively. We further denote the proportion of null and non-null group as $\pi_0$ and $\pi_1$, respectively. Then we model the observed $p$-values as \citep{chung2014gpa},
\textbf{
	\begin{equation}
	p_{j}\sim\begin{cases}
	U\left[0,1\right], & \gamma_{j}=0,\\
	Beta\left(\alpha,1\right), & \gamma_{j}=1,
	\end{cases}
	\end{equation}
}where $U[0,1]$ denotes the uniform distribution on [0,1] and $Beta(\alpha,1)$ is the beta distribution with parameter $(\alpha,1)$. We constrain $0<\alpha<1$ to model the fact that $p$-values from the non-null group tend to be closer to 0 rather than 1.

Suppose that we have collected not only the $p$-values of $M$ SNPs from GWAS, but also functional annotations of these SNPs. To incorporate information from functional annotations for prioritization of risk variants and detection of tissue-specific functions for a complex phenotype, we consider the following latent sparse mixed model:
\begin{equation}
\log\frac{\Pr\left(\gamma_{j}=1|\mathbf{Z}_{j},\mathbf{A}_{j}\right)}{\Pr\left(\gamma_{j}=0|\mathbf{Z}_{j},\mathbf{A}_{j}\right)}=\mathbf{Z}_{j}\mathbf{b}+\mathbf{A}_{j}\boldsymbol{\beta},\label{eq:logistic}
\end{equation}
where $\mathbf{Z}\in\mathbb{R}^{M\times\left(L+1\right)}$ is the
design matrix for fixed effects, comprised of an intercept and $L$ covariates, $\mathbf{b}\in\mathbb{R}^{L+1}$ is the vector of fixed effects, $\mathbf{A}\in\mathbb{R}^{M\times K}$ is the design matrix for random effects, $\boldsymbol{\beta}\in\mathbb{R}^{K}$ is the vector of random effects, and $K$ is the number of random effects. Both the $j$-th row of $\mathbf{Z}$ (i.e., $\mathbf{Z}_{j}$) and $\mathbf{A}$ (i.e., $\mathbf{A}_{j}$) corresponds to the $j$-th SNP. Note that $\gamma_{j}$ is a latent variable in model (\ref{eq:logistic}) but its corresponding $p_{j}$ is observed. This makes our model different from the standard generalized linear mixed model.

Now we partition functional annotations into two categories: genic category annotations and tissue-specific annotations. According to \citep{schork2013all}, genomic regions, such as exon, intron, 5'UTR and 3'UTR, are considered as genic category annotations. For tissue-specific annotations, we used epigenetic markers (H3k4me1, H3k4me3, H3k36me3, H3k27me3, H3k9me3, H3k27ac, H3k9ac, and DNase I Hypersensitivity) of multiple tissues from the Roadmap project. As we are more interested in the detection of tissue-specific results, we put genic category annotation data into $\mathbf{Z}$ and tissue-specific annotation data into $\mathbf{A}$, where each column of $\mathbf{Z}$ corresponds to a genic functional category and each column of $\mathbf{A}$ corresponds to a tissue-specific functional category. In the simplest case, the entries in $\mathbf{Z}$ and $\mathbf{A}$ are binary. For example, $Z_{jl} = 1$ means that the $j$-th SNP has a function in the $l$-th genic category and $Z_{jl} = 0$ otherwise. Our model also allows the entries in $\mathbf{Z}$ and $\mathbf{A}$ to be continuous variables, e.g., a score $Z_{jl}$ between 0 and 1 can be used to indicate the degree that the $j$-the SNP has a function in the $l$-th category. The closer to 1, the more likely it has a functional role. The entries in $\mathbf{A}$ are defined in the same way as those of $\mathbf{Z}$.

To adaptively select tissue-specific annotations, we assign a spike-slab prior on $\beta_{k}$:
\begin{equation}
\beta_{k}\sim\begin{cases}
N\left(\beta_{k}|0,\sigma^{2}\right), & \eta_{k}=1,\\
\delta_{0}\left(\beta_{k}\right), & \eta_{k}=0,
\end{cases}\label{eq:betaprior}
\end{equation}
where $N\left(\beta_{k}|0,\sigma^{2}\right)$ denotes the Gaussian distribution with mean $0$ and variance $\sigma^{2}$, $\delta_{0}$ denotes the Dirac delta function at zero, $\eta_{k}=1$ or $\eta_{k}=0$ means the $k$-th annotation is relevant or irrelevant to the given phenotype, respectively. Here $\eta_k$ is a Bernoulli variable with probability $\omega$ being 1:
\begin{equation}
\eta_k \sim \omega^{\eta_k} (1-\omega)^{1-\eta_k },
\end{equation}
where $\omega$ can be interpreted as the proportion of relevant annotations corresponding to this phenotype.

Let $\boldsymbol{\theta}=\left\{ \alpha,\mathbf{b},\sigma^{2},\omega\right\}$ be the collection of model parameters. The logarithm of the marginal likelihood can be written as
\begin{equation}
\log\Pr\left(\mathbf{p}|\mathbf{Z},\mathbf{A};\boldsymbol{\theta}\right)=\log\sum_{\boldsymbol{\gamma}}\sum_{\boldsymbol{\eta}}\int\Pr\left(\mathbf{p},\boldsymbol{\gamma},\boldsymbol{\beta},\boldsymbol{\eta}|\mathbf{Z},\mathbf{A};\boldsymbol{\theta}\right)d\boldsymbol{\beta},\label{eq:LL}
\end{equation}
where
\begin{equation}
\Pr\left(\mathbf{p},\boldsymbol{\gamma},\boldsymbol{\beta},\boldsymbol{\eta}|\mathbf{Z},\mathbf{A};\boldsymbol{\theta}\right)=\Pr\left(\mathbf{p}|\boldsymbol{\gamma};\alpha\right)\Pr\left(\boldsymbol{\gamma}|\mathbf{Z},\mathbf{A},\boldsymbol{\beta};\mathbf{b}\right)\Pr\left(\boldsymbol{\beta}|\boldsymbol{\eta};\sigma^{2}\right)\Pr\left(\boldsymbol{\eta}|\omega\right).
\end{equation}
Our goal is to maximize the marginal likelihood to obtain the estimation $\hat{\boldsymbol{\theta}}$ of $\boldsymbol{\theta}$ and compute the posterior
\begin{equation}
\Pr\left(\boldsymbol{\gamma},\boldsymbol{\beta},\boldsymbol{\eta}|\mathbf{p},\mathbf{Z},\mathbf{A};\hat{\boldsymbol{\theta}}\right)=\frac{\Pr\left(\mathbf{p},\boldsymbol{\gamma},\boldsymbol{\beta},\boldsymbol{\eta}|\mathbf{Z},\mathbf{A};\hat{\boldsymbol{\theta}}\right)}{\Pr\left(\mathbf{p}|\mathbf{Z},\mathbf{A};\hat{\boldsymbol{\theta}}\right)}.\label{eq:posterior}
\end{equation}
Then we can infer the risk SNPs and relevant tissue-specific functional annotations for this phenotype and calculate the false discovery rate.

\subsection{Algorithm}
Exact evaluation of posterior (\ref{eq:posterior}) is intractable. One difficulty is due to the sigmoid function resulting from the logistic model. The other comes from the spike-slab prior. To address this issue, we propose a variational EM algorithm for parameter estimation and posterior approximation.

Before starting the derivation of our algorithm, we first re-parametrize the spike-slab prior (\ref{eq:betaprior}) by introducing a new Gaussian variable $\tilde{\beta}_{k}\sim N\left(0,\sigma^{2}\right)$, then the product $\eta_{k}\tilde{\beta}_{k}$ has the same distribution with $\beta_{k}$ in model (\ref{eq:betaprior}). So model (\ref{eq:logistic}) can be written as
\begin{equation}
\log\frac{\Pr\left(\gamma_{j}=1|\mathbf{Z}_{j},\mathbf{A}_{j}\right)}{\Pr\left(\gamma_{j}=0|\mathbf{Z}_{j},\mathbf{A}_{j}\right)}=\mathbf{Z}_{j}\mathbf{b}+\sum_{k=1}^{K}A_{jk}\beta_{k}=\mathbf{Z}_{j}\mathbf{b}+\sum_{k=1}^{K}A_{jk}\eta_{k}\tilde{\beta}_{k}.
\end{equation}
Hence the complete-data likelihood $\Pr\left(\mathbf{p},\boldsymbol{\gamma},\boldsymbol{\beta},\boldsymbol{\eta}|\mathbf{Z},\mathbf{A};\boldsymbol{\theta}\right)$ can be re-written as
\begin{equation}
\Pr\left(\mathbf{p},\boldsymbol{\gamma},\tilde{\boldsymbol{\beta}},\boldsymbol{\eta}|\mathbf{Z},\mathbf{A};\boldsymbol{\theta}\right)=\Pr\left(\mathbf{p}|\boldsymbol{\gamma};\alpha\right)\Pr\left(\boldsymbol{\gamma}|\mathbf{Z},\mathbf{A},\tilde{\boldsymbol{\beta}},\boldsymbol{\eta};\mathbf{b}\right)\Pr\left(\tilde{\boldsymbol{\beta}},\boldsymbol{\eta}|\sigma^{2},\omega\right),
\end{equation}
where
\begin{equation}
\mathbf{\Pr\left(p|\gamma;\alpha\right)} = \prod_{j=1}^{M}\Pr\left(p_{j}|\gamma_{j};\alpha\right)=\prod_{j=1}^{M}\left(\alpha p_{j}^{\alpha-1}\right)^{\gamma_{j}},
\end{equation}
\begin{align}
& \Pr\left(\boldsymbol{\gamma}|\mathbf{Z},\mathbf{A},\tilde{\boldsymbol{\beta}},\boldsymbol{\eta};\mathbf{b}\right)\nonumber
=  \prod_{j=1}^{M}\Pr\left(\gamma_{j}|\mathbf{Z}_{j},\mathbf{A}_{j},\tilde{\boldsymbol{\beta}},\boldsymbol{\eta};\mathbf{b}\right)\nonumber \\
= & \prod_{j=1}^{M}e^{\gamma_{j}\left(\mathbf{Z}_{j}\mathbf{b}+\sum_{k}A_{jk}\eta_{k}\tilde{\beta}_{k}\right)}S\left(-\mathbf{Z}_{j}\mathbf{b}-\sum_{k=1}^{K}A_{jk}\eta_{k}\tilde{\beta}_{k}\right),\label{eq:gammaprior}
\end{align}
\begin{equation}
\Pr\left(\tilde{\boldsymbol{\beta}},\boldsymbol{\eta}|\sigma^{2},\omega\right)=\Pr\left(\tilde{\boldsymbol{\beta}}|\sigma^{2}\right)\Pr\left(\boldsymbol{\eta}|\omega\right)=\prod_{k=1}^{K}N\left(\tilde{\beta}_{k}|0,\sigma^{2}\right)\omega^{\eta_{k}}\left(1-\omega\right)^{1-\eta_{k}},\label{eq:betaetaprior}
\end{equation}
where $S\left(\cdot\right)$ is the sigmoid function and $S\left(x\right)=\left(1+e^{-x}\right)^{-1}$. With this reparameterization, we get rid of the Dirac delta function.

Due to the intractability caused by the sigmoid function inside integration (\ref{eq:LL}), we consider the JJ bound \citep{jaakkola2000bayesian}:
\begin{equation}
S\left(x\right)\ge S\left(\xi\right)\exp\left\{ \left(x-\xi\right)/2-\lambda\left(\xi\right)\left(x^{2}-\xi^{2}\right)\right\} ,\label{eq:JJbound}
\end{equation}
where $\lambda\left(\xi\right)=\frac{1}{2\xi}\left[S\left(\xi\right)-\frac{1}{2}\right]$ and the right-hand-side of the inequality (\ref{eq:JJbound}) is the JJ bound. Clearly, the JJ bound is in the exponential of a quadratic form. Applying this bound to (\ref{eq:gammaprior}), we can get a tractable lower bound of $\Pr\left(\boldsymbol{\gamma}|\mathbf{Z},\mathbf{A},\tilde{\boldsymbol{\beta}},\boldsymbol{\eta};\mathbf{b}\right)$ , denoted as $h\left(\boldsymbol{\gamma}|\mathbf{Z},\mathbf{A},\tilde{\boldsymbol{\beta}},\boldsymbol{\eta};\mathbf{b},\boldsymbol{\xi}\right)$, where $\boldsymbol{\xi}\in\mathbb{R}^{M}$ is variational parameter. Let $\boldsymbol{\Theta}=\left\{\alpha,\mathbf{b},\boldsymbol{\xi},\sigma^{2},\omega\right\}$. The lower bound of the complete-data likelihood is defined as
\begin{equation}
f\left(\mathbf{p},\boldsymbol{\gamma},\tilde{\boldsymbol{\beta}},\boldsymbol{\eta}|\mathbf{Z},\mathbf{A};\boldsymbol{\Theta}\right)=\Pr\left(\mathbf{p}|\boldsymbol{\gamma};\alpha\right)h\left(\boldsymbol{\gamma}|\mathbf{Z},\mathbf{A},\tilde{\boldsymbol{\beta}},\boldsymbol{\eta};\mathbf{b},\boldsymbol{\xi}\right)\Pr\left(\tilde{\boldsymbol{\beta}},\boldsymbol{\eta}|\sigma^{2},\omega\right).
\end{equation}

Next we derive the variational EM algorithm. Let $q\left(\boldsymbol{\gamma},\tilde{\boldsymbol{\beta}},\boldsymbol{\eta}\right)$ be an approximation of the posterior $\Pr\left(\boldsymbol{\gamma},\tilde{\boldsymbol{\beta}},\boldsymbol{\eta}|\mathbf{p},\mathbf{Z},\mathbf{A};\boldsymbol{\theta}\right)$. We can obtain a lower bound of the logarithm of the marginal likelihood
\begin{align}
& \log\Pr\left(\mathbf{p}|\mathbf{Z},\mathbf{A};\boldsymbol{\theta}\right)\nonumber \\
= & \log\sum_{\boldsymbol{\gamma}}\sum_{\boldsymbol{\eta}}\int\Pr\left(\mathbf{p},\boldsymbol{\gamma},\tilde{\boldsymbol{\beta}},\boldsymbol{\eta}|\mathbf{Z},\mathbf{A};\boldsymbol{\theta}\right)d\tilde{\boldsymbol{\beta}}\nonumber \\
\ge & \log\sum_{\boldsymbol{\gamma}}\sum_{\boldsymbol{\eta}}\int f\left(\mathbf{p},\boldsymbol{\gamma},\tilde{\boldsymbol{\beta}},\boldsymbol{\eta}|\mathbf{Z},\mathbf{A};\boldsymbol{\Theta}\right)d\tilde{\boldsymbol{\beta}}\nonumber \\
\ge & \sum_{\boldsymbol{\gamma}}\sum_{\boldsymbol{\eta}} q\left(\boldsymbol{\gamma},\tilde{\boldsymbol{\beta}},\boldsymbol{\eta}\right)\log\frac{f\left(\mathbf{p},\boldsymbol{\gamma},\tilde{\boldsymbol{\beta}},\boldsymbol{\eta}|\mathbf{Z},\mathbf{A};\boldsymbol{\Theta}\right)}{q\left(\boldsymbol{\gamma},\tilde{\boldsymbol{\beta}},\boldsymbol{\eta}\right)}d\tilde{\boldsymbol{\beta}}\nonumber \\
= & \mathbf{E}_{q}\left[\log f\left(\mathbf{p},\boldsymbol{\gamma},\tilde{\boldsymbol{\beta}},\boldsymbol{\eta}|\mathbf{Z},\mathbf{A};\boldsymbol{\Theta}\right)-\log q\left(\boldsymbol{\gamma},\tilde{\boldsymbol{\beta}},\boldsymbol{\eta}\right)\right]\nonumber \\
\triangleq & L\left(q\right),\label{eq:Lq}
\end{align}
where $L(q)$ is the lower bound. The first inequality is based on the JJ bound. The second inequality follows Jensen's inequality. To make it feasible to evaluate the lower bound, we use the mean-field theory and assume that $q\left(\boldsymbol{\gamma},\tilde{\boldsymbol{\beta}},\boldsymbol{\eta}\right)$ can be factorized as
\begin{equation}
q\left(\boldsymbol{\gamma},\tilde{\boldsymbol{\beta}},\boldsymbol{\eta}\right)=\left(\prod_{k=1}^{K}q\left(\tilde{\beta}_{k},\eta_{k}\right)\right)\left(\prod_{j=1}^{M}q\left(\gamma_{j}\right)\right),
\end{equation}
where $q\left(\tilde{\beta}_{k},\eta_{k}\right)=q\left(\tilde{\beta}_{k}|\eta_{k}\right)q\left(\eta_{k}\right)$. It turns out that $q\left(\boldsymbol{\gamma},\tilde{\boldsymbol{\beta}},\boldsymbol{\eta}\right)$ can be obtained analytically and thus the lower bound $L(q)$ can be exactly evaluated. By setting the derivative of $L(q)$ with respect to the parameters in $\boldsymbol{\Theta}$ be zero, we can obtain the updating equations for parameter estimation. The detailed derivation of the algorithm can be found in Section 1 of Supplementary Document.

We note that LSMM covers two special cases: (1) Two-groups model only (denoted as TGM), when all the coefficients in $\mathbf{b}$ (except the intercept term) and $\boldsymbol{\beta}$ are zero; (2) Two-groups model plus fixed effects model only (denoted as LFM for the abbreviation of latent fixed effect model), when all coefficients in $\boldsymbol{\beta}$ are zero. This motivates us developing a four-stage algorithm based on warm starts. More specifically, in the first stage, we run an EM algorithm to obtain the two parameters ($\alpha$ and the proportion of non-null group $\pi_{1}$) in the TGM. Then we use the estimated parameters as the starting point to run the second stage variational EM algorithm to fit the LFM and obtain the parameter $\alpha$, $\mathbf{b}$ and the posterior probability of $\boldsymbol{\gamma}$. In the third stage, we treat the obtained posterior as the value of $\boldsymbol{\gamma}$ and fit the logistic sparse mixed model to obtain the required initial value for the parameters in the next stage. Finally, in the fourth stage we run the above variational EM algorithm with the obtained parameters at the second and third stage until convergence. Since all the iterations are built upon the framework of EM algorithm, the lower bound is guaranteed to increase at each iteration. The details of the algorithm design are provided in Section 2 of Supplementary Document.

\subsection{Identification of risk SNPs and Detection of relevant tissue-specific functional annotations}
After the convergence of the variational EM algorithm, the approximated posterior of latent variables $\boldsymbol{\gamma}$ and $\boldsymbol{\eta}$ can be obtained. Using this information, we are able to prioritize risk SNPs and relevant tissue-specific functional annotations.

Risk SNPs are identified based on $q\left(\gamma_{j}=1\right)$, an approximation of the posterior probability that the $j$-th SNP is associated with this phenotype. Accordingly, we can calculate the approximated local false discovery rate $fdr_{j}=1-q\left(\gamma_{j}=1\right)$. To control the global false discovery rate (FDR), we sort SNPs by $fdr$ from the smallest to the largest and regard the $j$-th re-ordered SNP as a risk SNP if 
\begin{equation}
FDR_{(j)}=\frac{\sum_{i=1}^{j}fdr_{(j)}}{j}\le\tau,
\end{equation}
where $fdr_{(j)}$ is the $j$-th ordered $fdr$, $FDR_{(j)}$ is the corresponding global FDR, and $\tau$ is the threshold of global FDR. In simulations, we chose $\tau=0.1$.

Relevant tissue-specific functional annotations are inferred from $q\left(\eta_{k}=1\right)$, an approximation of the posterior probability that annotation $k$ is relevant to this phenotype. Similarly, we can calculate the approximated local false discovery rate $fdr_{k}=1-q\left(\eta_{k}=1\right)$ and convert it into the global false discovery rate. We can either control the local false discovery rate (e.g., $fdr_{k}\le0.1$) or global false discovery rate with $\tau=0.1$.

\section{Results}

\subsection{Simulation}
We conducted simulations to evaluate the performance of the proposed LSMM. The simulation data was generated as follows. The numbers of SNPs, fixed effects (genic category annotations) and random effects (tissue-specific functional annotations) were set to be $M=100,000$, $L=10$ and $K=500$ respectively. The entries in design matrices $Z_{jl}$ and $A_{jk}$ were generated from $Bernoulli\left(0.1\right)$, $j=1,...,M$, $l=1,...,L$ and $k=1,...,K$. Given the proportion of relevant tissue-specific functional annotations $\omega$, $\eta_{k}$ was drawn from $Bernoulli\left(\omega\right)$ and the corresponding nonzero entries of random effects $\boldsymbol{\beta}$ were simulated from $N\left(0,1\right)$. The first entry of the coefficients of fixed effects $\mathbf{b}$, i.e., the intercept in the logistic model, was fixed at $-2$ and other entries were generated from $N\left(0,1\right)$ and then kept fixed in multiple replications. After that, we simulated $\gamma_{j}$ from Bernoulli distribution with probability $S\left(\mathbf{Z}_{j}\mathbf{b}+\mathbf{A}_{j}\boldsymbol{\beta}\right)$, and then generated $p_{j}$ from $U\left[0,1\right]$ if $\gamma_{j}=0$ and $Beta\left(\alpha,1\right)$ otherwise.

\begin{figure}[!htbp]
	\centering
	\includegraphics[width=0.23\textwidth]{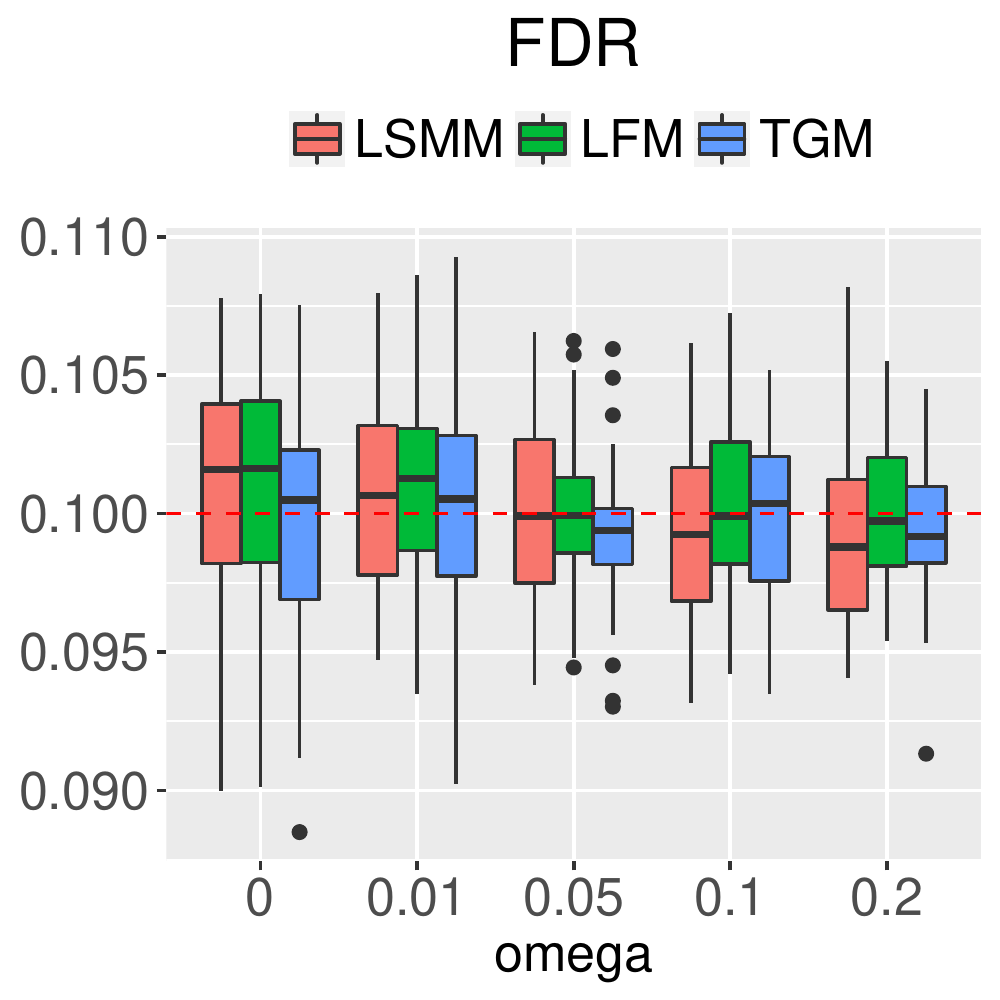}\quad
	\includegraphics[width=0.23\textwidth]{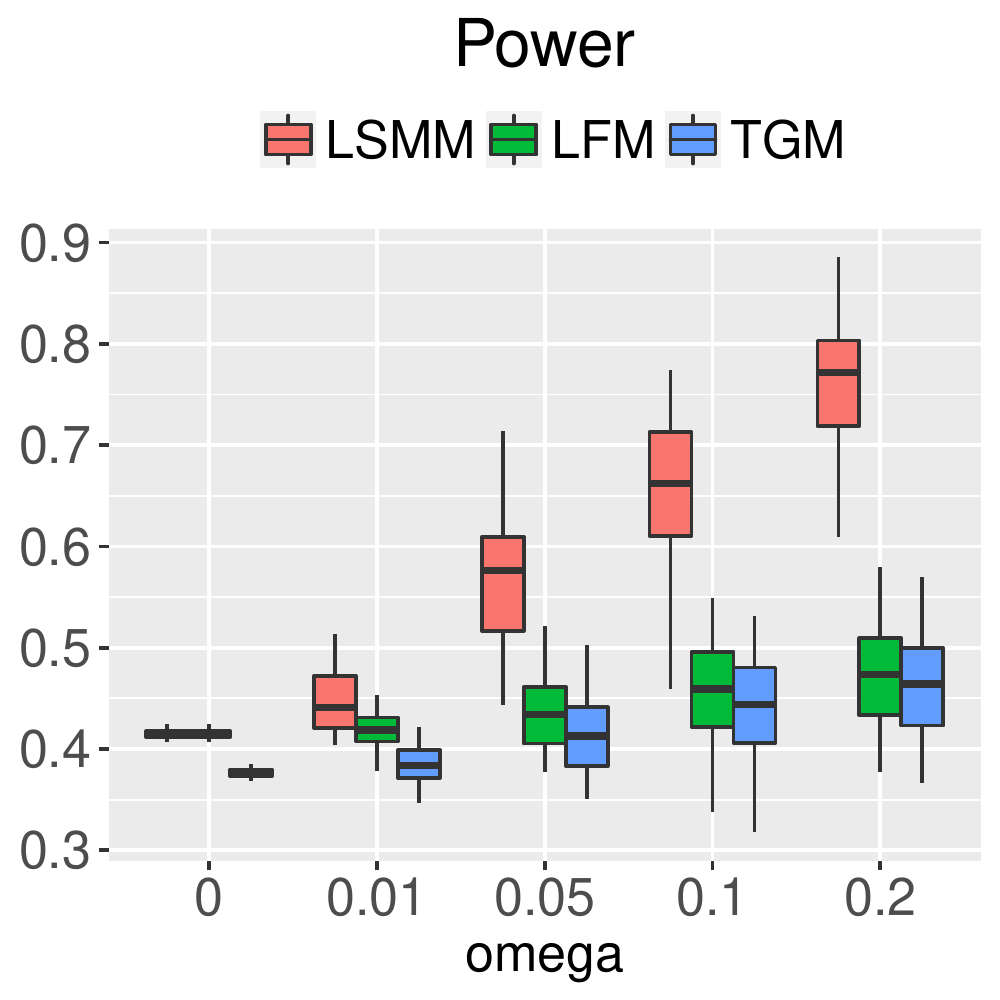}\quad
	\includegraphics[width=0.23\textwidth]{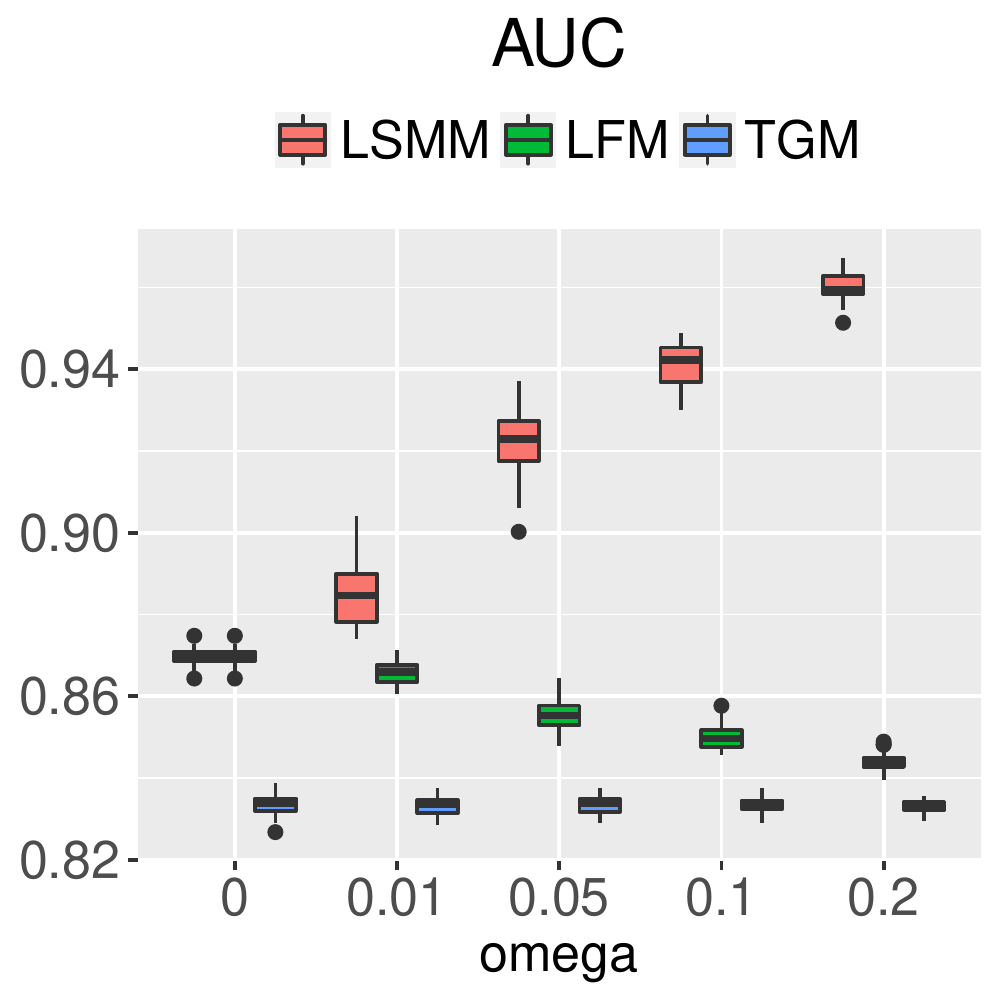}\quad
	\includegraphics[width=0.23\textwidth]{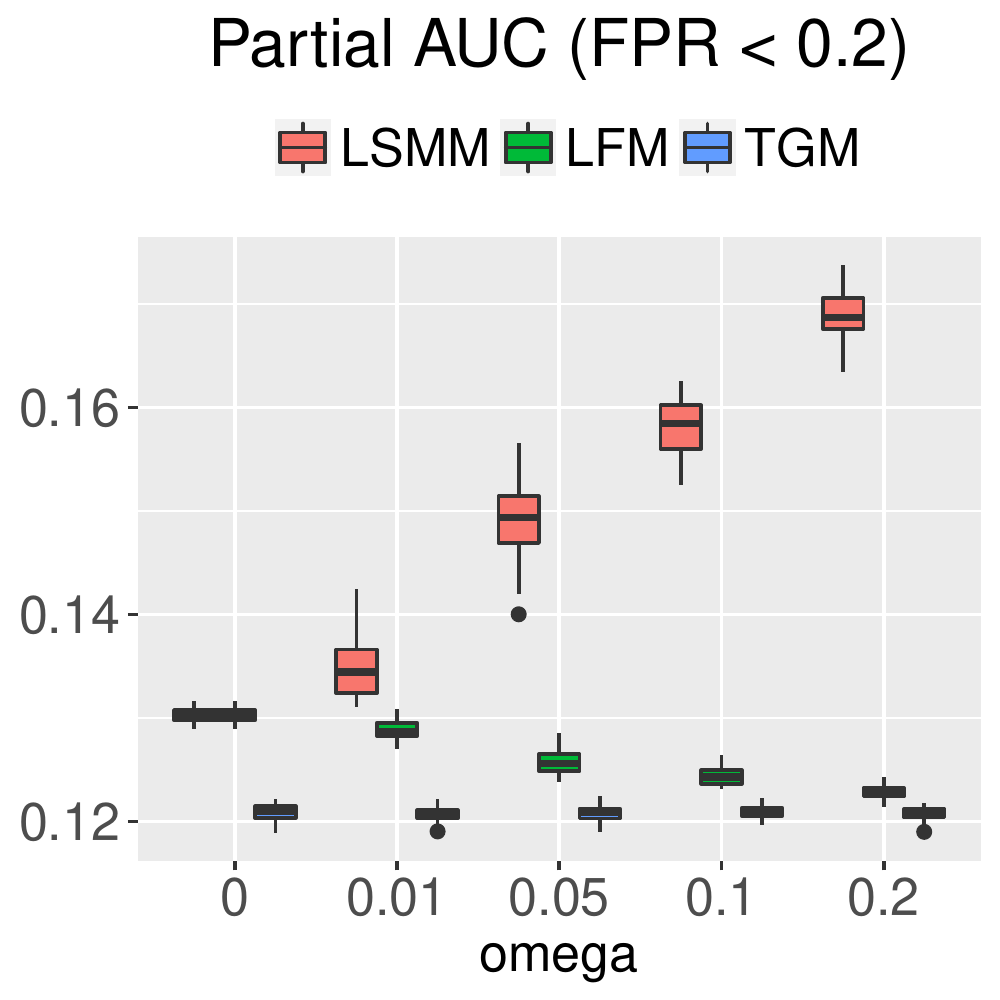}\\
	\caption{FDR, power, AUC and partial AUC of LSMM, LFM and TGM for identification of risk SNPs with $\alpha=0.2$ and $K=500$. We controlled global FDR at 0.1 to evaluate empirical FDR and power. The results are summarized from 50 replications.\label{fig:risk SNP}}
\end{figure}

We first evaluated the performance of LSMM in the identification of risk SNPs. We compared LSMM with two special cases, LFM (with fixed effects only) and TGM (without fixed effects and random effects). After prioritizing the risk SNPs using these methods, we made a comparison upon their empirical FDR, power, area under the receiver operating characteristic curve (AUC) and partial AUC. We varied the proportion of relevant random effects $\omega$ at $\left\{ 0,0.01,0.05,0.1,0.2\right\} $. Figure \ref{fig:risk SNP} shows the performance of these three models with $\alpha=0.2$ and $K=500$ (results for other scenarios are shown in Figures S2-S9 in Supplementary Document). As shown in Figure \ref{fig:risk SNP}, the empirical FDRs are indeed controlled at the nominal level ($\tau=0.1$) for all these models. For TGM and LFM, the powers increase as the proportion of relevant functional annotations $\omega$ increases. This is because a larger $\omega$ could result in an increasing proportion of non-null group for SNPs. However, the AUC and partial AUC of LFM slightly decrease because the estimates of fixed effects using LFM would become less accurate when the impact of functional annotations becomes larger. LSMM can adaptively select relevant functional annotations to improve its performance. As expected, it outperforms both TGM and LFM in terms of the power, AUC and partial AUC. One may wonder what if we do not do variable selection and simply treat the effects of all covariates as fixed effects. We evaluated this approach and found that, without variable selection, the FDR would be inflated when the GWAS signal is relatively weak (See Figure S10 in Supplementary Document). In addition, LSMM assumed independence among SNPs, which greatly facilitates the computation and inference of LSMM. We evaluated the impact of this assumption on LSMM. The details of the simulations are given in Section 3 of Supplementary Document. Because GWAS only aim to identify the local genomic region in LD with true risk genetic variants, it is reasonable to consider the identified SNPs not as false positives if they are in the flanking region of the true risk SNPs. In this sense, the results (Figure S1 in Supplementary Document) suggest that LSMM can provide a satisfactory FDR control.

Next we evaluated the performance of LSMM in the detection of relevant tissue-specific functional annotations in terms of the FDR, power, AUC and partial AUC. We varied the proportion of relevant tissue-specific functional annotations $\omega$ at $\left\{ 0.01,0.05,0.1,0.2\right\} $. The results with $\alpha=0.2$ and $K=500$ are given in Figure \ref{fig:relevant annotation} (results for other scenarios are shown in Figures S11-S18 in Supplementary Document). The empirical FDR is controlled at 0.1 with conservativeness. This is because the variational approach is adopted to approximate the posterior, e.g., the JJ bound and mean-field approximation. The performance of LSMM in the detection of relevant functional annotations depends on the signal strength of the GWAS data. When the signal of the GWAS data is relatively strong, i.e., $\alpha$ is relatively small, LSMM has a very good performance of detecting relevant functional annotations, as indicated by power, AUC and partial AUC. We also conducted the following simulations to examine the role of adjusting covariates (i.e., genic category annotations) using fixed effects for detecting relevant tissue-specific annotations. We consider the case that genic category annotations and some tissue-specific annotations are correlated and $\mathbf{b}$, the vector of coefficients corresponding to genic category annotations, is nonzero. Without adjusting genic category annotations, some irrelevant tissue-specific annotations will be falsely included in the model due to their correlation with genic category annotations. To verify this, we simulated a case that 10 genic category annotations and first 50 tissue-specific annotations are correlated with correlation coefficient varied at $\left\{ 0,0.2,0.4,0.6,0.8\right\}$ and the remaining annotations are generated independently. To simulate the design matrices for genic category and tissue-specific annotations, we first simulated $M$ samples from a multivariate normal distribution with the correlation matrix among annotations and then made a cutoff so that 10$\%$ of the entries would be 1 and the others be 0. The results are shown in Figure S19 in Supplementary Document. In the presence of correlation, as expected, a larger FDR of detecting relevant tissue-specific annotations is observed without adjusting genic category annotations.
\begin{figure}[!htbp]
	\centering
	\includegraphics[width=0.23\textwidth]{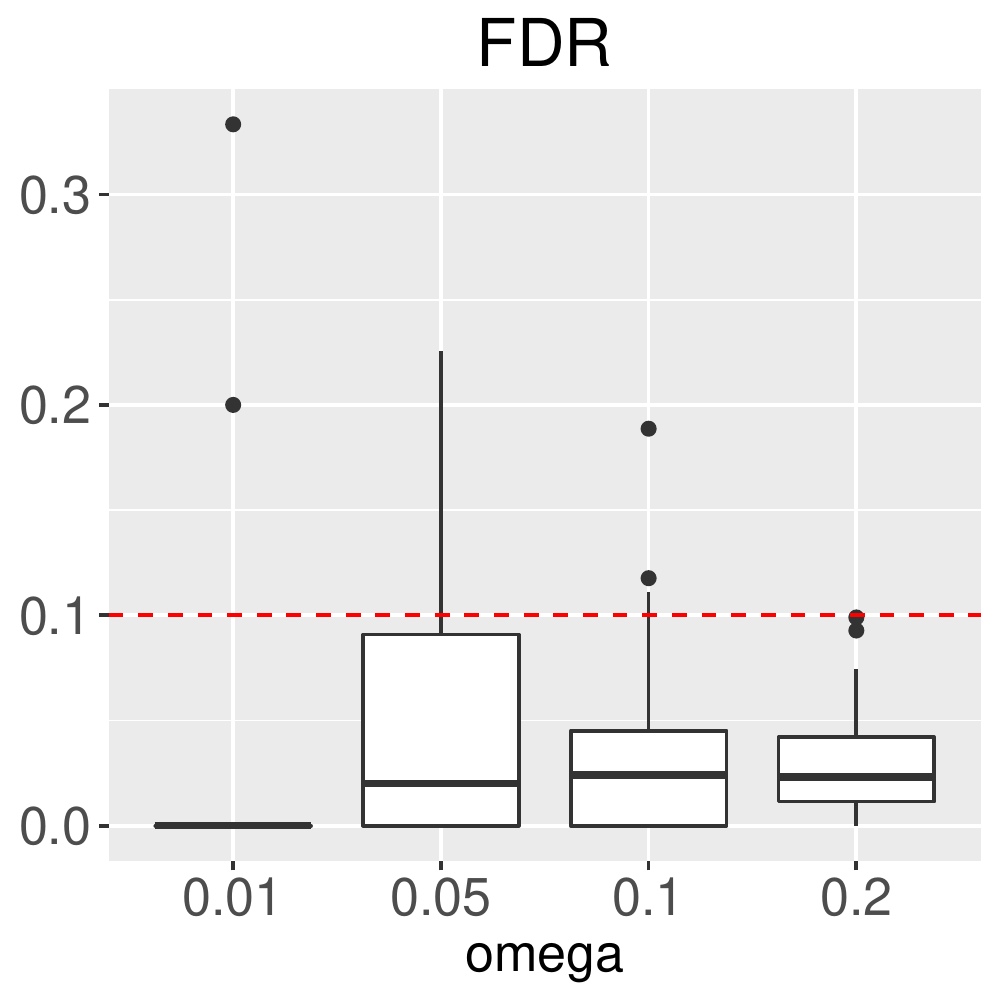}\quad
	\includegraphics[width=0.23\textwidth]{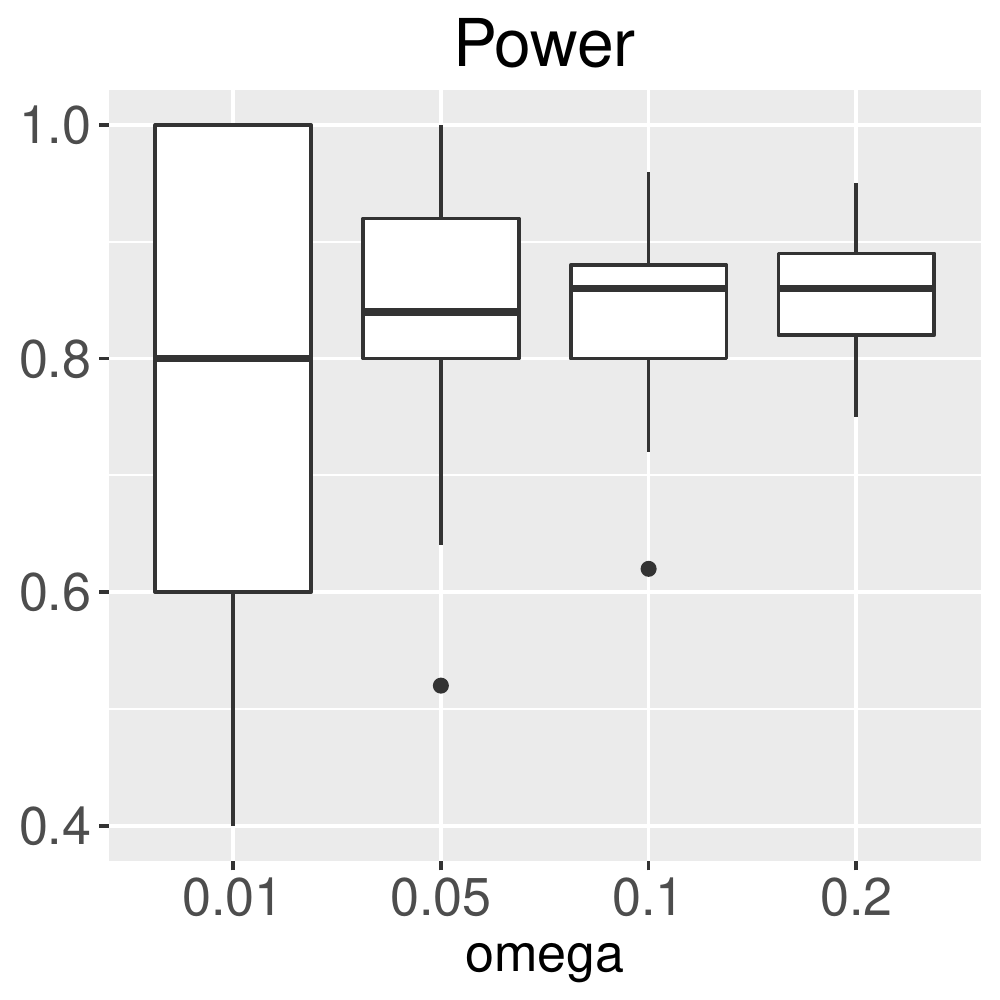}\quad
	\includegraphics[width=0.23\textwidth]{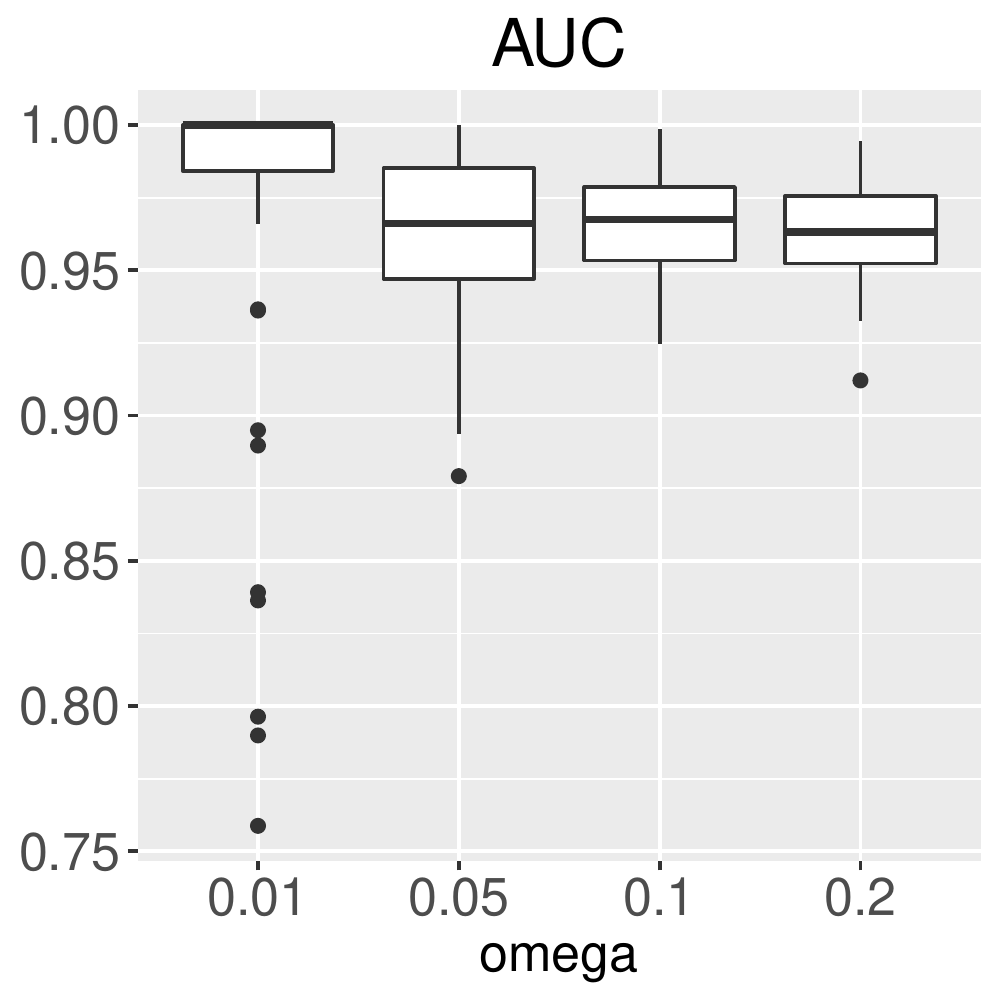}\quad
	\includegraphics[width=0.23\textwidth]{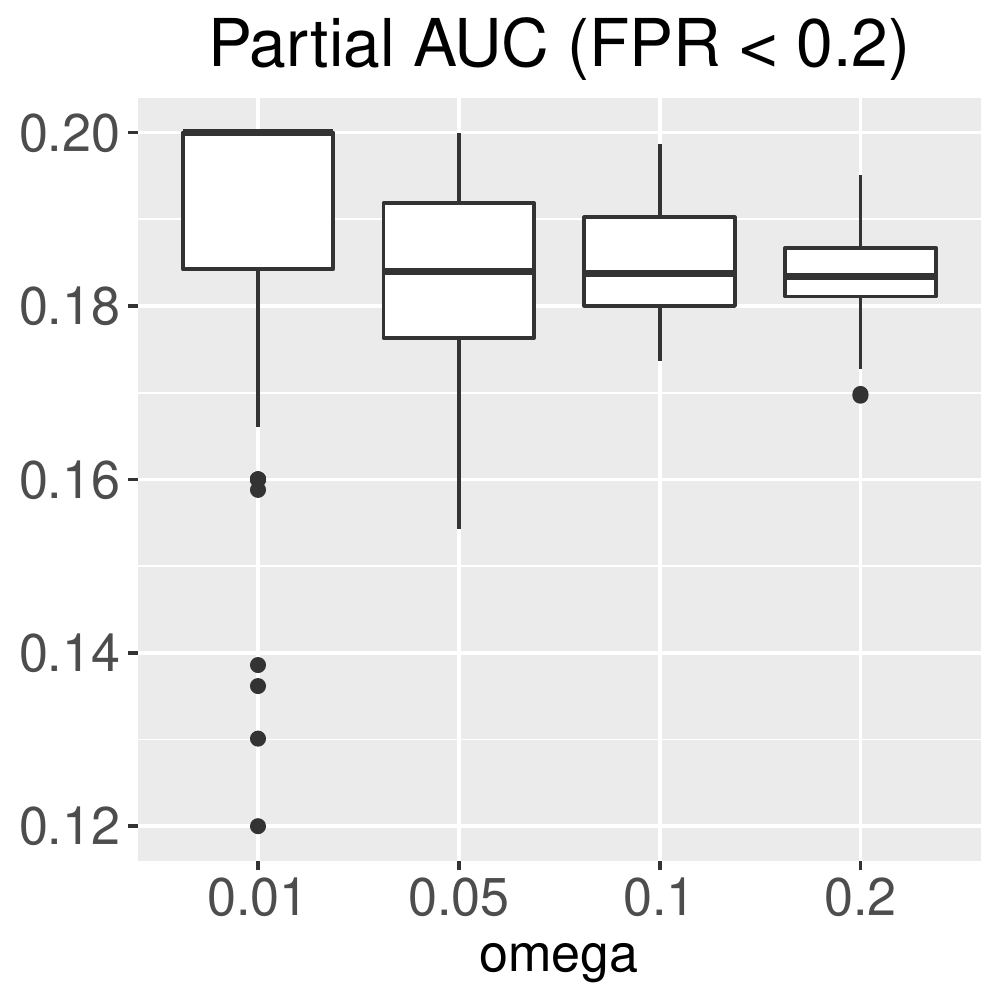}\\
	\caption{FDR, power, AUC and partial AUC of LSMM for detection of relevant annotations with $\alpha=0.2$ and $K=500$. We controlled global FDR at 0.1 to evaluate empirical FDR and power. The results are summarized from 50 replications.\label{fig:relevant annotation}}
\end{figure}

Regarding parameter estimation, LSMM provides a satisfactory estimate of $\alpha$, the parameter in Beta distribution (See Figures S31-S33 in Supplementary Document). When the signal strength of GWAS data is not very weak, the estimated fixed effects $\mathbf{b}$ (Figures S34-S44 in Supplementary Document) and the proportion of non-zero random effects $\omega$ (Figure S45 in Supplementary Document) are relatively accurate.

\begin{figure}[!htbp]
	\centering
	\includegraphics[width=0.3\textwidth]{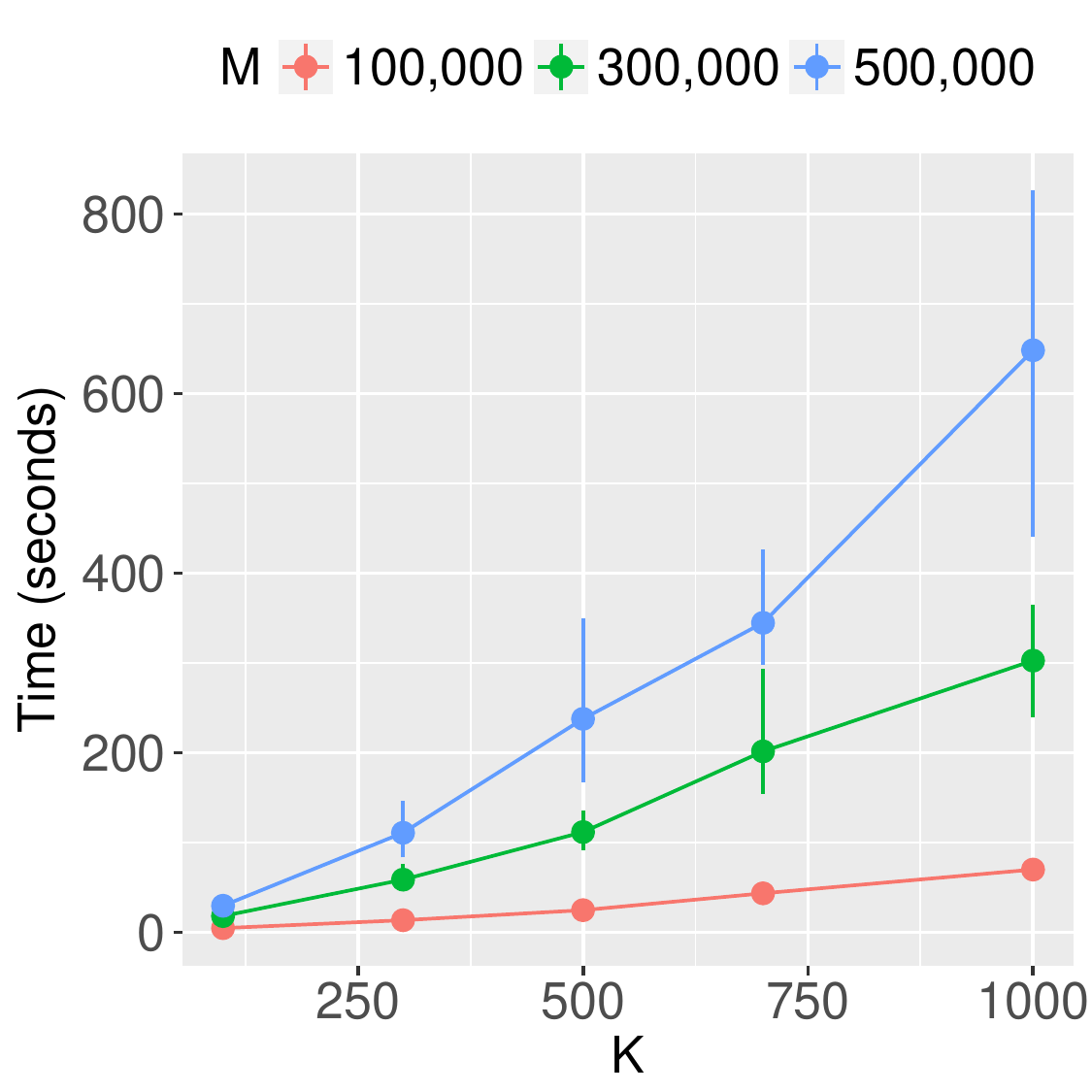}\quad
	\includegraphics[width=0.3\textwidth]{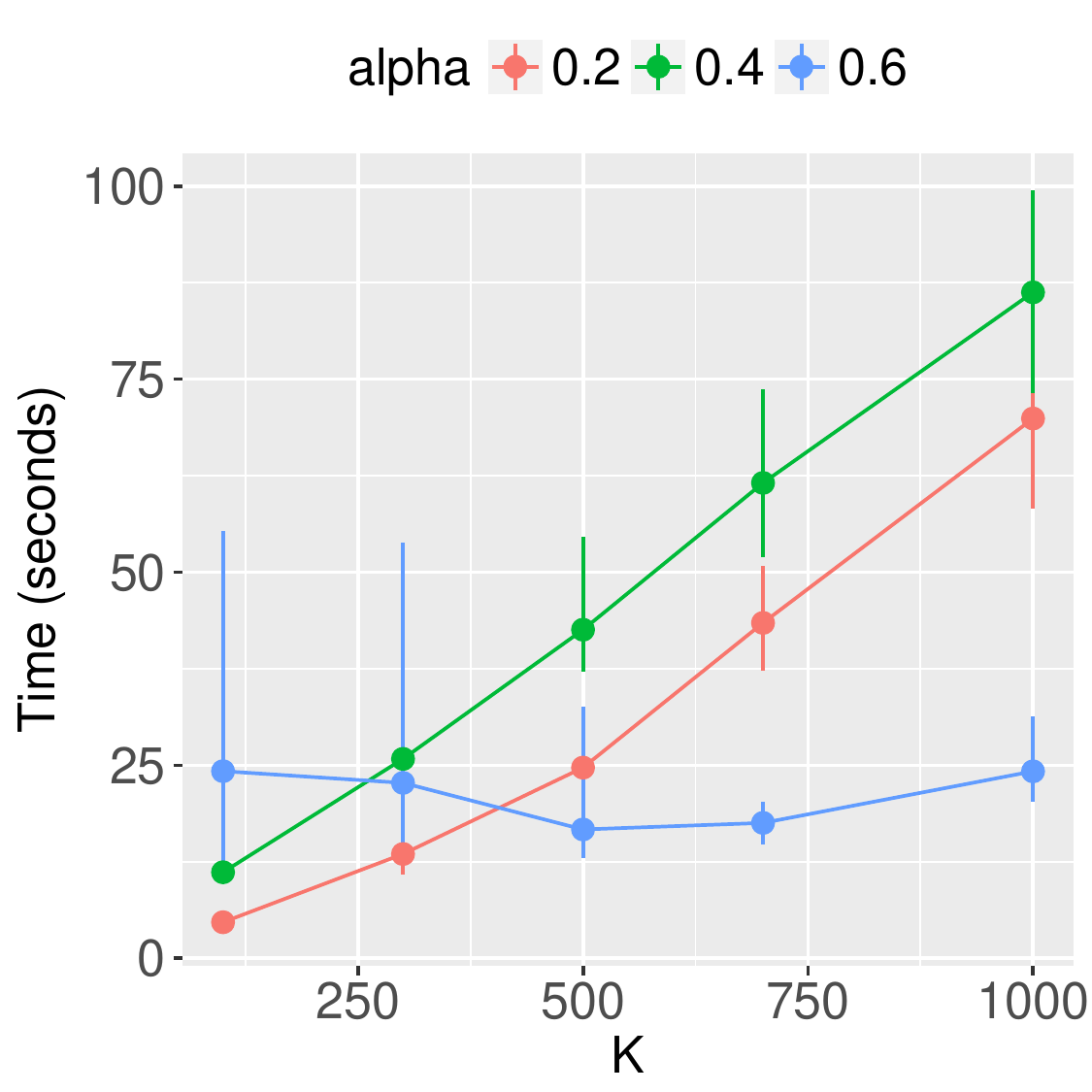}\\
	\caption{Computational time of LSMM. Left panel: We varied the number of SNPs $M$ and the number of random effects $K$, with $\alpha=0.2$. Right panel: We varied the number of random effects $K$ and the strength of GWAS signal $\alpha$ with $M=100,000$. The results are summarized from 10 replications.\label{fig:time_simulation}}
\end{figure}

The computational time of LSMM depends on the strength of GWAS signal, the number of SNPs and the number of random effects. The left panel of Figure \ref{fig:time_simulation} shows that the computational time is nearly linear with respect to $M$ and $K$ with $\alpha=0.2$. In the right panel, we fixed $M=100,000$ and varied $K$ and $\alpha$. When the GWAS signal is relatively weak, e.g., $\alpha=0.6$, the timings of LSMM remain the same for different scales of random effects. This is because LSMM adopts a warm-start strategy and its last two stages start from the estimates at the second stage (i.e., fixed effects only) and converge in a few iterations because the GWAS signal is too weak to provide information for updating the random effects.

\begin{figure}[!htbp]
	\centering
	\includegraphics[width=0.23\textwidth]{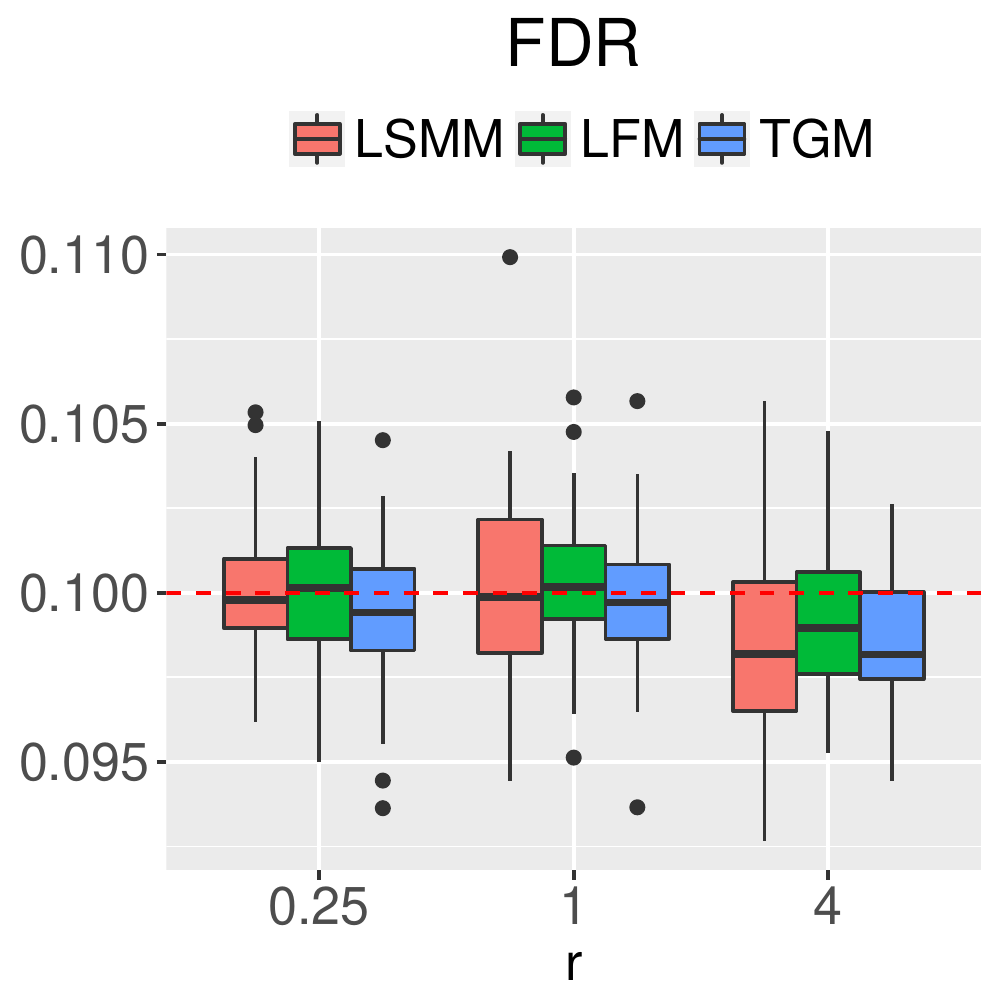}\quad
	\includegraphics[width=0.23\textwidth]{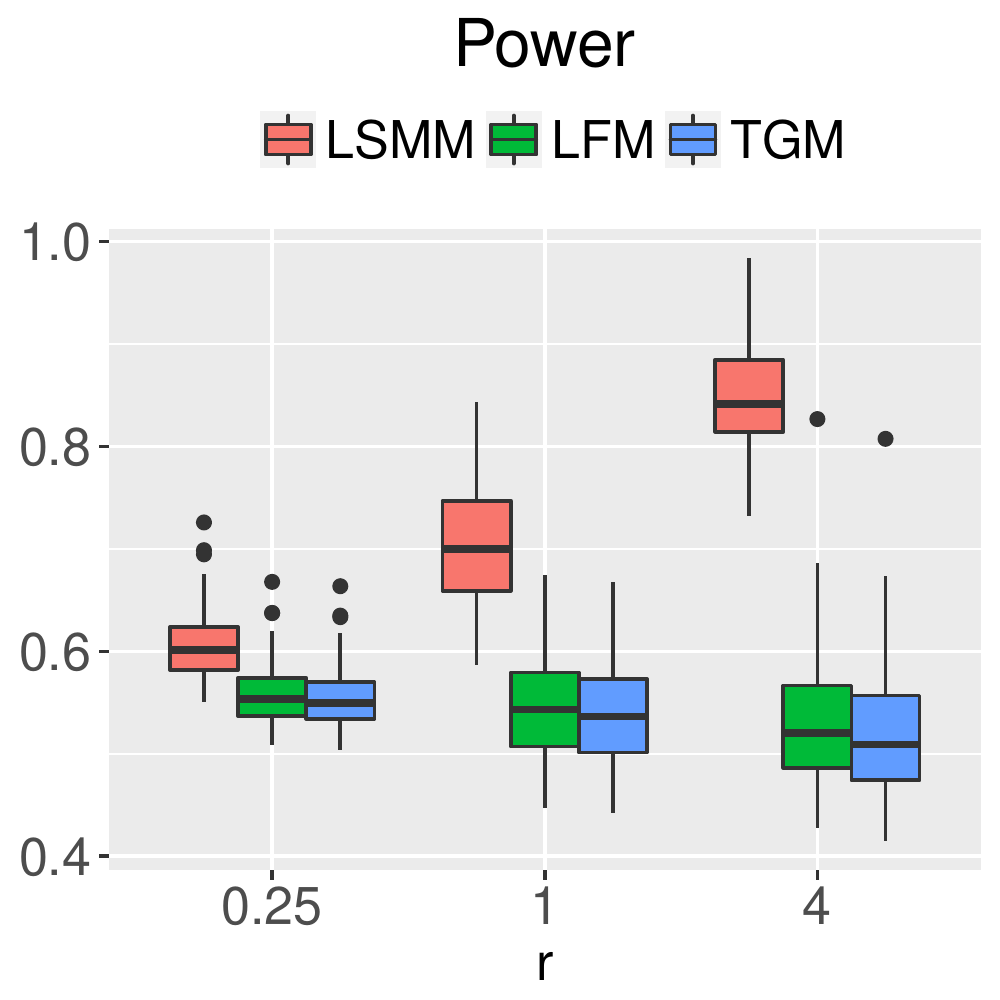}\quad
	\includegraphics[width=0.23\textwidth]{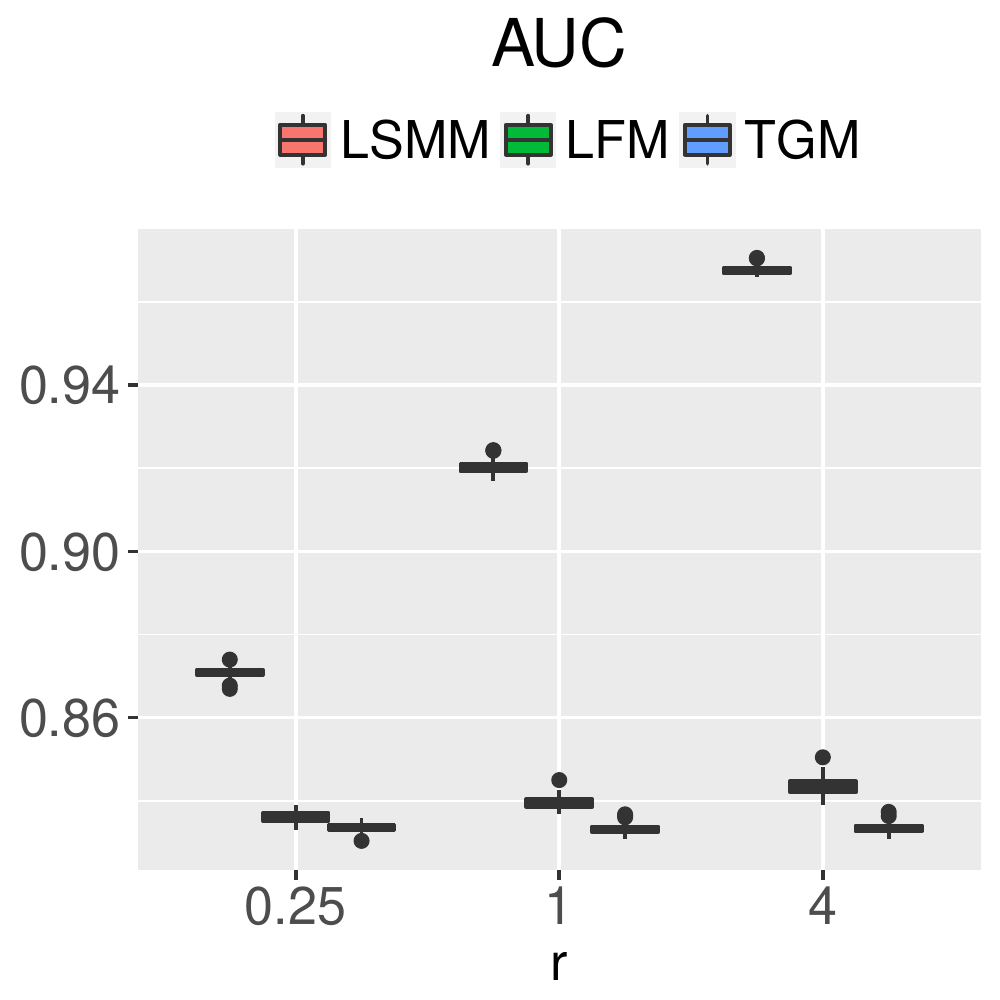}\quad
	\includegraphics[width=0.23\textwidth]{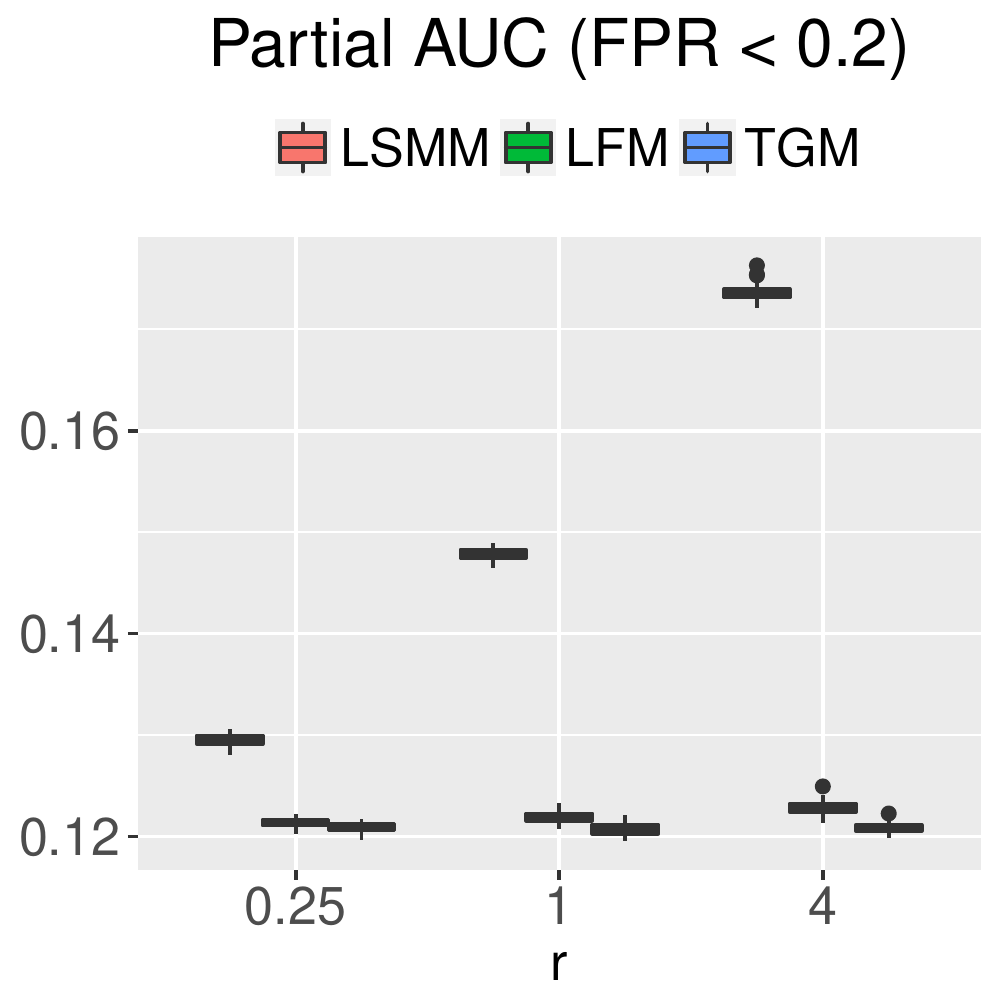}\\
	\caption{FDR, power, AUC and partial AUC of LSMM, LFM and TGM for identification of risk SNPs based on probit model (\ref{eq:probit}) with $K=500$. We controlled global FDR at 0.1 to evaluate empirical FDR and power. The results are summarized from 50 replications.\label{fig:probit SNP}}
\end{figure}

To test the robustness of LSMM, instead of using generative model (\ref{eq:logistic}), we conducted simulations based on probit model:
\begin{equation}
y_{j}=\mathbf{Z}_{j}\mathbf{b}+\mathbf{A}_{j}\boldsymbol{\beta}+e_{j},\label{eq:probit}
\end{equation}
where $e_{j}\sim N\left(0,\sigma_{e}^{2}\right)$. And we set $\gamma_{j}=1$ if $y_{j}>0$, $\gamma_{j}=0$ if $y_{j}\le0$. The first entry of the coefficients of fixed effects $\mathbf{b}$, i.e. the intercept term, was fixed at $-1$ and other entries were generated from $N\left(0,1\right)$ and fixed during multiple replications. We set $\alpha=0.2$ and varied the signal-noise ratio $r=\left\{ 4:1,1:1,1:4\right\}$. Figure \ref{fig:probit SNP} shows the performance in identification of risk SNPs when $K=500$. We note that FDRs are all well-controlled at the nominal level and LSMM shows the best performance in power, AUC and partial AUC. The advantages of LSMM over LFM and TGM is more apparent as the signal-noise ratio increases. The performance of LSMM in the detection of relevant functional annotations is provided in Figure \ref{fig:probit annotation}. Results for other scenarios are shown in Figures S20-S23 in Supplementary Document. Furthermore, we simulated the underlying distribution of $p$-values in non-null group from other distributions rather than the Beta distribution. The experimental results indicate that the FDR of LSMM is still well controlled at the nominal level, suggesting the robustness of LSMM and its potentially wide usage (results are shown in Figures S24-S26 in Supplementary Document). 
\begin{figure}[!htbp]
	\centering
	\includegraphics[width=0.23\textwidth]{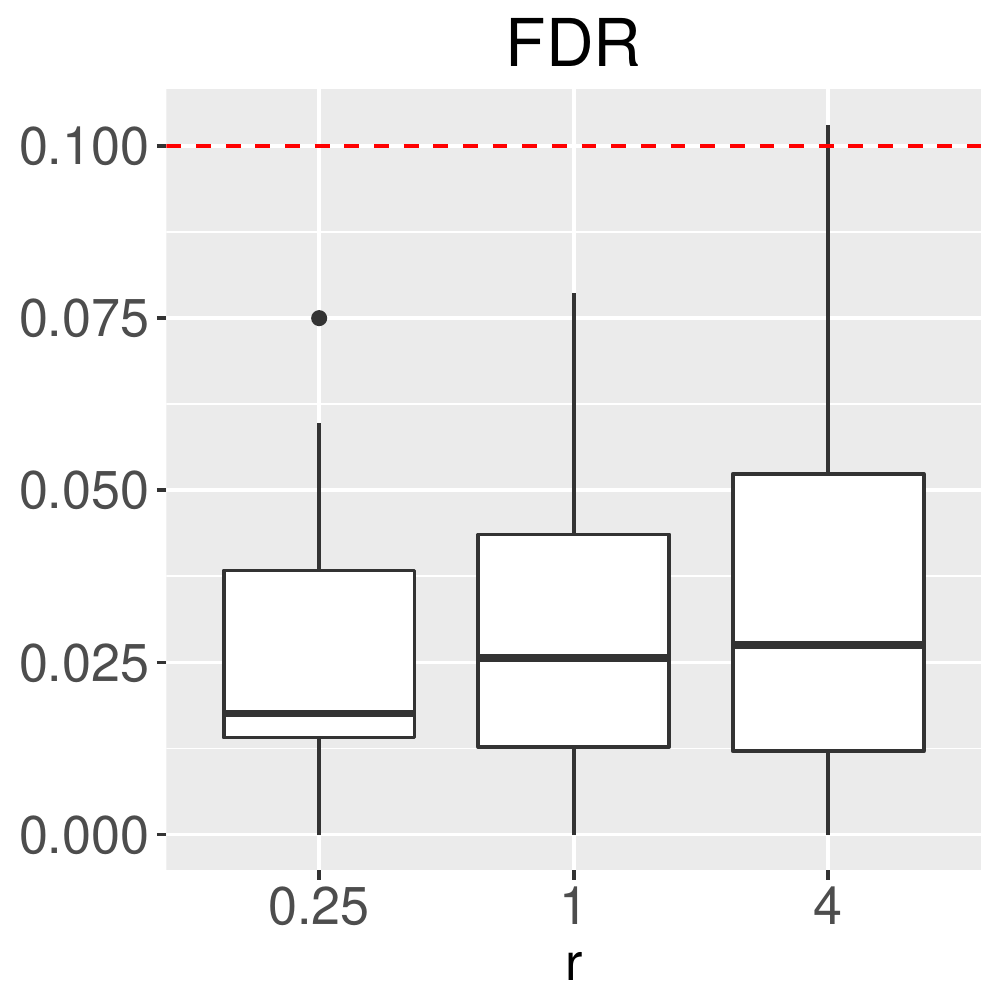}\quad
	\includegraphics[width=0.23\textwidth]{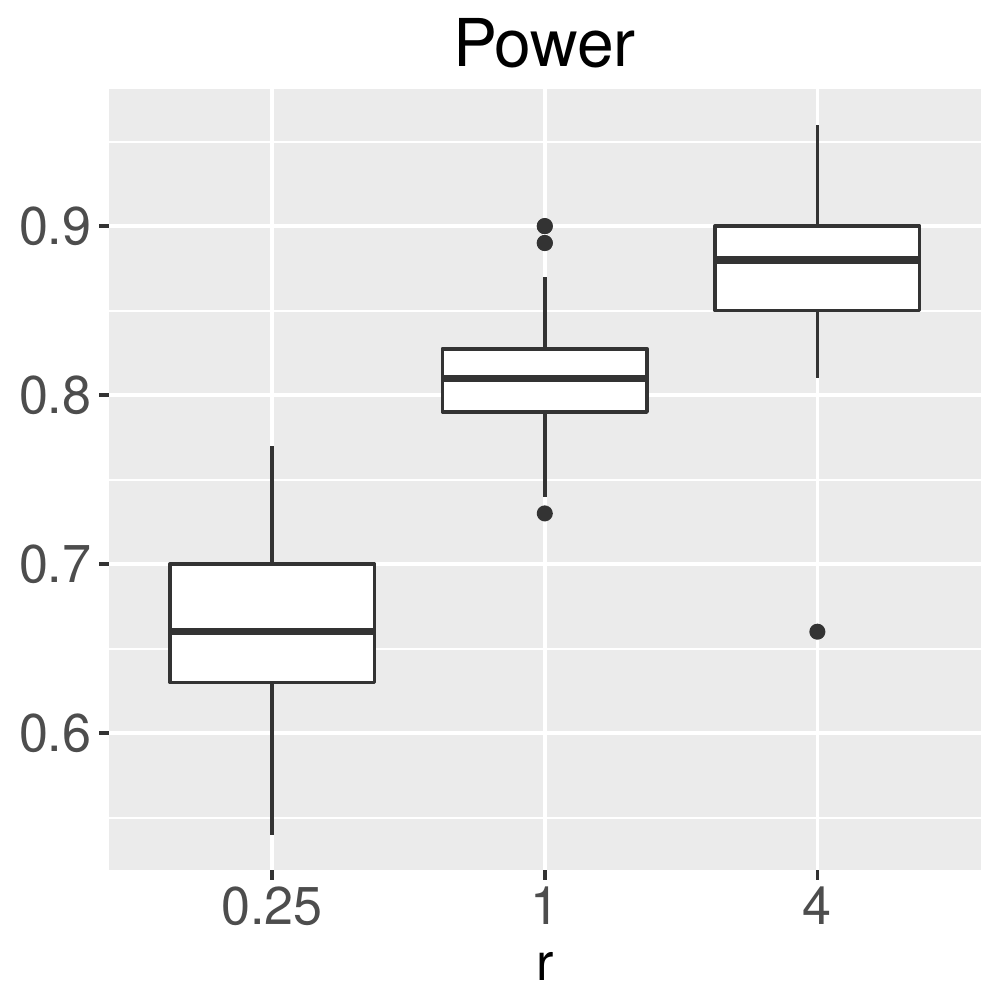}\quad
	\includegraphics[width=0.23\textwidth]{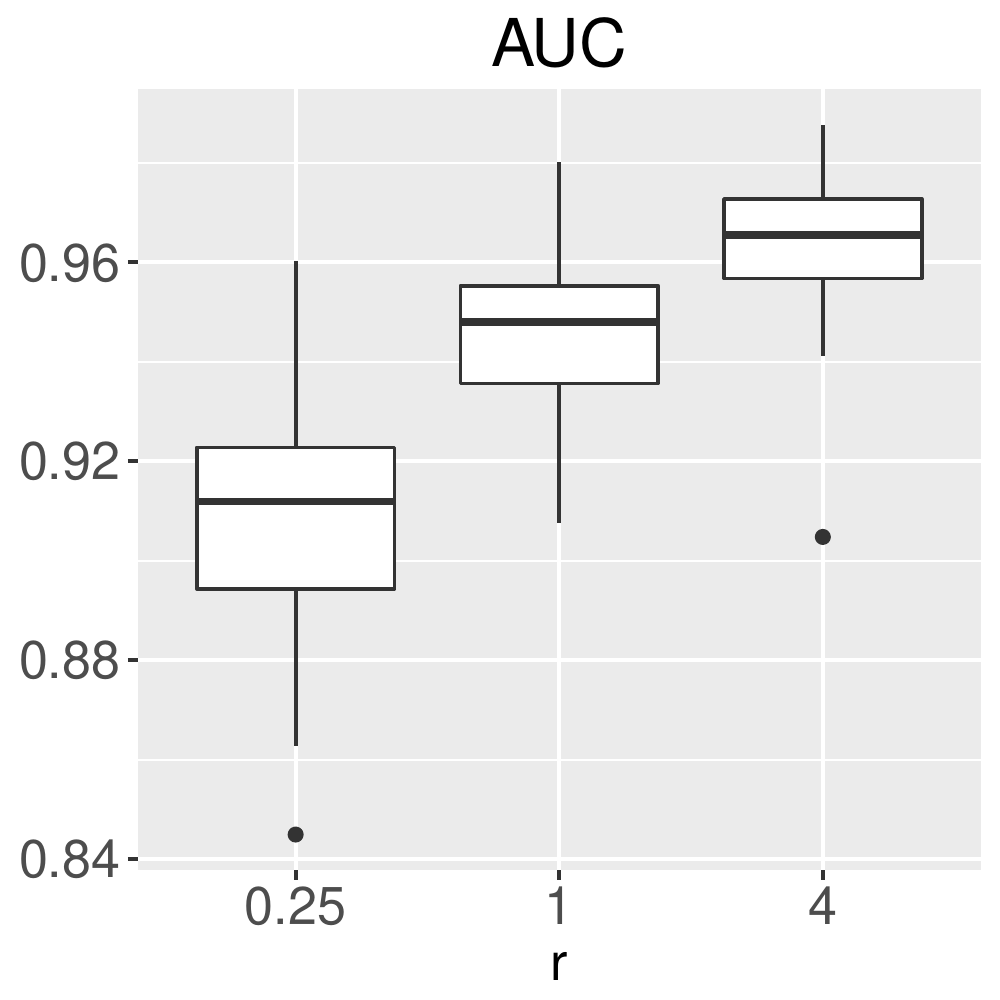}\quad
	\includegraphics[width=0.23\textwidth]{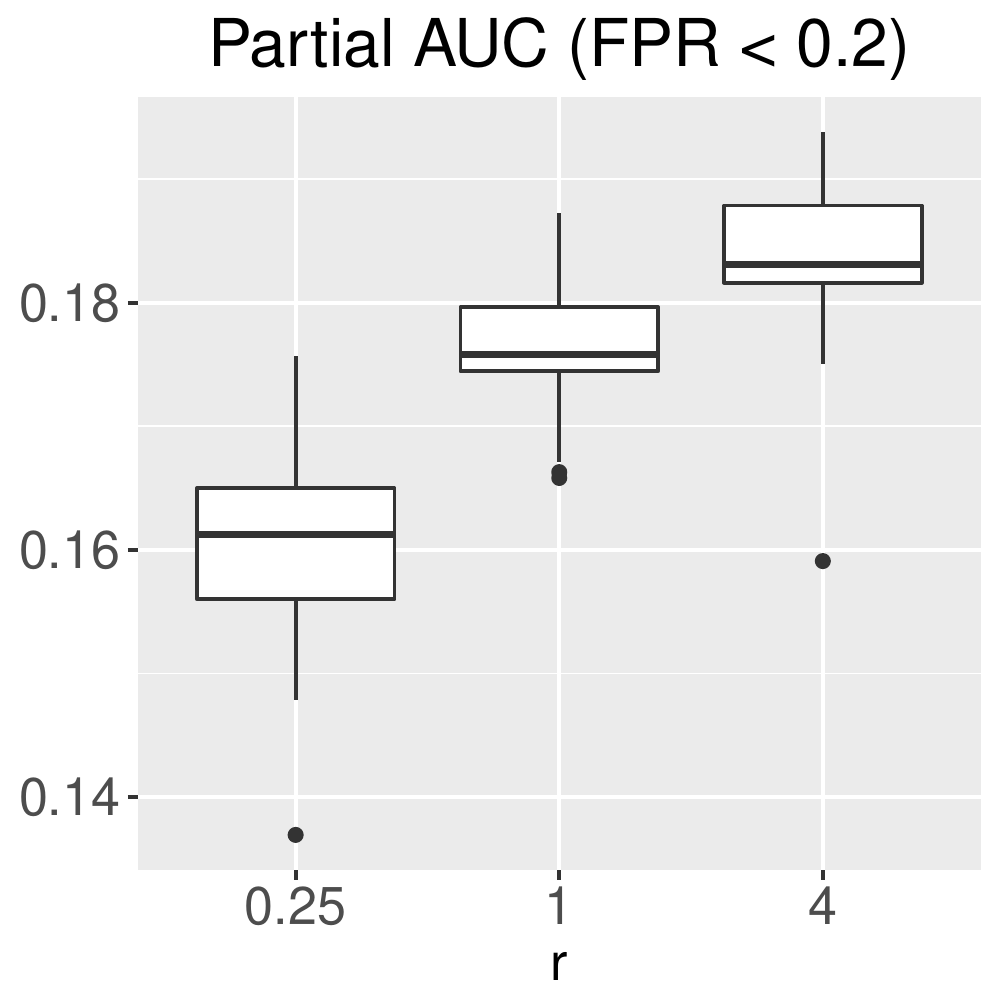}\\
	\caption{FDR, power, AUC and partial AUC of LSMM, LFM and TGM for detection of relevant annotations based on probit model with $K=500$. We controlled global FDR at 0.1 to evaluate empirical FDR and power. The results are summarized from 50 replications.\label{fig:probit annotation}}
\end{figure}

We compared LSMM with GPA in the identification of risk variants and detection of tissue-specific annotations. As LSMM can integrate both genic category and functional annotations, we compared GPA with LSMM without fixed effects (integrate functional annotations only) for a fair comparison. From the model setup, one main difference between GPA and LSMM is that GPA assumes conditional independence among annotations, whereas in LSMM we do not make this assumption. To check the influence of correlated functional annotations, we simulated a case that the first 10 functional annotations were correlated and all the others were independent. We set $\alpha=0.2$ and varied the correlation among annotations $corr$ at $\left\{ 0,0.2,0.4,0.6,0.8\right\}$. To simulate the design matrices for correlated functional annotations, we first simulated $M$ samples from a multivariate normal distribution with the correlation matrix among annotations and then made a cutoff so that 10$\%$ of the entries would be 1 and the others be 0. Figure \ref{fig:GPA} shows the results with $K=500$ (results for other scenarios are shown in Figures S27-S29 in Supplementary Document). We observe that the empirical FDRs of LSMM and LSMM without fixed effects are indeed controlled at 0.1, but the FDR of GPA inflates very much when annotations are correlated. As the FDR of GPA is not controlled, the power of GPA is not comparable to the other two models. According to the AUC and partial AUC, the performance of GPA becomes worse as the correlation among annotations increase, while the performance of LSMM is still stable and outstanding. It implies that LSMM is able to identify true relevant annotations among correlated misleading ones. We also conducted simulations to compare LSMM with cmfdr, a fully Bayesian approach to incorporate genic category annotations in GWAS using MCMC sampling algorithm. We find that cmfdr is not able to handle a large number of annotations and the MCMC sampling algorithm is very time-consuming. The result is shown in Figure S30 in Supplementary Document. Besides the computational time, we observe the empirical FDR of cmfdr is slightly inflated and its performance for prioritization of risk variants is inferior to LSMM in terms of AUC and partial AUC.
\begin{figure}[!htbp]
	\centering
	\includegraphics[width=0.23\textwidth]{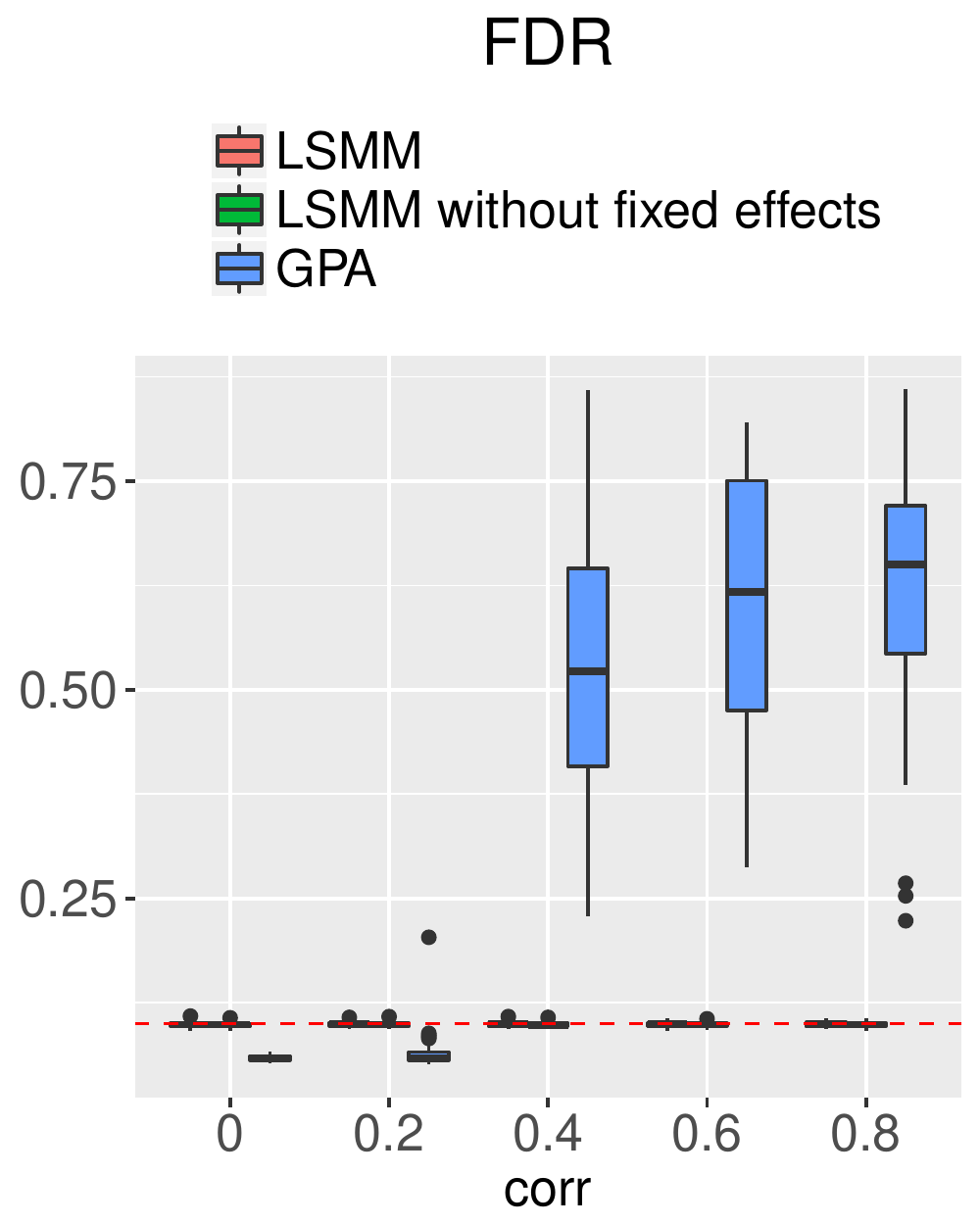}\quad
	\includegraphics[width=0.23\textwidth]{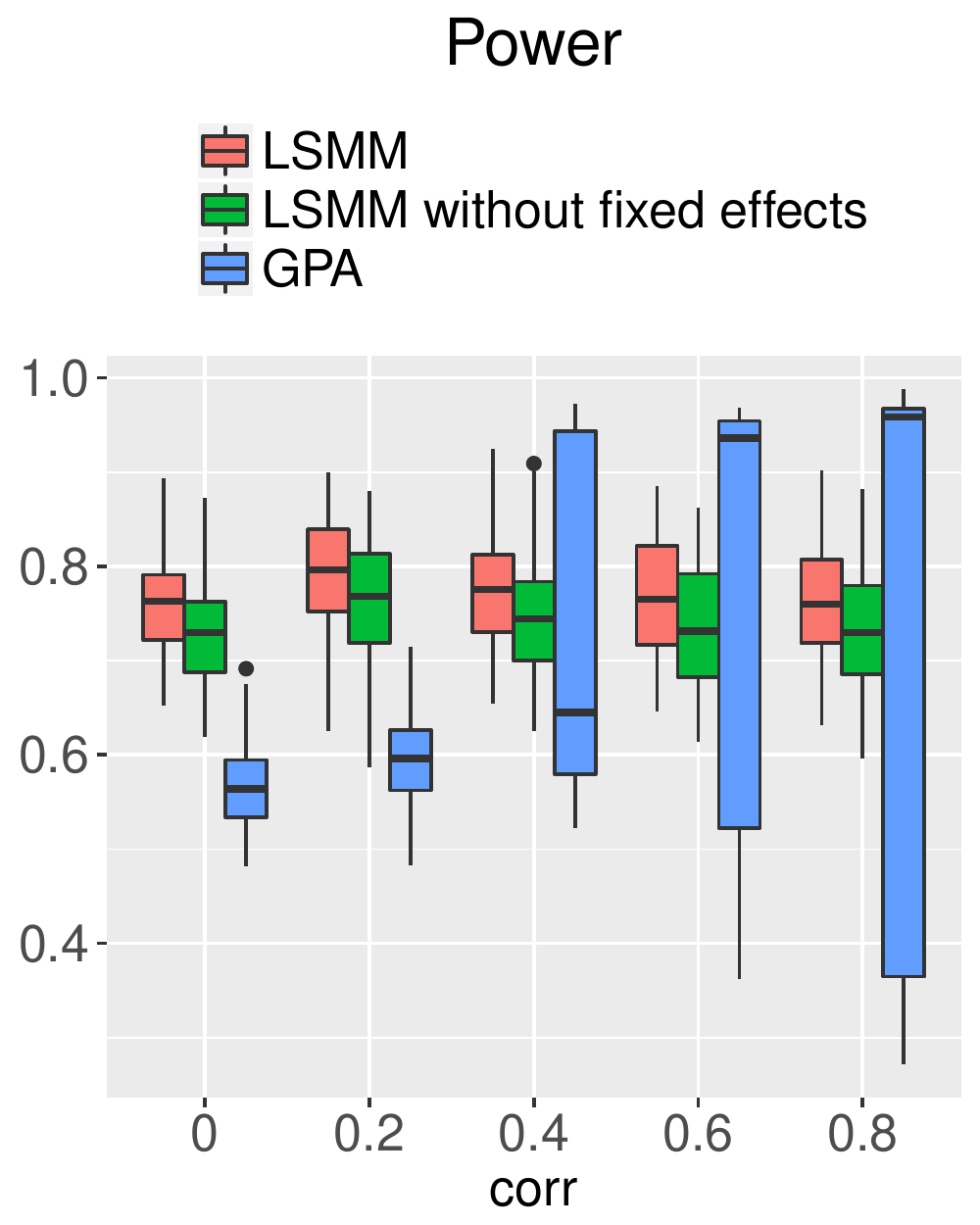}\quad
	\includegraphics[width=0.23\textwidth]{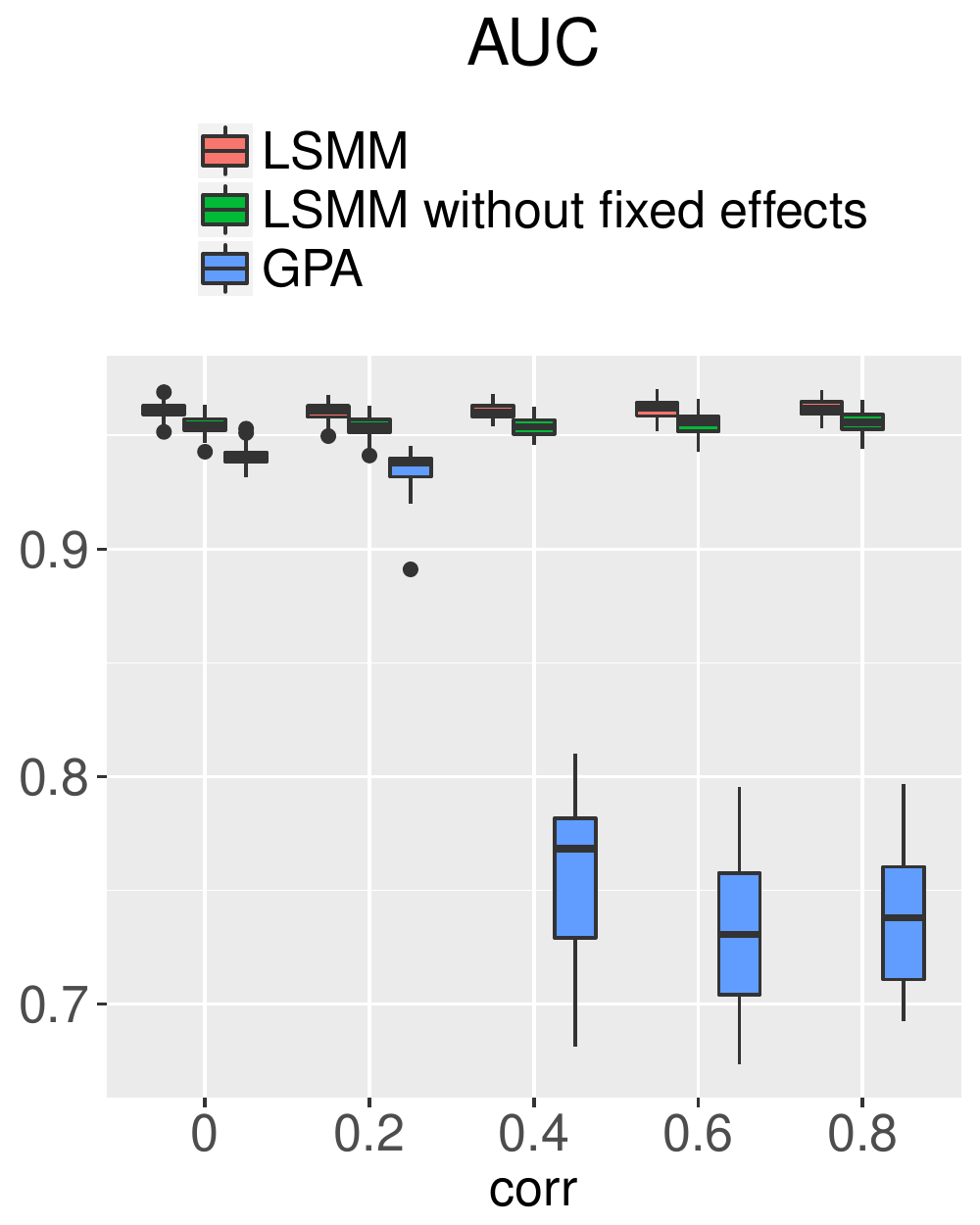}\quad
	\includegraphics[width=0.23\textwidth]{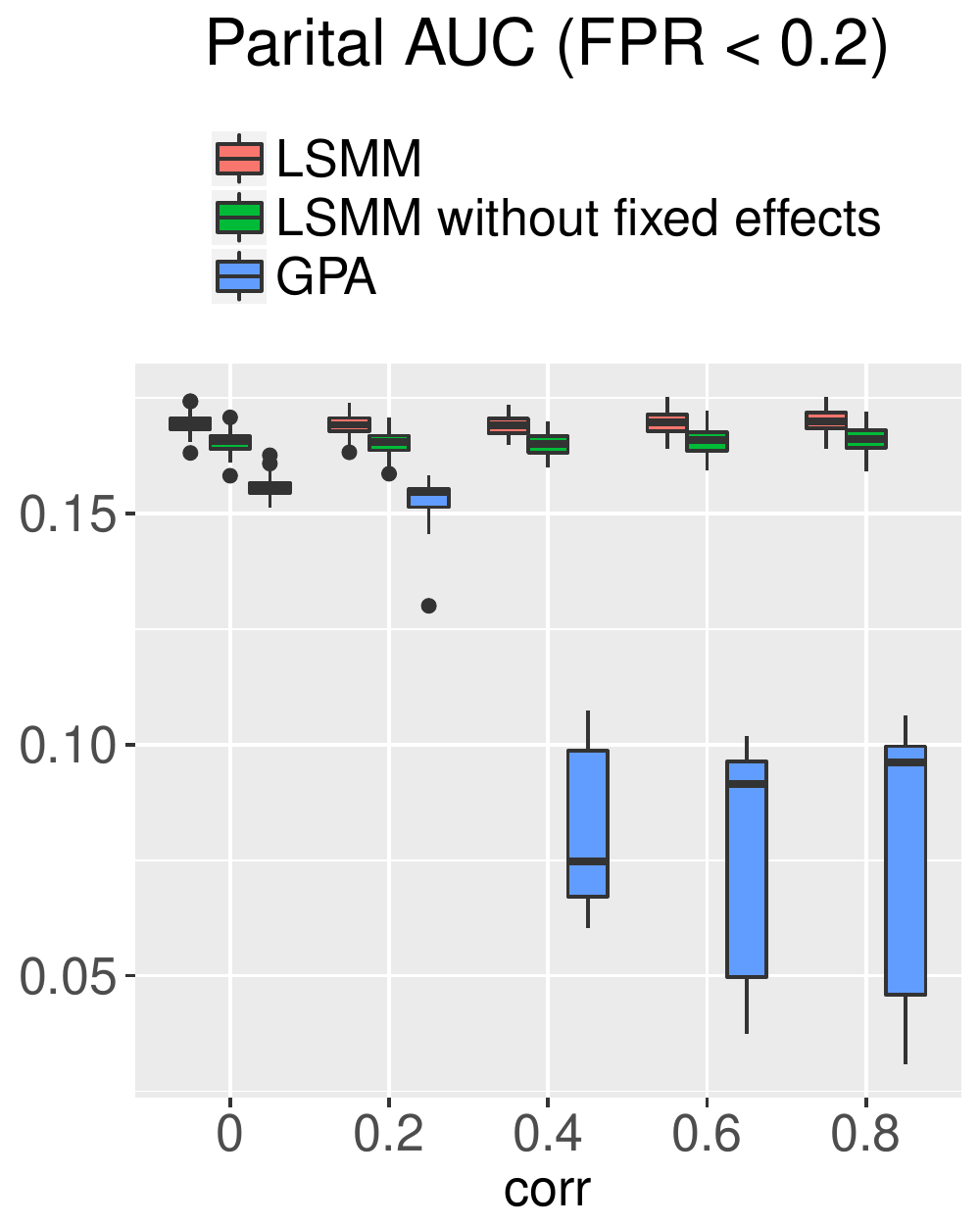}\\
	\caption{FDR, power, AUC and partial AUC of LSMM, LSMM without fixed effects and GPA for identification of risk SNPs with $K=500$. We controlled global FDR at 0.1 to evaluate empirical FDR and power. The results are summarized from 50 replications.\label{fig:GPA}}
\end{figure}

\subsection{Real Data Analysis}
We applied LSMM to analyze 30 GWAS of complex phenotypes. The source of the 30 GWAS is given in Table S2 in Supplementary Document. We used ANNOVAR \citep{wang2010annovar} to provide the genic category annotations: upstream, downstream, exonic, intergenic, intronic, ncRNA\_exonic, ncRNA\_intronic, UTR3 and UTR5, where ncRNA means variant overlaps a transcript without coding annotation in the gene definition. We obtained 127 tissue-specific functional annotations from GenoSkylinePlus \citep{lu2017systematic} (http://genocanyon.med.yale.edu/GenoSkyline). To avoid unusually large GWAS signals in the MHC region (Chromosome 6, 25Mb - 35Mb), we excluded the SNPs in this region.

\begin{figure}[!htbp]
	\centering
	\includegraphics[width=0.6\textwidth]{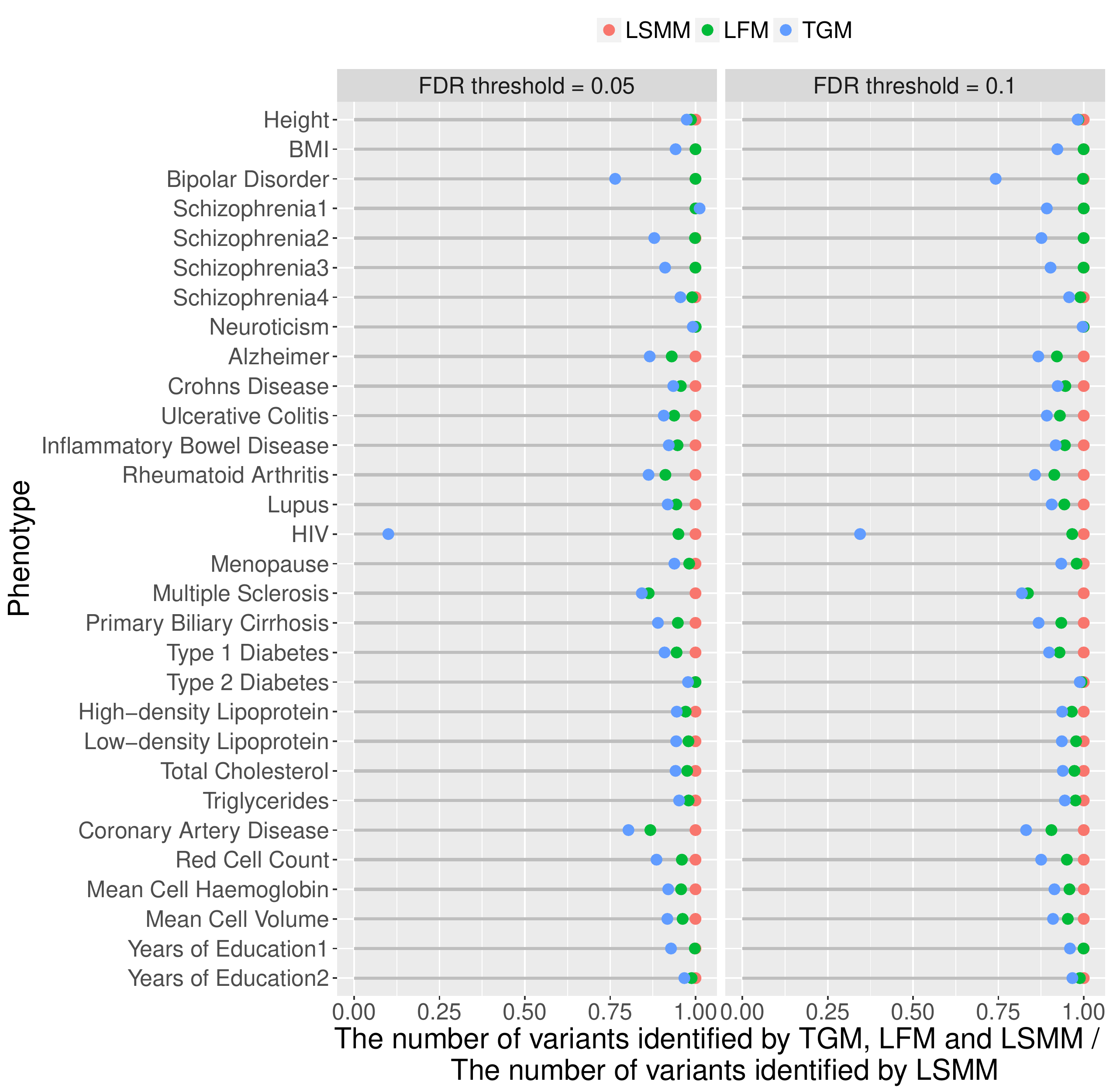}\\
	\caption{The number of risk variants identified by TGM, LFM and LSMM for 30 GWAS, under the same level of global FDR control (0.05 and 0.1). For visualization purpose, these numbers are normalized by dividing the corresponding number of variants identified by LSMM.\label{fig:NoSNPs}}
\end{figure}

\begin{figure*}[!htbp]
	\centering
	\includegraphics[width=0.76\textwidth]{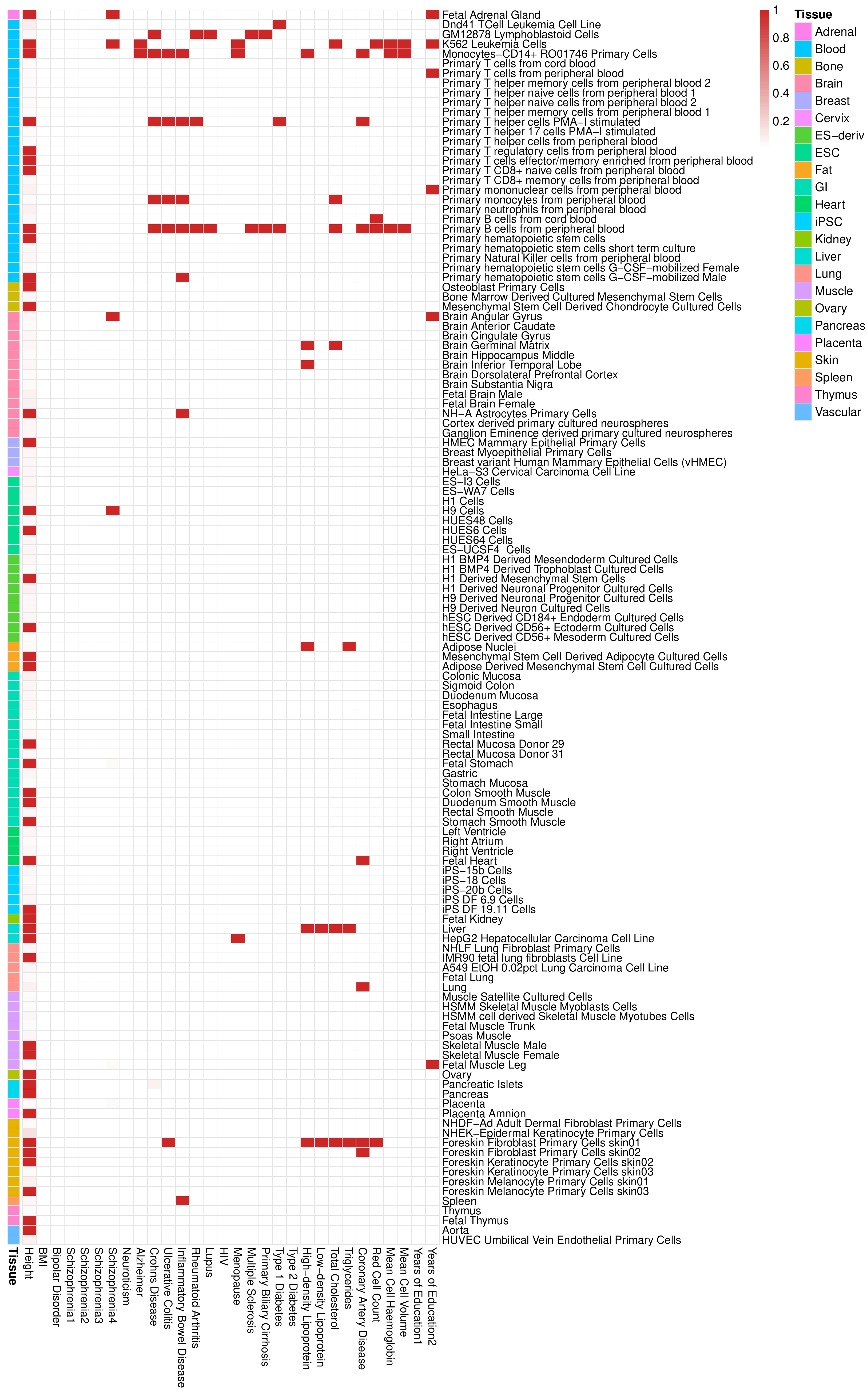}
	\caption{Relevant tissue-specific functional annotations for 30 GWAS.\label{fig:heatmap}}
\end{figure*}

We compared the number of identified risk SNPs using TGM, LFM and LSMM for 30 GWAS. Using LSMM as a reference, we calculated the ratio of the number of risk SNPs each method identified to that from LSMM under FDR thresholds $\tau=0.05$ and $\tau=0.1$. The results are shown in Figure \ref{fig:NoSNPs}. For detecting the relevant tissue-specific functional annotations, we controlled the local fdr at $0.1$. Figure \ref{fig:heatmap} shows the approximated posterior probability for annotations and phenotypes, where the darkness of the red entry implies the level of relevance between the corresponding tissue-specific functional annotation and the phenotype, the darker the more relevant.

Figure \ref{fig:NoSNPs} shows that LSMM can identify more risk variants than TGM and LFM, under the same level of FDR control. The differences between TGM and LFM is due to the impact of genic category annotations and the differences between LFM and LSMM can be attributed to tissue-specific functional annotations. For HIV and bipolar disorder, a clear improvement in the identification of risk SNPs can be found from TGM to LFM, reflecting a large enrichment of genic category annotations.
The contribution of tissue specific annotations can be clearly seen with the improvement from LFM to LSMM in several GWAS analyses, such as multiple sclerosis and coronary artery disease (CAD). For multiple sclerosis, genic category annotations do not show huge contributions, however, the contributions of tissue-specific annotations are substantial. As shown in Figure \ref{fig:heatmap}, its relevant tissue-specific annotations are related with immune system, GM12878 lymphoblastoid cells and primary B cells from peripheral blood. For CAD, both enrichment of genic category and tissue-specific annotations are estimated and its relevant cells are from a few different tissues, including blood, heart, lung and skin (See Figure \ref{fig:heatmap}). As a cardiovascular disease, it is reasonable to discover the relevance of these cells to CAD, and \citet{fernandez2016immune} has shown its relationship with immune system. The annotations in lung and skin we detected may provide some new insights about the disease.

Among the 30 GWAS, we analyzed four GWAS of schizophrenia with different sample sizes, Schizophrenia1 (9,379 cases and 7,736 controls), Schizophrenia2 (9,394 cases and 12,462 controls), Schizophrenia3 (13,833 cases and 18,310 controls ) and Schizophrenia4 (36,989 cases and 113,075 controls). The detailed results are summarized in Table S3 in Supplementary Document. The Manhattan plots using TGM and LSMM are provided in Figure S46 in Supplementary Document. Clearly, LSMM steadily improves over TGM and LFM in the analysis of schizophrenia, a highly polygenic trait, with different sample sizes. In particular, for Schizophrenia3, LSMM identified 1,492 risk variants which could not be identified by TGM. Interestingly, the majority of them (872 variants) can be re-identified in Schizophrenia4 using TGM. This indicates that LSMM has a better power in prioritizing risk variants than TGM. For Schizophrenia4, four tissue-specific functional annotations are detected. In our analysis, both genetic variants related to functions of brain cells (brain angular gyrus) and blood cells (K562 leukemia cells) are detected to be relevant. This evidence not only connects Schizophrenia with brain, but also suggests the biological link between Schizophrenia and immune system \citep{ripke2014biological}. We also analyzed two GWAS of years of education (Years of Education 1 and 2). Compared with Years of Eduction 1, the GWAS data set for Years of Education 2 is based on a larger sample size, and thus it enables LSMM to detect relevant functional annotations in brain and immune system. Our results are consistent with \citet{finucane2015partitioning}.

More findings about the relevance between tissue-specific annotations and GWAS are shown in Figure \ref{fig:heatmap}. Some are concordant with previous GWAS analyses. For example, we detect the functional annotation in liver to be relevant to the lipid-related phenotypes, including low-density lipoprotein, high-density lipoprotein, triglycerides and total cholesterol. Similar functional enrichment has been found by \citet{kundaje2015integrative, finucane2015partitioning} and \citet{lu2017systematic}.
For height, more than 40 tissue-specific functional annotations are detected to be relevant using LSMM, which reflects its highly polygenic genetic architecture. These relevant annotations include cells in bone, vascular and skeletal muscle which were also shown significant enrichments for height by \citet{finucane2015partitioning}. Recent research has linked some neurodegenerative diseases, which were believed to be more related to brain and neural system, to the immune system, such as Alzheimer's disease \citep{sims2017rare} and Parkinson's disease \citep{Sulzer2017}. For Alzheimer's disease, similar results have been found using LSMM. The relevant functional annotations are from blood cells, including monocytes-CD14+ and K562 leukemia cells.
For autoimmune diseases including Crohn's disease, ulcerative colitis, inflammatory bowel disease, rheumatoid arthritis, lupus, menopause, multiple sclerosis and primary biliary cirrhosis, the detected relevant functional annotations are mainly from the immune system and have many overlaps. Our results also provide the genomic level supports to previous medical literature, such as the relevance between spleen and inflammatory bowel disease \citep{muller1993splenic}, between liver and menopause \citep{mucci2001age}. The result also provides several new insights. Lipid-related phenotypes including high-density lipoprotein and total cholesterol are also relevant to functional annotations in immune system and brain. Additionally, annotations in immune system are considered relevant to blood-related phenotypes including red cell count, mean cell haemoglobin and mean cell volume. The foreskin fibroblast primary cells in skin are relevant to ulcerative colitis, four lipid-related phenotypes and red cell count.

Regarding the computational time, LSMM takes less than six minutes to handle each of the 30 GWAS datasets. We also recorded timings of cmfdr as a comparison. As cmfdr is not scalable to a large number of covariates, we only integrated the 9 genic category annotations in cmfdr. The MCMC algorithm was suggested \citep{zablocki2014covariate} to run with 5,000 burn-in and 20,000 main iterations. According to our estimates, cmfdr takes more than ten days for most phenotypes. The detailed timing results are shown in Figure S47 of Supplementary Document.

If we did not adjust the genic category annotation, more relevant tissue-specific functional annotations would be detected (results are shown in Figure S48 in Supplementary Document). It indicates that LSMM could adjust covariates' effects and provide a more reliable identification of relevant functional annotations.

\section{Conclusion}

We have presented a statistical approach, LSMM, to integrate genic category annotations and a large amount of tissue-specific functional annotations with GWAS data. LSMM can not only improve the statistical power in the identification of risk SNPs, but also infer relevant tissue-specific functional annotations to the phenotype, offering new insights to explore the genetic architecture of complex traits or diseases. Through comprehensive simulations and real data analysis of 30 GWAS, LSMM is shown to be statistically efficient and computationally scalable. As more annotation data will become publicly available in the future, we believe LSMM is widely useful for integrative analysis of genomic data.

\section*{Acknowledgement}
This work was supported in part by grant NO. 61501389 from National Science Funding of China, grants NO. 22302815, NO. 12316116 and NO. 12301417 from the Hong Kong Research Grant Council, startup grant R9405 from The Hong Kong University of Science and Technology, and Duke-NUS Medical School WBS: R-913-200-098-263, and MOE2016-T2-2-029 from Ministry of Eduction, Singapore.

\pagebreak{}

\newcommand{\beginsupplement}{%
	\setcounter{table}{0}
	\renewcommand{\thetable}{S\arabic{table}}%
	\setcounter{figure}{0}
	\renewcommand{\thefigure}{S\arabic{figure}}%
	\setcounter{equation}{0}
	\renewcommand{\theequation}{S\arabic{equation}}%
}
\beginsupplement

\section*{Supplementary Document}
\setcounter{section}{0}
\section{The variational EM algorithm}

\subsection*{E-step}

Let $\boldsymbol{\theta}=\left\{ \alpha,\mathbf{b},\sigma^{2},\omega\right\} $
be the collection of model parameters. The logarithm of the marginal
likelihood is

\[
\log\Pr\left(\mathbf{p}|\mathbf{Z},\mathbf{A};\boldsymbol{\theta}\right)=\log\sum_{\boldsymbol{\gamma}}\sum_{\boldsymbol{\eta}}\int\Pr\left(\mathbf{p},\boldsymbol{\gamma},\tilde{\boldsymbol{\beta}},\boldsymbol{\eta}|\mathbf{Z},\mathbf{A};\boldsymbol{\theta}\right)d\tilde{\boldsymbol{\beta}}.
\]

Using the signoid function denoted as $S\left(x\right)=\frac{1}{1+e^{-x}}$,
the complete-data likelihood can be written as

\[
\Pr\left(\mathbf{p},\boldsymbol{\gamma},\tilde{\boldsymbol{\beta}},\boldsymbol{\eta}|\mathbf{Z},\mathbf{A};\boldsymbol{\theta}\right)=\Pr\left(\mathbf{p}|\boldsymbol{\gamma};\alpha\right)\Pr\left(\boldsymbol{\gamma}|\mathbf{Z},\mathbf{A},\tilde{\boldsymbol{\beta}},\boldsymbol{\eta};\mathbf{b}\right)\Pr\left(\tilde{\boldsymbol{\beta}},\boldsymbol{\eta}|\sigma^{2},\omega\right),
\]
.where

\begin{eqnarray*}
	\mathbf{\Pr\left(p|\boldsymbol{\gamma};\alpha\right)} & = & \prod_{j=1}^{M}\Pr\left(p_{j}|\gamma_{j};\alpha\right)=\prod_{j=1}^{M}\left(\alpha p_{j}^{\alpha-1}\right)^{\gamma_{j}},\\
	\Pr\left(\boldsymbol{\gamma}|\mathbf{Z},\mathbf{A},\tilde{\boldsymbol{\beta}},\boldsymbol{\eta};\mathbf{b}\right) & = & \prod_{j=1}^{M}\Pr\left(\gamma_{j}|\mathbf{Z}_{j},\mathbf{A}_{j},\tilde{\boldsymbol{\beta}},\boldsymbol{\eta};\mathbf{b}\right)\\
	& = & \prod_{j=1}^{M}e^{\gamma_{j}\left(\mathbf{Z}_{j}\mathbf{b}+\sum_{k}A_{jk}\eta_{k}\tilde{\beta}_{k}\right)}S\left(-\mathbf{Z}_{j}\mathbf{b}-\sum_{k}A_{jk}\eta_{k}\tilde{\beta}_{k}\right),\\
	\Pr\left(\tilde{\boldsymbol{\beta}},\boldsymbol{\eta}|\sigma^{2},\omega\right) & = & \prod_{k=1}^{K}\Pr\left(\tilde{\beta}_{k},\eta_{k}|\sigma^{2},\omega\right)=\prod_{k=1}^{K}N\left(\tilde{\beta}_{k}|0,\sigma^{2}\right)\omega^{\eta_{k}}\left(1-\omega\right)^{1-\eta_{k}}.
\end{eqnarray*}

We can use JJ bound \citep{jaakkola2000bayesian} to get the tractable
lower bound of $\Pr\left(\boldsymbol{\gamma}|\mathbf{Z},\mathbf{A},\tilde{\boldsymbol{\beta}},\boldsymbol{\eta};\mathbf{b}\right)$
which is denoted by $h\left(\boldsymbol{\gamma}|\mathbf{Z},\mathbf{A},\tilde{\boldsymbol{\beta}},\boldsymbol{\eta};\mathbf{b},\boldsymbol{\xi}\right)$:

\begin{eqnarray*}
	&  & \Pr\left(\gamma_{j}|\mathbf{Z}_{j},\mathbf{A}_{j},\tilde{\boldsymbol{\beta}},\boldsymbol{\eta};\mathbf{b}\right)\\
	& = & e^{\gamma_{j}\left(\mathbf{Z}_{j}\mathbf{b}+\sum_{k}A_{jk}\eta_{k}\tilde{\beta}_{k}\right)}S\left(-\mathbf{Z}_{j}\mathbf{b}-\sum_{k}A_{jk}\eta_{k}\tilde{\beta}_{k}\right)\\
	& \ge & e^{\gamma_{j}\left(\mathbf{Z}_{j}\mathbf{b}+\sum_{k}A_{jk}\eta_{k}\tilde{\beta}_{k}\right)}S\left(\xi_{j}\right)\exp\left(-\lambda\left(\xi_{j}\right)\left(\left(\mathbf{Z}_{j}\mathbf{b}+\sum_{k}A_{jk}\eta_{k}\tilde{\beta}_{k}\right)^{2}-\xi_{j}^{2}\right)-\frac{\mathbf{Z}_{j}\mathbf{b}+\sum_{k}A_{jk}\eta_{k}\tilde{\beta}_{k}+\xi_{j}}{2}\right)\\
	& = & h\left(\gamma_{j}|\mathbf{Z}_{j},\mathbf{A}_{j},\tilde{\boldsymbol{\beta}},\boldsymbol{\eta};\mathbf{b},\xi_{j}\right),
\end{eqnarray*}
where 

\[
\lambda\left(\xi_{j}\right)=\frac{1}{2\xi_{j}}\left(S\left(\xi_{j}\right)-\frac{1}{2}\right).
\]

Let $\boldsymbol{\Theta}=\left\{ \alpha,\mathbf{b},\boldsymbol{\xi},\sigma^{2},\omega\right\} $.
Then

\[
f\left(\mathbf{p},\boldsymbol{\gamma},\tilde{\boldsymbol{\beta}},\boldsymbol{\eta}|\mathbf{Z},\mathbf{A};\boldsymbol{\Theta}\right)=\Pr\left(\mathbf{p}|\boldsymbol{\gamma};\alpha\right)h\left(\boldsymbol{\gamma}|\mathbf{Z},\mathbf{A},\tilde{\boldsymbol{\beta}},\boldsymbol{\eta};\mathbf{b},\boldsymbol{\xi}\right)\Pr\left(\tilde{\boldsymbol{\beta}},\boldsymbol{\eta}|\sigma^{2},\omega\right)
\]
is a lower bound of complete-data likelihood.

Next, let $q\left(\boldsymbol{\gamma},\tilde{\boldsymbol{\beta}},\boldsymbol{\eta}\right)$
be an approximation of the posterior $\Pr\left(\boldsymbol{\gamma},\tilde{\boldsymbol{\beta}},\boldsymbol{\eta}|\mathbf{p},\mathbf{Z},\mathbf{A};\boldsymbol{\theta}\right)$.
Then we can obtain a lower bound of the logarithm of the marginal
likelihood:

\begin{eqnarray*}
	&  & \log\Pr\left(\mathbf{p}|\mathbf{Z},\mathbf{A};\boldsymbol{\theta}\right)\\
	& = & \log\sum_{\boldsymbol{\gamma}}\sum_{\boldsymbol{\eta}}\int\Pr\left(\mathbf{p},\boldsymbol{\gamma},\tilde{\boldsymbol{\beta}},\boldsymbol{\eta}|\mathbf{Z},\mathbf{A};\boldsymbol{\theta}\right)d\tilde{\boldsymbol{\beta}}\\
	& \ge & \log\sum_{\boldsymbol{\gamma}}\sum_{\boldsymbol{\eta}}\int f\left(\mathbf{p},\boldsymbol{\gamma},\tilde{\boldsymbol{\beta}},\boldsymbol{\eta}|\mathbf{Z},\mathbf{A};\boldsymbol{\Theta}\right)d\tilde{\boldsymbol{\beta}}\\
	& \ge & \sum_{\boldsymbol{\gamma}}\sum_{\boldsymbol{\eta}}\intop q\left(\boldsymbol{\gamma},\tilde{\boldsymbol{\beta}},\boldsymbol{\eta}\right)\log\frac{f\left(\mathbf{p},\boldsymbol{\gamma},\tilde{\boldsymbol{\beta}},\boldsymbol{\eta}|\mathbf{Z},\mathbf{A};\boldsymbol{\Theta}\right)}{q\left(\boldsymbol{\gamma},\tilde{\boldsymbol{\beta}},\boldsymbol{\eta}\right)}d\tilde{\boldsymbol{\beta}}\\
	& = & \mathbf{E}_{q}\left[\log f\left(\mathbf{p},\boldsymbol{\gamma},\tilde{\boldsymbol{\beta}},\boldsymbol{\eta}|\mathbf{Z},\mathbf{A};\boldsymbol{\Theta}\right)-\log q\left(\boldsymbol{\gamma},\tilde{\boldsymbol{\beta}},\boldsymbol{\eta}\right)\right]\\
	& \triangleq & L\left(q\right),
\end{eqnarray*}
where $L(q)$ is the lower bound. The second inequality follows Jensen's
inequality. And 

\begin{eqnarray*}
	&  & \log\Pr\left(\mathbf{p}|\boldsymbol{\gamma},\alpha\right)\\
	& = & \sum_{j=1}^{M}\left(\gamma_{j}\left(\log\alpha+\left(\alpha-1\right)\log p_{j}\right)\right),\\
	\\
	&  & \log h\left(\boldsymbol{\gamma}|\mathbf{Z},\mathbf{A},\tilde{\boldsymbol{\beta}},\boldsymbol{\eta},\mathbf{b},\boldsymbol{\xi}\right)\\
	& = & \sum_{j=1}^{M}\left(\gamma_{j}\left(\mathbf{Z}_{j}\mathbf{b}+\sum_{k}A_{jk}\eta_{k}\tilde{\beta}_{k}\right)+\log S\left(\xi_{j}\right)\right)\\
	& + & \sum_{j=1}^{M}\left(-\lambda\left(\xi_{j}\right)\left(\left(\mathbf{Z}_{j}\mathbf{b}+\sum_{k}A_{jk}\eta_{k}\tilde{\beta}_{k}\right)^{2}-\xi_{j}^{2}\right)-\left(\mathbf{Z}_{j}\mathbf{b}+\sum_{k}A_{jk}\eta_{k}\tilde{\beta}_{k}+\xi_{j}\right)/2\right),\\
	\\
	&  & \log\Pr\left(\tilde{\boldsymbol{\beta}},\boldsymbol{\eta}|\sigma^{2},\omega\right)\\
	& = & -\frac{1}{2\sigma^{2}}\sum_{k=1}^{K}\tilde{\beta}_{k}^{2}-\frac{K}{2}\log\left(2\pi\sigma^{2}\right)+\sum_{k=1}^{K}\eta_{k}\log\omega+\sum_{k=1}^{K}\left(1-\eta_{k}\right)\log\left(1-\omega\right).
\end{eqnarray*}

To make it feasible to evaluate the lower bound, we assume that $q\left(\boldsymbol{\gamma},\tilde{\boldsymbol{\beta}},\boldsymbol{\eta}\right)$
can be factorized as

\[
q\left(\boldsymbol{\gamma},\tilde{\boldsymbol{\beta}},\boldsymbol{\eta}\right)=\left(\prod_{k=1}^{K}q\left(\tilde{\beta}_{k},\eta_{k}\right)\right)\left(\prod_{j=1}^{M}q\left(\gamma_{j}\right)\right),
\]
where $q\left(\tilde{\beta}_{k},\eta_{k}\right)=q\left(\tilde{\beta}_{k}|\eta_{k}\right)q\left(\eta_{k}\right)$,$q\left(\gamma_{j}=1\right)=\pi_{j}$,
$q\left(\eta_{k}=1\right)=\omega_{k}$.

We can obtain an approximation according to the mean-field method:

\begin{eqnarray*}
	&  & \log q\left(\tilde{\beta}_{i},\eta_{i}\right)\\
	& = & \mathbf{E}_{k\ne i}\mathbf{E}_{\boldsymbol{\gamma}}\left[\log f\left(\mathbf{p},\boldsymbol{\gamma},\tilde{\boldsymbol{\beta}},\boldsymbol{\eta}|\mathbf{Z},\mathbf{A};\boldsymbol{\Theta}\right)\right]\\
	& = & \left(-\frac{1}{2\sigma^{2}}-\sum_{j=1}^{M}\lambda\left(\xi_{j}\right)A_{ji}^{2}\eta_{i}^{2}\right)\tilde{\beta}_{i}^{2}\\
	&  & +\sum_{j=1}^{M}\left(\left(\pi_{j}-\frac{1}{2}-2\lambda\left(\xi_{j}\right)\mathbf{Z}_{j}\mathbf{b}\right)A_{ji}-2\lambda\left(\xi_{j}\right)A_{ji}\sum_{k\ne i}A_{jk}\mathbf{E}_{k}\left[\eta_{k}\tilde{\beta}_{k}\right]\right)\eta_{i}\tilde{\beta}_{i}\\
	&  & +\eta_{i}\log\omega+\left(1-\eta_{i}\right)\log\left(1-\omega\right)+const,
\end{eqnarray*}
where the expectation is taken under the distribtion $q\left(\boldsymbol{\boldsymbol{\gamma}}\right)$
and $q\left(\tilde{\beta}_{-i},\eta_{-i}\right)=\prod_{k\ne i}q\left(\tilde{\beta}_{k},\eta_{k}\right)$. 

When $\eta_{i}=1$, we have

\begin{eqnarray*}
	&  & \log q\left(\tilde{\beta}_{i}|\eta_{i}=1\right)\\
	& = & \left(-\frac{1}{2\sigma^{2}}-\sum_{j=1}^{M}\lambda\left(\xi_{j}\right)A_{ji}^{2}\right)\tilde{\beta}_{i}^{2}\\
	&  & +\sum_{j=1}^{M}\left(\left(\pi_{j}-\frac{1}{2}-2\lambda\left(\xi_{j}\right)\mathbf{Z}_{j}\mathbf{b}\right)A_{ji}-2\lambda\left(\xi_{j}\right)A_{ji}\sum_{k\ne i}A_{jk}\mathbf{E}_{k}\left[\eta_{k}\tilde{\beta}_{k}\right]\right)\tilde{\beta}_{i}+const,
\end{eqnarray*}
where $\mathbf{E}_{k}$ denotes the expectation under $q\left(\tilde{\beta}_{k},\eta_{k}\right)$,
and the constant doesn't depend on $\tilde{\beta}_{i}$. Because $\log q\left(\tilde{\beta}_{i}|\eta_{i}=1\right)$
is a quadratic form, 

\[
q\left(\tilde{\beta}_{i}|\eta_{i}=1\right)=N\left(\mu_{i},s_{i}^{2}\right),
\]
where 

\begin{eqnarray*}
	\mu_{i} & = & s_{i}^{2}\sum_{j=1}^{M}\left(\pi_{j}-\frac{1}{2}-2\lambda\left(\xi_{j}\right)\left(\mathbf{Z}_{j}\mathbf{b}+\sum_{k\ne i}A_{jk}\mathbf{E}_{k}\left[\eta_{k}\tilde{\beta}_{k}\right]\right)A_{ji}\right),\\
	s_{i}^{2} & = & \frac{\sigma^{2}}{1+2\sigma^{2}\sum_{j=1}^{M}\lambda\left(\xi_{j}\right)A_{ji}^{2}}.
\end{eqnarray*}

When $\eta_{i}=0$, we have

\begin{eqnarray*}
	\log q\left(\tilde{\beta}_{i}|\eta_{i}=0\right) & = & -\frac{1}{2\sigma^{2}}\tilde{\beta}_{i}^{2}+const.
\end{eqnarray*}
So

\begin{eqnarray*}
	q\left(\tilde{\beta}_{i}|\eta_{i}=0\right) & = & N\left(0,\sigma^{2}\right).
\end{eqnarray*}

Therefore we have

\[
q\left(\tilde{\beta}_{i},\eta_{i}\right)=\left[\omega_{i}N\left(\mu_{i},s_{i}^{2}\right)\right]^{\eta_{i}}\left[\left(1-\omega_{i}\right)N\left(0,\sigma^{2}\right)\right]^{1-\eta_{i}}.
\]

Now we evaluate the variational lower bound $L\left(q\right)$.

\begin{eqnarray*}
	&  & \mathbf{E}_{q}\left[\log\Pr\left(\mathbf{p}|\boldsymbol{\gamma},\alpha\right)\right]\\
	& = & \sum_{j=1}^{M}\left(\pi_{j}\left(\log\alpha+\left(\alpha-1\right)\log p_{j}\right)\right),\\
	\\
	&  & \mathbf{E}_{q}\left[\log h\left(\boldsymbol{\gamma}|\mathbf{Z},\mathbf{A},\tilde{\boldsymbol{\beta}},\boldsymbol{\eta},\mathbf{b},\boldsymbol{\xi}\right)\right]\\
	& = & \sum_{j=1}^{M}\left(\pi_{j}\left(\mathbf{Z}_{j}\mathbf{b}+\sum_{k}A_{jk}\omega_{k}\mu_{k}\right)+\log S\left(\xi_{j}\right)-\lambda\left(\xi_{j}\right)\left(\left(\mathbf{Z}_{j}\mathbf{b}+\sum_{k}A_{jk}\omega_{k}\mu_{k}\right)^{2}-\xi_{j}^{2}\right)\right)\\
	&  & +\sum_{j=1}^{M}\left(-\left(\mathbf{Z}_{j}\mathbf{b}+\sum_{k}A_{jk}\omega_{k}\mu_{k}+\xi_{j}\right)/2+\lambda\left(\xi_{j}\right)\sum_{k}A_{jk}^{2}\omega_{k}^{2}\mu_{k}^{2}-\lambda\left(\xi_{j}\right)\sum_{k}A_{jk}^{2}\omega_{k}\left(s_{k}^{2}+\mu_{k}^{2}\right)\right),\\
	\\
	&  & \mathbf{E}_{q}\left[\log\Pr\left(\tilde{\boldsymbol{\beta}},\boldsymbol{\eta}|\sigma^{2},\omega\right)\right]\\
	& = & -\frac{1}{2\sigma^{2}}\sum_{k=1}^{K}\left(\omega_{k}\left(s_{k}^{2}+\mu_{k}^{2}\right)+\left(1-\omega_{k}\right)\sigma^{2}\right)-\frac{K}{2}\log\left(2\pi\sigma^{2}\right)+\sum_{k=1}^{K}\omega_{k}\log\omega+\sum_{k=1}^{K}\left(1-\omega_{k}\right)\log\left(1-\omega\right),\\
	\\
	&  & -\mathbf{E}_{q}\left[\log q\left(\boldsymbol{\gamma},\tilde{\boldsymbol{\beta}},\boldsymbol{\eta}\right)\right]\\
	& = & \sum_{k=1}^{K}\left(\frac{1}{2}\omega_{k}\left(\log s_{k}^{2}-\log\sigma^{2}\right)-\omega_{k}\log\omega_{k}-\left(1-\omega_{k}\right)\log\left(1-\omega_{k}\right)\right)+\frac{K}{2}\log\sigma^{2}+\frac{K}{2}+\frac{K}{2}\log\left(2\pi\right)\\
	&  & -\sum_{j=1}^{M}\left(\pi_{j}\log\pi_{j}+\left(1-\pi_{j}\right)\log\left(1-\pi_{j}\right)\right).
\end{eqnarray*}

We set the partial derivative of the lower bound $L(q)$ w.r.t to
$\omega_{k},\pi_{j}$ and $\xi_{j}$ be 0 to get the variational parameters
$\omega_{k},\pi_{j}$ and $\xi_{j}$:

\begin{eqnarray*}
	\omega_{k} & = & \frac{1}{1+\exp\left(-u_{k}\right)},\textrm{ where }u_{k}=\log\frac{\omega}{1-\omega}+\frac{1}{2}\log\frac{s_{k}^{2}}{\sigma^{2}}+\frac{\mu_{k}^{2}}{2s_{k}^{2}},\\
	v_{j} & = & \log\alpha+\left(\alpha-1\right)\log p_{j}+\mathbf{Z}_{j}\mathbf{b}+\sum_{k=1}^{K}A_{jk}\omega_{k}\mu_{k},\\
	\xi_{j}^{2} & = & \left(\mathbf{Z}_{j}\mathbf{b}+\sum_{k}A_{jk}\omega_{k}\mu_{k}\right)^{2}+\sum_{k}A_{jk}^{2}\left(\omega_{k}\left(s_{k}^{2}+\mu_{k}^{2}\right)-\omega_{k}^{2}\mu_{k}^{2}\right).
\end{eqnarray*}

The variational lower bound $L(q)$ is 

\begin{eqnarray*}
	&  & L(q)\\
	& = & \sum_{j=1}^{M}\left(\pi_{j}\left(\log\alpha+\left(\alpha-1\right)\log p_{j}\right)\right)\\
	&  & +\sum_{j=1}^{M}\left(\pi_{j}\left(\mathbf{Z}_{j}\mathbf{b}+\sum_{k}A_{jk}\omega_{k}\mu_{k}\right)+\log S\left(\xi_{j}\right)-\lambda\left(\xi_{j}\right)\left(\left(\beta_{0}+\sum_{k}A_{jk}\omega_{k}\mu_{k}\right)^{2}-\xi_{j}^{2}\right)\right)\\
	&  & +\sum_{j=1}^{M}\left(-\left(\mathbf{Z}_{j}\mathbf{b}+\sum_{k}A_{jk}\omega_{k}\mu_{k}+\xi_{j}\right)/2+\lambda\left(\xi_{j}\right)\sum_{k}A_{jk}^{2}\omega_{k}^{2}\mu_{k}^{2}-\lambda\left(\xi_{j}\right)\sum_{k}A_{jk}^{2}\omega_{k}\left(s_{k}^{2}+\mu_{k}^{2}\right)\right)\\
	&  & -\frac{1}{2\sigma^{2}}\sum_{k=1}^{K}\left(\omega_{k}\left(s_{k}^{2}+\mu_{k}^{2}\right)-\omega_{k}\sigma^{2}\right)+\sum_{k=1}^{K}\omega_{k}\log\omega+\sum_{k=1}^{K}\left(1-\omega_{k}\right)\log\left(1-\omega\right)\\
	&  & +\sum_{k=1}^{K}\left(\frac{1}{2}\omega_{k}\left(\log s_{k}^{2}-\log\sigma^{2}\right)-\omega_{k}\log\omega_{k}-\left(1-\omega_{k}\right)\log\left(1-\omega_{k}\right)\right)\\
	&  & -\sum_{j=1}^{M}\left(\pi_{j}\log\pi_{j}+\left(1-\pi_{j}\right)\log\left(1-\pi_{j}\right)\right).
\end{eqnarray*}

\subsection*{{\normalsize{}M-step}}

Now we update $\alpha$, $\mathbf{b}$, $\sigma^{2}$, $\omega$.
We set the partial derivative of $L(q)$ w.r.t the parameters to be
0 and get

\begin{eqnarray*}
	\alpha & = & -\frac{\sum_{j=1}^{M}\pi_{j}}{\sum_{j=1}^{M}\pi_{j}\log p_{j}},\\
	\sigma^{2} & = & \frac{\sum_{k=1}^{K}\omega_{k}\left(s_{k}^{2}+\mu_{k}^{2}\right)}{\sum_{k=1}^{K}\omega_{k}},\\
	\omega & = & \frac{1}{K}\sum_{k=1}^{K}\omega_{k},
\end{eqnarray*}
and use Newton's method to update $\mathbf{b}$:

\[
\mathbf{b}=\mathbf{b}_{old}-\mathbf{H}^{-1}\mathbf{g},
\]
where

\begin{eqnarray*}
	\mathbf{g} & = & \sum_{j=1}^{M}\mathbf{Z}_{j}^{T}\left(\pi_{j}-2\lambda\left(\xi_{j}\right)\left(\mathbf{Z}_{j}\mathbf{b}+\sum_{k}A_{jk}\omega_{k}\mu_{k}\right)-\frac{1}{2}\right),\\
	\mathbf{H} & = & -2\mathbf{Z}_{j}^{T}\lambda\left(\xi_{j}\right)\mathbf{Z}_{j}.
\end{eqnarray*}

\subsection*{{\normalsize{}Implementation}}
\begin{itemize}
	\item Initialize $\alpha$, $\sigma^{2}$, $\omega$, $\mathbf{b}$, $\left\{ \omega_{k},\mu_{k}\right\} _{k=1,...K}$,
	$\left\{ \xi_{j},\pi_{j}\right\} _{j=1,...,M}$. Let $\tilde{y}=\sum_{k}A_{jk}\omega_{k}\mu_{k}$.
	\item E-step: For $i=1,...,K$, first obtain $\tilde{y}_{i}=\tilde{y}-A_{ji}\omega_{i}\mu_{i}$,
	and then update $\mu_{i},s_{i}^{2},\omega_{i}$ and $\tilde{y}$ as
	follows
	
	\begin{eqnarray*}
		s_{i}^{2} & = & \frac{\sigma^{2}}{1+2\sigma^{2}\sum_{j=1}^{M}\lambda\left(\xi_{j}\right)A_{ji}^{2}},\\
		\mu_{i} & = & s_{i}^{2}\sum_{j=1}^{M}\left(\left(\pi_{j}-\frac{1}{2}-2\lambda\left(\xi_{j}\right)\left(\mathbf{Z}_{j}\mathbf{b}+\tilde{y}_{i}\right)\right)A_{ji}\right),\\
		\omega_{i} & = & \frac{1}{1+\exp\left(-u_{i}\right)},\textrm{ where }u_{i}=\log\frac{\omega}{1-\omega}+\frac{1}{2}\log\frac{s_{i}^{2}}{\sigma^{2}}+\frac{\mu_{i}^{2}}{2s_{i}^{2}},\\
		\tilde{y} & = & \tilde{y}_{i}+A_{ji}\omega_{i}\mu_{i}.
	\end{eqnarray*}
	
	Then for $j=1,...,M$, update $\pi_{j},\xi_{j}$ as follows
	
	\begin{eqnarray*}
		\pi_{j} & = & \frac{1}{1+\exp\left(-v_{j}\right)},\textrm{ where }v_{j}=\log\alpha+\left(\alpha-1\right)\log p_{j}+\mathbf{Z}_{j}\mathbf{b}+\tilde{y},\\
		\xi_{j}^{2} & = & \left(\mathbf{Z}_{j}\mathbf{b}+\tilde{y}\right)^{2}+\sum_{k}A_{jk}^{2}\left(\omega_{k}\left(s_{k}^{2}+\mu_{k}^{2}\right)-\omega_{k}^{2}\mu_{k}^{2}\right).
	\end{eqnarray*}
	
	Calculate $L\left(q\right)$:
	
	\begin{eqnarray*}
		&  & L(q)\\
		& = & \sum_{j=1}^{M}\pi_{j}\left(\log\alpha+\left(\alpha-1\right)\log p_{j}\right)-\sum_{j=1}^{M}\left(\pi_{j}\log\pi_{j}+\left(1-\pi_{j}\right)\log\left(1-\pi_{j}\right)\right)\\
		&  & +\sum_{j=1}^{M}\left(\pi_{j}\left(\mathbf{Z}_{j}\mathbf{b}+\tilde{y}\right)+\log S\left(\xi_{j}\right)-\frac{\mathbf{Z}_{j}\mathbf{b}+\tilde{y}+\xi_{j}}{2}\right)\\
		&  & -\frac{1}{2\sigma^{2}}\sum_{k=1}^{K}\left(\omega_{k}\left(s_{k}^{2}+\mu_{k}^{2}\right)-\omega_{k}\sigma^{2}\right)+\sum_{k=1}^{K}\omega_{k}\log\omega+\sum_{k=1}^{K}\left(1-\omega_{k}\right)\log\left(1-\omega\right)\\
		&  & +\sum_{k=1}^{K}\left(\frac{1}{2}\omega_{k}\left(\log s_{k}^{2}-\log\sigma^{2}\right)-\omega_{k}\log\omega_{k}-\left(1-\omega_{k}\right)\log\left(1-\omega_{k}\right)\right).
	\end{eqnarray*}
	
	\item M-step
	
	\begin{eqnarray*}
		\alpha & = & -\frac{\sum_{j=1}^{M}\pi_{j}}{\sum_{j=1}^{M}\pi_{j}\log p_{j}},\\
		\sigma^{2} & = & \frac{\sum_{k=1}^{K}\omega_{k}\left(s_{k}^{2}+\mu_{k}^{2}\right)}{\sum_{k=1}^{K}\omega_{k}},\\
		\omega & = & \frac{1}{K}\sum_{k=1}^{K}\omega_{k},\\
		\mathbf{g} & = & -\sum_{j=1}^{M}\mathbf{Z}_{j}^{T}\left(\pi_{j}-2\lambda\left(\xi_{j}\right)\left(\mathbf{Z}_{j}\mathbf{b}+\tilde{y}\right)-\frac{1}{2}\right),\\
		\mathbf{H} & = & 2\sum_{j=1}^{M}\lambda\left(\xi_{j}\right)\mathbf{Z}_{j}^{T}\mathbf{Z}_{j},\\
		\mathbf{b} & = & \mathbf{b}_{old}-\mathbf{H}^{-1}\mathbf{g}.
	\end{eqnarray*}
	
	\item Evaluate $L(q)$ to track the convergence of the algorithm.
\end{itemize}

\section{Details of the proposed algorithm}

\subsection*{Stage 1: Two-groups model (TGM)}

Suppose we have the $p$-values of $M$ SNPs for a given a phenotype.
Let $\gamma_{j}$ be the latent variables indicating whether the $j$-th
SNP is associated with this phenotype. Here $\gamma_{j}=0$ means
unassociated and $\gamma_{j}=1$ means associated. Then we have the
following two-groups model: 

\textbf{
	\[
	p_{j}\sim\begin{cases}
	U\left[0,1\right], & \gamma_{j}=0,\\
	Beta\left(\alpha,1\right), & \gamma_{j}=1,
	\end{cases}
	\]
}where $\mathbf{p}\in\mathbb{R}^{M}$ are the $p$-values, $0<\alpha<1$
and $\Pr\left(\gamma_{j}=1\right)=\pi_{1}$. 

We can use EM algorithm to compute the posterior and parameter estimation.

Let $\boldsymbol{\theta}=\left\{ \alpha,\pi_{1}\right\} $ be the
collection of model parameters. The logarithm of the marginal likelihood
is
\[
\log\Pr\left(\mathbf{p}|\boldsymbol{\theta}\right)=\log\sum_{\boldsymbol{\gamma}}\Pr\left(\mathbf{p},\boldsymbol{\gamma}|\boldsymbol{\theta}\right)=\log\sum_{\boldsymbol{\gamma}}\Pr\left(\mathbf{p}|\boldsymbol{\gamma};\alpha\right)\Pr\left(\boldsymbol{\gamma}|\pi_{1}\right),
\]
where 
\begin{eqnarray*}
	\Pr\left(\mathbf{p}|\boldsymbol{\gamma};\alpha\right) & = & \prod_{j=1}^{M}\Pr\left(p_{j}|\gamma_{j};\alpha\right)=\prod_{j=1}^{M}\left(\alpha p_{j}^{\alpha-1}\right)^{\gamma_{j}},\\
	\Pr\left(\boldsymbol{\gamma}|\pi_{1}\right) & = & \prod_{j=1}^{M}\pi_{1}^{\gamma_{j}}\left(1-\pi_{1}\right)^{1-\gamma_{j}}.
\end{eqnarray*}

In the E step, we compute the posterior: 
\[
\tilde{\gamma}_{j}=q\left(\gamma_{j}=1\right)=\frac{\pi_{1}\alpha p_{j}^{\alpha-1}}{\pi_{1}\alpha p_{j}^{\alpha-1}+1-\pi_{1}},
\]
and get the Q function:

\begin{eqnarray*}
	Q & = & \mathbf{E}_{q}\left[\log\Pr\left(\mathbf{p}|\boldsymbol{\gamma};\alpha\right)+\log\Pr\left(\boldsymbol{\gamma}|\pi_{1}\right)\right]\\
	& = & \sum_{j=1}^{M}\tilde{\gamma}_{j}\left(\log\alpha+\left(\alpha-1\right)\log p_{j}+\log\pi_{1}\right)+\sum_{j=1}^{M}\left(1-\tilde{\gamma}_{j}\right)\log\left(1-\pi_{1}\right).
\end{eqnarray*}

The incomplete log likelihood can be evaluated as:
\begin{eqnarray*}
	L & = & \sum_{j=1}^{M}\tilde{\gamma}_{j}\left(\log\alpha+\left(\alpha-1\right)\log p_{j}+\log\pi_{1}-\log\tilde{\gamma}_{j}\right)+\sum_{j=1}^{M}\left(1-\tilde{\gamma}_{j}\right)\left(\log\left(1-\pi_{1}\right)-\log\left(1-\tilde{\gamma}_{j}\right)\right).
\end{eqnarray*}

In the M step, we update $\alpha$ and $\pi_{1}$ by maximizing the
Q function. We have 

\begin{eqnarray*}
	\alpha & = & -\frac{\sum_{j=1}^{M}\tilde{\gamma}_{j}}{\sum_{j=1}^{M}\tilde{\gamma}_{j}\log p_{j}},\\
	\pi_{1} & = & \frac{1}{M}\sum_{j=1}^{M}\tilde{\gamma}_{j}.
\end{eqnarray*}

\subsubsection*{Algorithm:}

Input: $\mathbf{p}$, Initialize: $\alpha=0.1$, $\pi_{1}=0.1$, Output:
$\alpha$, $\pi_{1}$, $\left\{ \tilde{\gamma}_{j}\right\} _{j=1,...,M}$.
\begin{itemize}
	\item Initialize $\alpha=0.1$, $\pi_{1}=0.1$. 
	\item E-step: For $j=1,...,M$, calculate $\tilde{\gamma}_{j}$ as follows
	
	\[
	\tilde{\gamma}_{j}=\frac{\pi_{1}\alpha p_{j}^{\alpha-1}}{\pi_{1}\alpha p_{j}^{\alpha-1}+1-\pi_{1}}.
	\]
	
	Calculate $L$:
	
	\begin{eqnarray*}
		L & = & \sum_{j=1}^{M}\tilde{\gamma}_{j}\left(\log\alpha+\left(\alpha-1\right)\log p_{j}+\log\pi_{1}-\log\tilde{\gamma}_{j}\right)+\sum_{j=1}^{M}\left(1-\tilde{\gamma}_{j}\right)\left(\log\left(1-\pi_{1}\right)-\log\left(1-\tilde{\gamma}_{j}\right)\right).
	\end{eqnarray*}
	
	\item M-step:
	
	\begin{eqnarray*}
		\alpha & = & -\frac{\sum_{j=1}^{M}\tilde{\gamma}_{j}}{\sum_{j=1}^{M}\tilde{\gamma}_{j}\log p_{j}},\\
		\pi_{1} & = & \frac{1}{M}\sum_{j=1}^{M}\tilde{\gamma}_{j}.
	\end{eqnarray*}
	
	\item Check convergence.
\end{itemize}

\subsection*{Stage 2: Latent fixed-effect model (LFM)}

Suppose we have the $p$-values of $M$ SNPs for a given a phenotype.
Similarly, we assume 

\textbf{
	\[
	p_{j}\sim\begin{cases}
	U\left[0,1\right], & \gamma_{j}=0,\\
	Beta\left(\alpha,1\right), & \gamma_{j}=1,
	\end{cases}
	\]
}where $\mathbf{p}\in\mathbb{R}^{M}$ are the $p$-values, $\gamma_{j}=1$
indicates the $j$-th is associated with this phenotype and $\gamma_{j}=0$
otherwise, and $0<\alpha<1$.

To integrate more information, we consider the logistic fixed-effect
model: 

\[
\log\frac{\Pr\left(\gamma_{j}=1|\mathbf{Z}_{j}\right)}{\Pr\left(\gamma_{j}=0|\mathbf{Z}_{j}\right)}=\mathbf{Z}_{j}\mathbf{b},
\]
where $\mathbf{Z}\in\mathbb{R}^{M\times\left(L+1\right)}$ and $\mathbf{b}=\left[b_{0},b_{1},b_{2},...,b_{L}\right]^{T}$
is an unknown vector of fixed effects, $L$ is the number of covariates. 

We can use EM algorithm to compute the posterior and parameter estimation.

Let $\boldsymbol{\theta}=\left\{ \alpha,\mathbf{b}\right\} $ be the
collection of model parameters. The complete data likelihood can be
written as
\[
\Pr\left(\mathbf{p},\boldsymbol{\gamma}|\mathbf{Z};\boldsymbol{\theta}\right)=\Pr\left(\mathbf{p}|\boldsymbol{\gamma};\alpha\right)\Pr\left(\boldsymbol{\gamma}|\mathbf{Z};\mathbf{b}\right),
\]
where 
\begin{eqnarray*}
	\Pr\left(\mathbf{p}|\boldsymbol{\gamma};\alpha\right) & = & \prod_{j=1}^{M}\Pr\left(p_{j}|\gamma_{j};\alpha\right)=\prod_{j=1}^{M}\left(\alpha p_{j}^{\alpha-1}\right)^{\gamma_{j}},\\
	\Pr\left(\boldsymbol{\gamma}|\mathbf{Z};\mathbf{b}\right) & = & \prod_{j=1}^{M}e^{\gamma_{j}\mathbf{Z}_{j}\mathbf{b}}S\left(-\mathbf{Z}_{j}\mathbf{b}\right).
\end{eqnarray*}

In the E step, we compute the posterior:
\[
\tilde{\gamma}_{j}=q\left(\gamma_{j}=1\right)=\frac{e^{\mathbf{Z}_{j}\mathbf{b}}\alpha p_{j}^{\alpha-1}}{e^{\mathbf{Z}_{j}\mathbf{b}}\alpha p_{j}^{\alpha-1}+1},
\]
and get the Q function:

\[
Q=\sum_{j=1}^{M}\tilde{\gamma}_{j}\left(\log\alpha+\left(\alpha-1\right)\log p_{j}+\mathbf{Z}_{j}\mathbf{b}\right)+\sum_{j=1}^{M}\log S\left(-\mathbf{Z}_{j}\mathbf{b}\right).
\]

The incomplete log likelihood can be evaluated as:
\begin{eqnarray*}
	L & = & \sum_{j=1}^{M}\tilde{\gamma}_{j}\left(\log\alpha+\left(\alpha-1\right)\log p_{j}+\mathbf{Z}_{j}\mathbf{b}-\log\tilde{\gamma}_{j}\right)-\sum_{j=1}^{M}\left(1-\tilde{\gamma}_{j}\right)\log\left(1-\tilde{\gamma}_{j}\right)+\sum_{j=1}^{M}\log S\left(-\mathbf{Z}_{j}\mathbf{b}\right).
\end{eqnarray*}

In the M step, we update $\alpha$ by maximizing the Q function. We
have 

\[
\alpha=-\frac{\sum_{j=1}^{M}\tilde{\gamma}_{j}}{\sum_{j=1}^{M}\tilde{\gamma}_{j}\log p_{j}}.
\]

We use Newton's method to update $\mathbf{b}$:

\[
\mathbf{b}=\mathbf{b}_{old}-\mathbf{H}^{-1}\mathbf{g},
\]
where 

\begin{eqnarray*}
	\mathbf{g} & = & \sum_{j=1}^{M}\left(-\tilde{\gamma}_{j}+S\left(\mathbf{Z}_{j}\mathbf{b}\right)\right)\mathbf{Z}_{j},\\
	\mathbf{H} & = & \sum_{j=1}^{M}S\left(\mathbf{Z}_{j}\mathbf{b}\right)S\left(-\mathbf{Z}_{j}\mathbf{b}\right)\mathbf{Z}_{j}^{T}\mathbf{Z}_{j}.
\end{eqnarray*}

\subsubsection*{Algorithm:}

Input: $\mathbf{p}$, $\mathbf{Z}$, $\alpha$, $b_{0}=\log\frac{\pi_{1}}{1-\pi_{1}}$,
Output: $\alpha$, $\mathbf{b}$, $\left\{ \tilde{\gamma}_{j}\right\} _{j=1,...,M}$.
\begin{itemize}
	\item Initialize $\alpha$, $\mathbf{b}=\left(b_{0},0,...,0\right)^{T}$. 
	\item E-step: For $j=1,...,M$, calculate $\tilde{\gamma}_{j}$ as follows
	
	\[
	\tilde{\gamma}_{j}=q\left(\gamma_{j}=1\right)=\frac{e^{\mathbf{Z}_{j}\mathbf{b}}\alpha p_{j}^{\alpha-1}}{e^{\mathbf{Z}_{j}\mathbf{b}}\alpha p_{j}^{\alpha-1}+1}.
	\]
	
	Calculate $L$:
	
	\begin{eqnarray*}
		L & = & \sum_{j=1}^{M}\tilde{\gamma}_{j}\left(\log\alpha+\left(\alpha-1\right)\log p_{j}+\mathbf{Z}_{j}\mathbf{b}-\log\tilde{\gamma}_{j}\right)-\sum_{j=1}^{M}\left(1-\tilde{\gamma}_{j}\right)\log\left(1-\tilde{\gamma}_{j}\right)+\sum_{j=1}^{M}\log S\left(-\mathbf{Z}_{j}\mathbf{b}\right).
	\end{eqnarray*}
	
	\item M-step
	
	\begin{eqnarray*}
		\alpha & = & -\frac{\sum_{j=1}^{M}\pi_{j}}{\sum_{j=1}^{M}\pi_{j}\log p_{j}},\\
		\mathbf{g} & = & \sum_{j=1}^{M}\left(-\tilde{\gamma}_{j}+S\left(\mathbf{Z}_{j}\mathbf{b}\right)\right)\mathbf{Z}_{j},\\
		\mathbf{H} & = & \sum_{j=1}^{M}S\left(\mathbf{Z}_{j}\mathbf{b}\right)S\left(-\mathbf{Z}_{j}\mathbf{b}\right)\mathbf{Z}_{j}^{T}\mathbf{Z}_{j},\\
		\mathbf{b} & = & \mathbf{b}_{old}-\mathbf{H}^{-1}\mathbf{g}.
	\end{eqnarray*}
	
	\item Check convergence.
\end{itemize}

\subsection*{Stage 3: Logistic sparse mixed model}

Suppose we the latent states $\boldsymbol{\gamma}$ of $M$ SNPs for
a given phenotype is given. We consider a logistic mixed model:

\[
\log\frac{\Pr\left(\gamma_{j}=1|\mathbf{Z}_{j},\mathbf{A}_{j}\right)}{\Pr\left(\gamma_{j}=0|\mathbf{Z}_{j},\mathbf{A}_{j}\right)}=\mathbf{Z}_{j}\mathbf{b}+\mathbf{A}_{j}\boldsymbol{\beta}=\sum_{l=0}^{L}\mathbf{Z}_{jl}b_{l}+\sum_{k=1}^{K}A_{jk}\beta_{k},
\]
where $\mathbf{Z}\in\mathbb{R}^{M\times(L+1)}$, $A\in\mathbb{R}^{M\times K}$,
$\mathbf{b}=\left[b_{0},b_{1},b_{2},...,b_{L}\right]^{T}$ is an unknown
vector of fixed effects, $\boldsymbol{\beta}=\left[\beta_{1},\beta_{2},...,\beta_{K}\right]^{T}$
is a unknown vector of random effects with a sprike-slab prior: 

\[
\beta_{k}\sim\begin{cases}
N\left(0,\sigma^{2}\right), & \eta_{k}=1,\\
\delta_{0}, & \eta_{k}=0,
\end{cases}
\]
where $\eta_{k}$ is another latent variable with $\Pr\left(\eta_{k}=1\right)=\omega$.
Here $\eta_{k}=1$ means the $k$-th annotation is relevant to this
phenotype and $\eta_{k}=0$ otherwise.

To handle the Dirac function, we reparemeterize the spike-slab prior
as $\tilde{\beta}_{k}\sim N\left(0,\sigma^{2}\right),$ then $\beta_{k}=\eta_{k}\tilde{\beta}_{k}.$

We can use variational EM algorithm to compute the posterior and parameter
estimation.

Let $\boldsymbol{\theta}=\left\{ \alpha,\mathbf{b},\sigma^{2},\omega\right\} $
be the collection of model parameters. Using the sigmoid function
denoted as $S\left(x\right)=\frac{1}{1+e^{-x}}$, the complete data
likelihood can be written as
\[
\Pr\left(\boldsymbol{\gamma},\tilde{\boldsymbol{\beta}},\boldsymbol{\eta}|\mathbf{Z},\mathbf{A};\boldsymbol{\theta}\right)=\Pr\left(\boldsymbol{\gamma}|\mathbf{Z},\mathbf{A},\tilde{\boldsymbol{\beta}},\boldsymbol{\eta};\mathbf{b}\right)\Pr\left(\tilde{\boldsymbol{\beta}},\boldsymbol{\eta}|\sigma^{2},\omega\right),
\]
where

\begin{eqnarray*}
	\Pr\left(\boldsymbol{\gamma}|\mathbf{Z},\mathbf{A},\tilde{\boldsymbol{\beta}},\boldsymbol{\eta};\mathbf{b}\right) & = & \prod_{j=1}^{M}\Pr\left(\gamma_{j}|\mathbf{Z}_{j},\mathbf{A}_{j},\tilde{\boldsymbol{\beta}},\boldsymbol{\eta};\mathbf{b}\right)\\
	& = & \prod_{j=1}^{M}e^{\gamma_{j}\left(\mathbf{Z}_{j}\mathbf{b}+\sum_{k}A_{jk}\eta_{k}\tilde{\beta}_{k}\right)}S\left(-\mathbf{Z}_{j}\mathbf{b}-\sum_{k}A_{jk}\eta_{k}\tilde{\beta}_{k}\right),\\
	\Pr\left(\tilde{\boldsymbol{\beta}},\boldsymbol{\eta}|\sigma^{2},\omega\right) & = & \prod_{k=1}^{K}\Pr\left(\tilde{\beta}_{k},\eta_{k}|\sigma^{2},\omega\right)=\prod_{k=1}^{K}N\left(\tilde{\beta}_{k}|0,\sigma^{2}\right)\omega^{\eta_{k}}\left(1-\omega\right)^{1-\eta_{k}}.
\end{eqnarray*}

We can use JJ bound \citep{jaakkola2000bayesian} to bound the sigmoid
function by 

\[
S\left(x\right)\ge S\left(\xi\right)\exp\left\{ \left(x-\xi\right)/2-\lambda\left(\xi\right)\left(x^{2}-\xi^{2}\right)\right\} ,
\]
where $\lambda\left(\xi\right)=\frac{1}{2\xi}\left[S\left(\xi\right)-\frac{1}{2}\right]$.
Using this bound, we have a tractable lower bound of $\Pr\left(\boldsymbol{\gamma}|\mathbf{Z},\mathbf{A},\tilde{\boldsymbol{\beta}},\boldsymbol{\eta};\mathbf{b}\right)$
which is denoted by $h\left(\boldsymbol{\gamma}|\mathbf{Z},\mathbf{A},\tilde{\boldsymbol{\beta}},\boldsymbol{\eta};\mathbf{b},\boldsymbol{\xi}\right)$:

\begin{eqnarray*}
	&  & h\left(\gamma_{j}|\mathbf{Z}_{j},\mathbf{A}_{j},\tilde{\boldsymbol{\beta}},\boldsymbol{\eta};\mathbf{b},\xi_{j}\right)\\
	& = & e^{\gamma_{j}\left(\mathbf{Z}_{j}\mathbf{b}+\sum_{k}A_{jk}\eta_{k}\tilde{\beta}_{k}\right)}S\left(\xi_{j}\right)\exp\left(-\lambda\left(\xi_{j}\right)\left(\left(\mathbf{Z}_{j}\mathbf{b}+\sum_{k}A_{jk}\eta_{k}\tilde{\beta}_{k}\right)^{2}-\xi_{j}^{2}\right)-\frac{\mathbf{Z}_{j}\mathbf{b}+\sum_{k}A_{jk}\eta_{k}\tilde{\beta}_{k}+\xi_{j}}{2}\right).
\end{eqnarray*}

Next, Let $q\left(\tilde{\boldsymbol{\beta}},\boldsymbol{\eta}\right)$
be an approximation of the posterior $\Pr\left(\tilde{\boldsymbol{\beta}},\boldsymbol{\eta}|\mathbf{Z},\mathbf{A};\boldsymbol{\theta}\right)$.
Then we can obtain a lower bound of the logarithm of the marginal
likelihood:

\begin{eqnarray*}
	&  & \log\Pr\left(\boldsymbol{\gamma}|\mathbf{Z},\mathbf{A};\boldsymbol{\theta}\right)\\
	& = & \log\sum_{\boldsymbol{\eta}}\int\Pr\left(\boldsymbol{\gamma},\tilde{\boldsymbol{\beta}},\boldsymbol{\eta}|\mathbf{Z},\mathbf{A};\boldsymbol{\theta}\right)d\tilde{\boldsymbol{\beta}}\\
	& = & \log\sum_{\boldsymbol{\eta}}\int\Pr\left(\boldsymbol{\gamma}|\mathbf{Z},\mathbf{A},\tilde{\boldsymbol{\beta}},\boldsymbol{\eta};\mathbf{b}\right)\Pr\left(\tilde{\boldsymbol{\beta}},\boldsymbol{\eta}|\sigma^{2},\omega\right)d\tilde{\boldsymbol{\beta}}\\
	& \ge & \log\sum_{\boldsymbol{\eta}}\int h\left(\boldsymbol{\gamma}|\mathbf{Z},\mathbf{A},\tilde{\boldsymbol{\beta}},\boldsymbol{\eta};\mathbf{b},\boldsymbol{\xi}\right)\Pr\left(\tilde{\boldsymbol{\beta}},\boldsymbol{\eta}|\sigma^{2},\omega\right)d\tilde{\boldsymbol{\beta}}\\
	& \ge & \sum_{\boldsymbol{\eta}}\int q\left(\tilde{\boldsymbol{\beta}},\boldsymbol{\eta}\right)\log\frac{h\left(\boldsymbol{\gamma}|\mathbf{Z},\mathbf{A},\tilde{\boldsymbol{\beta}},\boldsymbol{\eta};\mathbf{b},\boldsymbol{\xi}\right)\Pr\left(\tilde{\boldsymbol{\beta}},\boldsymbol{\eta}|\sigma^{2},\omega\right)}{q\left(\tilde{\boldsymbol{\beta}},\boldsymbol{\eta}\right)}d\tilde{\boldsymbol{\beta}}\\
	& = & \mathbf{E}_{q}\left[\log h\left(\boldsymbol{\gamma}|\mathbf{Z},\mathbf{A},\tilde{\boldsymbol{\beta}},\boldsymbol{\eta};\mathbf{b},\boldsymbol{\xi}\right)+\log\Pr\left(\tilde{\boldsymbol{\beta}},\boldsymbol{\eta}|\sigma^{2},\omega\right)-\log q\left(\tilde{\boldsymbol{\beta}},\boldsymbol{\eta}\right)\right]\\
	& \triangleq & L\left(q\right),
\end{eqnarray*}
where $L(q)$ is the lower bound. The second inequality follows Jensen's
inequality. We can maximize $L(q)$ instead of the marginal likelihood
to get parameter estimations. To make it feasible to evaluate the
lower bound, we assume that $q\left(\tilde{\boldsymbol{\beta}},\boldsymbol{\eta}\right)$
can be factorized as
\[
q\left(\tilde{\beta},\eta\right)=\prod_{k=1}^{K}q\left(\tilde{\beta}_{k},\eta_{k}\right)=\prod_{k=1}^{K}q\left(\tilde{\beta}_{k}|\eta_{k}\right)q\left(\eta_{k}\right),
\]
where $q\left(\eta_{k}=1\right)=\omega_{k}$.

We can obtain an approximation according to the mean-field method:

\begin{eqnarray*}
	\log q\left(\tilde{\beta}_{i},\eta_{i}\right) & = & \mathbf{E}_{k\ne i}\left[\log h\left(\boldsymbol{\gamma}|\mathbf{Z},\mathbf{A},\tilde{\boldsymbol{\beta}},\boldsymbol{\eta},\mathbf{b},\boldsymbol{\xi}\right)+\log\Pr\left(\tilde{\boldsymbol{\beta}},\boldsymbol{\eta}|\sigma^{2},\omega\right)\right],
\end{eqnarray*}
where the expectation is taken under the distribtion $q\left(\tilde{\beta}_{-i},\eta_{-i}\right)=\prod_{k\ne i}q\left(\tilde{\beta}_{k},\eta_{k}\right)$.
Then we have

\[
q\left(\tilde{\beta}_{i},\eta_{i}\right)=\left[\omega_{i}N\left(\mu_{i},s_{i}^{2}\right)\right]^{\eta_{i}}\left[\left(1-\omega_{i}\right)N\left(0,\sigma^{2}\right)\right]^{1-\eta_{i}},
\]
where 

\begin{eqnarray*}
	\mu_{i} & = & s_{i}^{2}\sum_{j=1}^{M}\left(\pi_{j}-\frac{1}{2}-2\lambda\left(\xi_{j}\right)\left(\mathbf{Z}_{j}\mathbf{b}+\sum_{k\ne i}A_{jk}\mathbf{E}_{k}\left[\eta_{k}\tilde{\beta}_{k}\right]\right)\right)A_{ji},\\
	s_{i}^{2} & = & \frac{\sigma^{2}}{1+2\sigma^{2}\sum_{j=1}^{M}\lambda\left(\xi_{j}\right)A_{ji}^{2}}.
\end{eqnarray*}

Then we maximize $L\left(q\right)$ with respect to $\omega_{k}$
and $\xi_{j}$ and get 

\begin{eqnarray*}
	\omega_{k} & = & \frac{1}{1+\exp\left(-u_{k}\right)},\textrm{ where }u_{k}=\log\frac{\omega}{1-\omega}+\frac{1}{2}\log\frac{s_{k}^{2}}{\sigma^{2}}+\frac{\mu_{k}^{2}}{2s_{k}^{2}},\\
	\xi_{j}^{2} & = & \left(\mathbf{Z}_{j}\mathbf{b}+\sum_{k}A_{jk}\omega_{k}\mu_{k}\right)^{2}+\sum_{k}A_{jk}^{2}\left(\omega_{k}\left(s_{k}^{2}+\mu_{k}^{2}\right)-\omega_{k}^{2}\mu_{k}^{2}\right).
\end{eqnarray*}

Now we have evaluate $L(q)$:

\begin{eqnarray*}
	&  & L(q)\\
	& = & \sum_{j=1}^{M}\left(\gamma_{j}\left(\mathbf{Z}_{j}\mathbf{b}+\sum_{k}A_{jk}\omega_{k}\mu_{k}\right)+\log S\left(\xi_{j}\right)-\lambda\left(\xi_{j}\right)\left(\left(\mathbf{Z}_{j}\mathbf{b}+\sum_{k}A_{jk}\omega_{k}\mu_{k}\right)^{2}-\xi_{j}^{2}\right)\right)\\
	&  & +\sum_{j=1}^{M}\left(-\left(\mathbf{Z}_{j}\mathbf{b}+\sum_{k}A_{jk}\omega_{k}\mu_{k}+\xi_{j}\right)/2+\lambda\left(\xi_{j}\right)\sum_{k}A_{jk}^{2}\omega_{k}^{2}\mu_{k}^{2}-\lambda\left(\xi_{j}\right)\sum_{k}A_{jk}^{2}\omega_{k}\left(s_{k}^{2}+\mu_{k}^{2}\right)\right)\\
	&  & -\frac{1}{2\sigma^{2}}\sum_{k=1}^{K}\left(\omega_{k}\left(s_{k}^{2}+\mu_{k}^{2}\right)-\omega_{k}\sigma^{2}\right)+\sum_{k=1}^{K}\omega_{k}\log\omega+\sum_{k=1}^{K}\left(1-\omega_{k}\right)\log\left(1-\omega\right)\\
	&  & +\sum_{k=1}^{K}\left(\frac{1}{2}\omega_{k}\left(\log s_{k}^{2}-\log\sigma^{2}\right)-\omega_{k}\log\omega_{k}-\left(1-\omega_{k}\right)\log\left(1-\omega_{k}\right)\right).
\end{eqnarray*}

With $q\left(\boldsymbol{\gamma},\tilde{\boldsymbol{\beta}},\boldsymbol{\eta}\right)$
obtained, we can evaluate the lower bound and then update the model
parameters by maximizing $L(q)$. 

In the M step, we update $\sigma^{2}$ and $\omega$ by maximizing
$L(q)$. We have 

\begin{eqnarray*}
	\sigma^{2} & = & \frac{\sum_{k=1}^{K}\omega_{k}\left(s_{k}^{2}+\mu_{k}^{2}\right)}{\sum_{k=1}^{K}\omega_{k}},\\
	\omega & = & \frac{1}{K}\sum_{k=1}^{K}\omega_{k}.
\end{eqnarray*}

We use Newton's method to update $\mathbf{b}$:

\[
\mathbf{b}=\mathbf{b}_{old}-\mathbf{H}^{-1}\mathbf{g},
\]
where 

\begin{eqnarray*}
	\mathbf{g} & = & -\sum_{j=1}^{M}\mathbf{Z}_{j}^{T}\left(\gamma_{j}-2\lambda\left(\xi_{j}\right)\left(\mathbf{Z}_{j}\mathbf{b}+\sum_{k}A_{jk}\omega_{k}\mu_{k}\right)-\frac{1}{2}\right),\\
	\mathbf{H} & = & 2\sum_{j=1}^{M}\lambda\left(\xi_{j}\right)\mathbf{Z}_{j}^{T}\mathbf{Z}_{j}.
\end{eqnarray*}

\subsubsection*{Algorithm:}

Input: $\mathbf{Z}$, $\mathbf{A}$, $\left\{ \gamma_{j}=\tilde{\gamma}_{j}\right\} _{j=1,...,M}$,
$\mathbf{b}$, Initialize: $\sigma^{2}=1$, $\omega=0.5$, $\left\{ \omega_{k}=0,\mu_{k}=0\right\} _{k=1,...K}$,
$\boldsymbol{\xi}=\mathbf{Zb}$, Output: $\mathbf{b}$, $\boldsymbol{\xi}$,
$\sigma^{2}$, $\omega$, $\left\{ \omega_{k},\mu_{k}\right\} _{k=1,...K}$.
\begin{itemize}
	\item Initialize $\mathbf{b}$, $\boldsymbol{\xi}=\mathbf{Zb}$, $\sigma^{2}=1$,
	$\omega=0.5$, $\left\{ \omega_{k}=0,\mu_{k}=0\right\} _{k=1,...K}$.
	Let $\tilde{y}=\sum_{k}A_{jk}\omega_{k}\mu_{k}$.
	\item E-step: For $i=1,...,K$, first obtain $\tilde{y}_{i}=\tilde{y}-A_{ji}\omega_{i}\mu_{i}$,
	and then update $\mu_{i},s_{i}^{2},\omega_{i}$ and $\tilde{y}$ as
	follows
	
	\begin{eqnarray*}
		s_{i}^{2} & = & \frac{\sigma^{2}}{1+2\sigma^{2}\sum_{j=1}^{M}\lambda\left(\xi_{j}\right)A_{ji}^{2}},\\
		\mu_{i} & = & s_{i}^{2}\sum_{j=1}^{M}\left(\left(\gamma_{j}-\frac{1}{2}-2\lambda\left(\xi_{j}\right)\left(\mathbf{Z}_{j}\mathbf{b}+\tilde{y}_{i}\right)\right)A_{ji}\right),\\
		\omega_{i} & = & \frac{1}{1+\exp\left(-u_{i}\right)},\textrm{ where }u_{i}=\log\frac{\omega}{1-\omega}+\frac{1}{2}\log\frac{s_{i}^{2}}{\sigma^{2}}+\frac{\mu_{i}^{2}}{2s_{i}^{2}},\\
		\tilde{y} & = & \tilde{y}_{i}+A_{ji}\omega_{i}\mu_{i}.
	\end{eqnarray*}
	
	Then for $j=1,...,M$, update $\xi_{j}$ as follows
	
	\begin{eqnarray*}
		\xi_{j}^{2} & = & \left(\mathbf{Z}_{j}\mathbf{b}+\tilde{y}\right)^{2}+\sum_{k}A_{jk}^{2}\left(\omega_{k}\left(s_{k}^{2}+\mu_{k}^{2}\right)-\omega_{k}^{2}\mu_{k}^{2}\right).
	\end{eqnarray*}
	
	Calculate $L\left(q\right)$:
	
	\begin{eqnarray*}
		&  & L(q)\\
		& = & \sum_{j=1}^{M}\left(\gamma_{j}\left(\mathbf{Z}_{j}\mathbf{b}+\tilde{y}\right)+\log S\left(\xi_{j}\right)-\frac{\mathbf{Z}_{j}\mathbf{b}+\tilde{y}+\xi_{j}}{2}\right)\\
		&  & -\frac{1}{2\sigma^{2}}\sum_{k=1}^{K}\left(\omega_{k}\left(s_{k}^{2}+\mu_{k}^{2}\right)-\omega_{k}\sigma^{2}\right)+\sum_{k=1}^{K}\omega_{k}\log\omega+\sum_{k=1}^{K}\left(1-\omega_{k}\right)\log\left(1-\omega\right)\\
		&  & +\sum_{k=1}^{K}\left(\frac{1}{2}\omega_{k}\left(\log s_{k}^{2}-\log\sigma^{2}\right)-\omega_{k}\log\omega_{k}-\left(1-\omega_{k}\right)\log\left(1-\omega_{k}\right)\right).
	\end{eqnarray*}
	
	\item M-step
	
	\begin{eqnarray*}
		\mathbf{g} & = & -\sum_{j=1}^{M}\mathbf{Z}_{j}^{T}\left(\pi_{j}-2\lambda\left(\xi_{j}\right)\left(\mathbf{Z}_{j}\mathbf{b}+\tilde{y}\right)-\frac{1}{2}\right),\\
		\mathbf{H} & = & 2\sum_{j=1}^{M}\lambda\left(\xi_{j}\right)\mathbf{Z}_{j}^{T}\mathbf{Z}_{j},\\
		\mathbf{b} & = & \mathbf{b}_{old}-\mathbf{H}^{-1}\mathbf{g},\\
		\sigma^{2} & = & \frac{\sum_{k=1}^{K}\omega_{k}\left(s_{k}^{2}+\mu_{k}^{2}\right)}{\sum_{k=1}^{K}\omega_{k}},\\
		\omega & = & \frac{1}{K}\sum_{k=1}^{K}\omega_{k}.
	\end{eqnarray*}
	
	\item Check convergence.
\end{itemize}

\subsection*{Stage 4: LSMM}

Input: $\mathbf{p}$, $\mathbf{Z}$, $\mathbf{A}$, $\alpha$,$\mathbf{b}$,
$\boldsymbol{\xi}$, $\sigma^{2}$, $\omega$, $\left\{ \omega_{k},\mu_{k}\right\} _{k=1,...K}$,
Initialize: $\left\{ \pi_{j}=\tilde{\gamma}_{j}\right\} _{j=1,...,M}$,
Output: $\alpha$,$\mathbf{b}$, $\sigma^{2}$, $\omega$, $\left\{ \omega_{k},\beta_{k}=\mu_{k}\omega_{k}\right\} _{k=1,...K}$,
$\left\{ \pi_{j}\right\} _{j=1,...,M}$

\subsubsection*{Algorithm:}
\begin{itemize}
	\item Initialize $\alpha$, $\sigma^{2}$, $\omega$, $\mathbf{b}$, $\left\{ \omega_{k},\mu_{k}\right\} _{k=1,...K}$,
	$\left\{ \xi_{j},\pi_{j}\right\} _{j=1,...,M}$. Let $\tilde{y}=\sum_{k}A_{jk}\omega_{k}\mu_{k}$.
	\item E-step: For $i=1,...,K$, first obtain $\tilde{y}_{i}=\tilde{y}-A_{ji}\omega_{i}\mu_{i}$,
	and then update $\mu_{i},s_{i}^{2},\omega_{i}$ and $\tilde{y}$ as
	follows
	
	\begin{eqnarray*}
		s_{i}^{2} & = & \frac{\sigma^{2}}{1+2\sigma^{2}\sum_{j=1}^{M}\lambda\left(\xi_{j}\right)A_{ji}^{2}},\\
		\mu_{i} & = & s_{i}^{2}\sum_{j=1}^{M}\left(\left(\pi_{j}-\frac{1}{2}-2\lambda\left(\xi_{j}\right)\left(\mathbf{Z}_{j}\mathbf{b}+\tilde{y}_{i}\right)\right)A_{ji}\right),\\
		\omega_{i} & = & \frac{1}{1+\exp\left(-u_{i}\right)},\textrm{ where }u_{i}=\log\frac{\omega}{1-\omega}+\frac{1}{2}\log\frac{s_{i}^{2}}{\sigma^{2}}+\frac{\mu_{i}^{2}}{2s_{i}^{2}},\\
		\tilde{y} & = & \tilde{y}_{i}+A_{ji}\omega_{i}\mu_{i}.
	\end{eqnarray*}
	
	Then for $j=1,...,M$, update $\pi_{j},\xi_{j}$ as follows
	
	\begin{eqnarray*}
		\pi_{j} & = & \frac{1}{1+\exp\left(-v_{j}\right)},\textrm{ where }v_{j}=\log\alpha+\left(\alpha-1\right)\log p_{j}+\mathbf{Z}_{j}\mathbf{b}+\tilde{y},\\
		\xi_{j}^{2} & = & \left(\mathbf{Z}_{j}\mathbf{b}+\tilde{y}\right)^{2}+\sum_{k}A_{jk}^{2}\left(\omega_{k}\left(s_{k}^{2}+\mu_{k}^{2}\right)-\omega_{k}^{2}\mu_{k}^{2}\right).
	\end{eqnarray*}
	
	Calculate $L\left(q\right)$:
	
	\begin{eqnarray*}
		&  & L(q)\\
		& = & \sum_{j=1}^{M}\pi_{j}\left(\log\alpha+\left(\alpha-1\right)\log p_{j}\right)-\sum_{j=1}^{M}\left(\pi_{j}\log\pi_{j}+\left(1-\pi_{j}\right)\log\left(1-\pi_{j}\right)\right)\\
		&  & +\sum_{j=1}^{M}\left(\pi_{j}\left(\mathbf{Z}_{j}\mathbf{b}+\tilde{y}\right)+\log S\left(\xi_{j}\right)-\frac{\mathbf{Z}_{j}\mathbf{b}+\tilde{y}+\xi_{j}}{2}\right)\\
		&  & -\frac{1}{2\sigma^{2}}\sum_{k=1}^{K}\left(\omega_{k}\left(s_{k}^{2}+\mu_{k}^{2}\right)-\omega_{k}\sigma^{2}\right)+\sum_{k=1}^{K}\omega_{k}\log\omega+\sum_{k=1}^{K}\left(1-\omega_{k}\right)\log\left(1-\omega\right)\\
		&  & +\sum_{k=1}^{K}\left(\frac{1}{2}\omega_{k}\left(\log s_{k}^{2}-\log\sigma^{2}\right)-\omega_{k}\log\omega_{k}-\left(1-\omega_{k}\right)\log\left(1-\omega_{k}\right)\right).
	\end{eqnarray*}
	
	\item M-step
	
	\begin{eqnarray*}
		\alpha & = & -\frac{\sum_{j=1}^{M}\pi_{j}}{\sum_{j=1}^{M}\pi_{j}\log p_{j}},\\
		\sigma^{2} & = & \frac{\sum_{k=1}^{K}\omega_{k}\left(s_{k}^{2}+\mu_{k}^{2}\right)}{\sum_{k=1}^{K}\omega_{k}},\\
		\omega & = & \frac{1}{K}\sum_{k=1}^{K}\omega_{k},\\
		\mathbf{g} & = & -\sum_{j=1}^{M}\mathbf{Z}_{j}^{T}\left(\pi_{j}-2\lambda\left(\xi_{j}\right)\left(\mathbf{Z}_{j}\mathbf{b}+\tilde{y}\right)-\frac{1}{2}\right),\\
		\mathbf{H} & = & 2\sum_{j=1}^{M}\lambda\left(\xi_{j}\right)\mathbf{Z}_{j}^{T}\mathbf{Z}_{j},\\
		\mathbf{b} & = & \mathbf{b}_{old}-\mathbf{H}^{-1}\mathbf{g}.
	\end{eqnarray*}
	
	\item Evaluate $L(q)$ to track the convergence of the algorithm.
\end{itemize}

\section{Simulation study for evaluating the LD effects on LSMM}

To study the influence of LD effects on our LSMM, we used the observed
genotype data (1,500 individuals from the 1958 British Birth Cohort
(58C)) from WTCCC (The Wellcome Trust Case Control Consortium, \citeyear{2007}).
For simplicity, we only consider 23874 SNPs in chromosome 1 after
quality control. We simulated a risk SNP every 1000 SNPs. So we had
24 risk SNPs. We assumed the 24 risk SNPs can explain 5\% phenotypic
variance. We used GCTA to simulation phenotypes and used PLINK to
get $p$-values for SNPs. Then we applied LSMM and detect risk SNPs. 

As the presence of LD effects, SNPs in a local genomic region would
be correlated and detection of risk SNPs would be difficult. We are
just expected to identify the region which contains the risk SNPs.
Here we used different distance threshold to define the region around
true risk SNPs. The identified risk SNPs which in the region of true
risk SNPs were considered as true positive. 

\begin{figure}[H]
	\begin{centering}
		\includegraphics[scale=0.4]{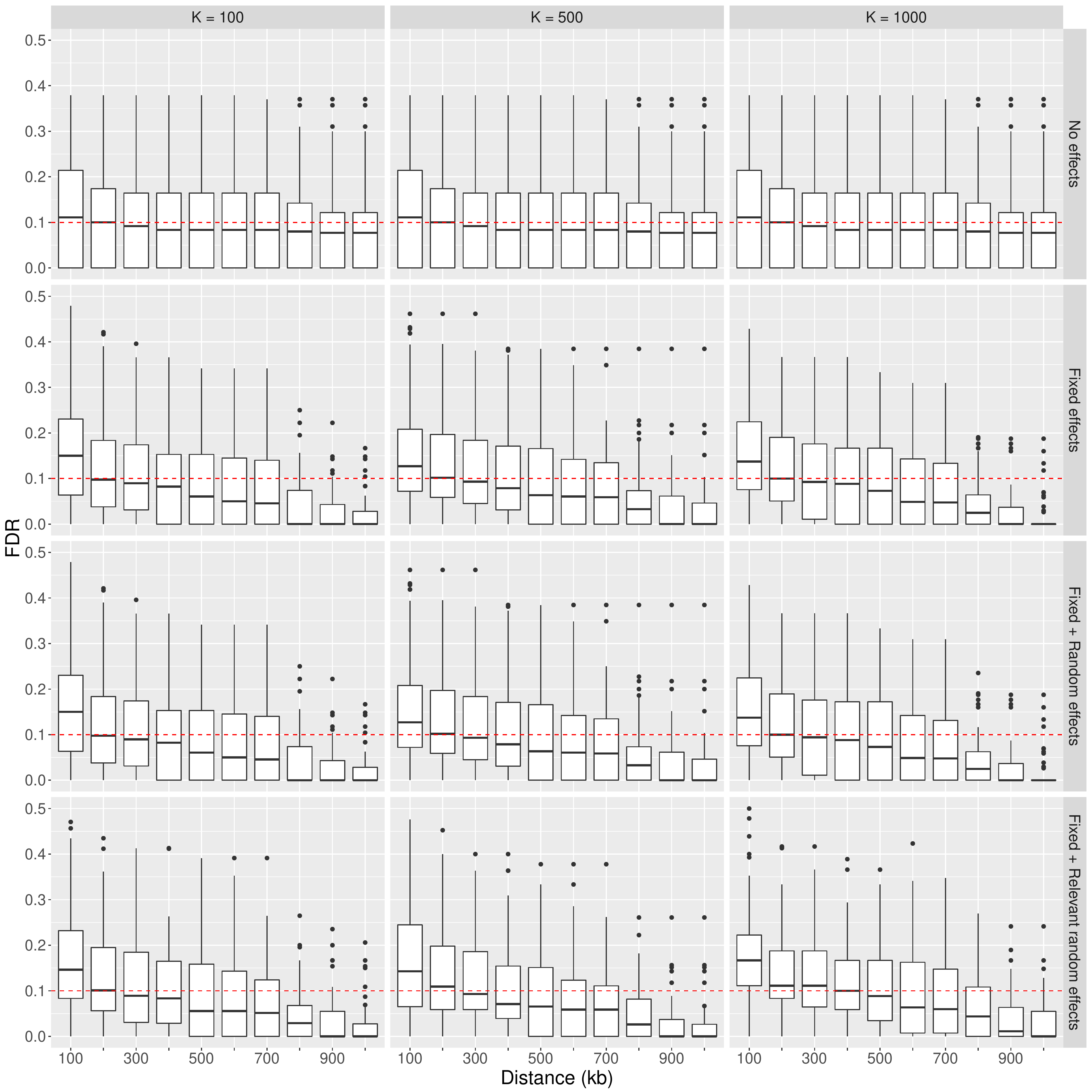}
		\par\end{centering}
	\caption{FDR of LSMM for identification of risk SNPs with different distance
		thresholds. The red line indicates the threshold of global FDR $\tau=0.1$.
		\label{fig:LD}}
\end{figure}

We considered four cases. The first case, no effects, means we only
used the $p$-values and didn't use fixed effects and random effects.
In the second case, fixed effects, we only add 10 fixed effects. In
the fixed effects, SNPs within 1Mb of true risk SNPs are annotated
with a probability of 0.6. In the third case, fixed + random effects,
we further add 100 random effects in which SNPs are annotated randomly.
In the fourth case, fixed + relevant random effects, we assume 20\%
of random effects are relevant to the phenotype and SNPs within 1Mb
of true risk SNPs are annotated with a probability of 0.6 in the relevant
random effects. The results of observed FDR were shown in Figure \ref{fig:LD}
based on 50 simulations. In the first case, when we used no effects,
the observed FDR was quite stable at 0.1. When we added fixed effects
and random effects, the observed FDR was just inflated a little with
the smallest distance threshold and became conservative as the distance
threshold increased. As a result, we believe that LSMM can provide
a satisfactory FDR control in detecting a local genomic region of
risk SNPs.

\section{More simulation results for different settings}

\subsection{Performance in identification of risk SNPs}

\begin{figure}[H]
	\begin{centering}
		\includegraphics[scale=0.36]{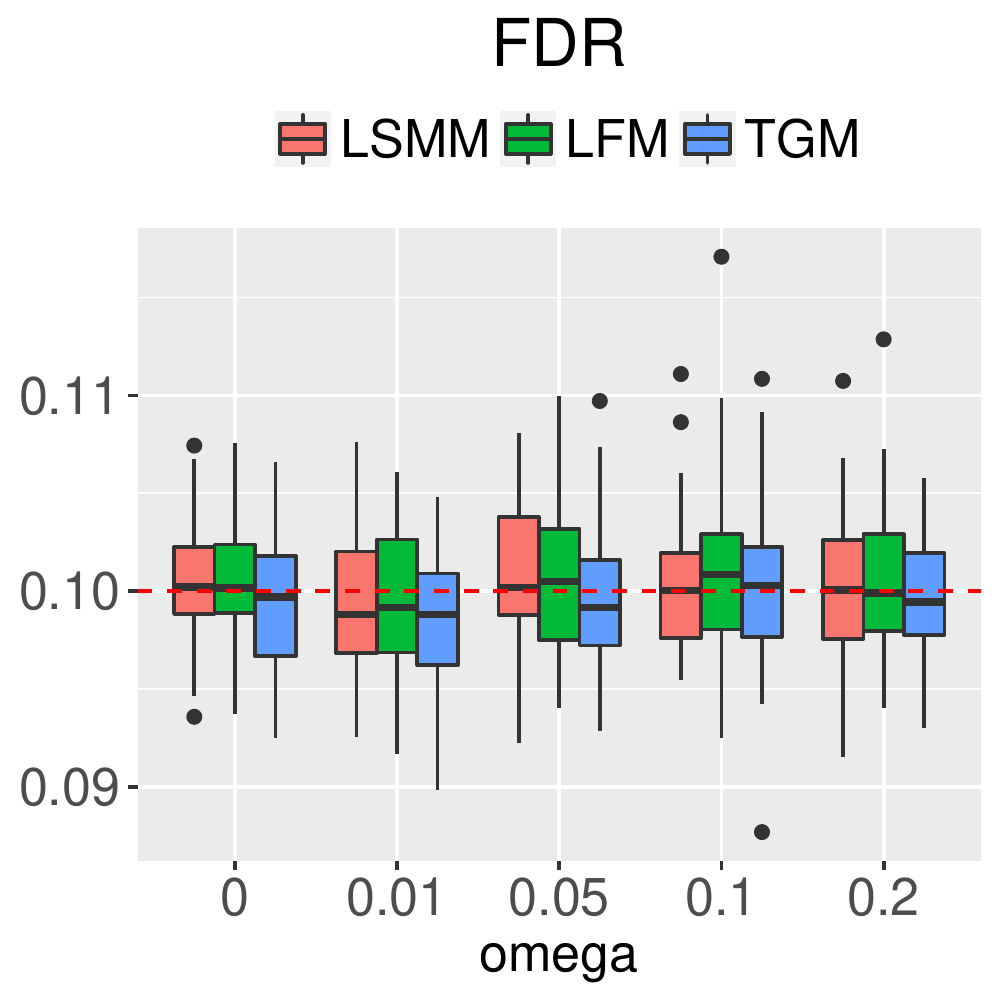}\includegraphics[scale=0.36]{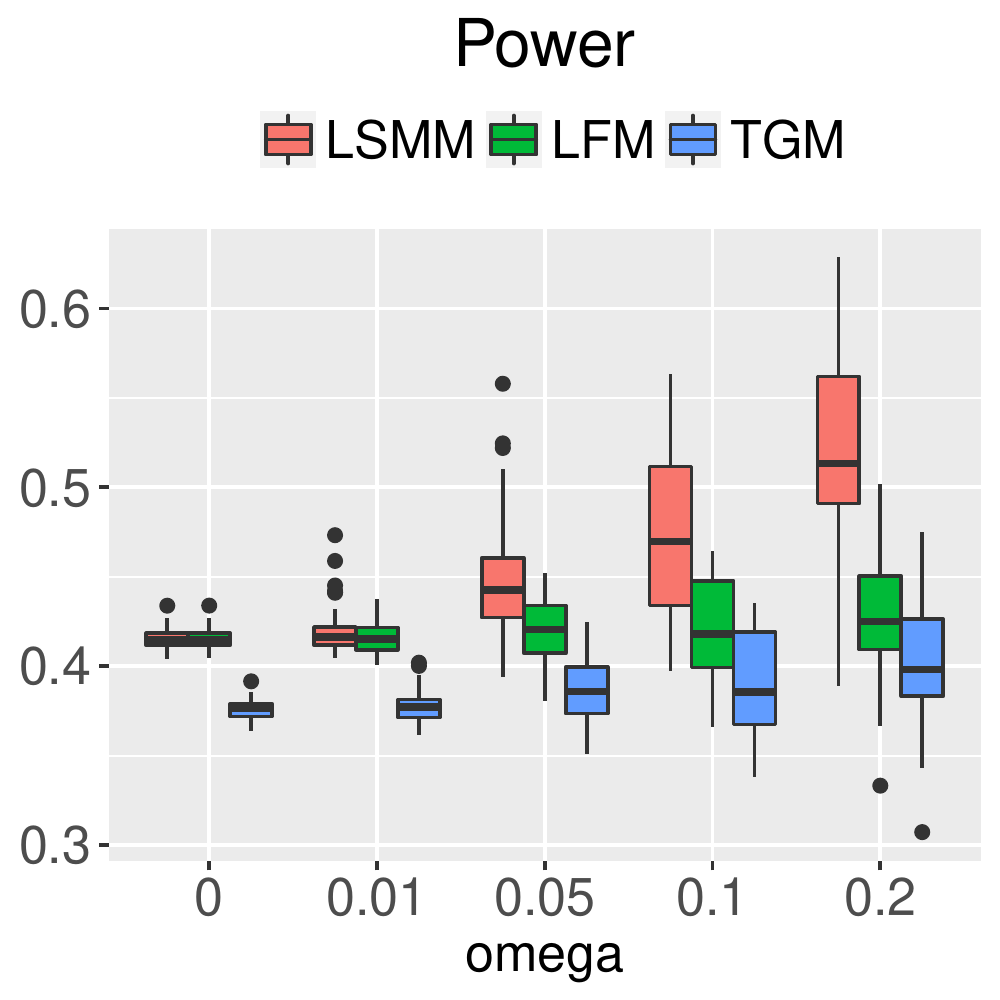}\includegraphics[scale=0.36]{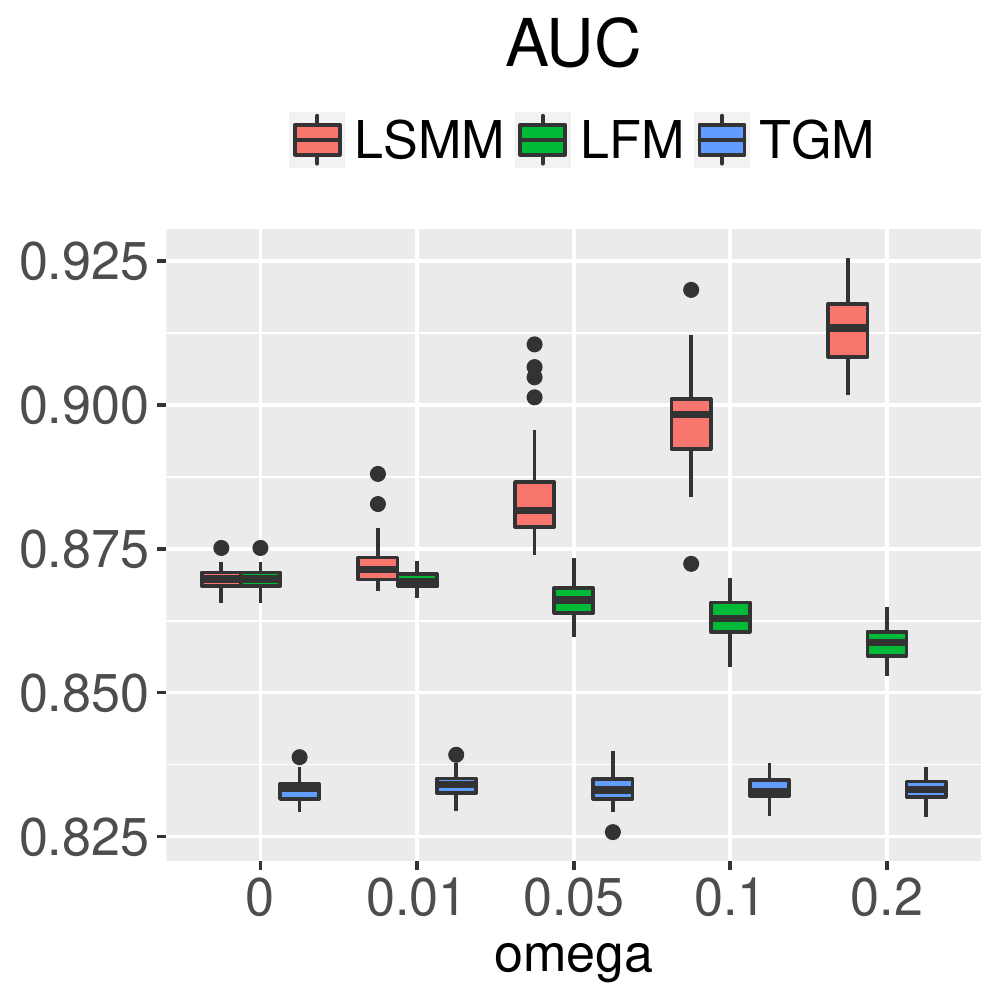}\includegraphics[scale=0.36]{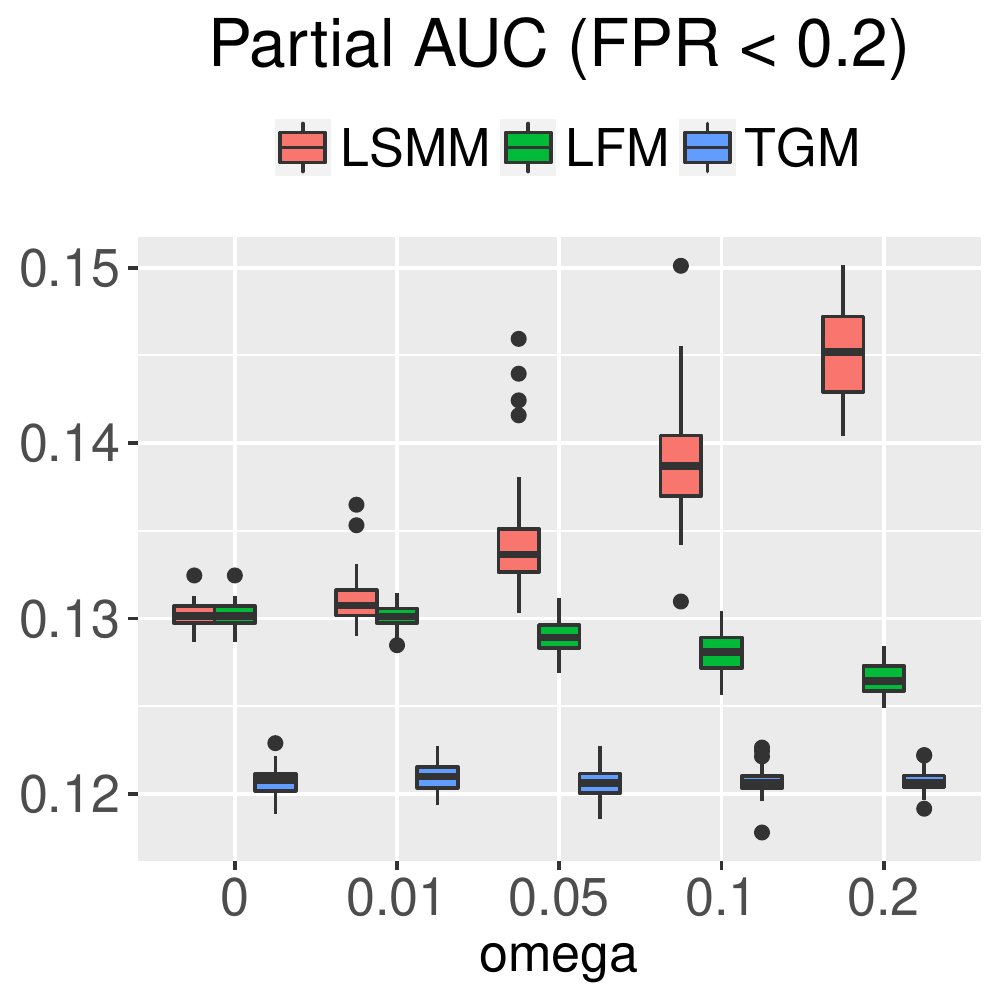}
		\par\end{centering}
	\caption{FDR, power, AUC and partial AUC of LSMM, LFM and TGM for identification
		of risk SNPs with $\alpha=0.2$ and $K=100$. We controlled global
		FDR at 0.1 to evaluate empirical FDR and power. The results are summarized
		from 50 replications.}
	
\end{figure}

\begin{figure}[H]
	\begin{centering}
		\includegraphics[scale=0.36]{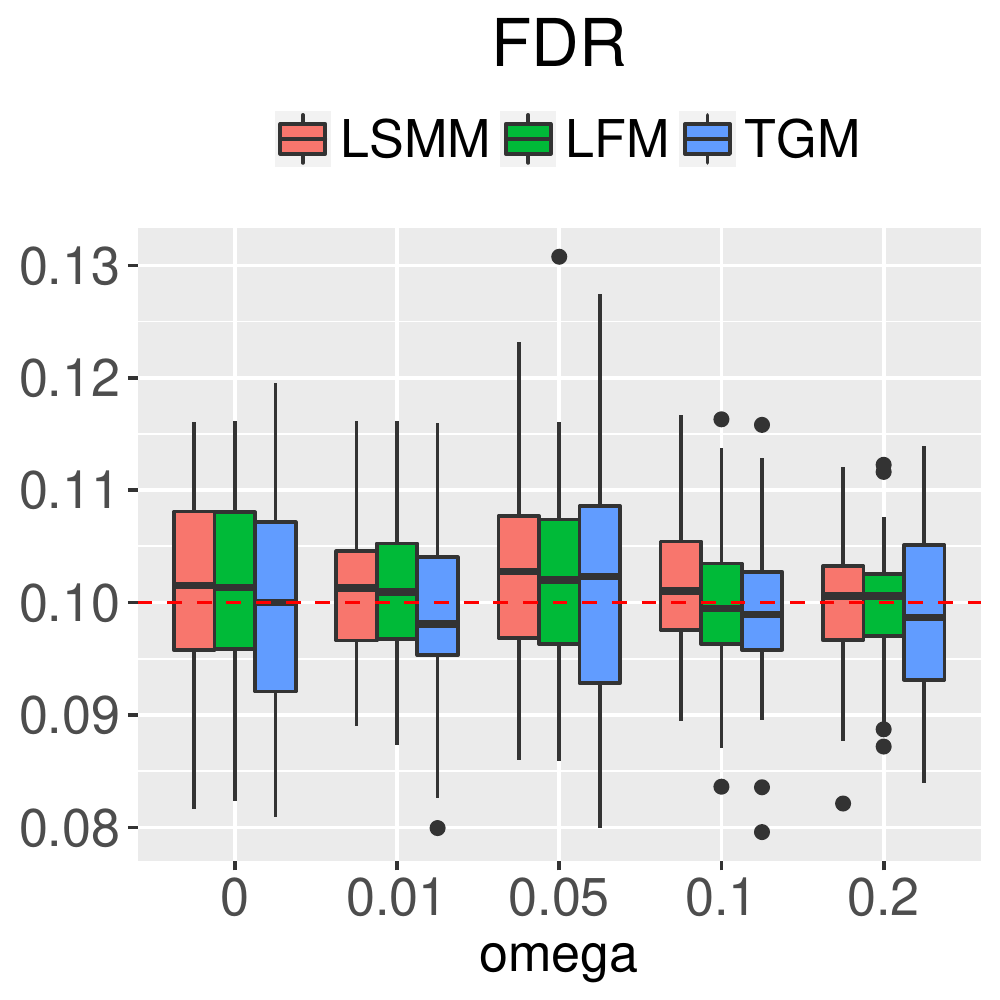}\includegraphics[scale=0.36]{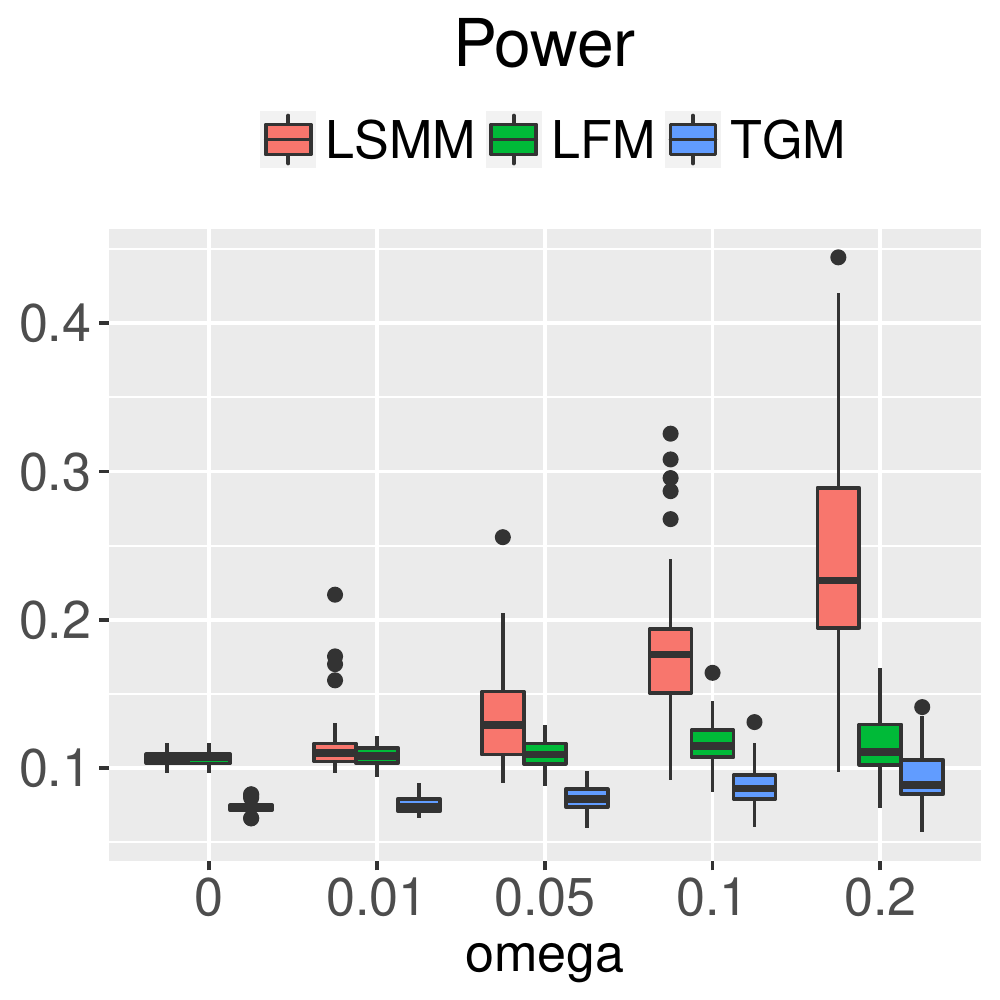}\includegraphics[scale=0.36]{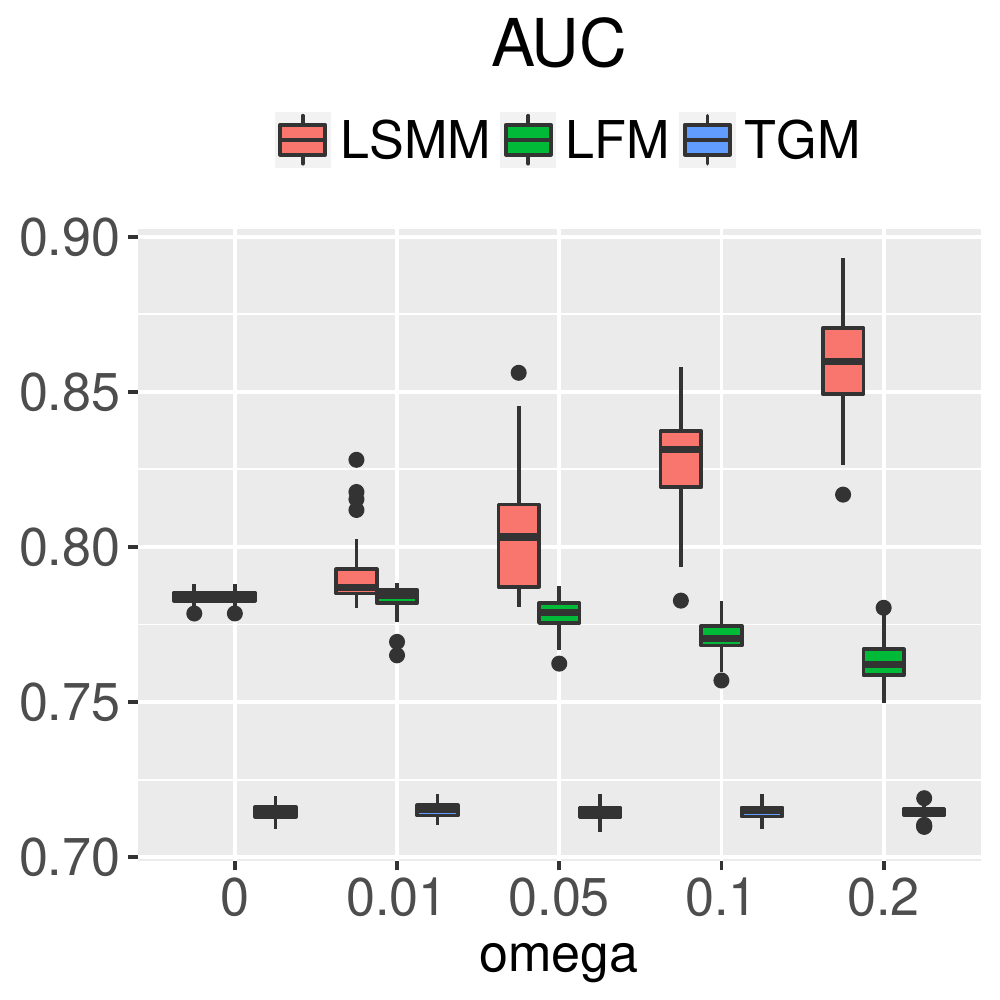}\includegraphics[scale=0.36]{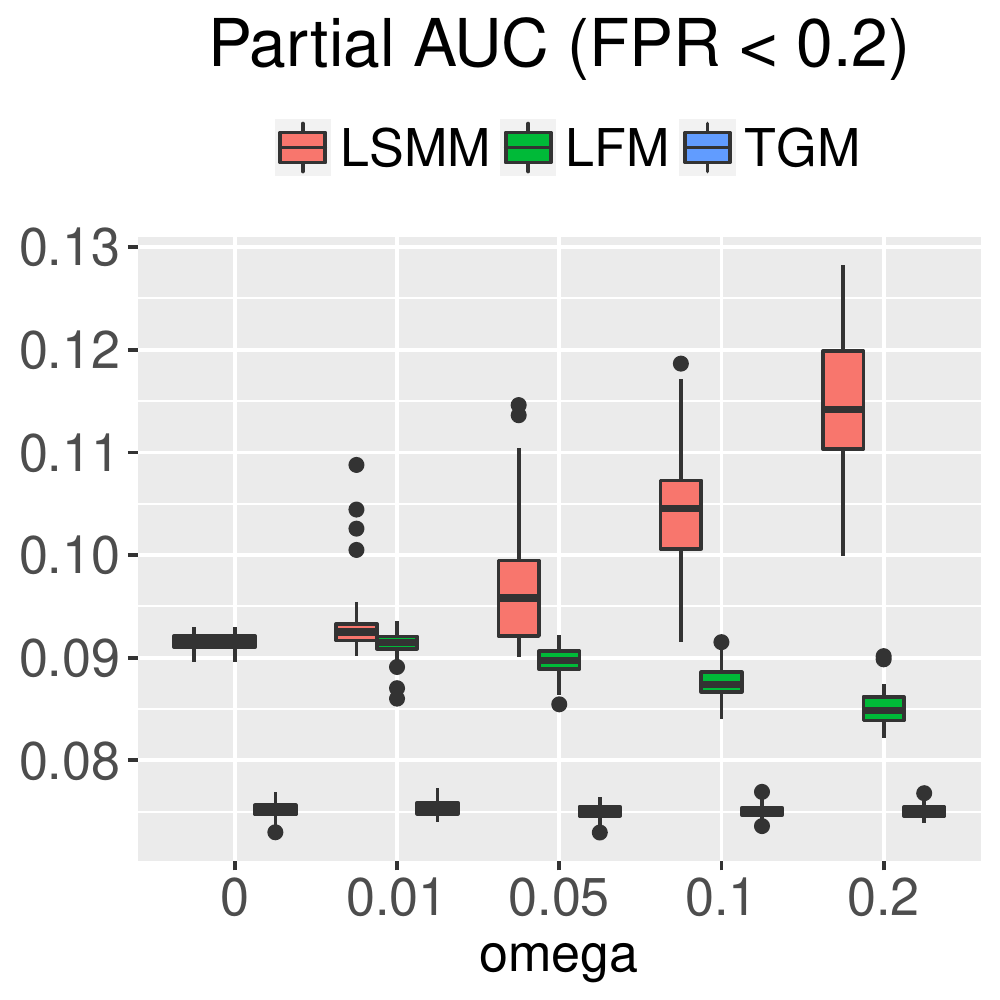}
		\par\end{centering}
	\caption{FDR, power, AUC and partial AUC of LSMM, LFM and TGM for identification
		of risk SNPs with $\alpha=0.4$ and $K=100$. We controlled global
		FDR at 0.1 to evaluate empirical FDR and power. The results are summarized
		from 50 replications.}
\end{figure}

\begin{figure}[H]
	\begin{centering}
		\includegraphics[scale=0.36]{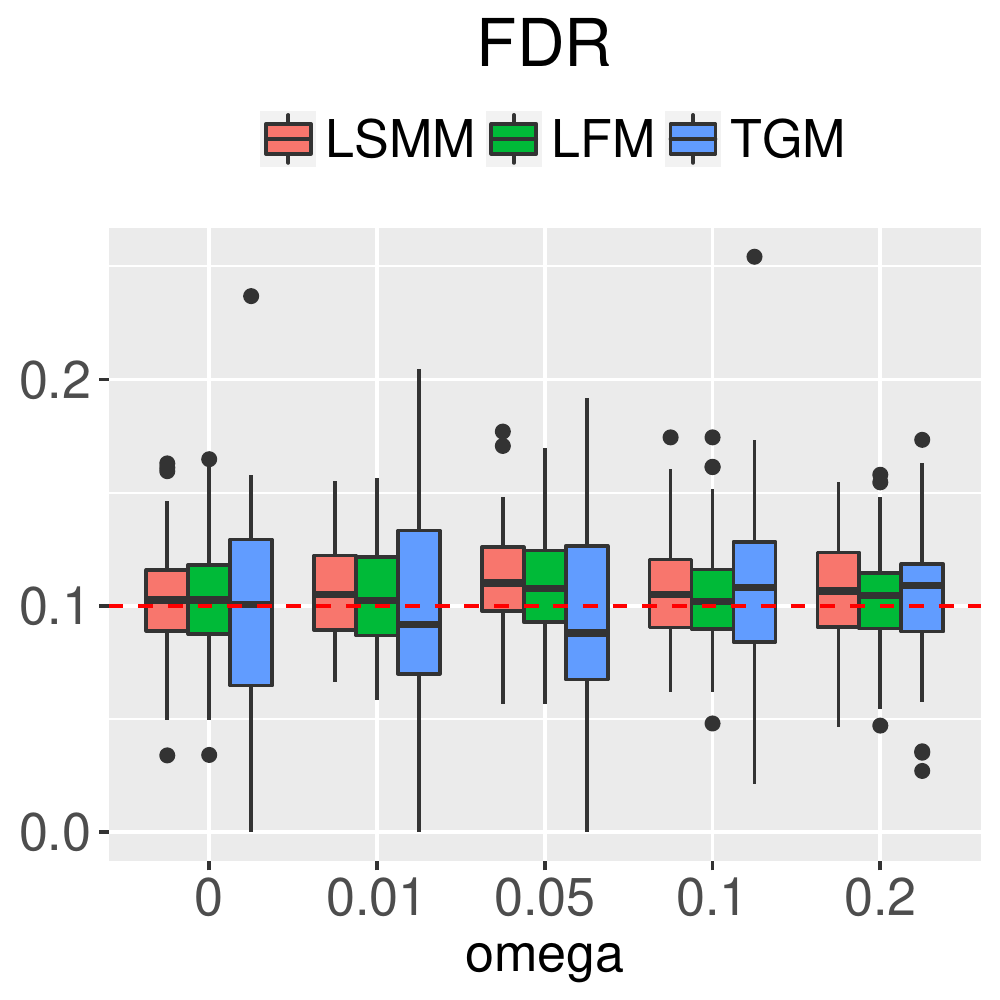}\includegraphics[scale=0.36]{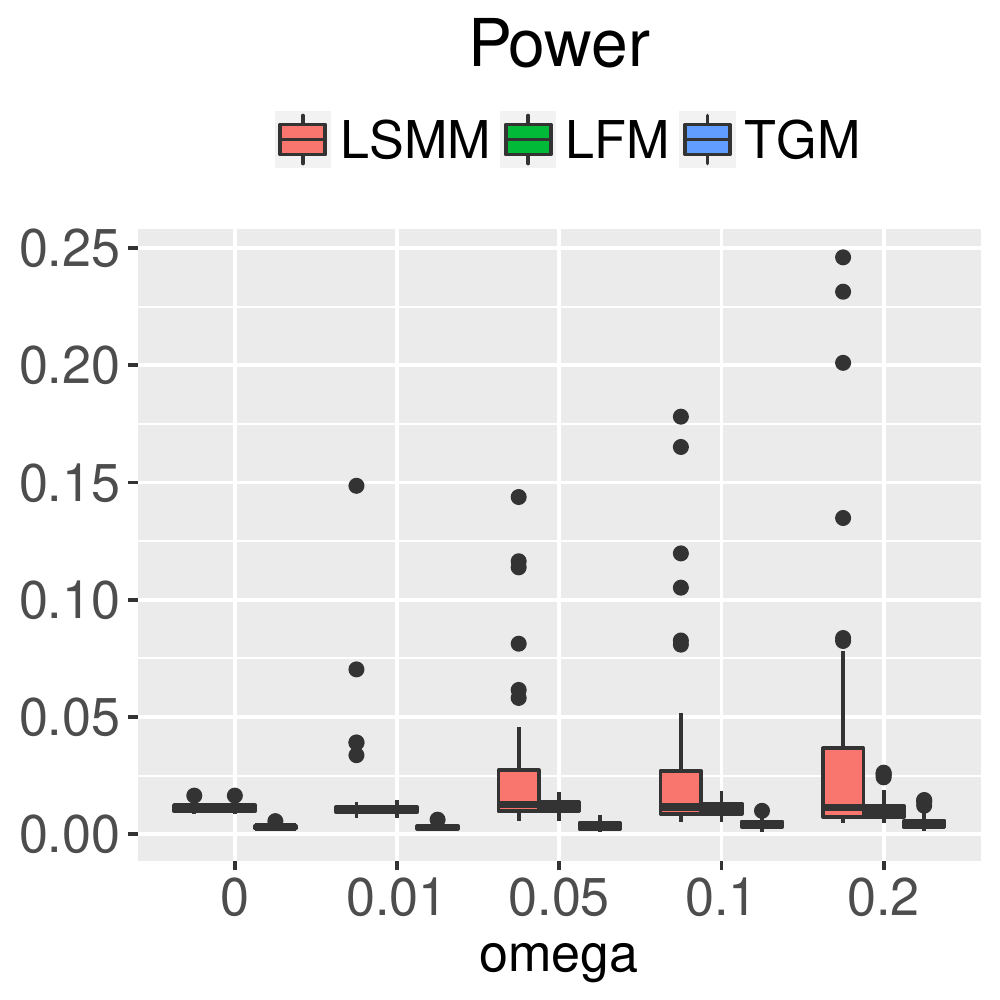}\includegraphics[scale=0.36]{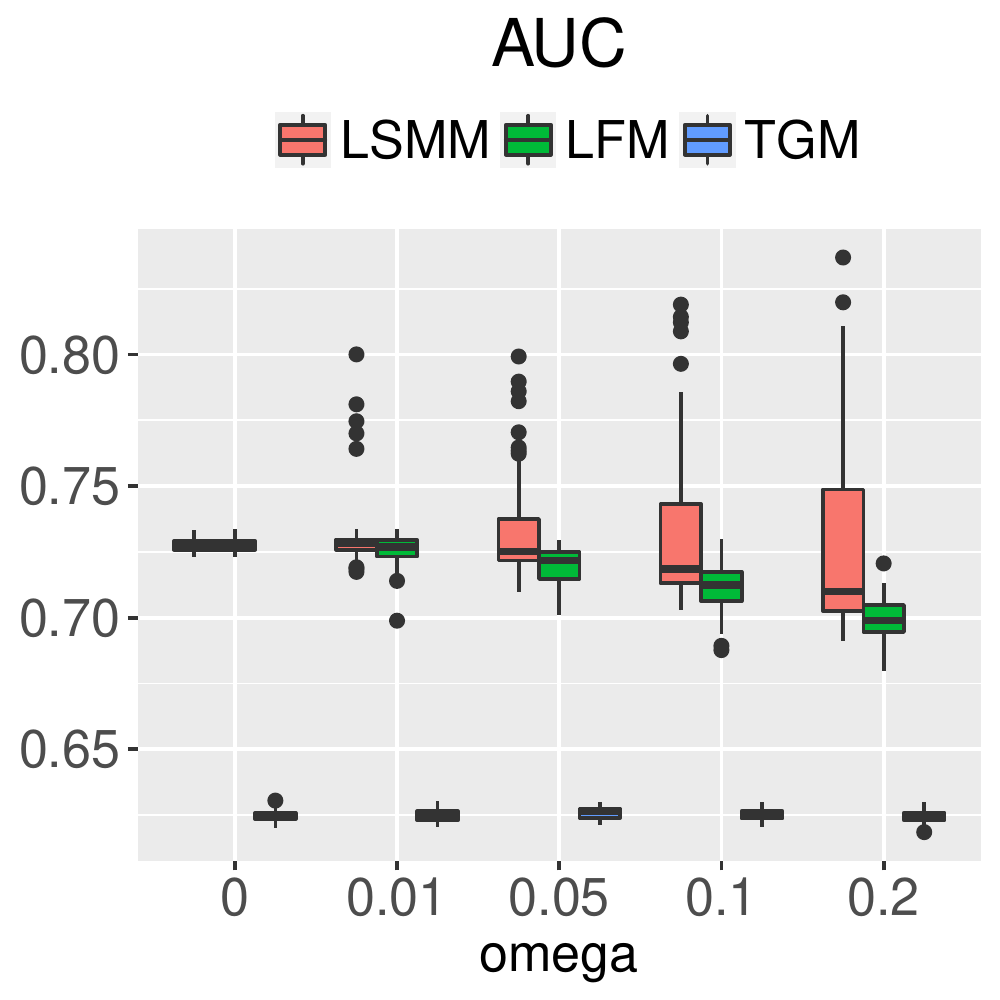}\includegraphics[scale=0.36]{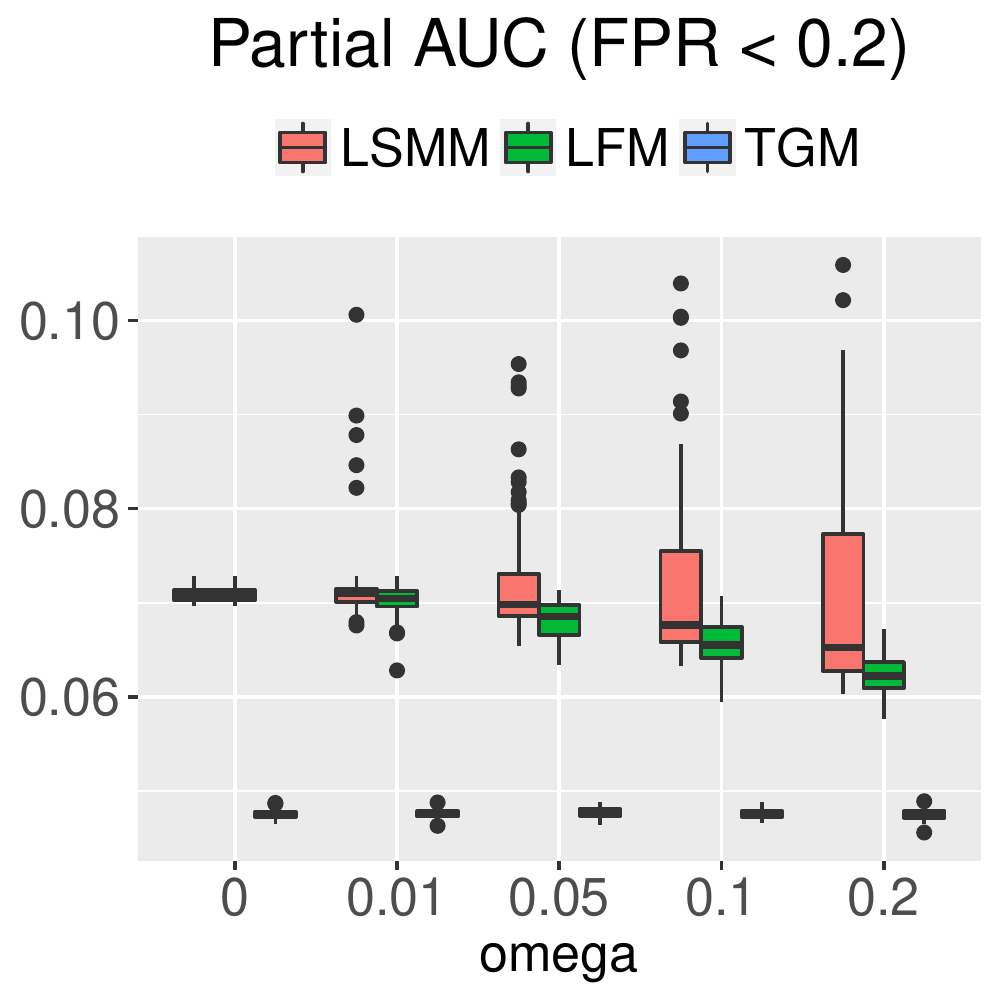}
		\par\end{centering}
	\caption{FDR, power, AUC and partial AUC of LSMM, LFM and TGM for identification
		of risk SNPs with $\alpha=0.6$ and $K=100$. We controlled global
		FDR at 0.1 to evaluate empirical FDR and power. The results are summarized
		from 50 replications.}
\end{figure}

\begin{figure}[H]
	\begin{centering}
		\includegraphics[scale=0.36]{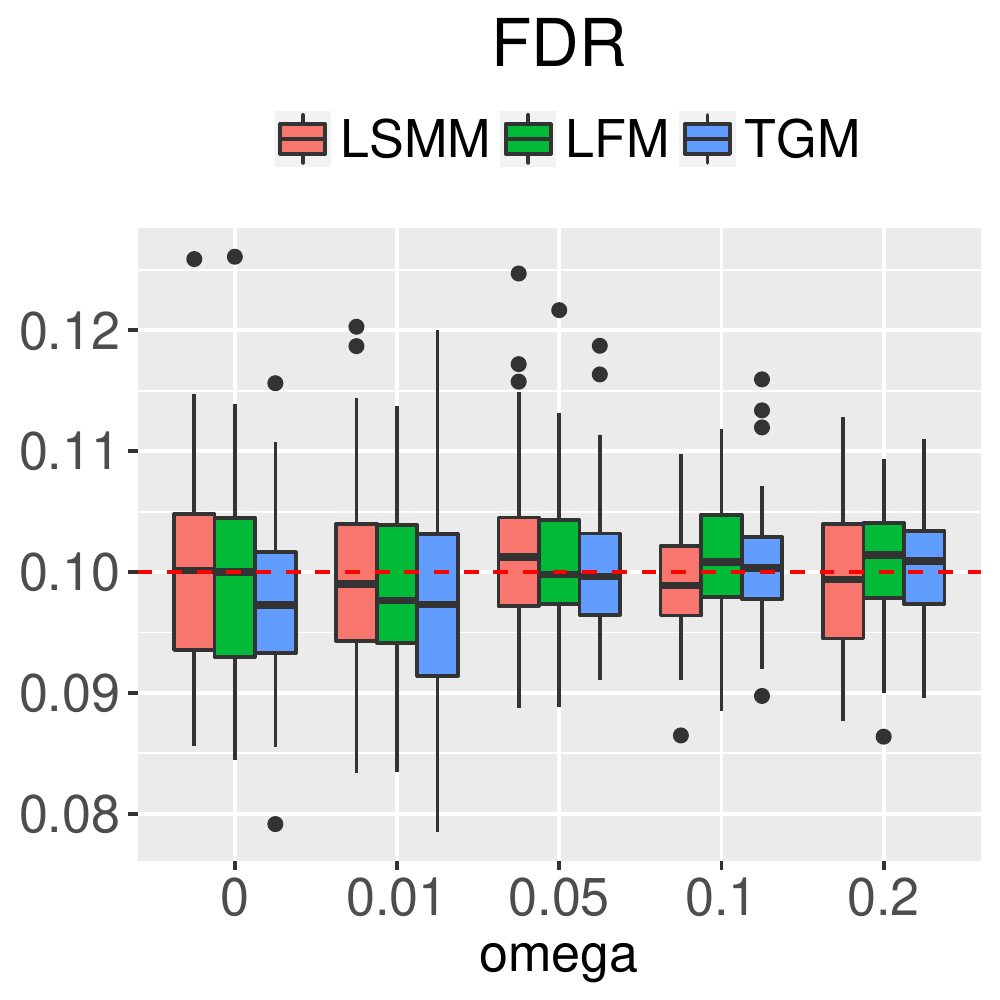}\includegraphics[scale=0.36]{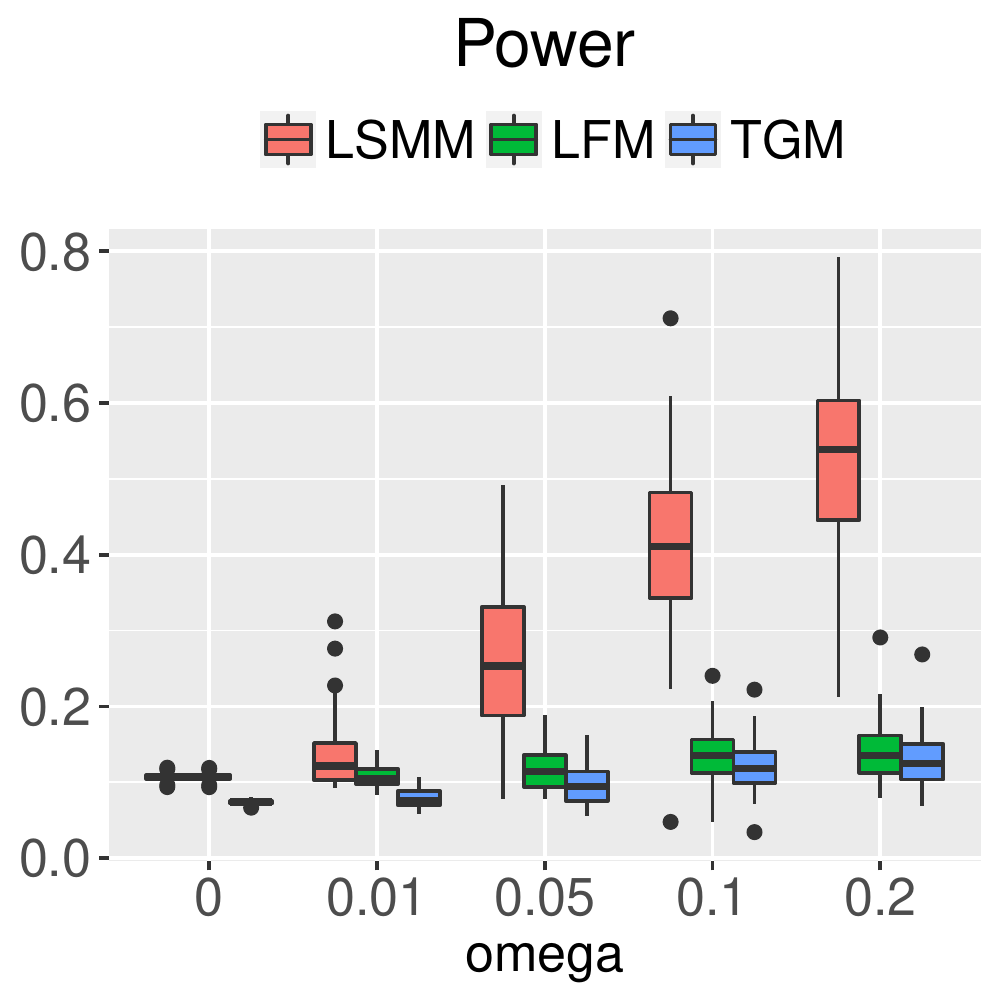}\includegraphics[scale=0.36]{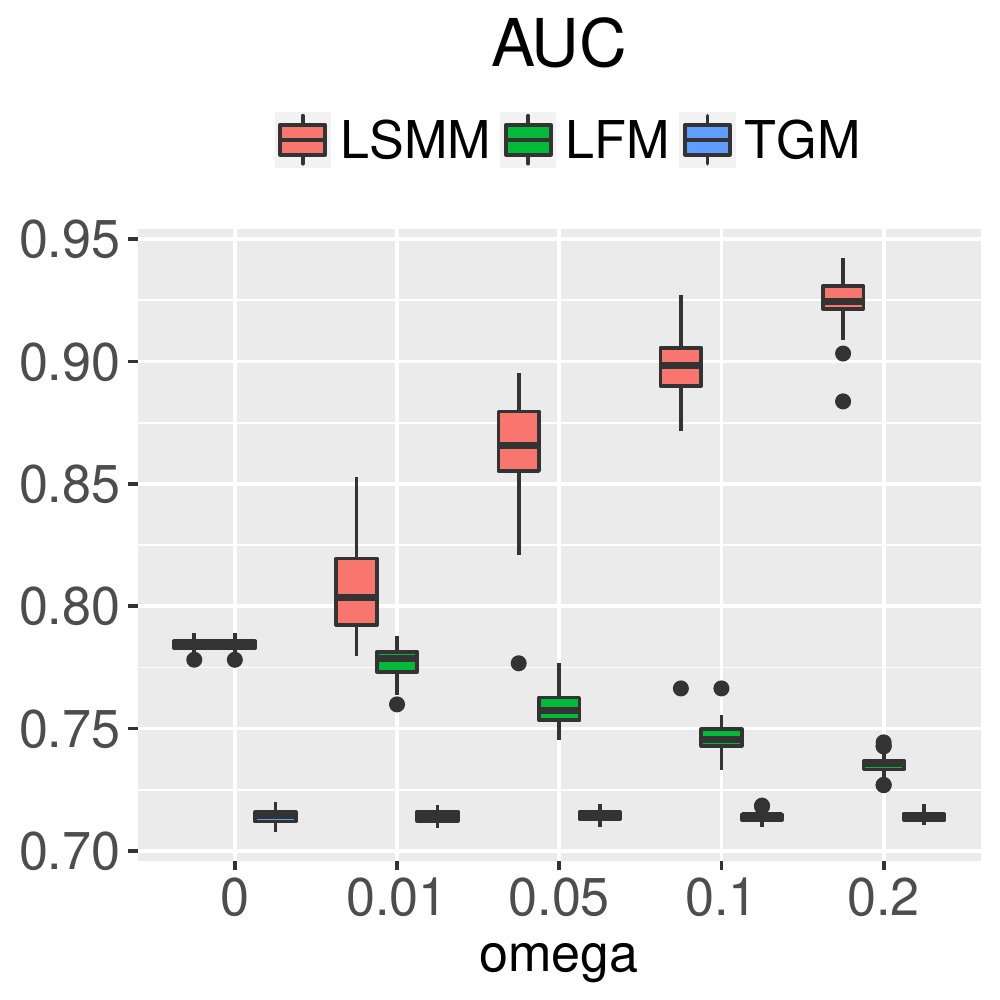}\includegraphics[scale=0.36]{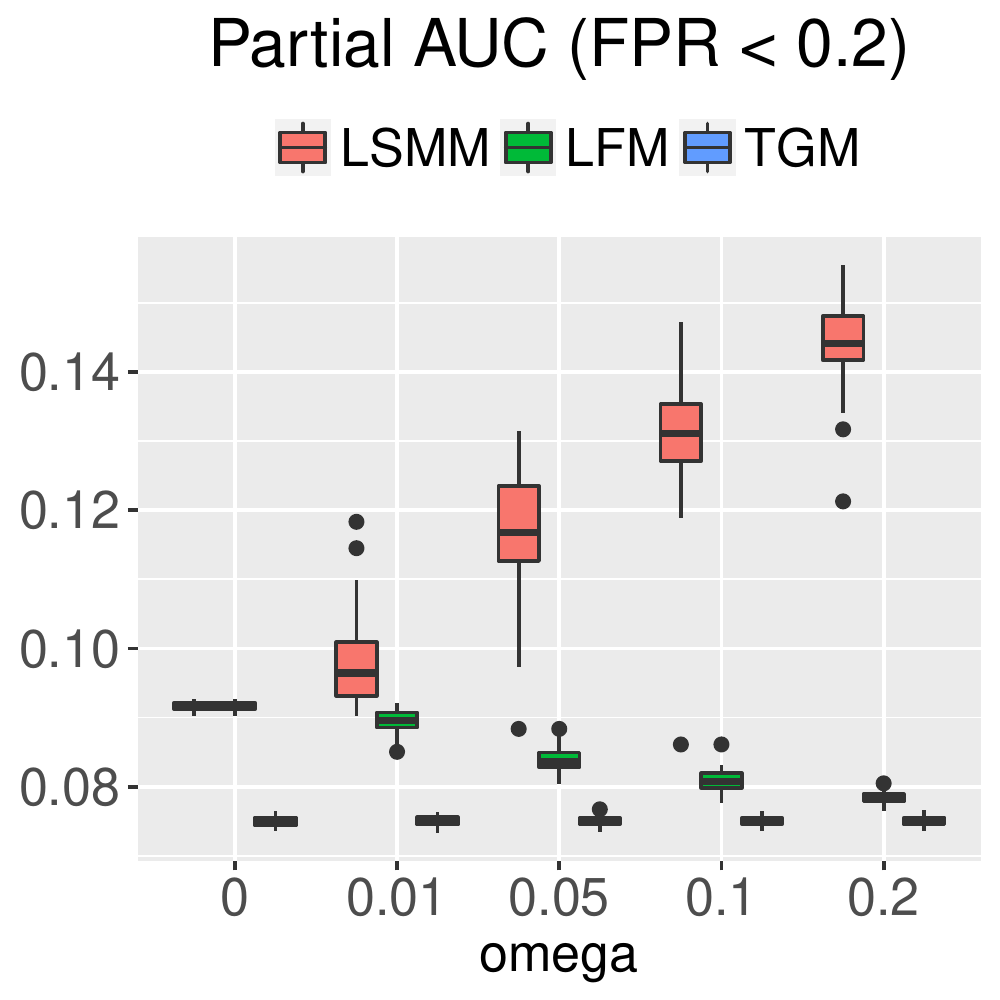}
		\par\end{centering}
	\caption{FDR, power, AUC and partial AUC of LSMM, LFM and TGM for identification
		of risk SNPs with $\alpha=0.4$ and $K=500$. We controlled global
		FDR at 0.1 to evaluate empirical FDR and power. The results are summarized
		from 50 replications.}
\end{figure}

\begin{figure}[H]
	\begin{centering}
		\includegraphics[scale=0.36]{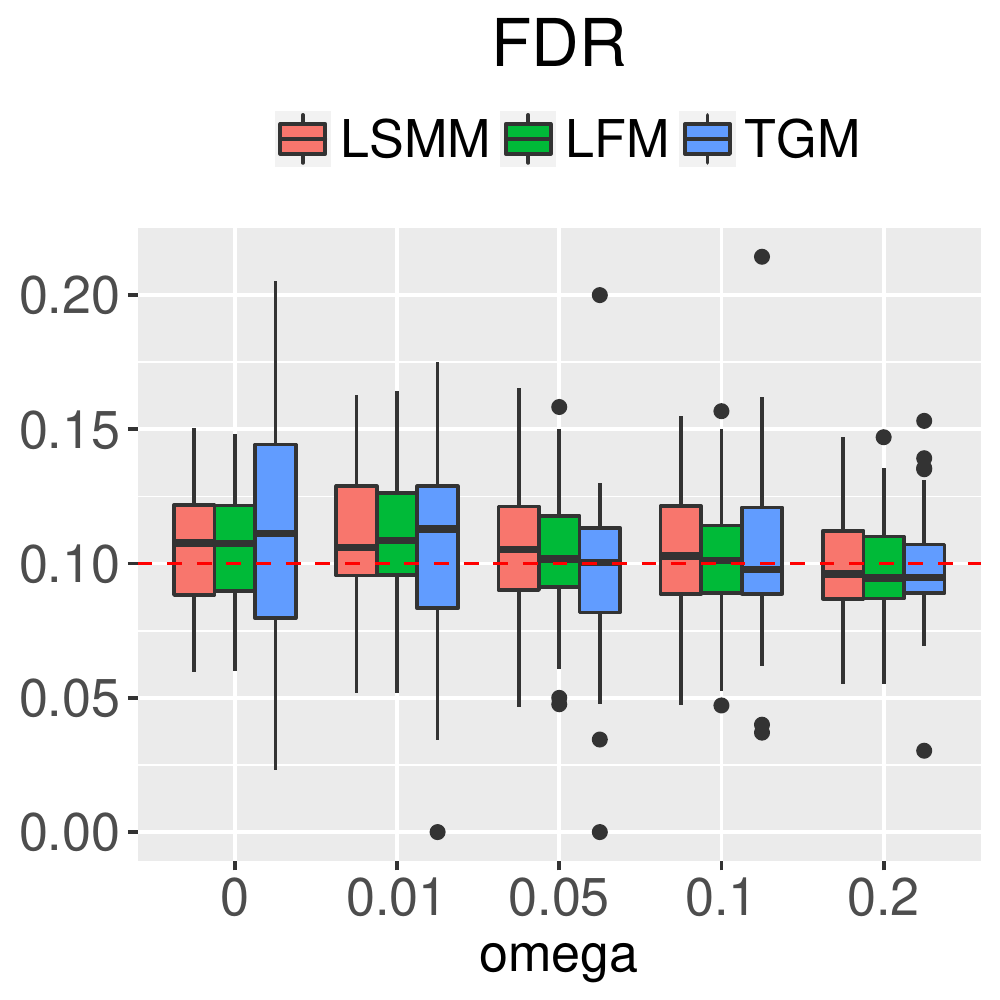}\includegraphics[scale=0.36]{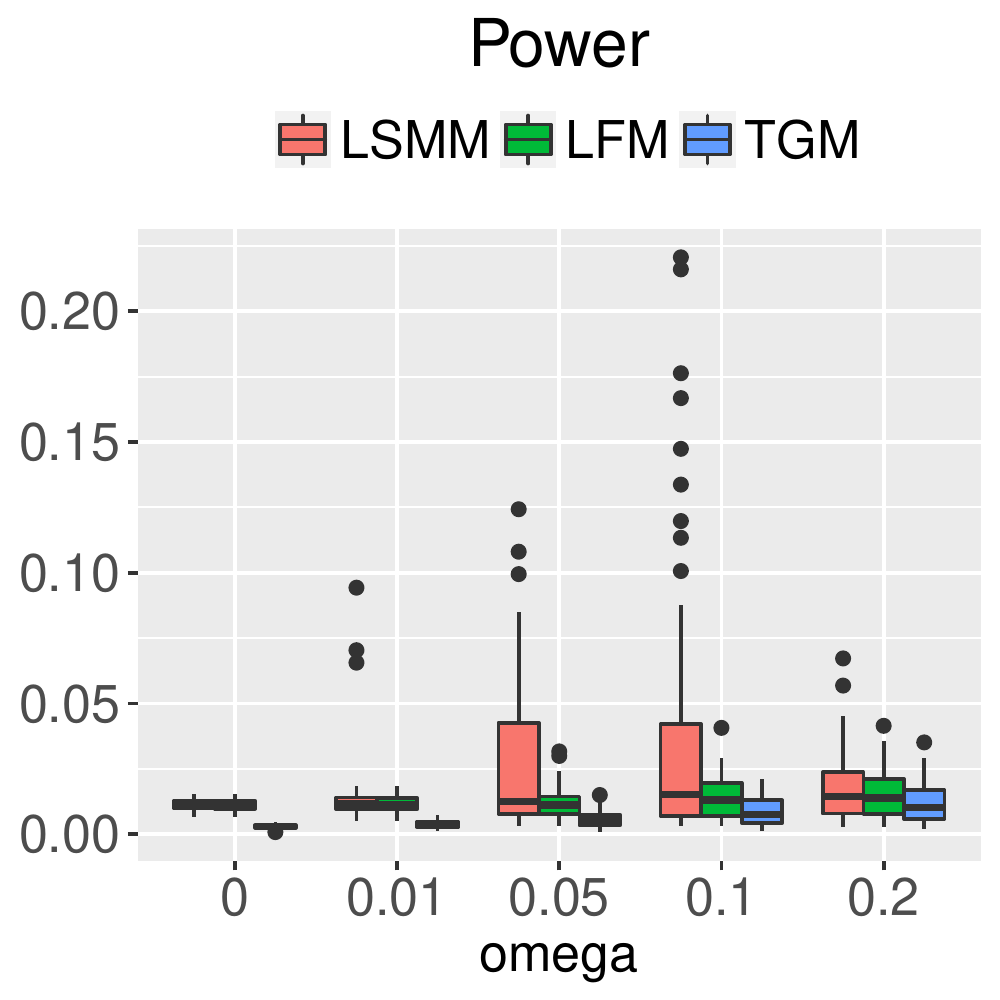}\includegraphics[scale=0.36]{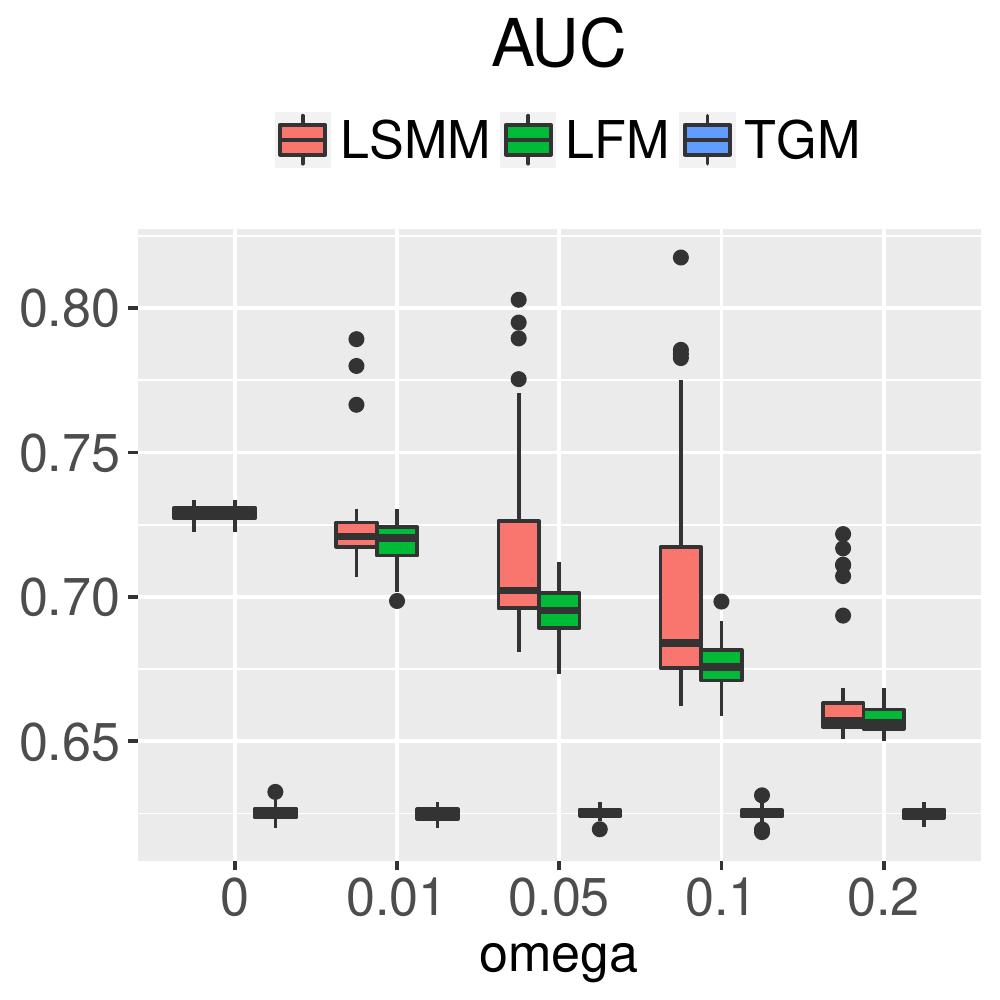}\includegraphics[scale=0.36]{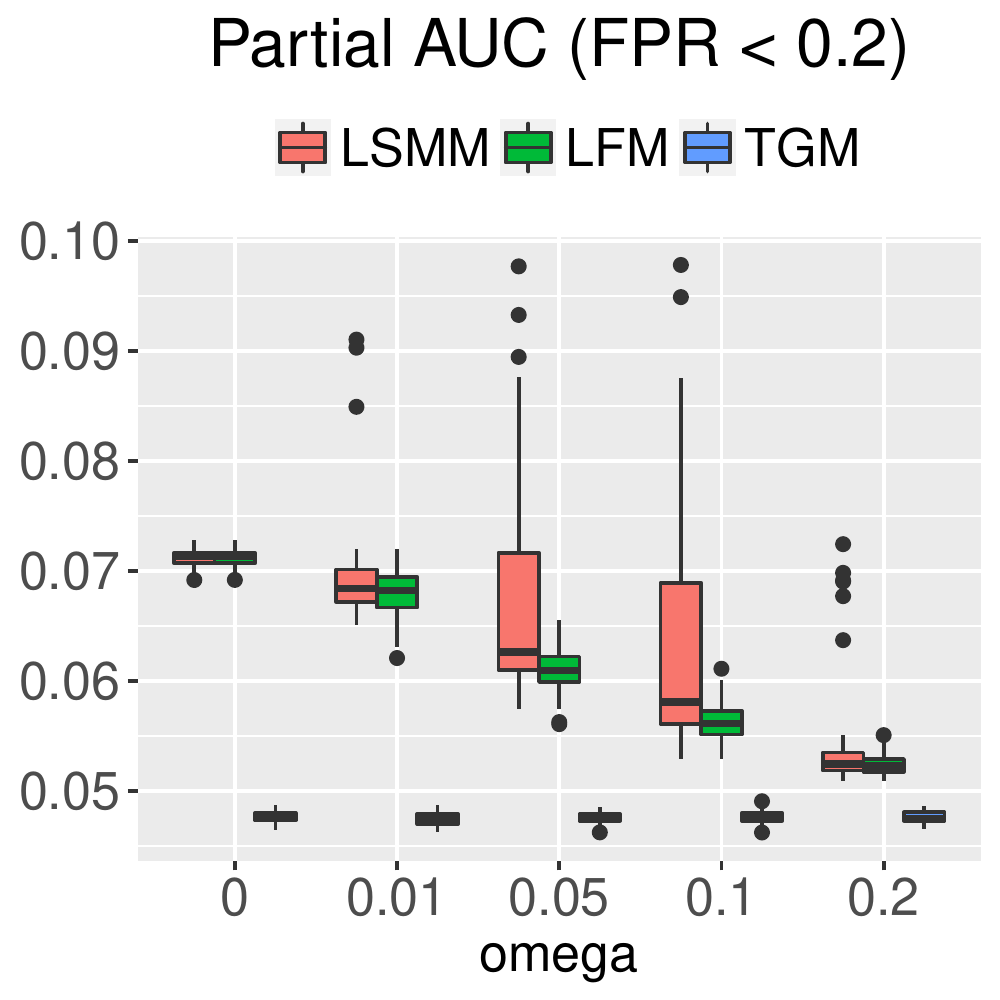}
		\par\end{centering}
	\caption{FDR, power, AUC and partial AUC of LSMM, LFM and TGM for identification
		of risk SNPs with $\alpha=0.6$ and $K=500$. We controlled global
		FDR at 0.1 to evaluate empirical FDR and power. The results are summarized
		from 50 replications.}
\end{figure}

\begin{figure}[H]
	\begin{centering}
		\includegraphics[scale=0.36]{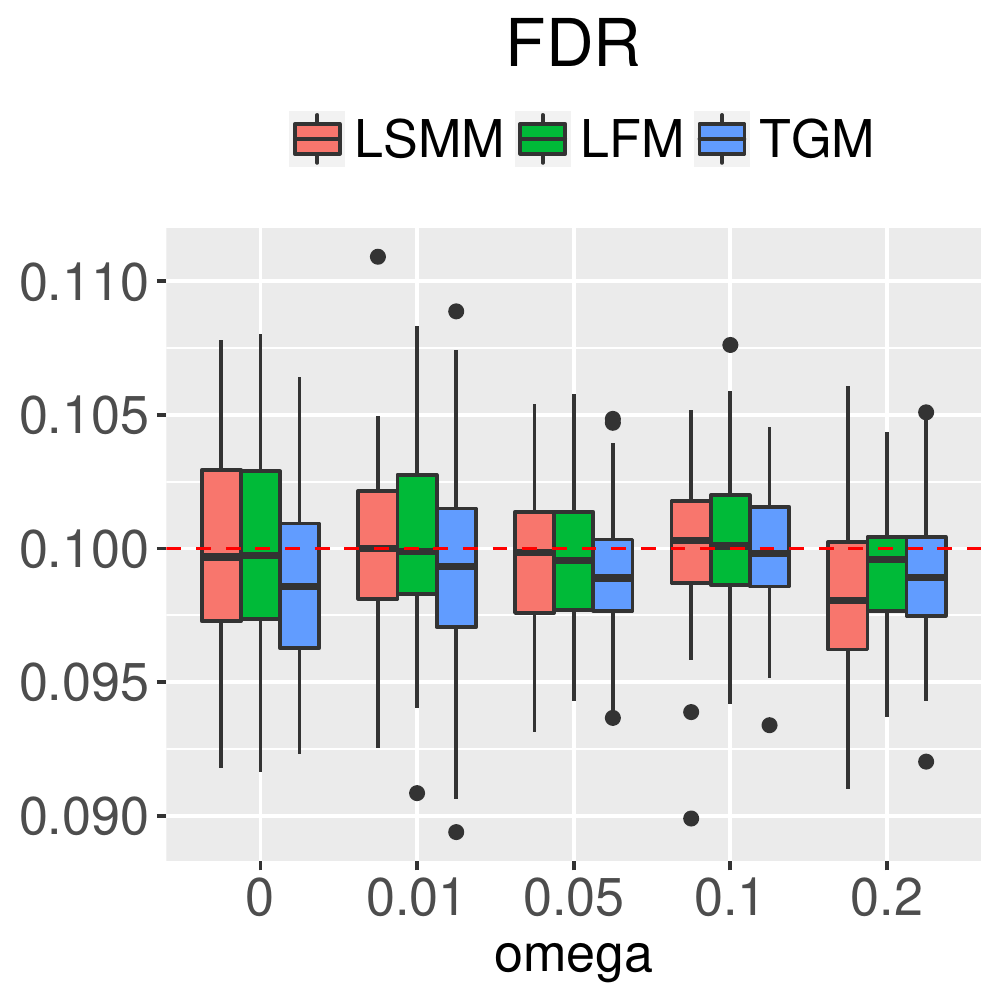}\includegraphics[scale=0.36]{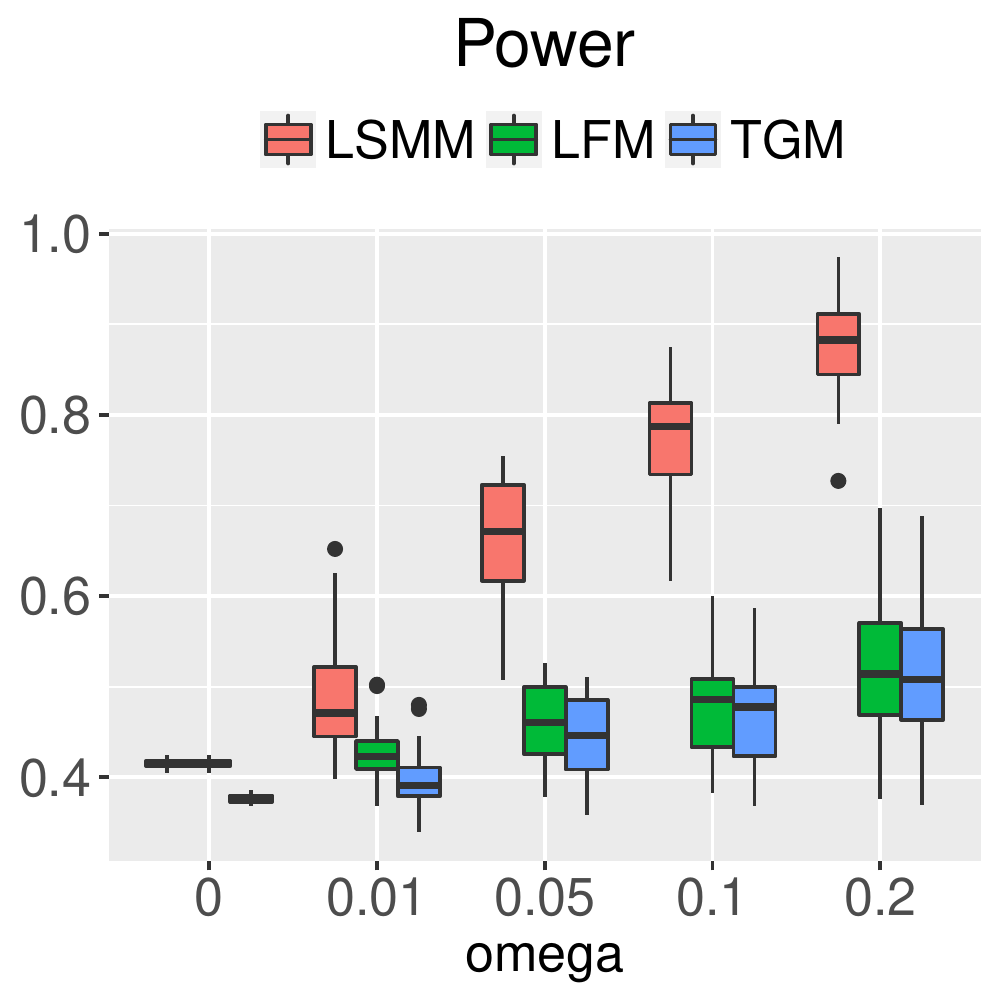}\includegraphics[scale=0.36]{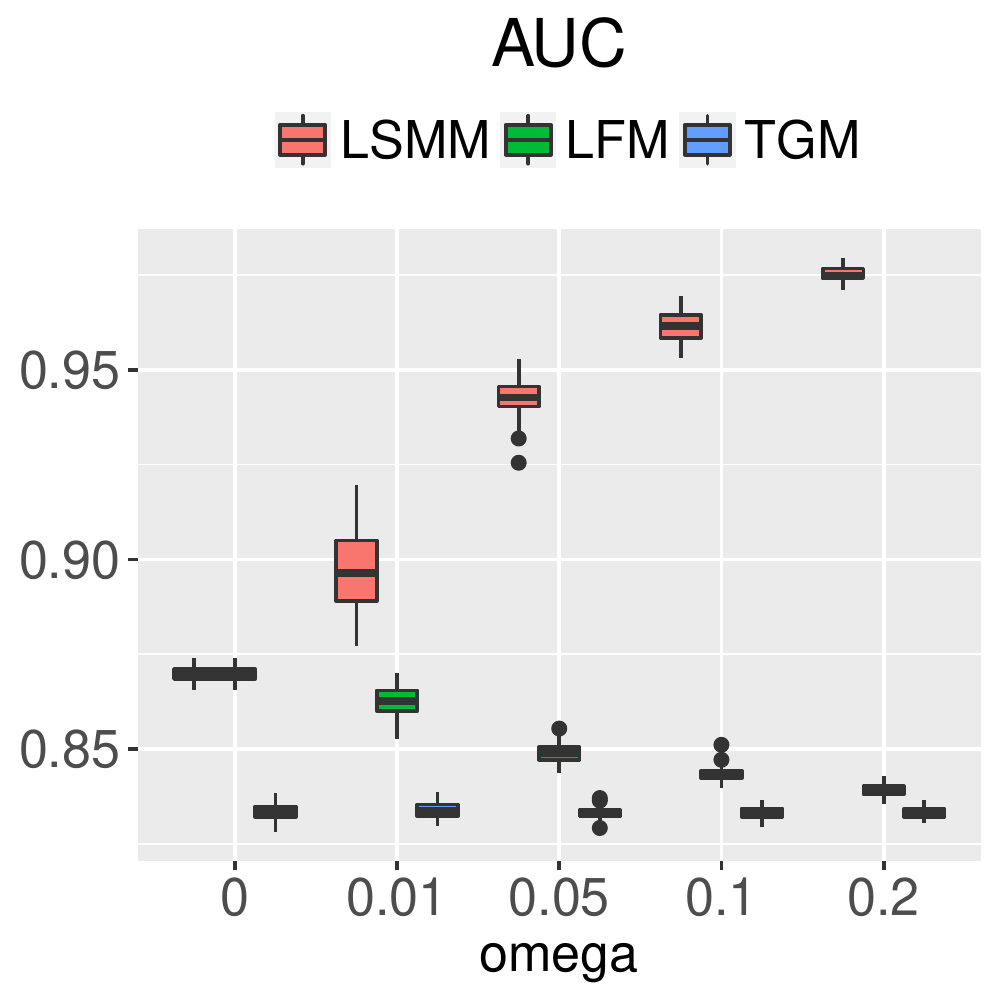}\includegraphics[scale=0.36]{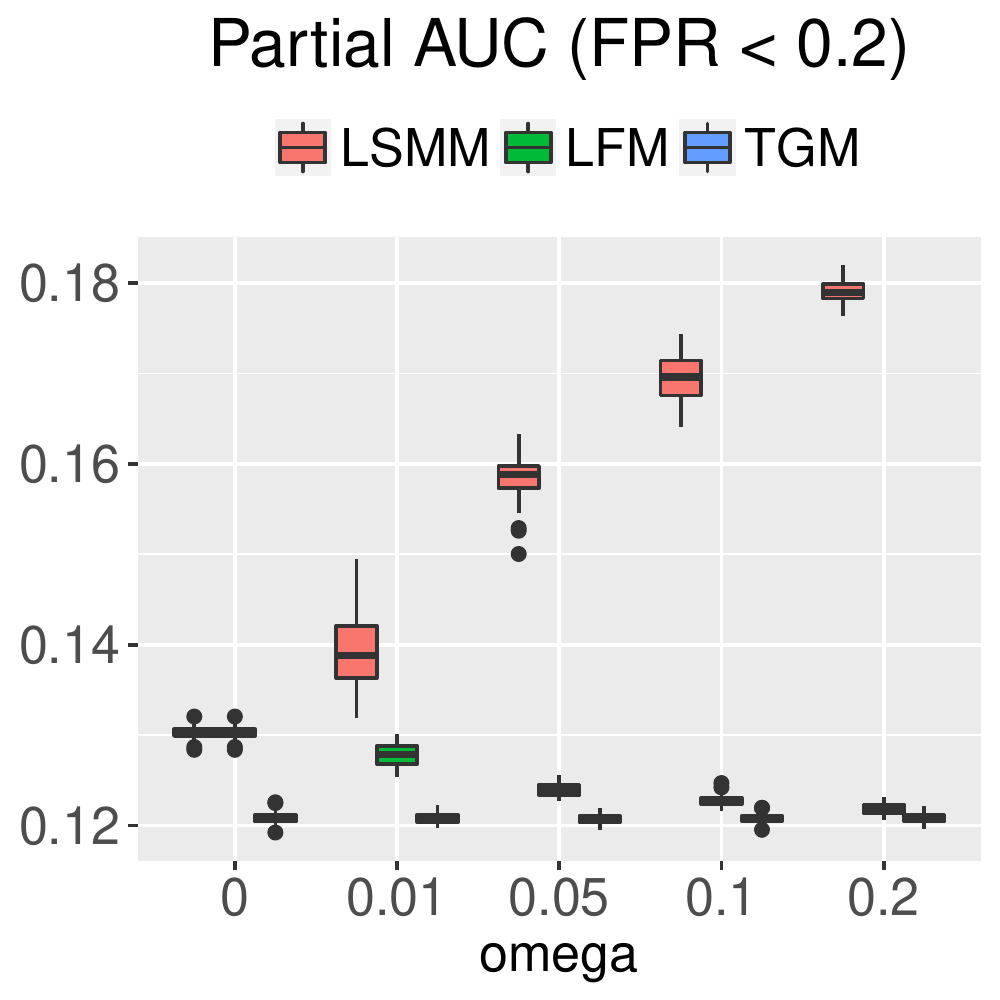}
		\par\end{centering}
	\caption{FDR, power, AUC and partial AUC of LSMM, LFM and TGM for identification
		of risk SNPs with $\alpha=0.2$ and $K=1000$. We controlled global
		FDR at 0.1 to evaluate empirical FDR and power. The results are summarized
		from 50 replications.}
\end{figure}

\begin{figure}[H]
	\begin{centering}
		\includegraphics[scale=0.36]{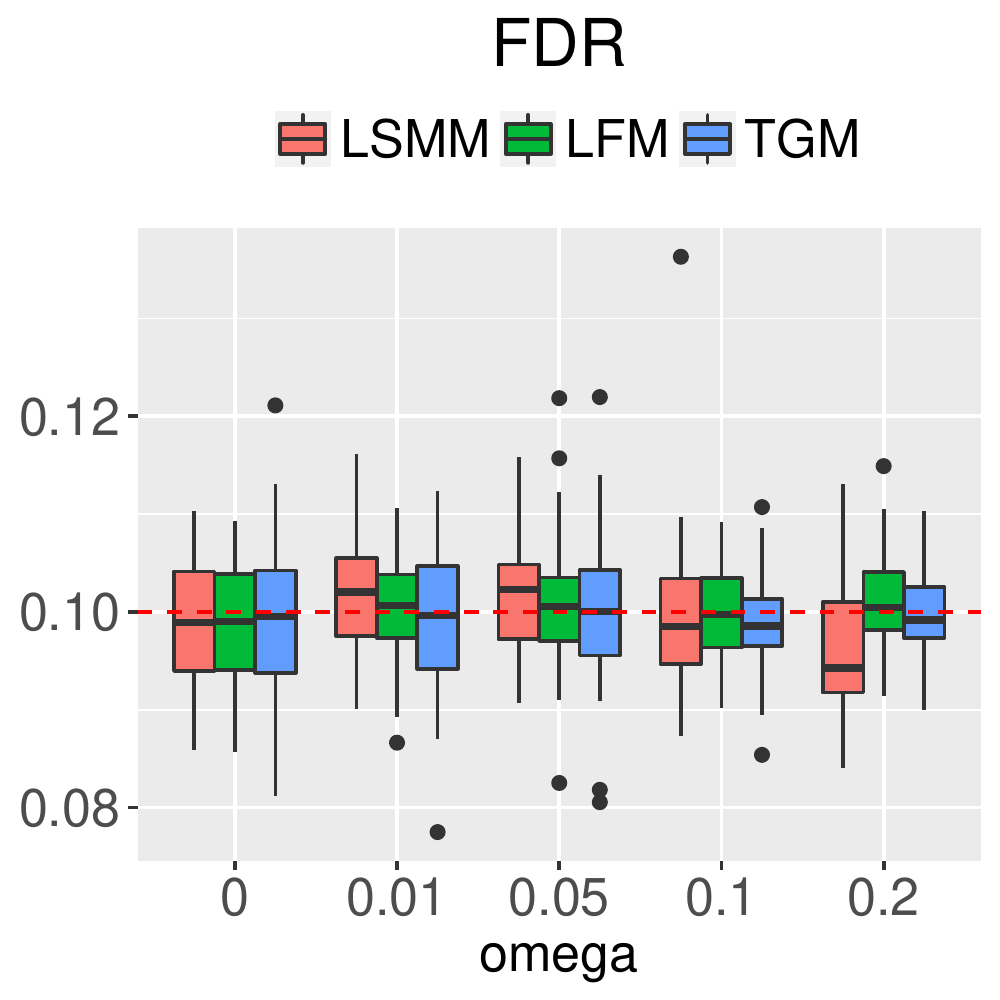}\includegraphics[scale=0.36]{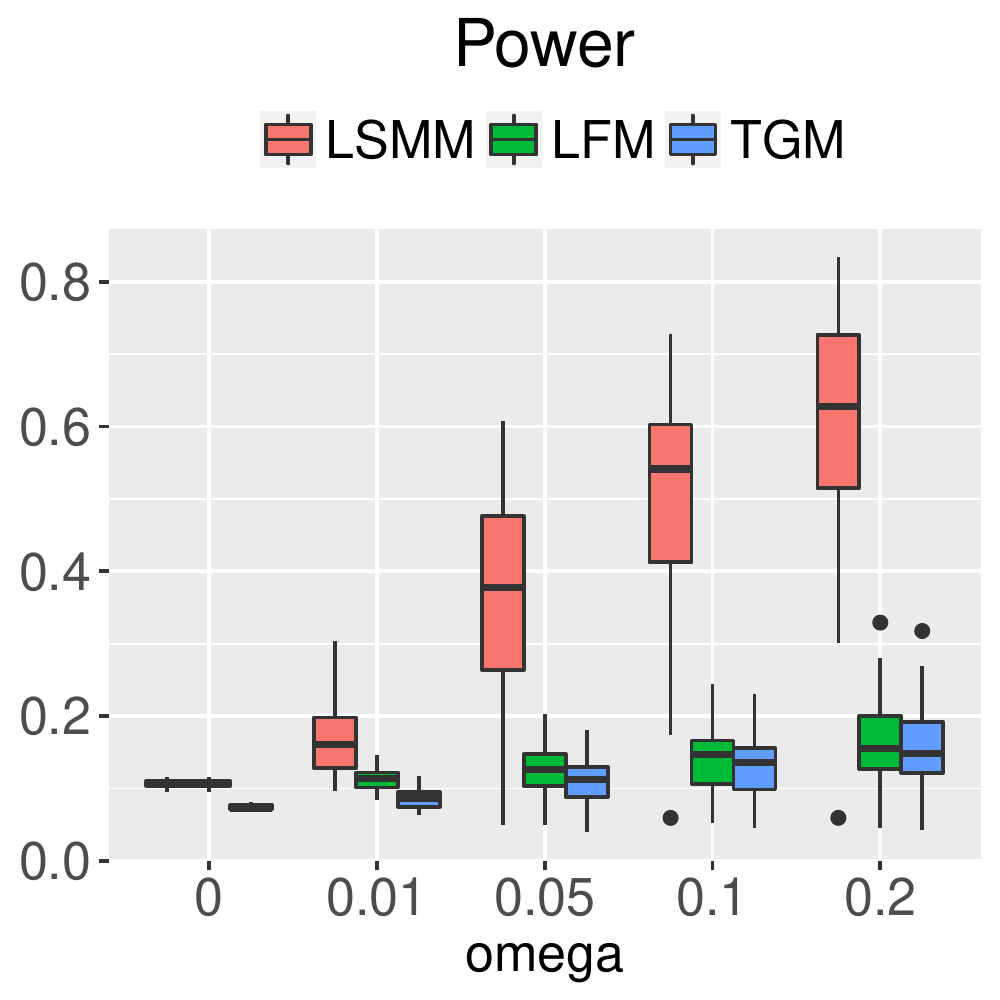}\includegraphics[scale=0.36]{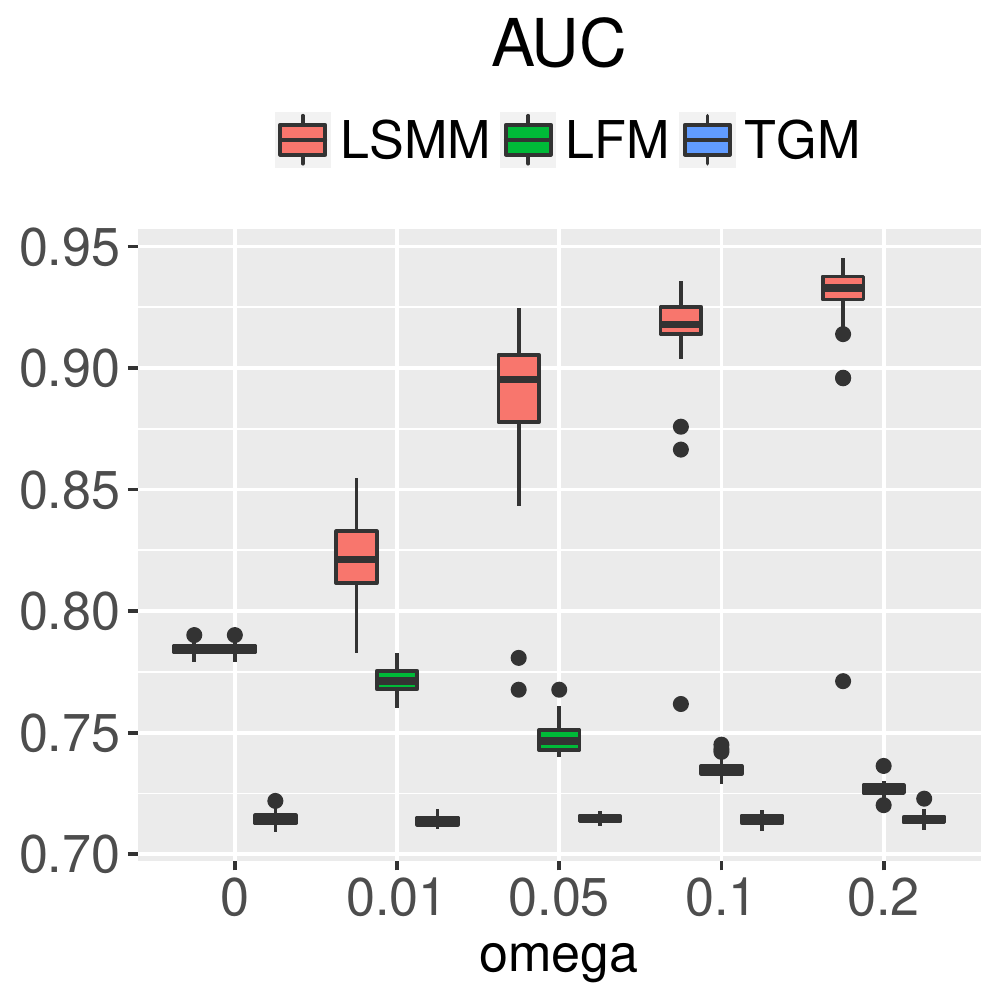}\includegraphics[scale=0.36]{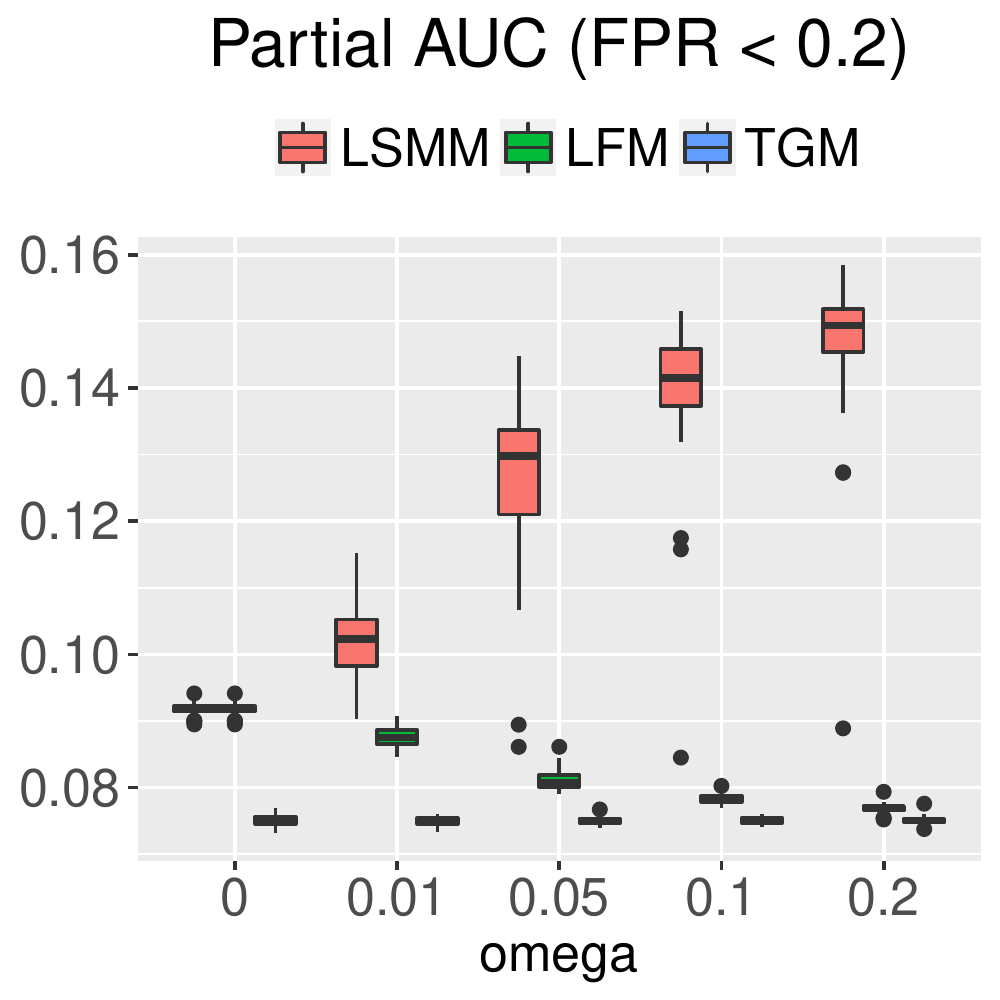}
		\par\end{centering}
	\caption{FDR, power, AUC and partial AUC of LSMM, LFM and TGM for identification
		of risk SNPs with $\alpha=0.4$ and $K=1000$. We controlled global
		FDR at 0.1 to evaluate empirical FDR and power. The results are summarized
		from 50 replications.}
\end{figure}

\begin{figure}[H]
	\begin{centering}
		\includegraphics[scale=0.36]{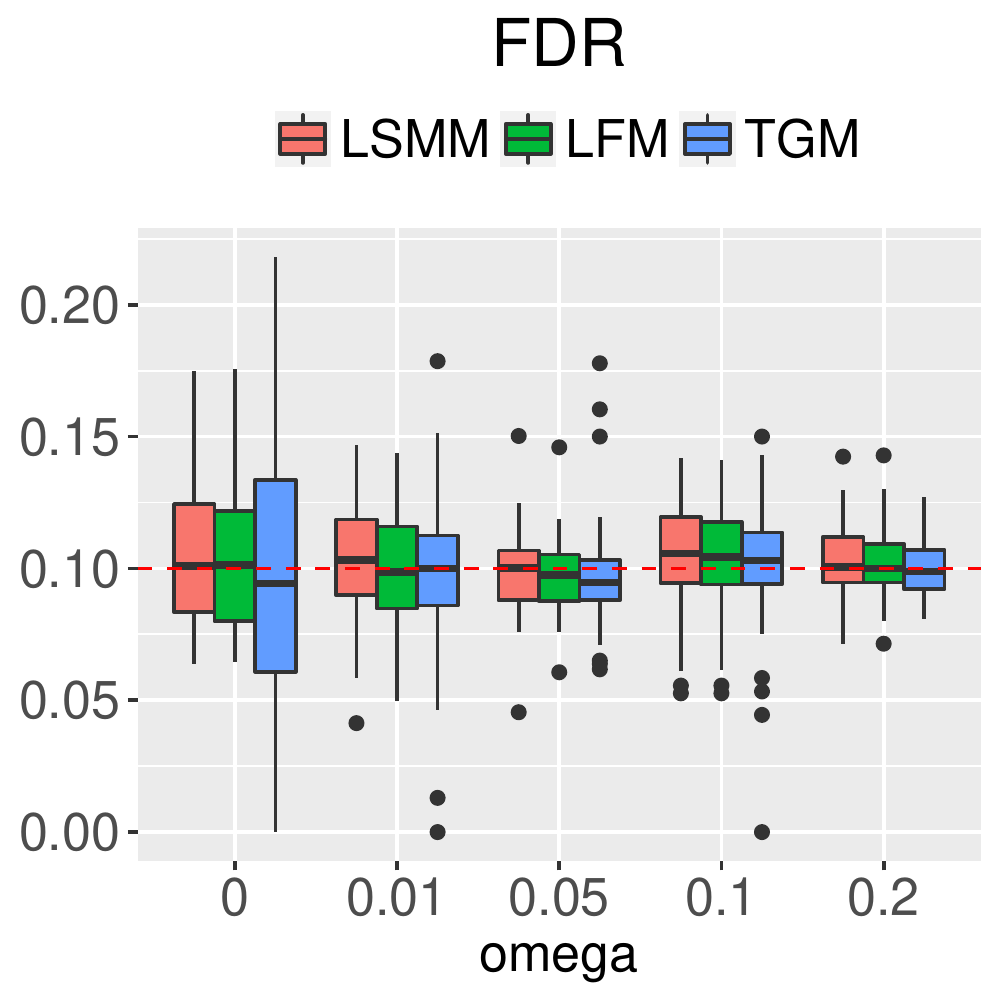}\includegraphics[scale=0.36]{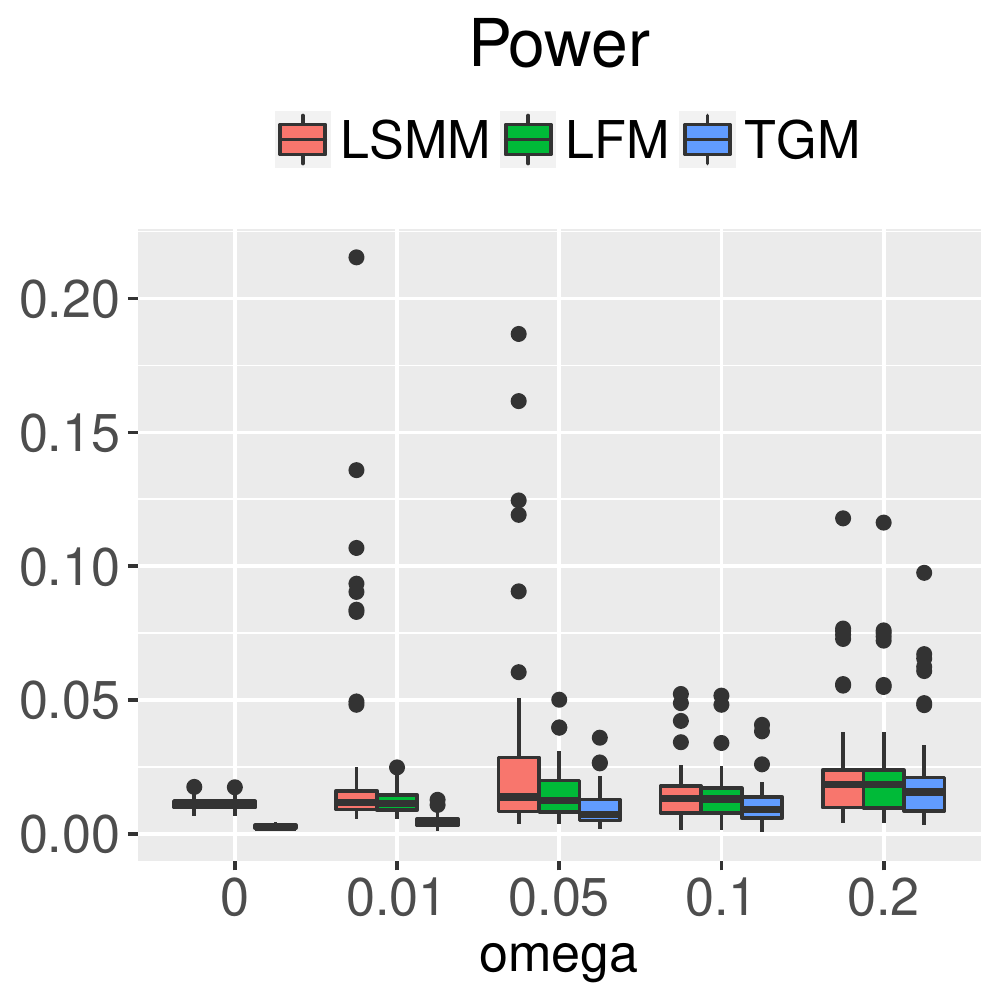}\includegraphics[scale=0.36]{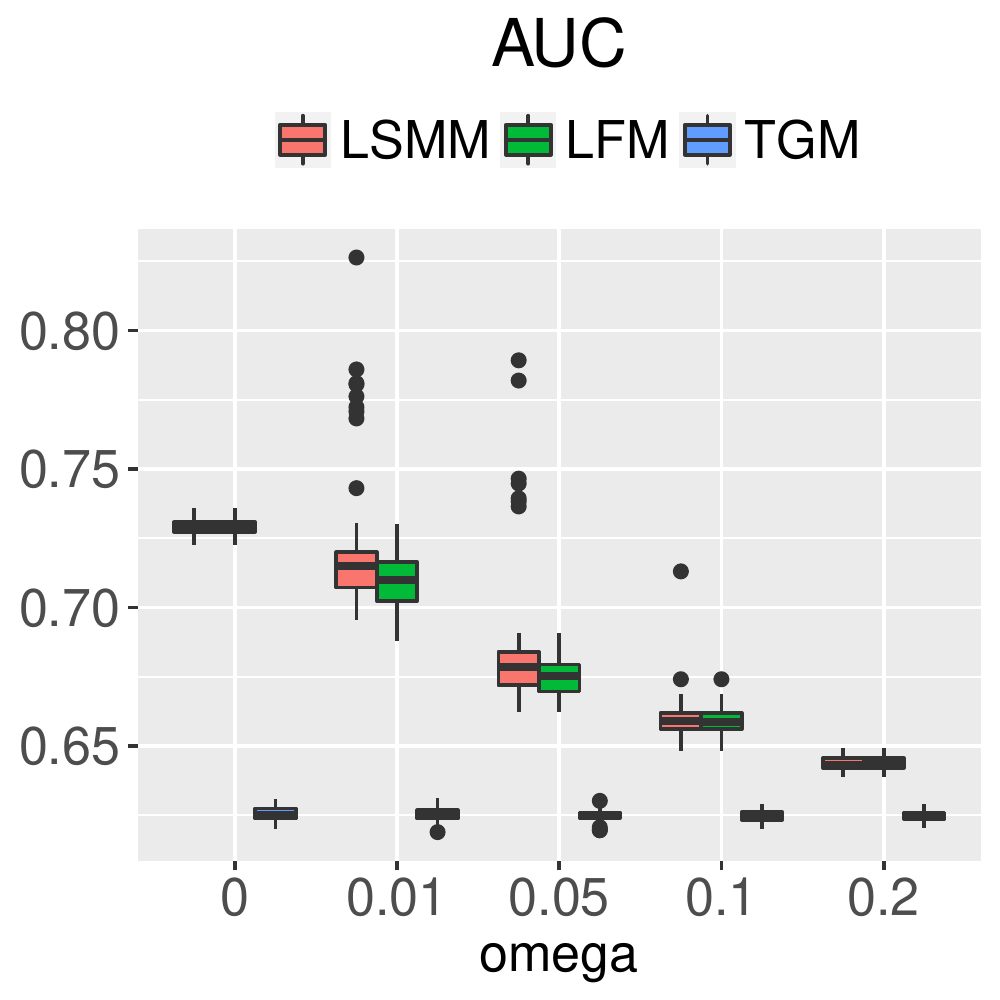}\includegraphics[scale=0.36]{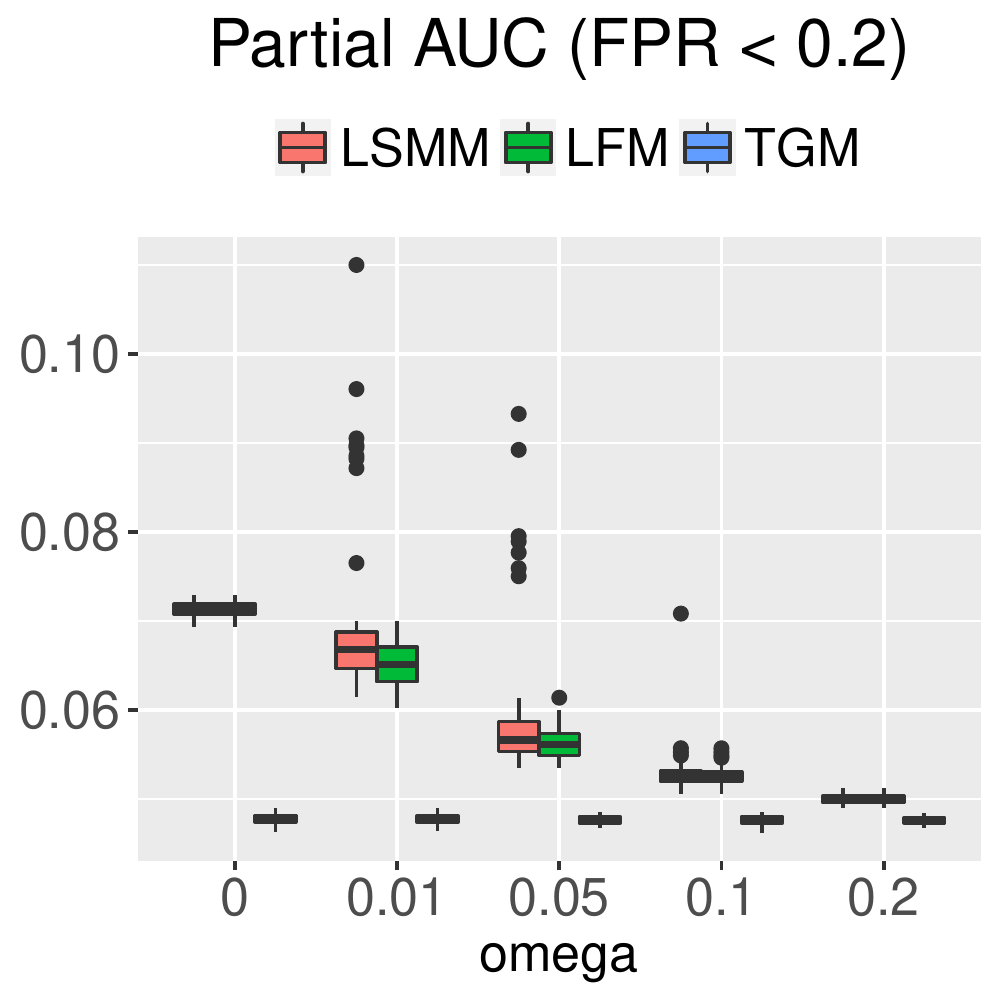}
		\par\end{centering}
	\caption{FDR, power, AUC and partial AUC of LSMM, LFM and TGM for identification
		of risk SNPs with $\alpha=0.6$ and $K=1000$. We controlled global
		FDR at 0.1 to evaluate empirical FDR and power. The results are summarized
		from 50 replications.}
\end{figure}

\subsection{Performance in identification of risk SNPs if treat all covariates
	as fixed effects}

\begin{figure}[H]
	\begin{centering}
		\includegraphics[scale=0.5]{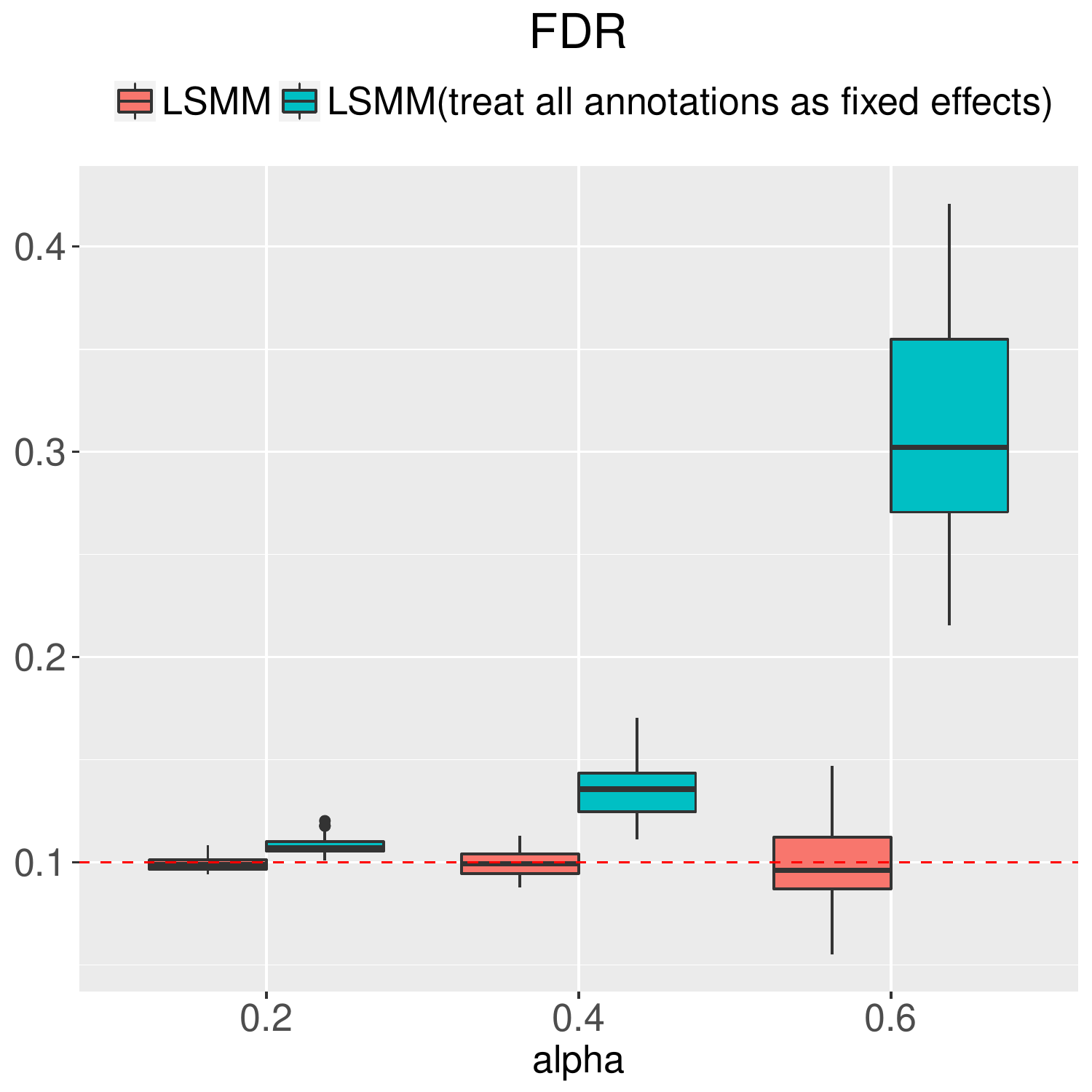}
		\par\end{centering}
	\caption{FDR of LSMM and LSMM (treat all covariates as fixed effects) for identification
		of risk SNPs with $K=500$. We controlled global FDR at 0.1 to evaluate
		empirical FDR and power. The results are summarized from 50 replications.}
	
\end{figure}

\subsection{Performance in identification of relevant annotations}

\begin{figure}[H]
	\begin{centering}
		\includegraphics[scale=0.36]{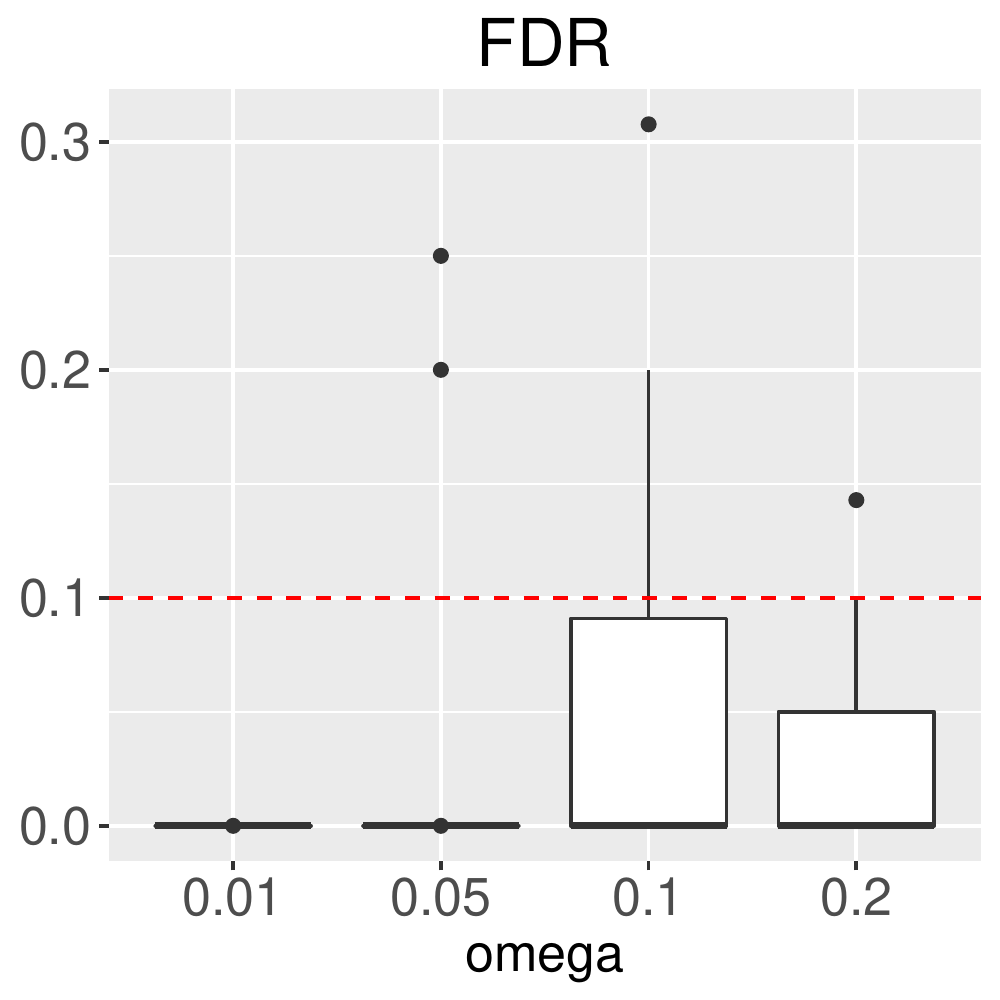}\includegraphics[scale=0.36]{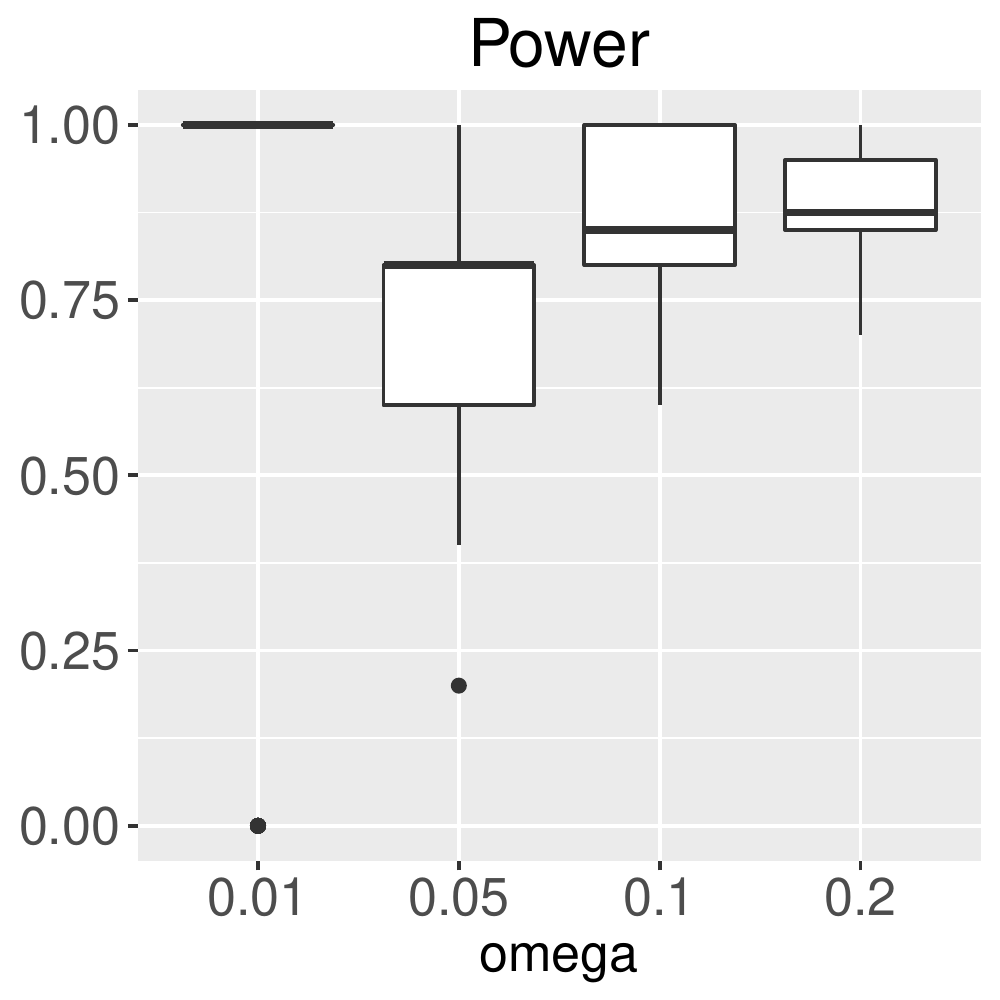}\includegraphics[scale=0.36]{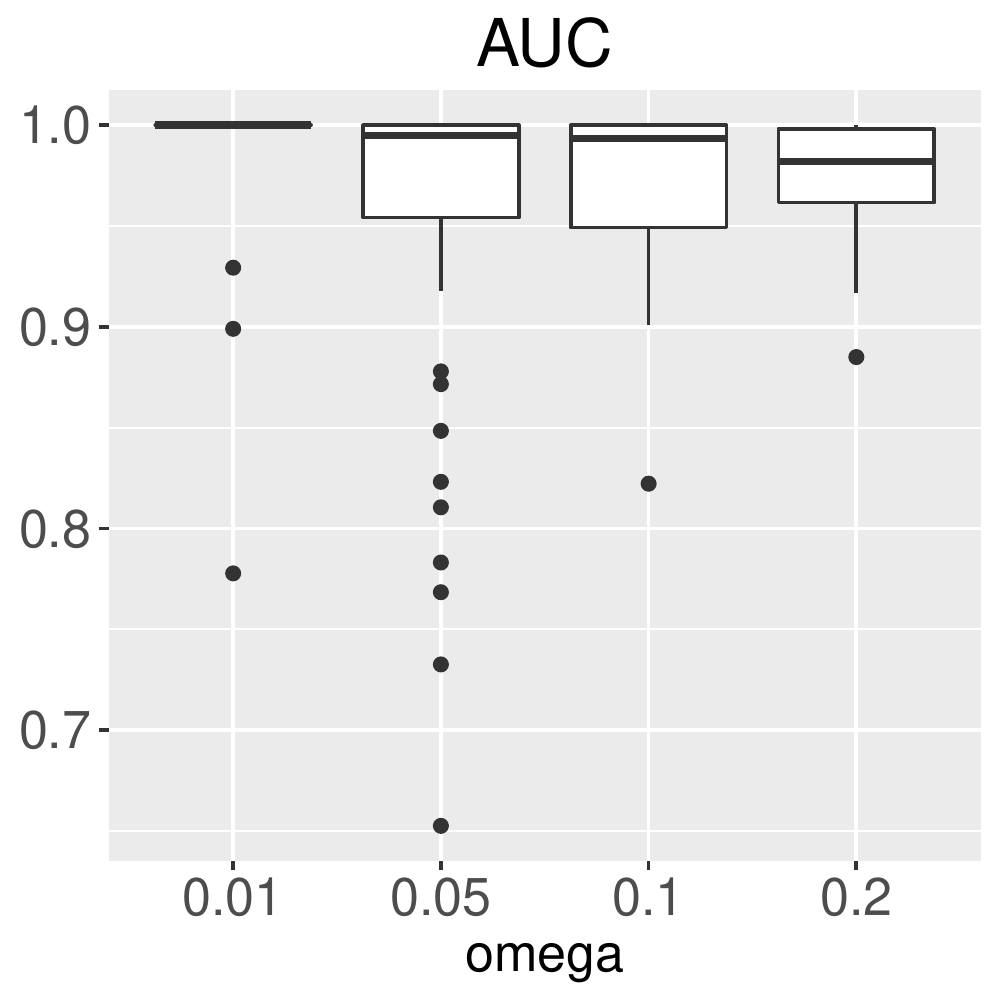}\includegraphics[scale=0.36]{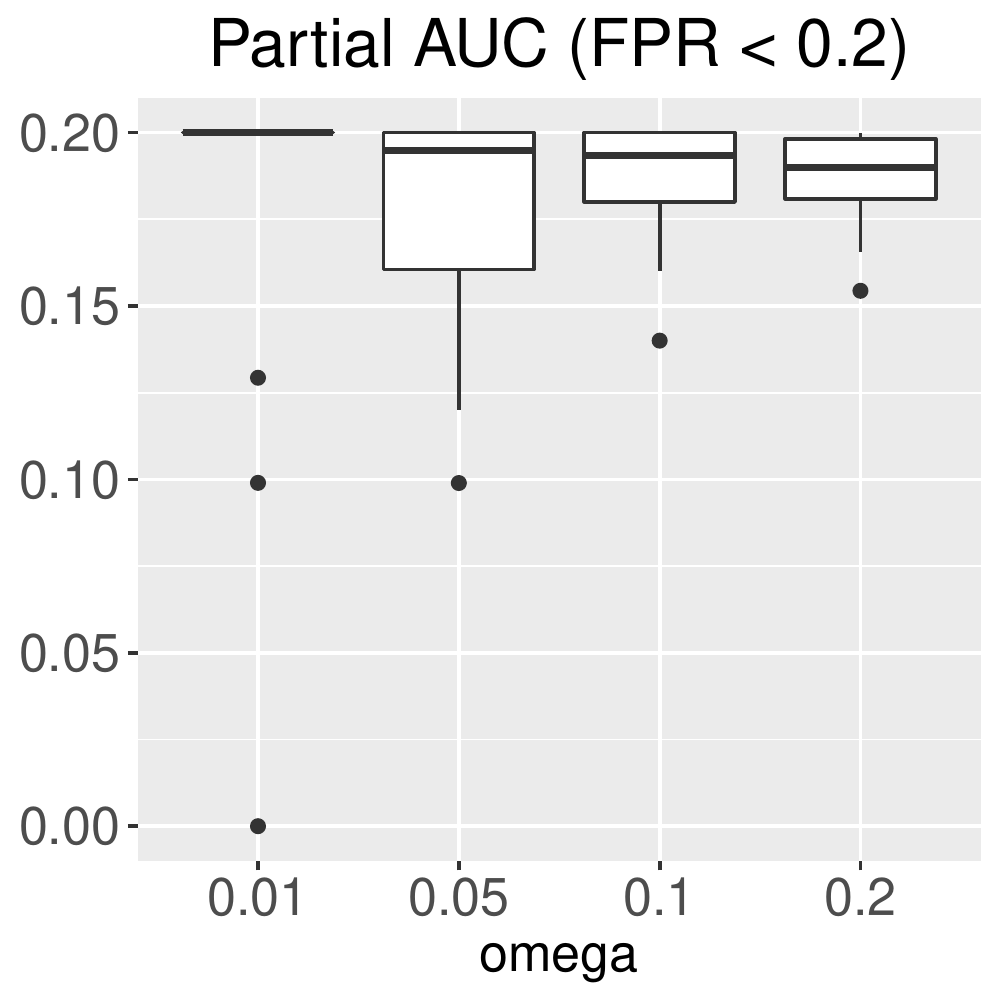}
		\par\end{centering}
	\caption{FDR, power, AUC and partial AUC of LSMM for detection of relevant
		annotations with $\alpha=0.2$ and $K=100$. We controlled global
		FDR at 0.1 to evaluate empirical FDR and power. The results are summarized
		from 50 replications.}
	
\end{figure}

\begin{figure}[H]
	\begin{centering}
		\includegraphics[scale=0.36]{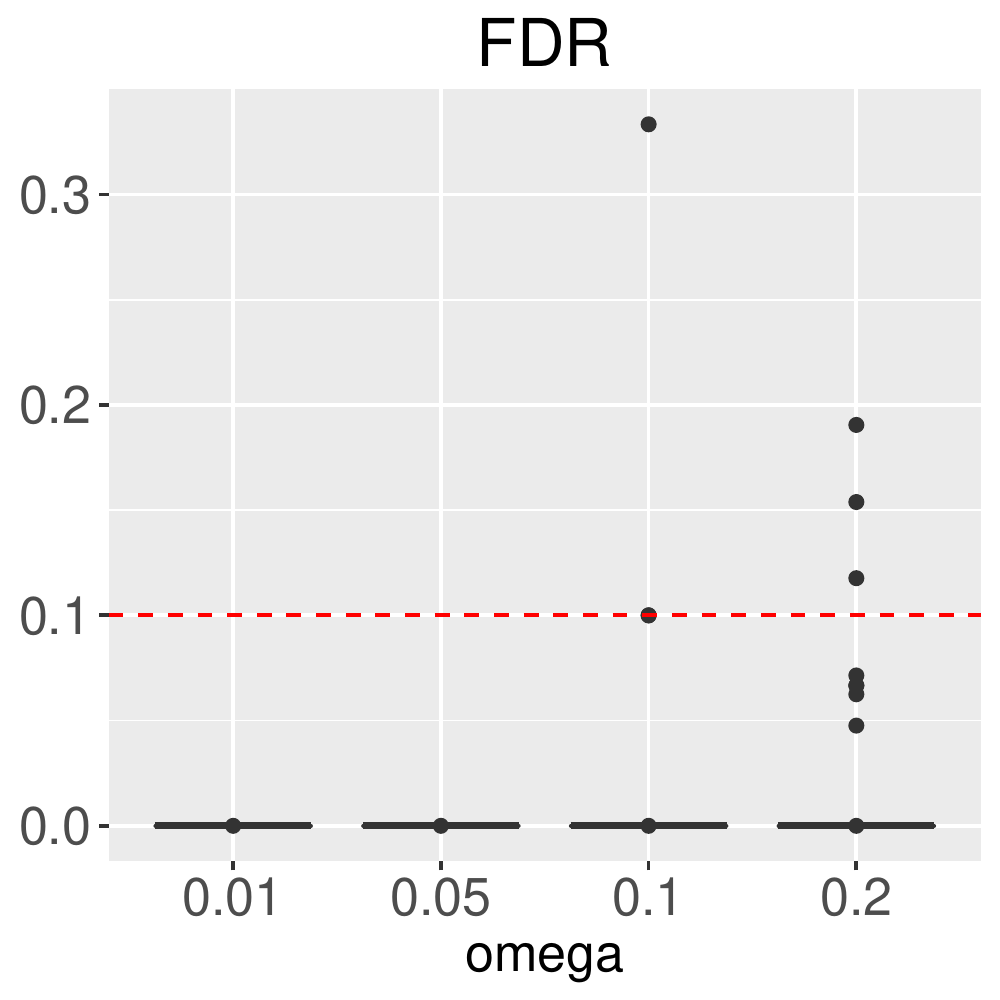}\includegraphics[scale=0.36]{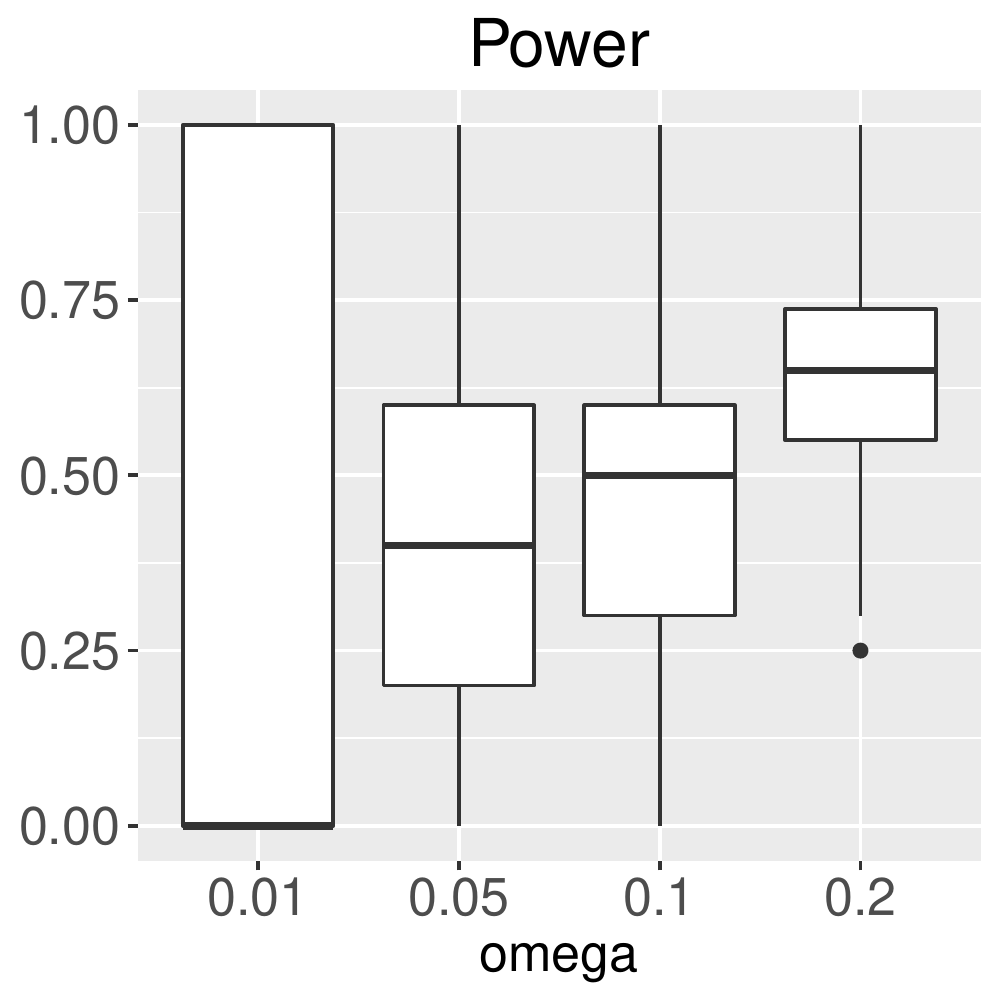}\includegraphics[scale=0.36]{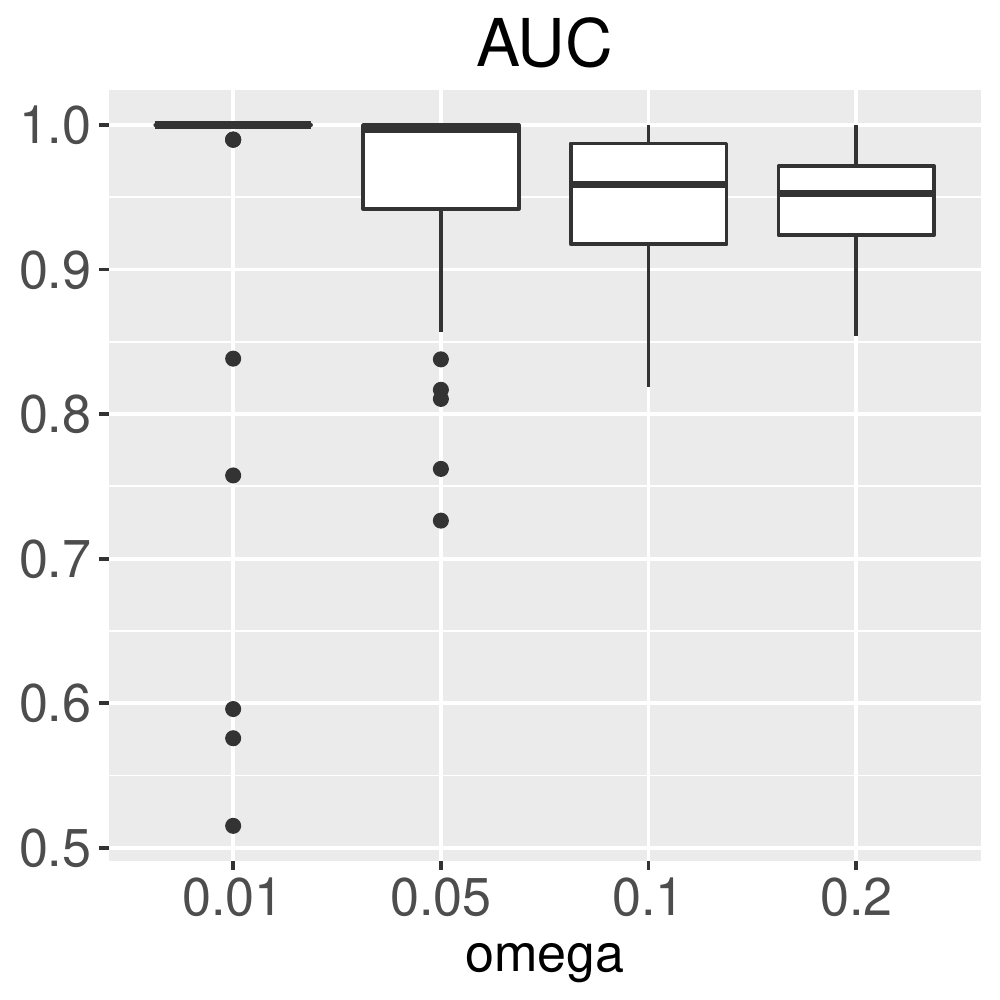}\includegraphics[scale=0.36]{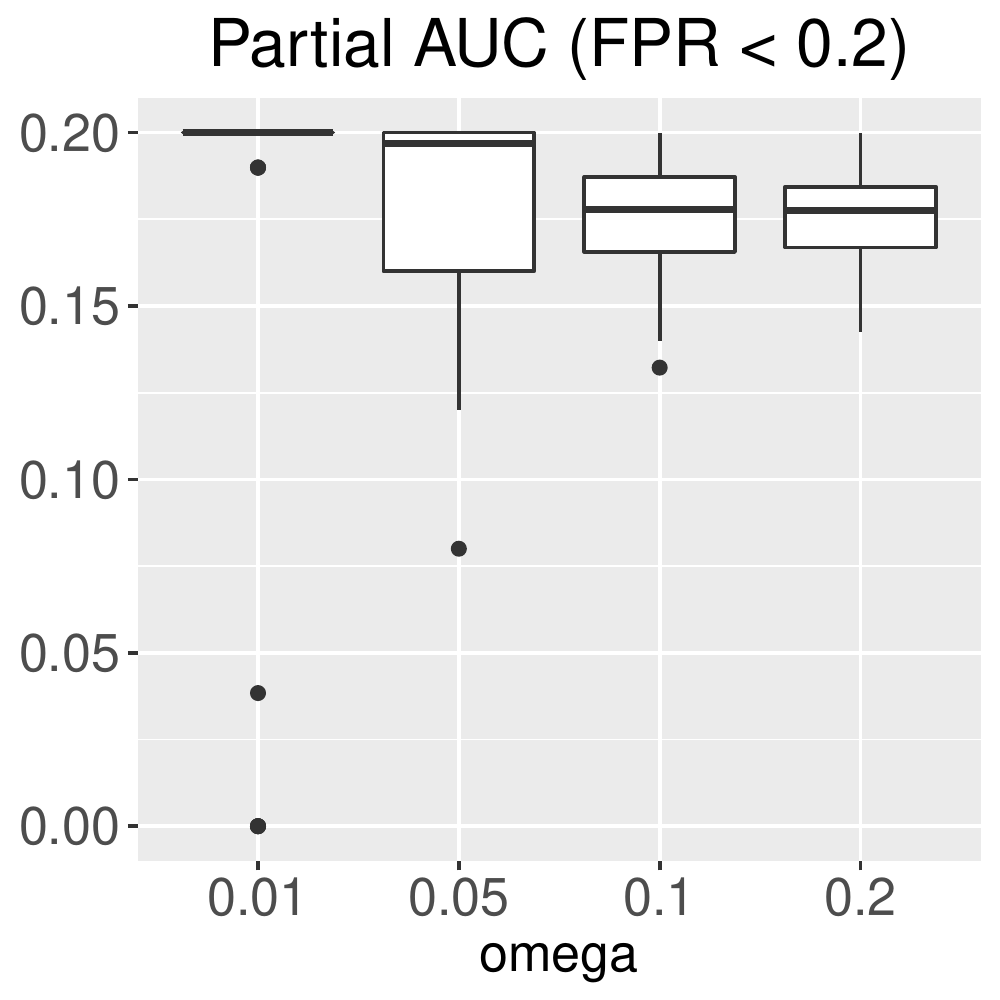}
		\par\end{centering}
	\caption{FDR, power, AUC and partial AUC of LSMM for detection of relevant
		annotations with $\alpha=0.4$ and $K=100$. We controlled global
		FDR at 0.1 to evaluate empirical FDR and power. The results are summarized
		from 50 replications.}
\end{figure}

\begin{figure}[H]
	\begin{centering}
		\includegraphics[scale=0.36]{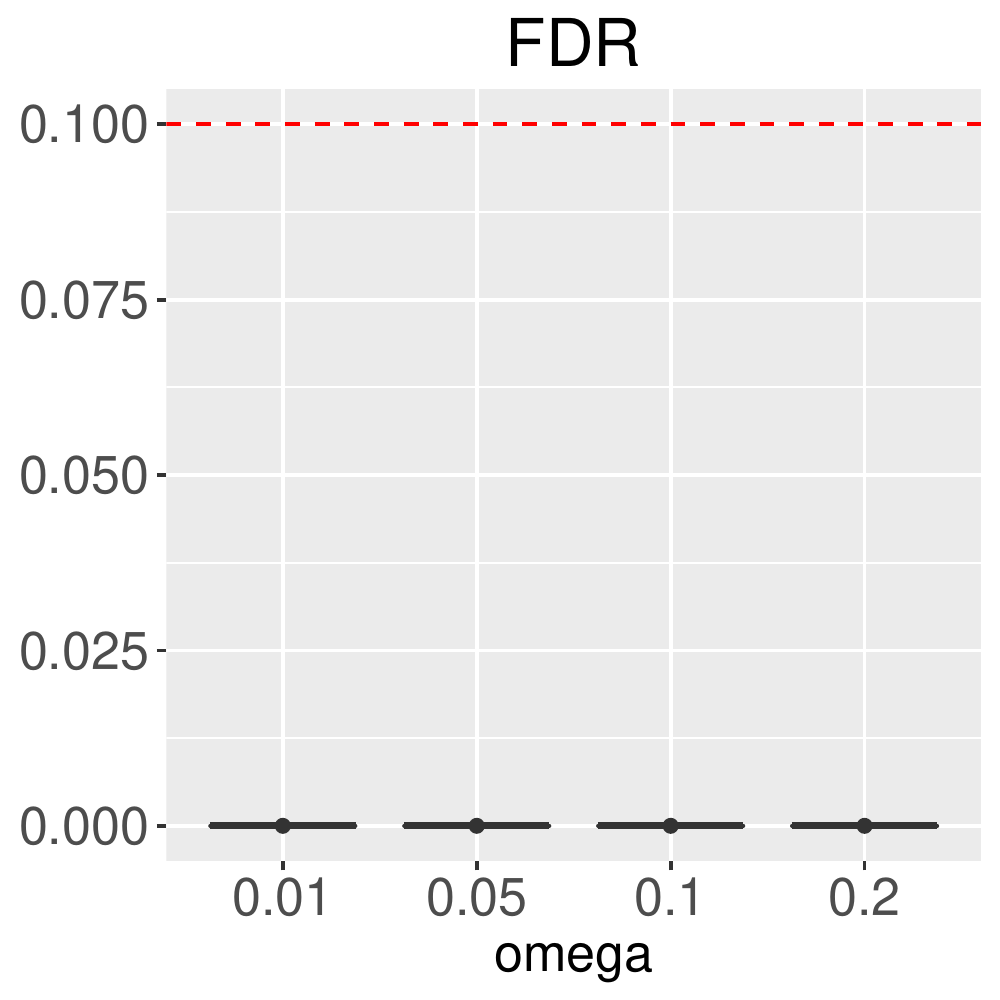}\includegraphics[scale=0.36]{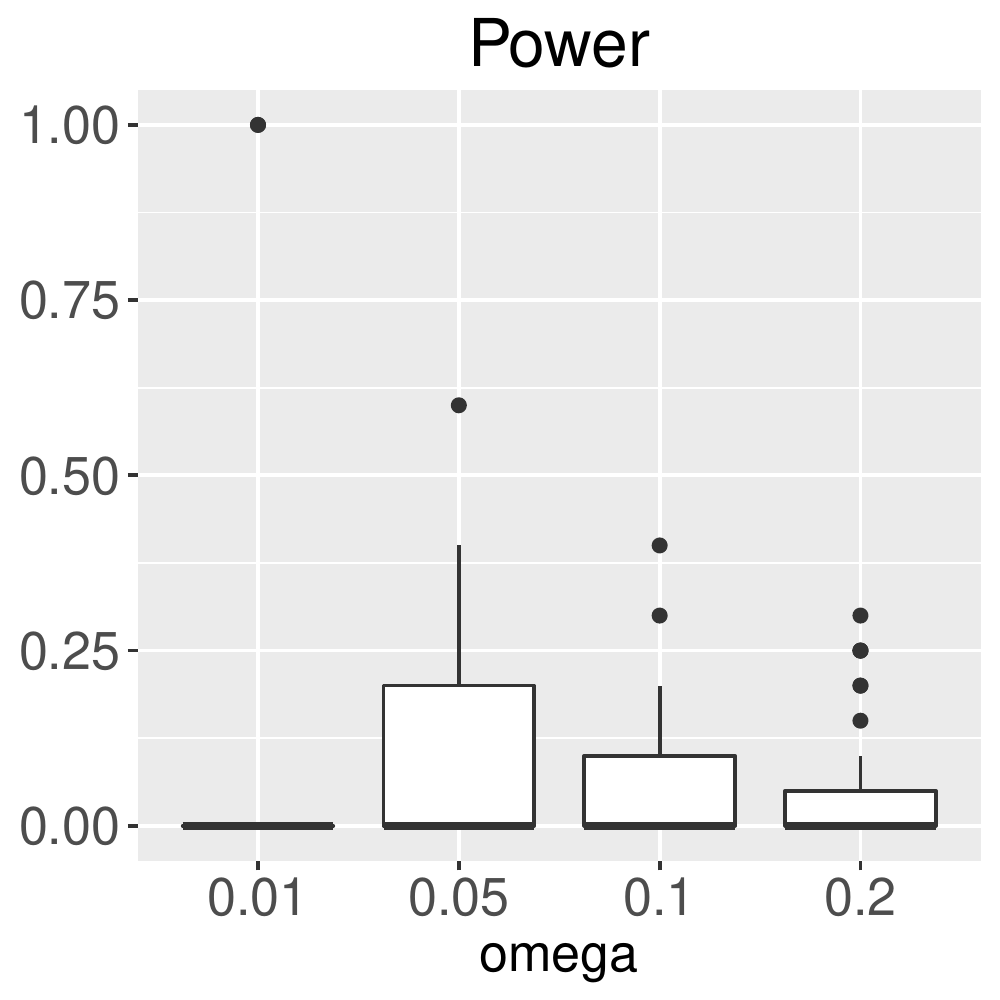}\includegraphics[scale=0.36]{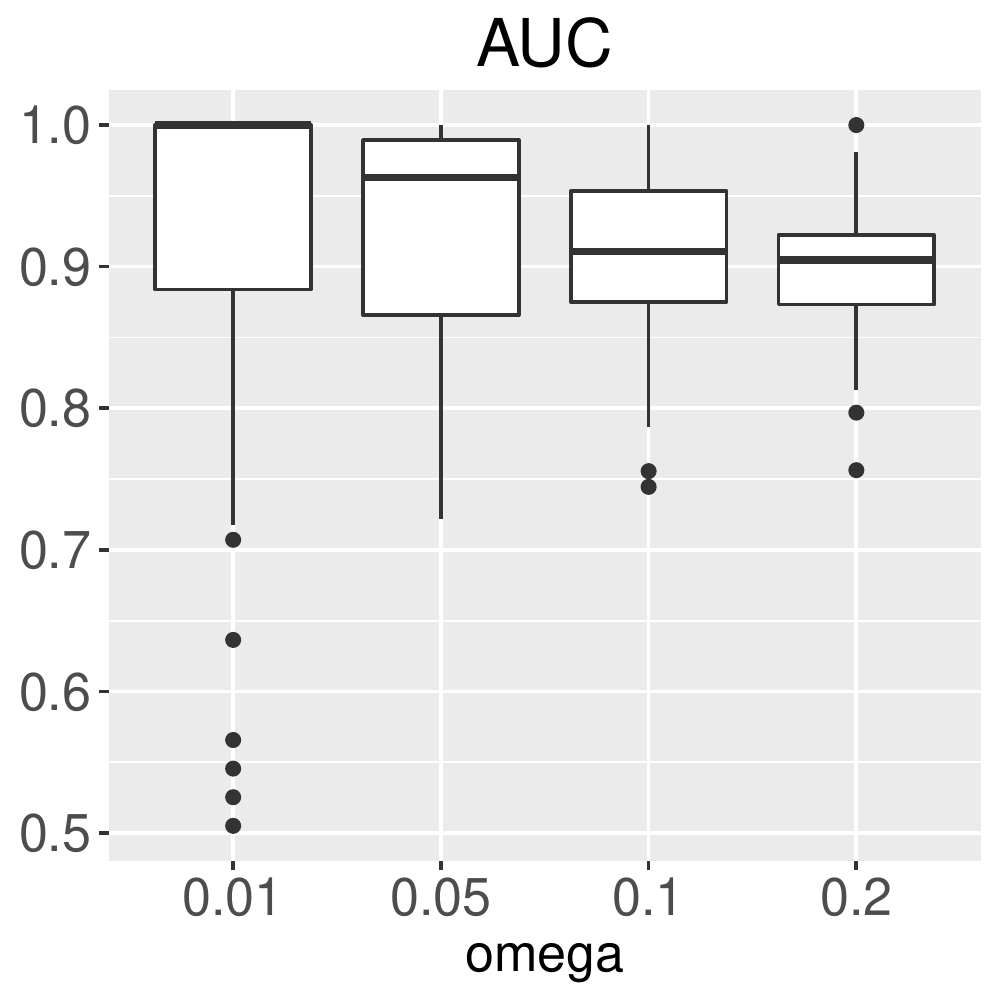}\includegraphics[scale=0.36]{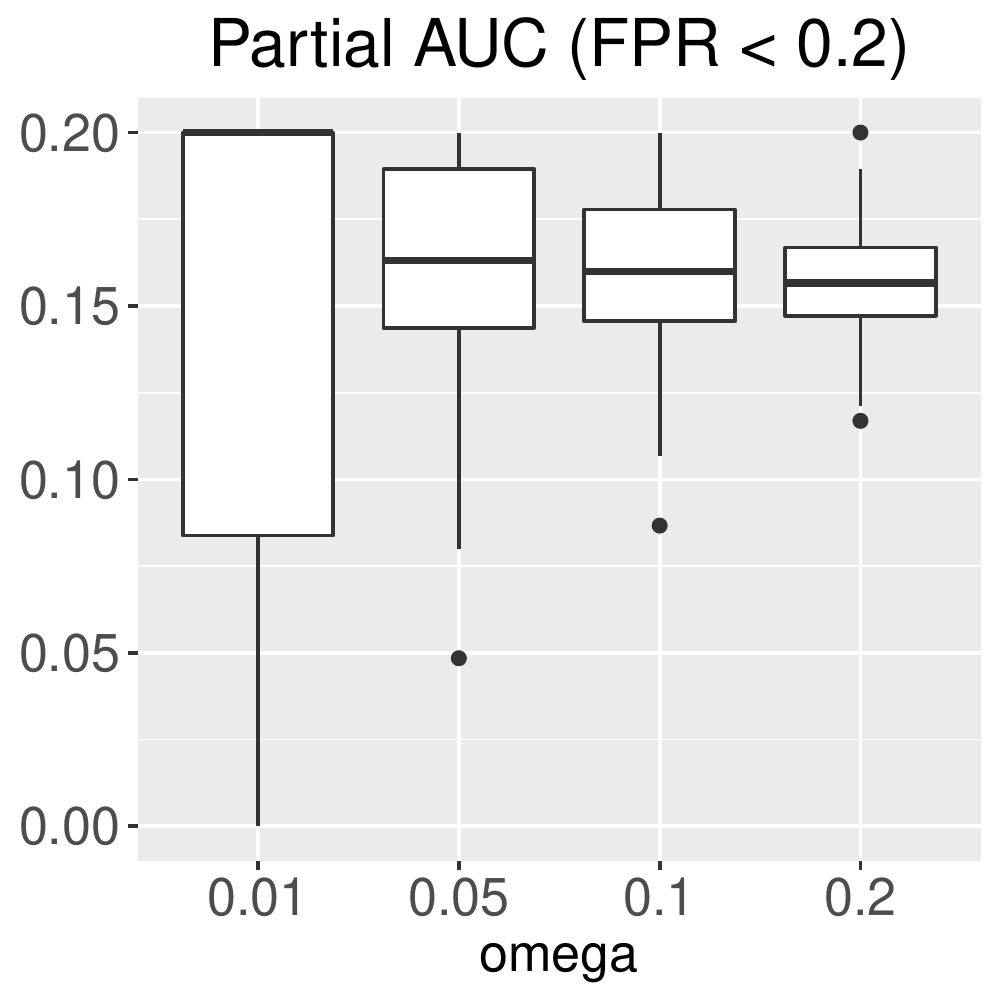}
		\par\end{centering}
	\caption{FDR, power, AUC and partial AUC of LSMM for detection of relevant
		annotations with $\alpha=0.6$ and $K=100$. We controlled global
		FDR at 0.1 to evaluate empirical FDR and power. The results are summarized
		from 50 replications.}
\end{figure}

\begin{figure}[H]
	\begin{centering}
		\includegraphics[scale=0.36]{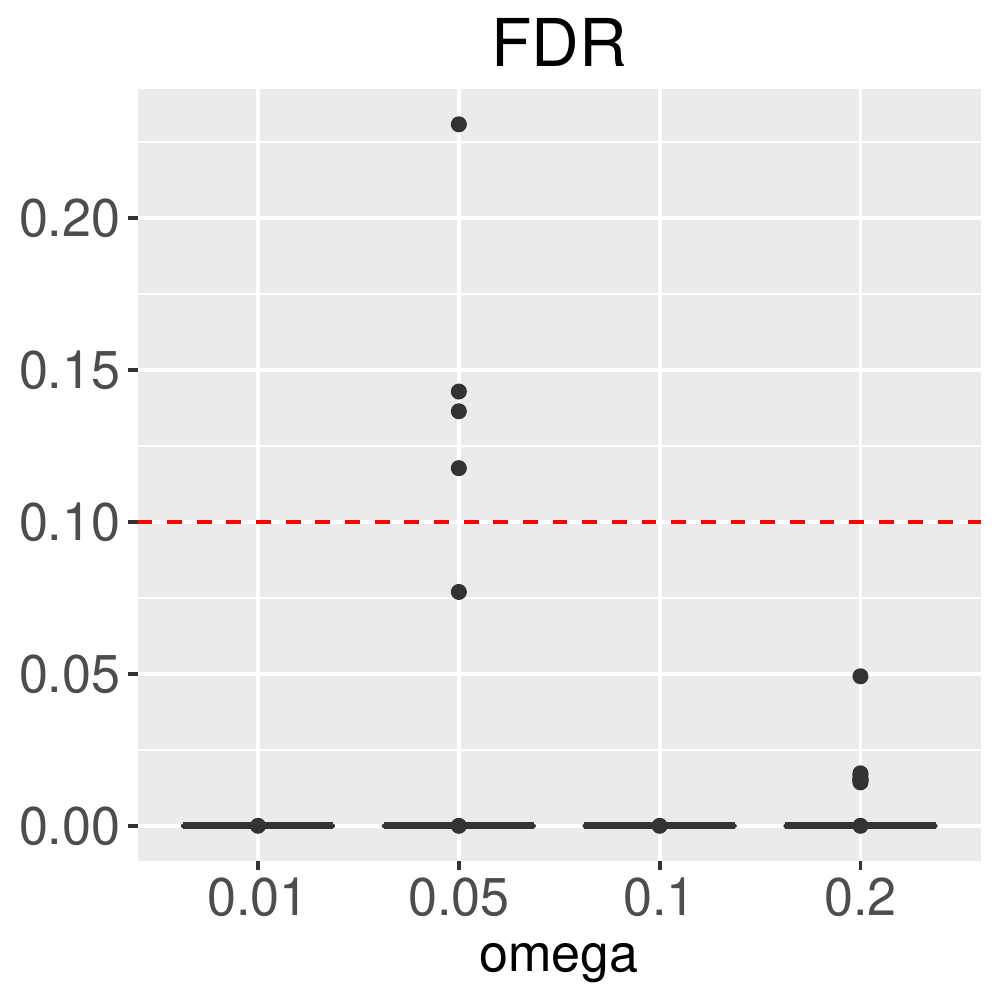}\includegraphics[scale=0.36]{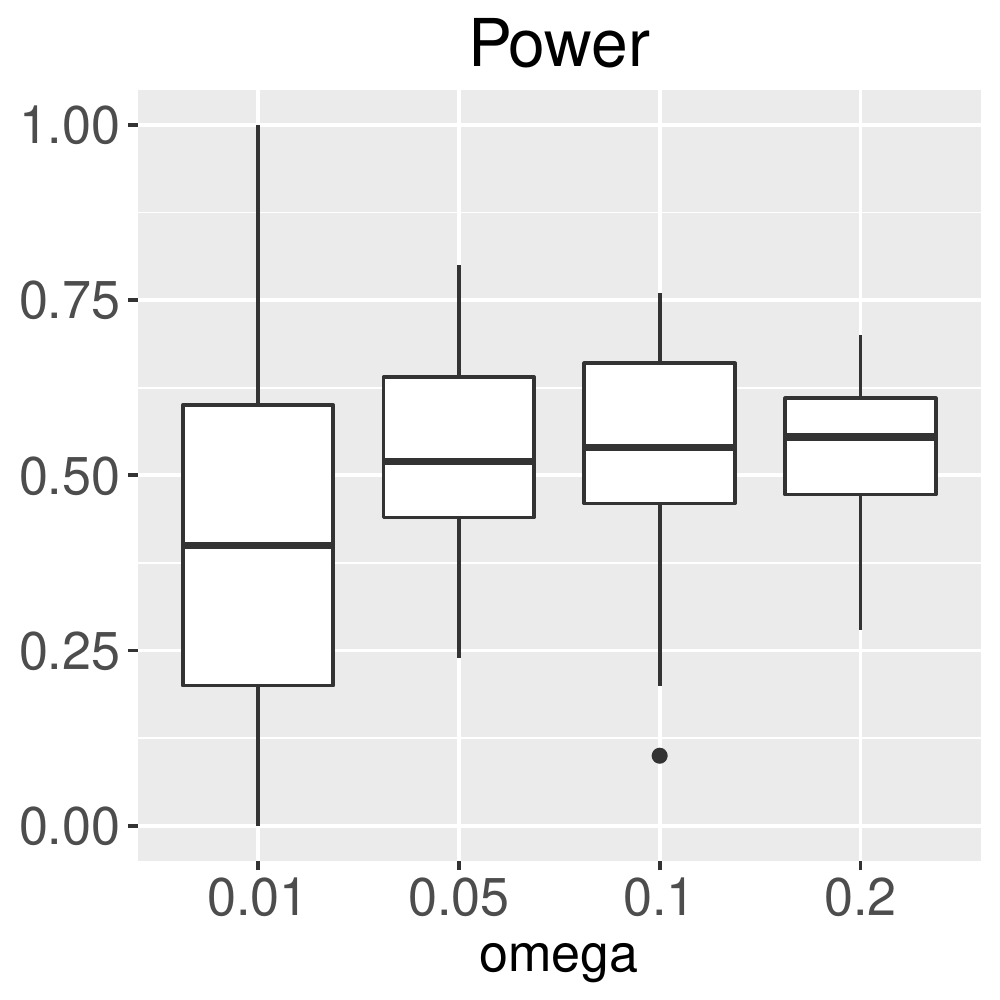}\includegraphics[scale=0.36]{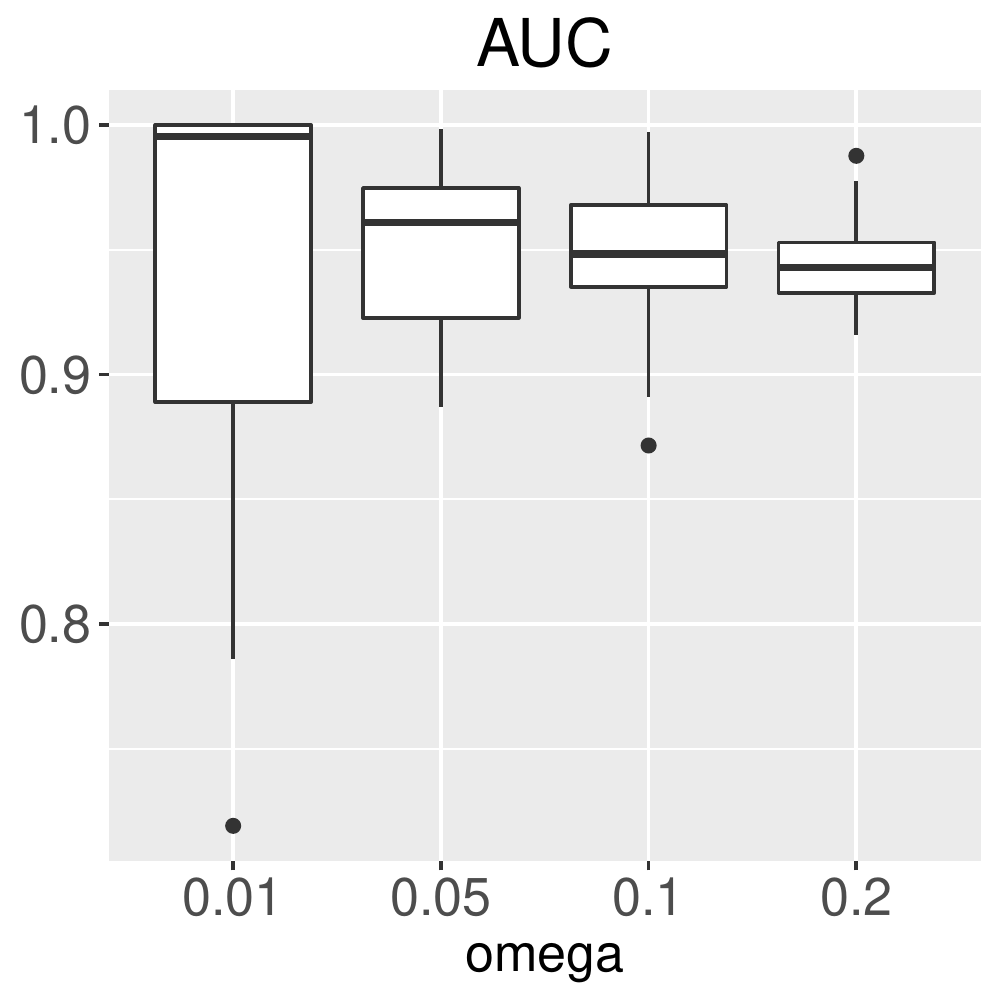}\includegraphics[scale=0.36]{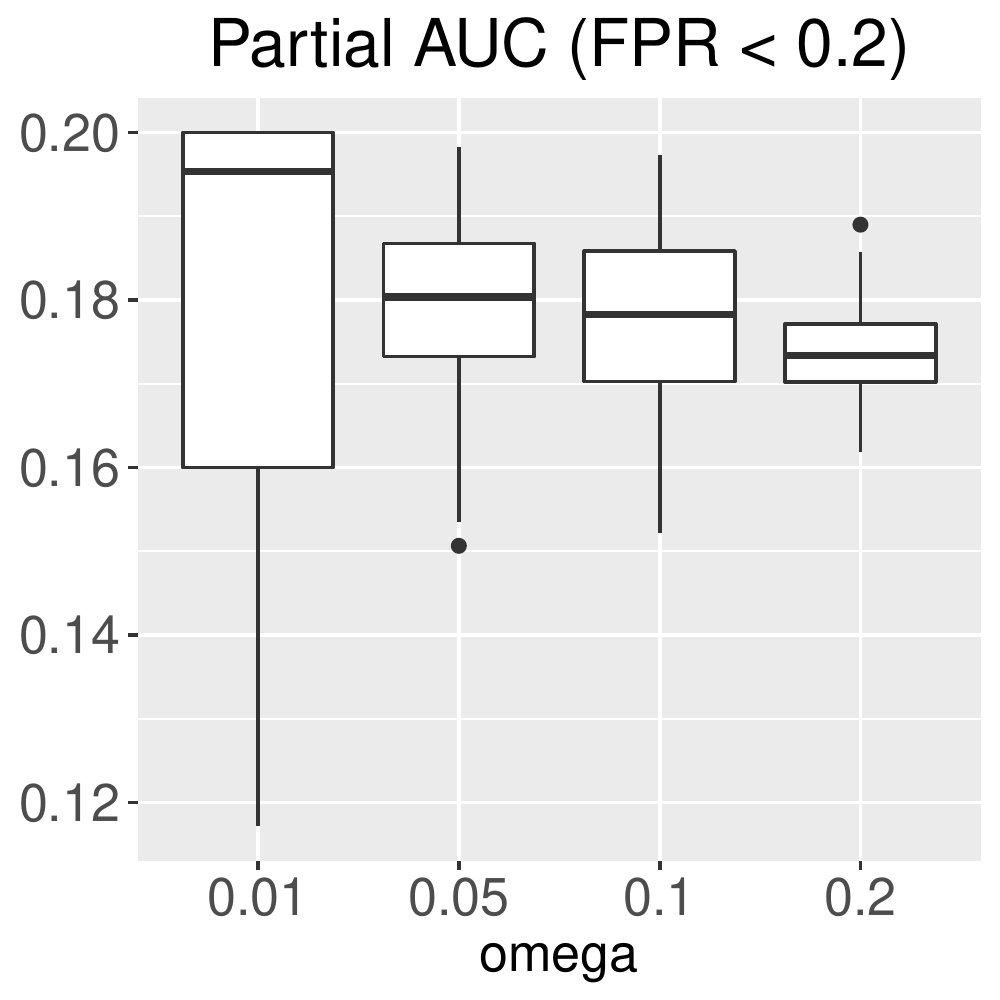}
		\par\end{centering}
	\caption{FDR, power, AUC and partial AUC of LSMM for detection of relevant
		annotations with $\alpha=0.4$ and $K=500$. We controlled global
		FDR at 0.1 to evaluate empirical FDR and power. The results are summarized
		from 50 replications.}
\end{figure}

\begin{figure}[H]
	\begin{centering}
		\includegraphics[scale=0.36]{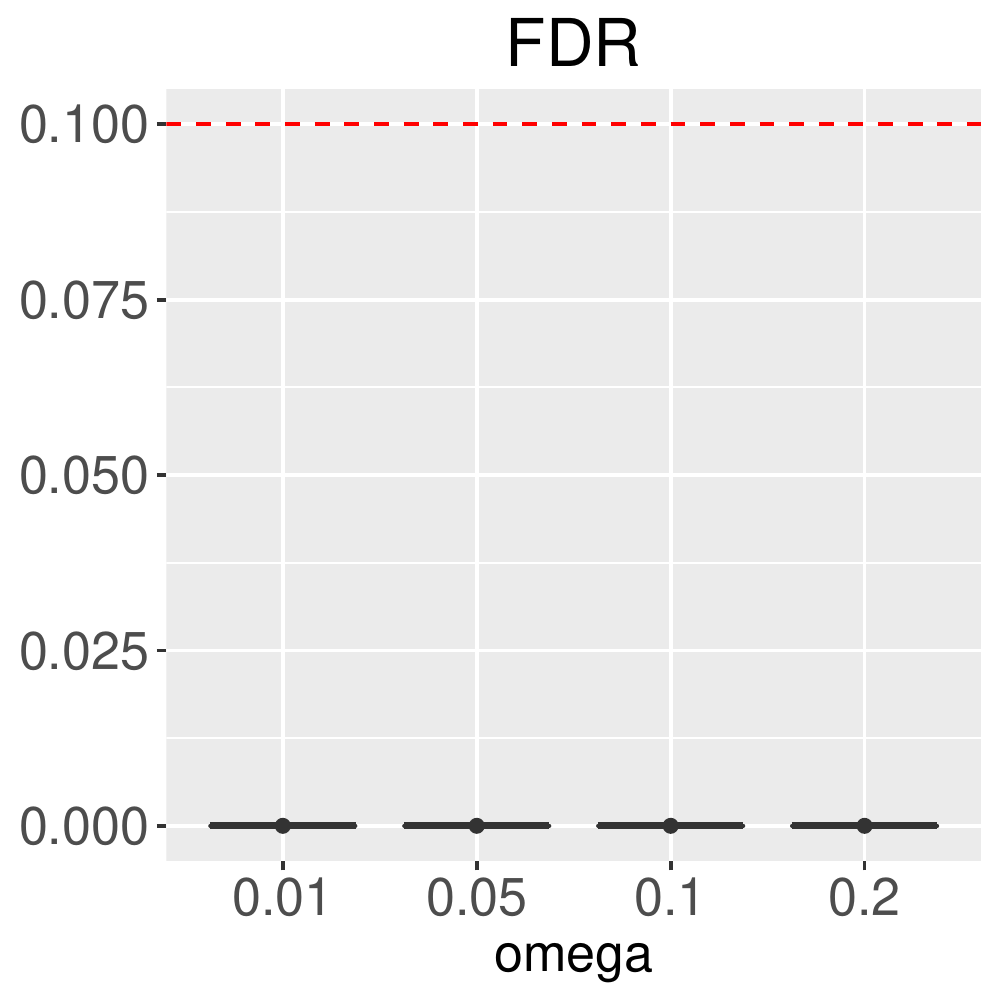}\includegraphics[scale=0.36]{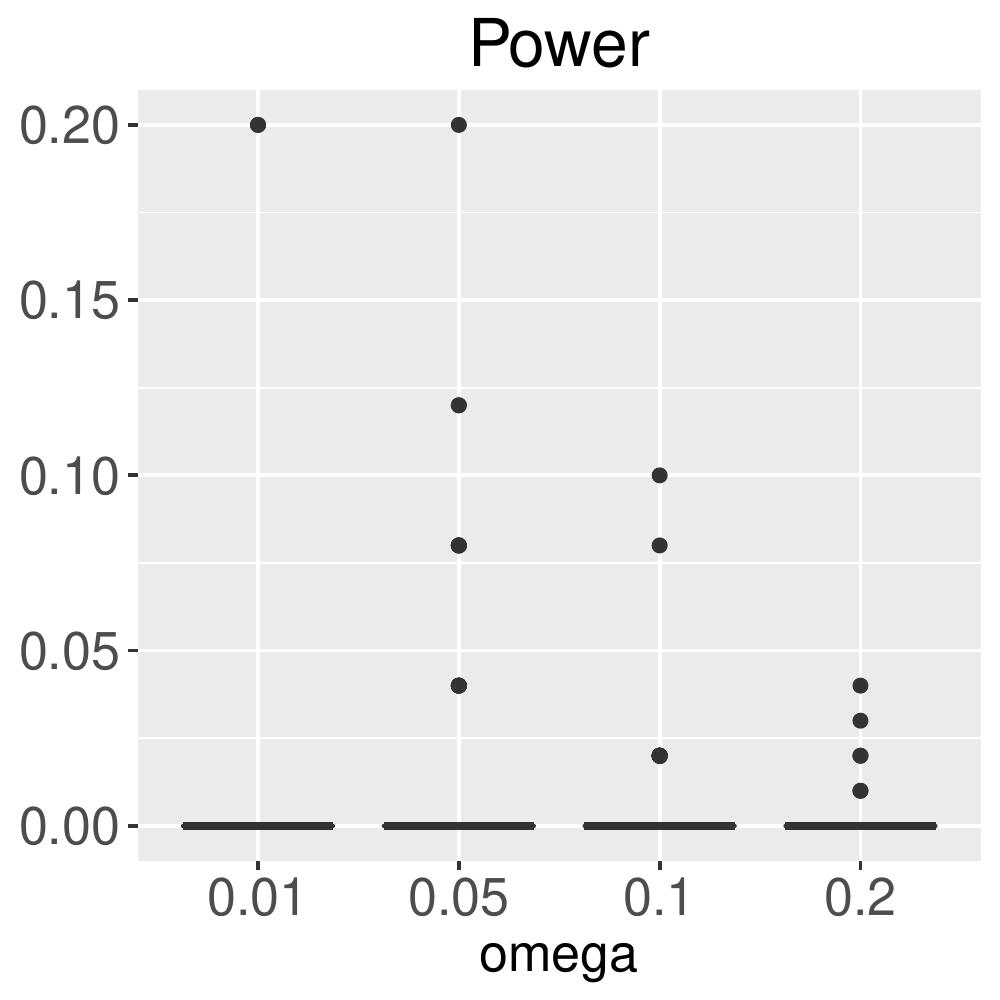}\includegraphics[scale=0.36]{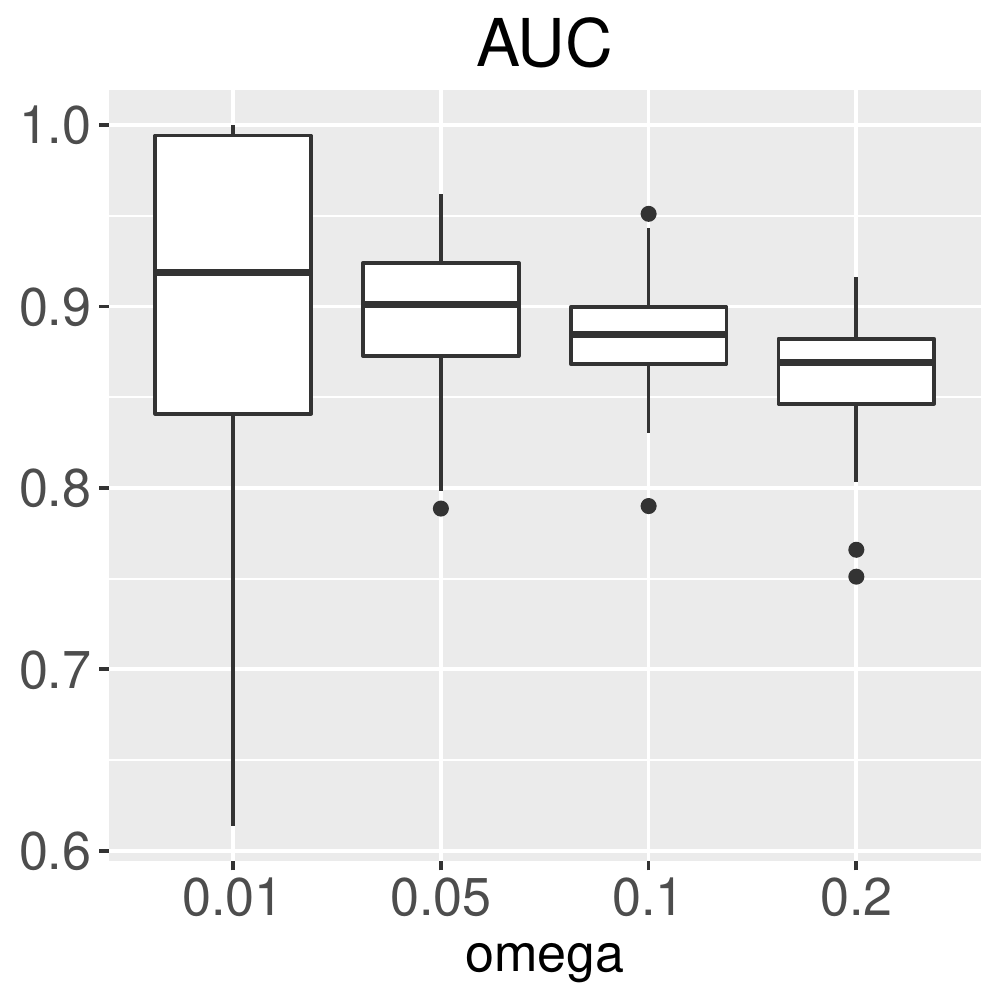}\includegraphics[scale=0.36]{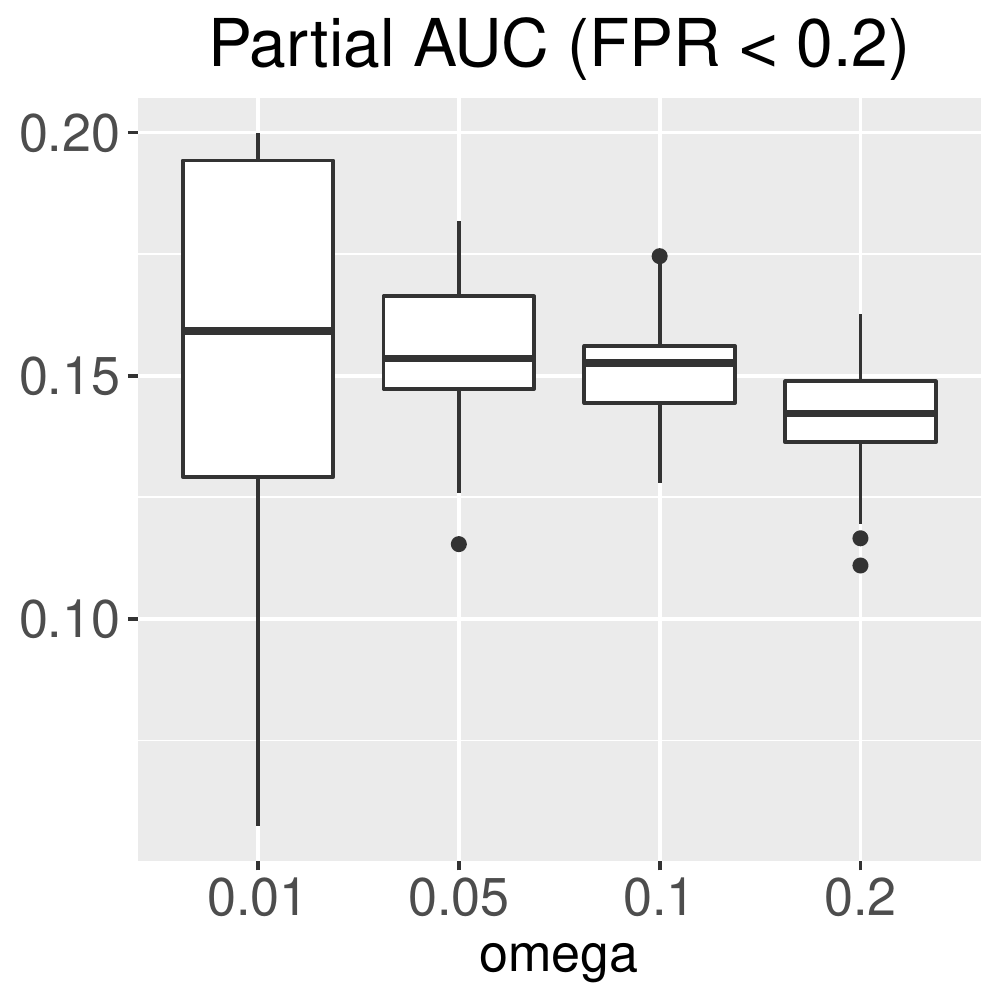}
		\par\end{centering}
	\caption{FDR, power, AUC and partial AUC of LSMM for detection of relevant
		annotations with $\alpha=0.6$ and $K=500$. We controlled global
		FDR at 0.1 to evaluate empirical FDR and power. The results are summarized
		from 50 replications.}
\end{figure}

\begin{figure}[H]
	\begin{centering}
		\includegraphics[scale=0.36]{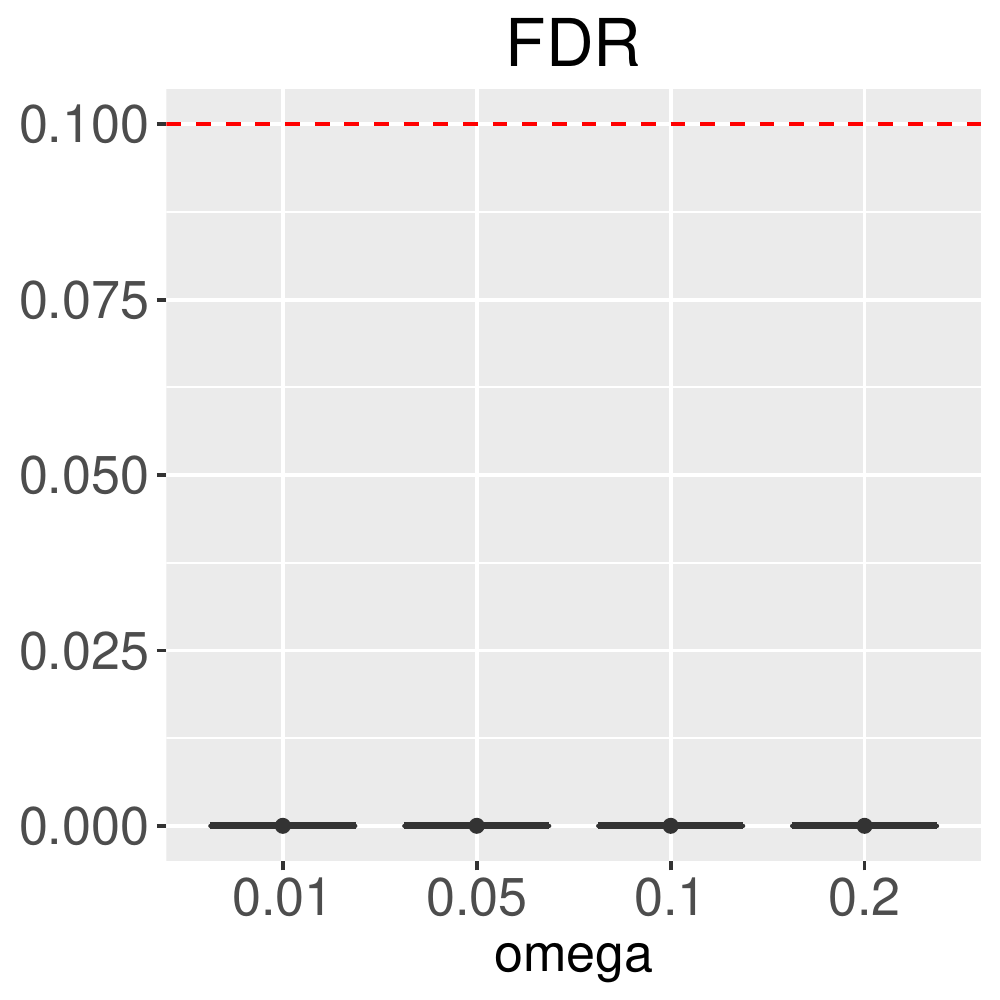}\includegraphics[scale=0.36]{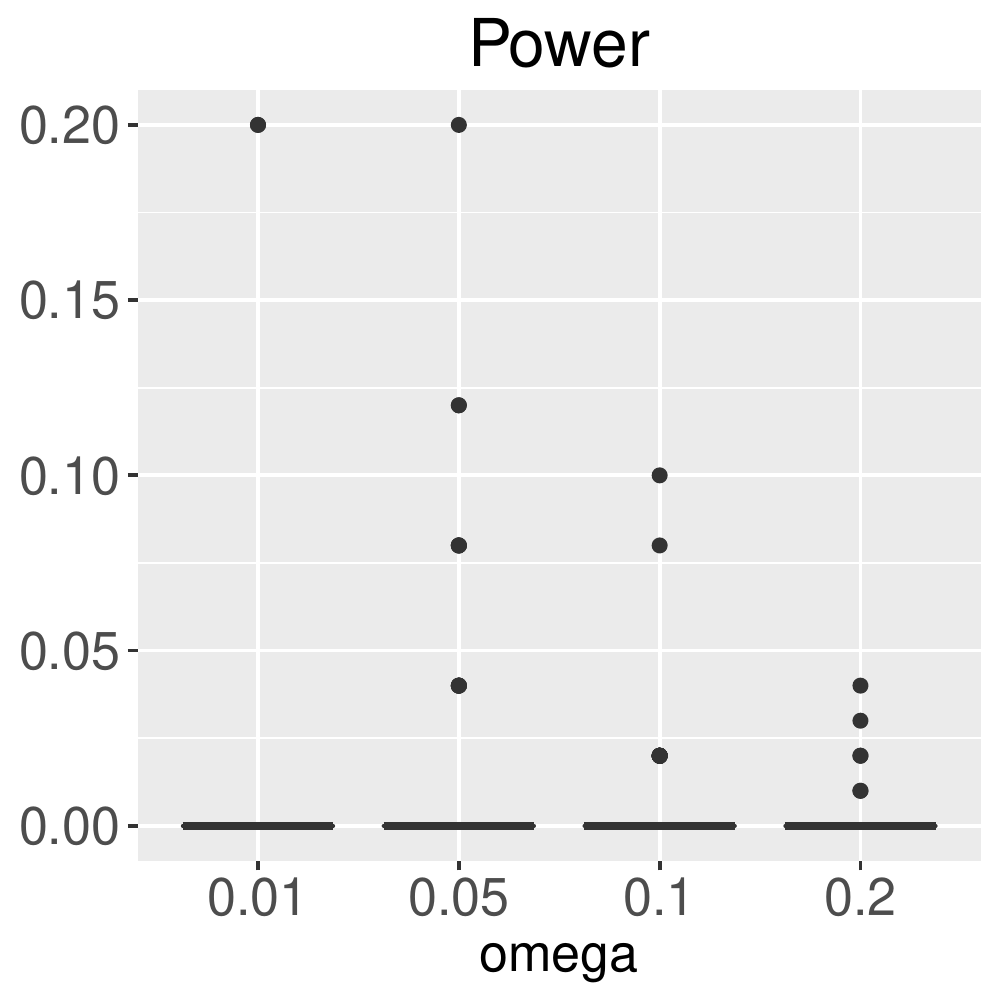}\includegraphics[scale=0.36]{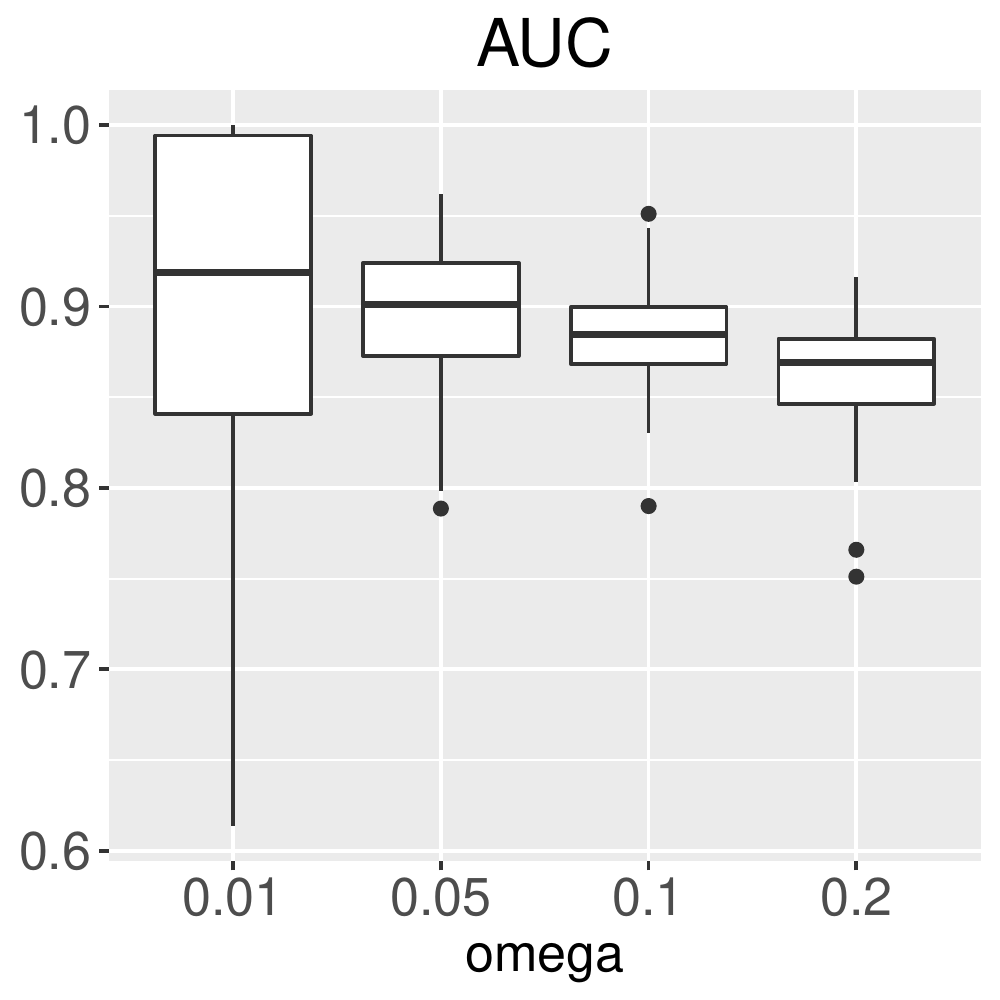}\includegraphics[scale=0.36]{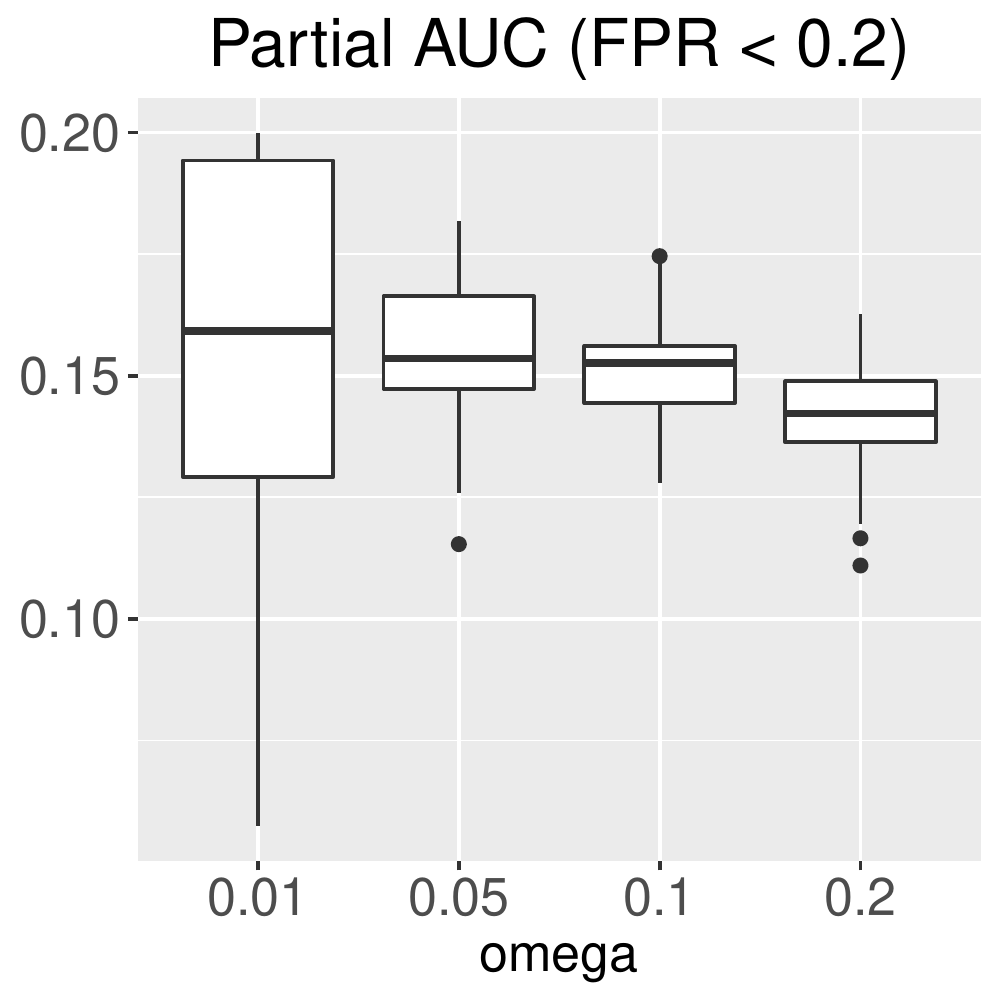}
		\par\end{centering}
	\caption{FDR, power, AUC and partial AUC of LSMM for detection of relevant
		annotations with $\alpha=0.2$ and $K=1000$. We controlled global
		FDR at 0.1 to evaluate empirical FDR and power. The results are summarized
		from 50 replications.}
\end{figure}

\begin{figure}[H]
	\begin{centering}
		\includegraphics[scale=0.36]{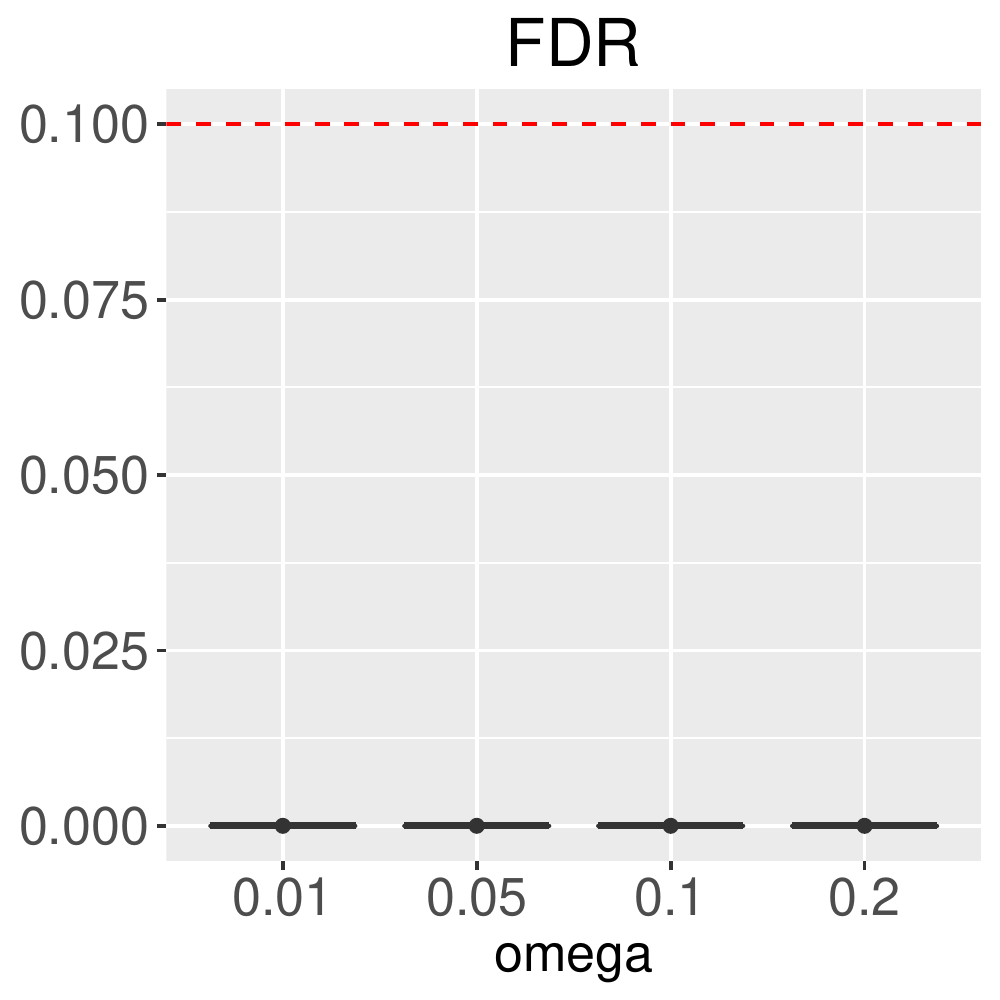}\includegraphics[scale=0.36]{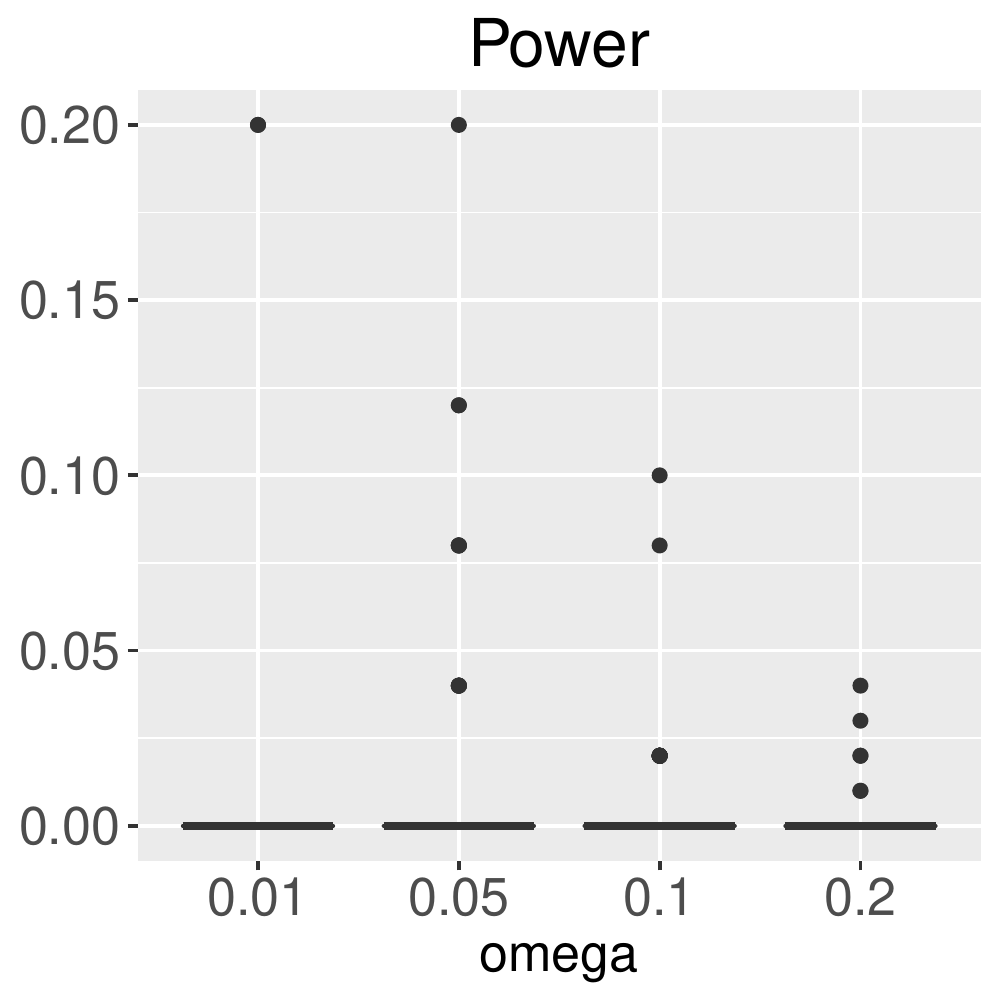}\includegraphics[scale=0.36]{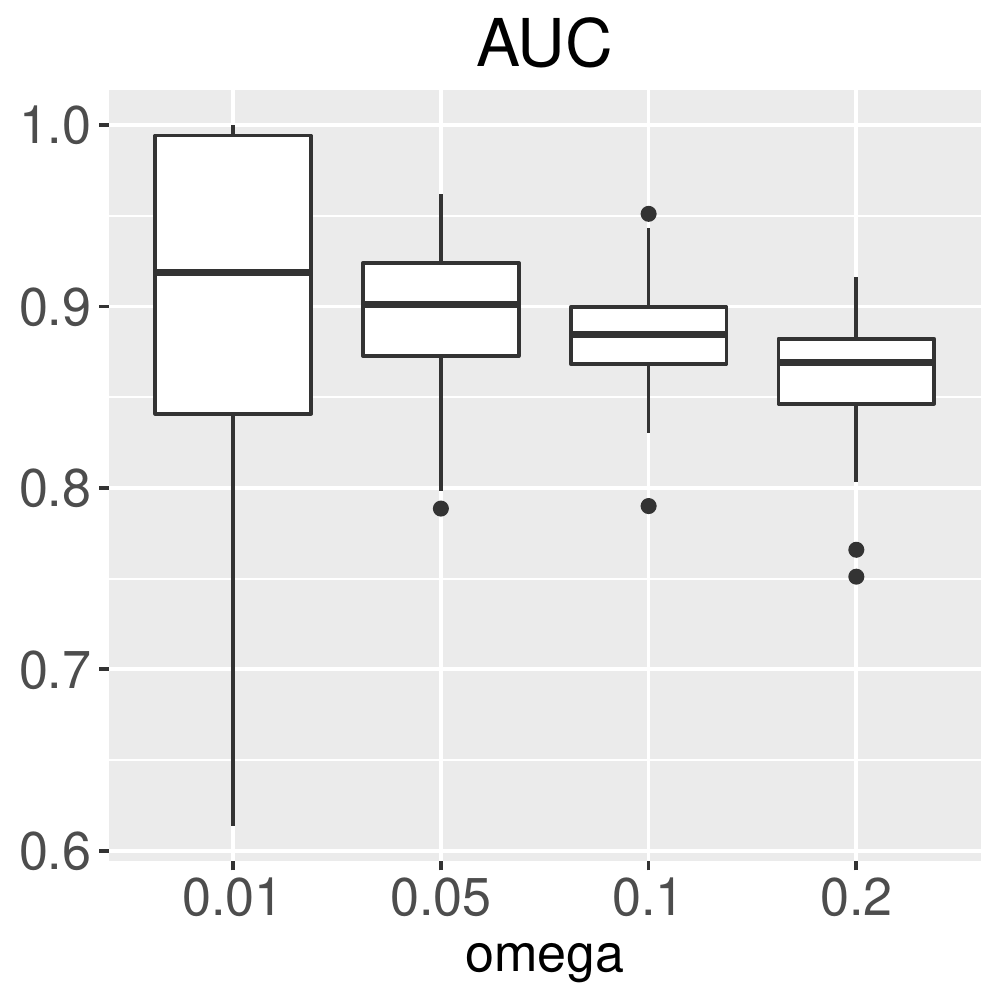}\includegraphics[scale=0.36]{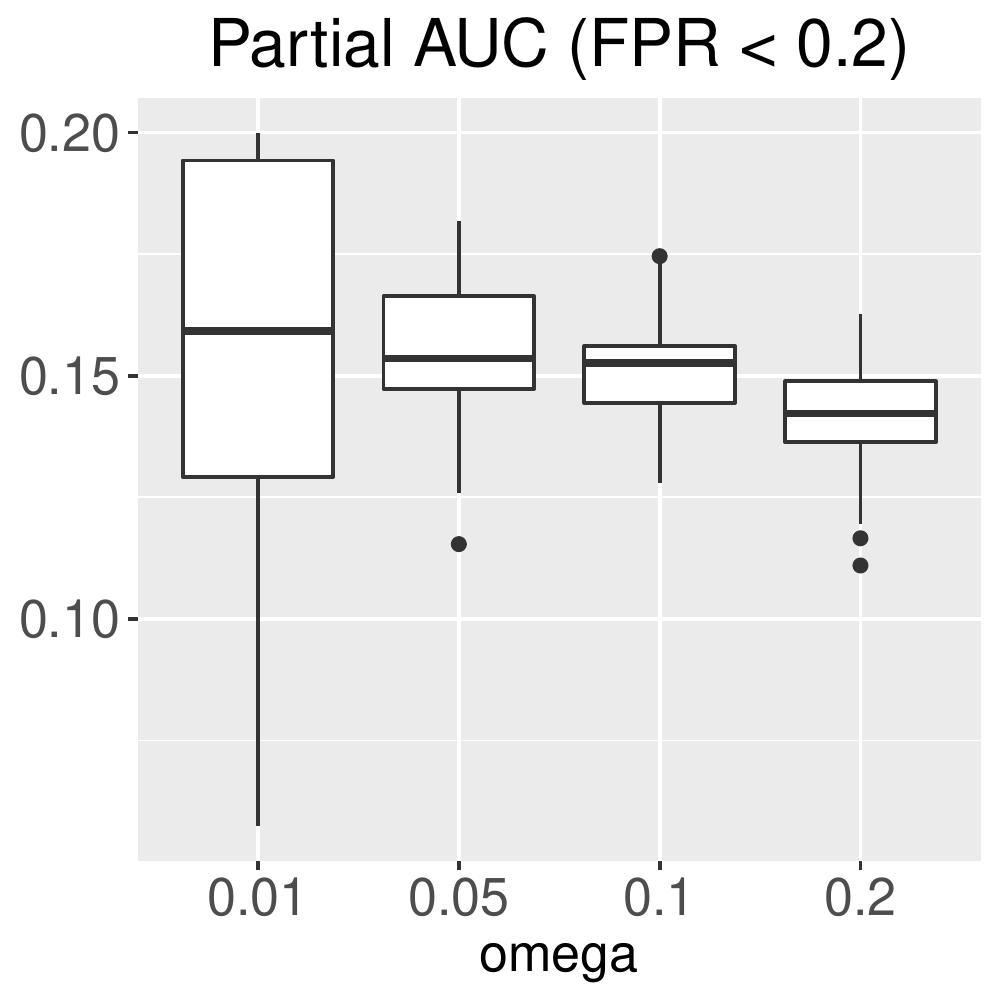}
		\par\end{centering}
	\caption{FDR, power, AUC and partial AUC of LSMM for detection of relevant
		annotations with $\alpha=0.4$ and $K=1000$. We controlled global
		FDR at 0.1 to evaluate empirical FDR and power. The results are summarized
		from 50 replications.}
\end{figure}

\begin{figure}[H]
	\begin{centering}
		\includegraphics[scale=0.36]{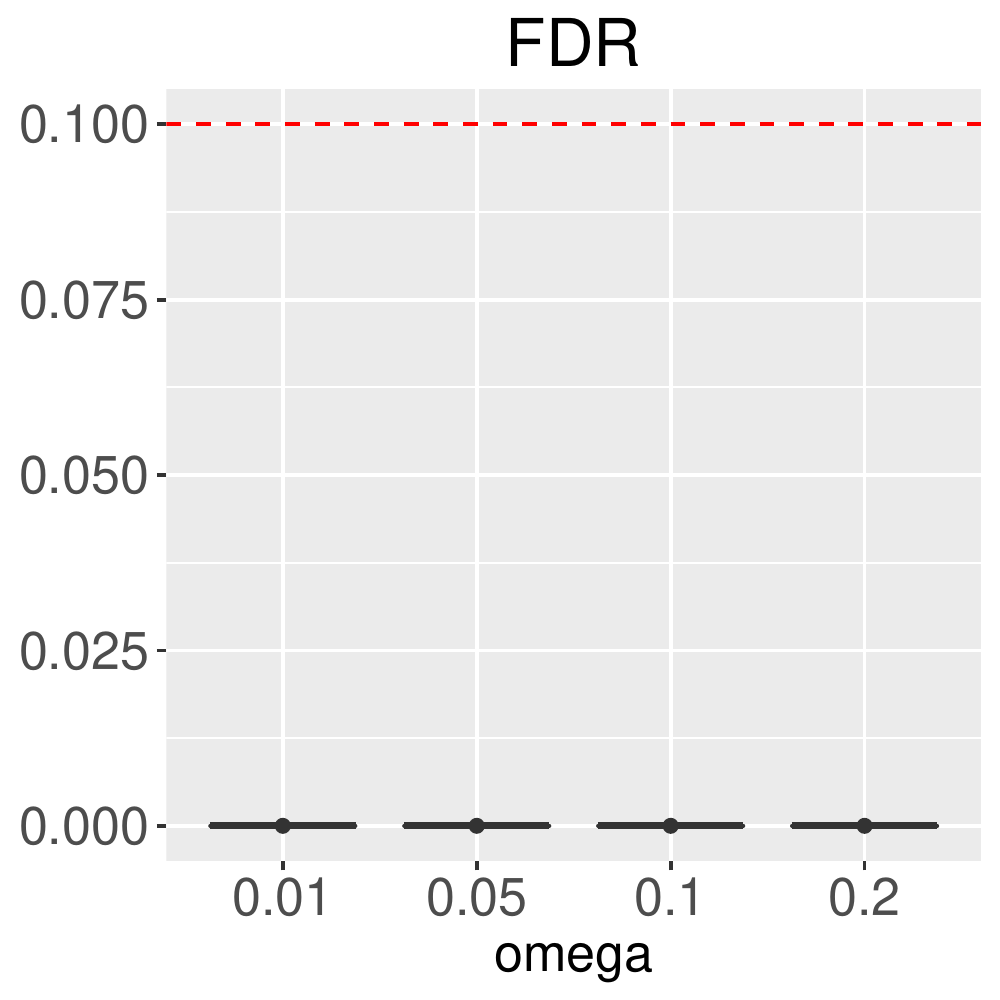}\includegraphics[scale=0.36]{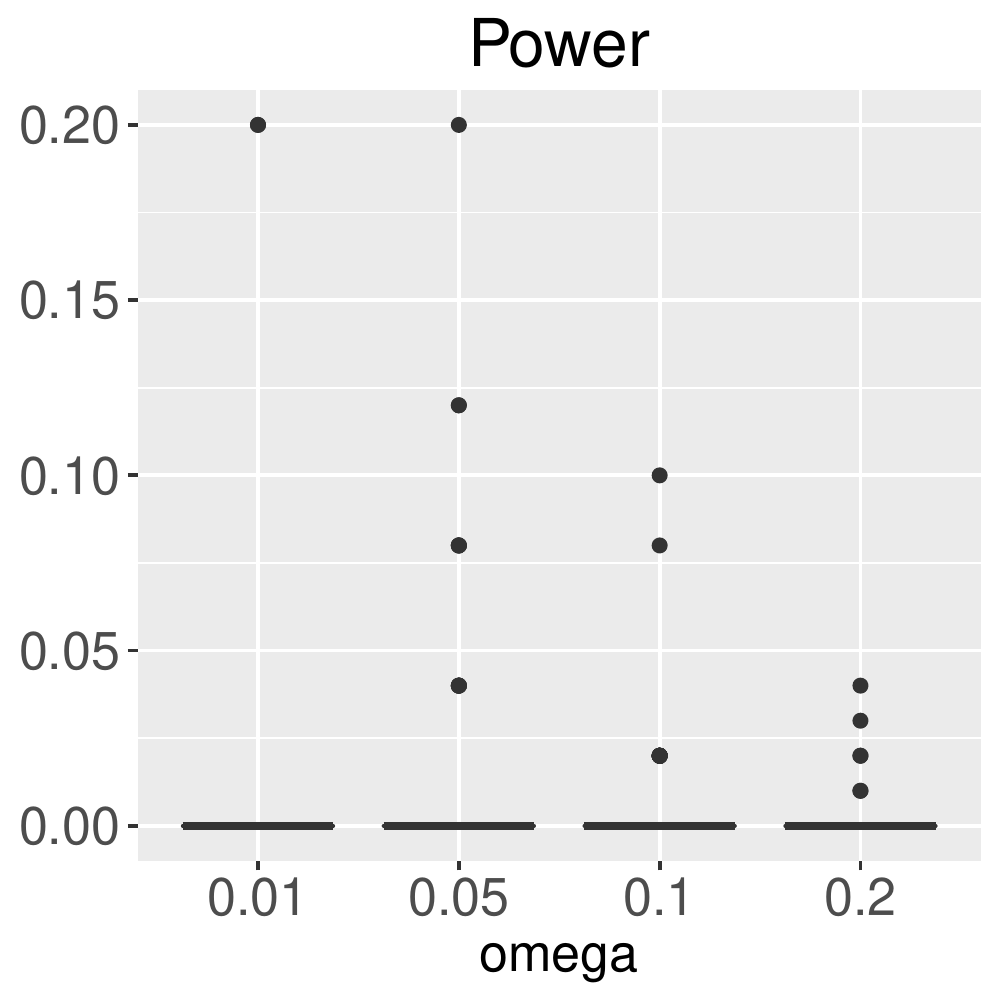}\includegraphics[scale=0.36]{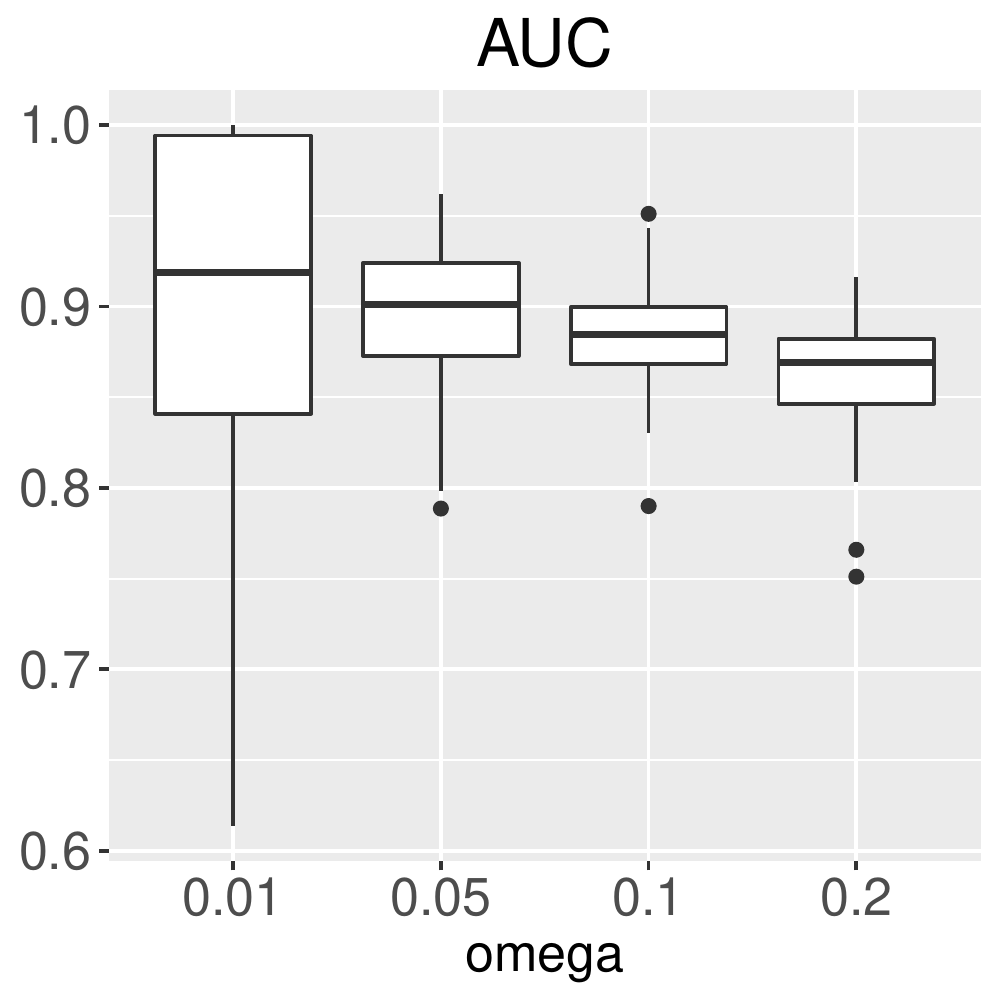}\includegraphics[scale=0.36]{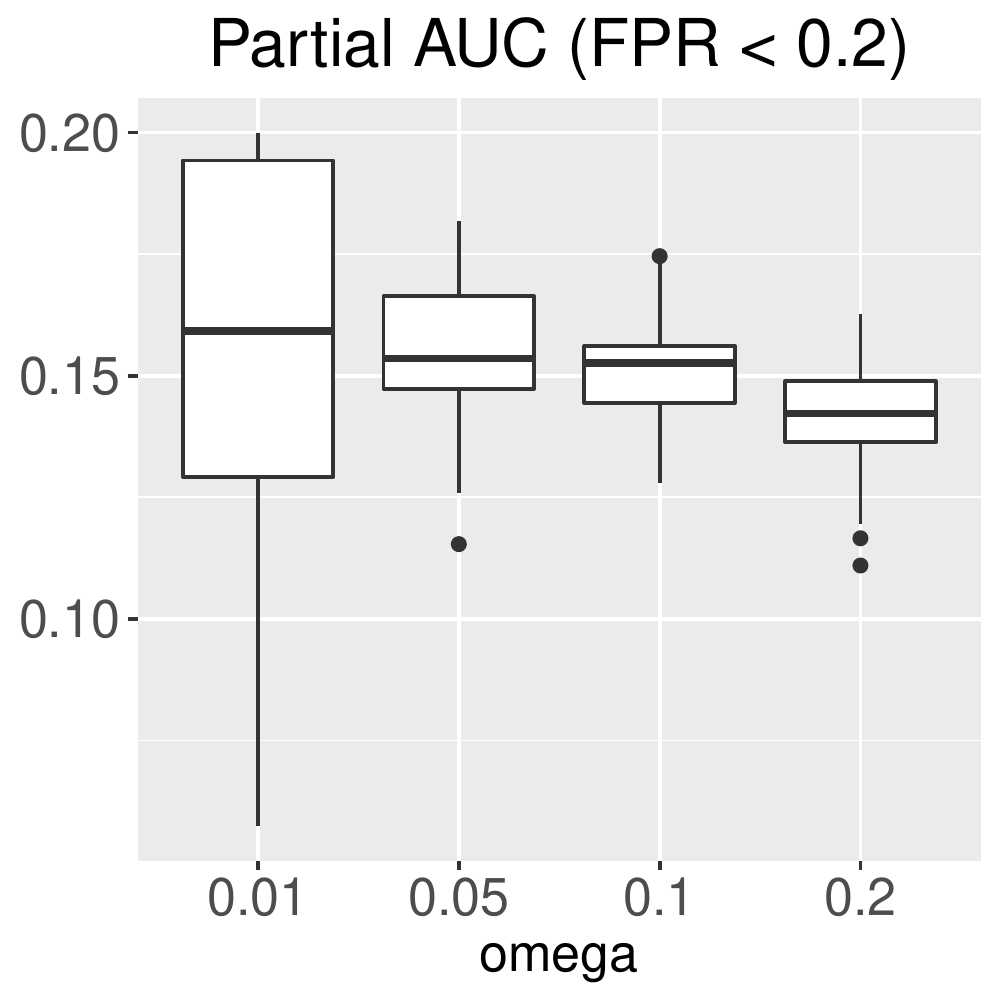}
		\par\end{centering}
	\caption{FDR, power, AUC and partial AUC of LSMM for detection of relevant
		annotations with $\alpha=0.6$ and $K=1000$. We controlled global
		FDR at 0.1 to evaluate empirical FDR and power. The results are summarized
		from 50 replications.}
\end{figure}

\subsection{Performance in identification of relevant annotations when fixed
	effects and random effects are not independent}

\begin{figure}[H]
	\begin{centering}
		\includegraphics[scale=0.7]{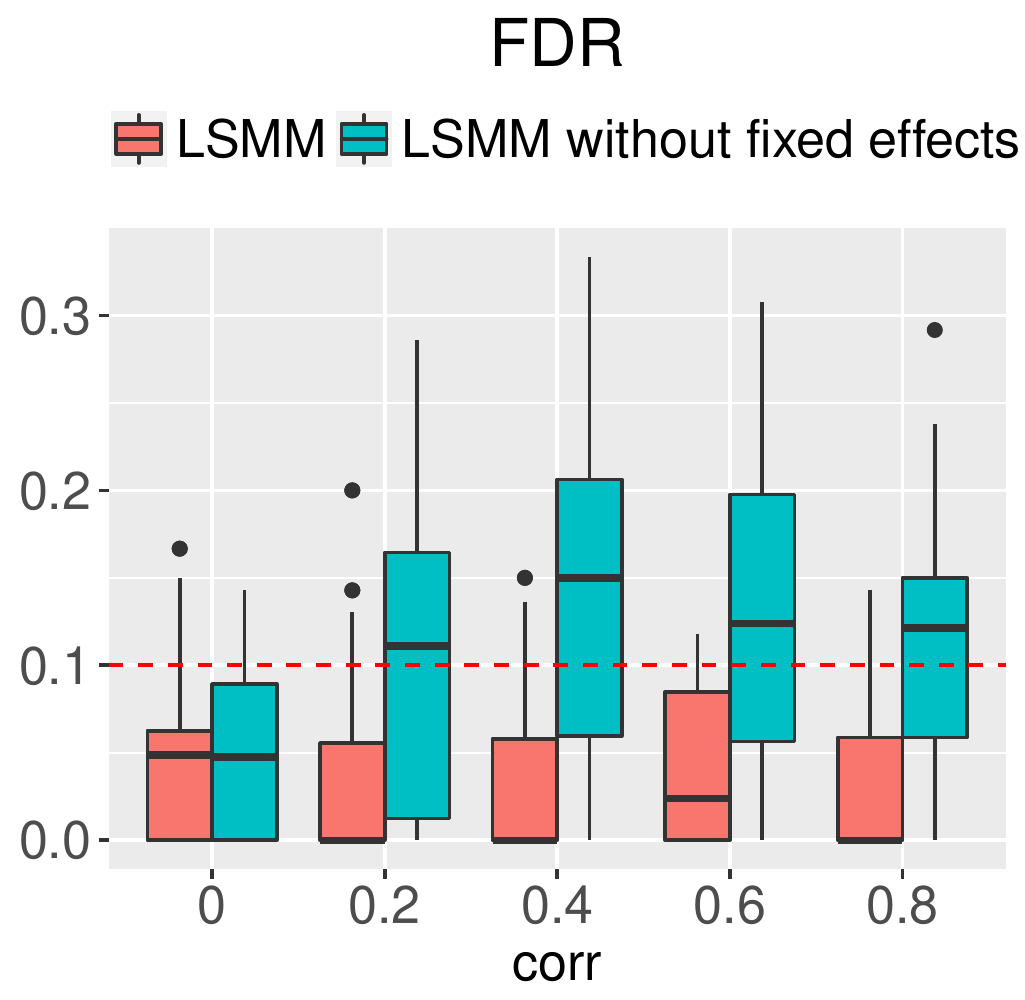}
		\par\end{centering}
	\caption{FDR of LSMM and LSMM without fixed effects for detection of relevant
		annotations with $K=100$. We controlled global FDR at 0.1 to evaluate
		empirical FDR and power. The results are summarized from 50 replications.}
\end{figure}

\subsection{Simulations based on probit model}

\begin{figure}[H]
	\begin{centering}
		\includegraphics[scale=0.36]{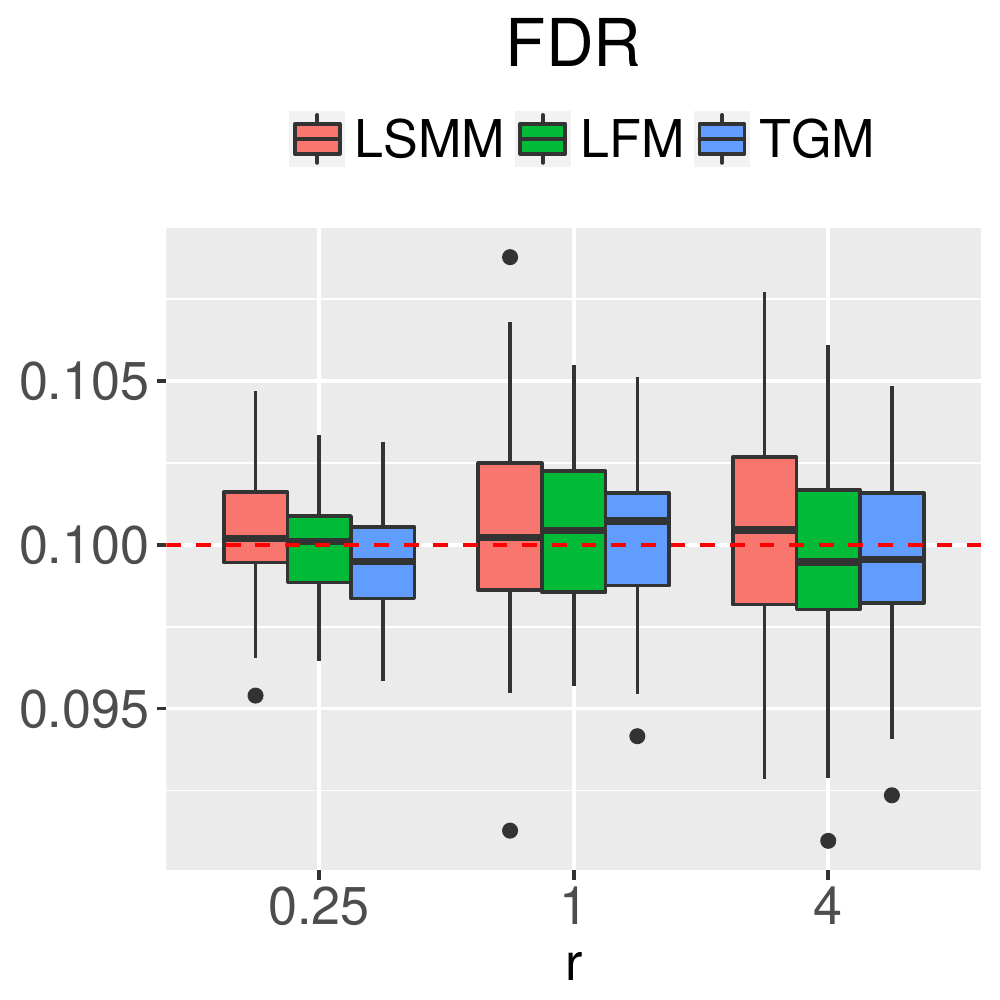}\includegraphics[scale=0.36]{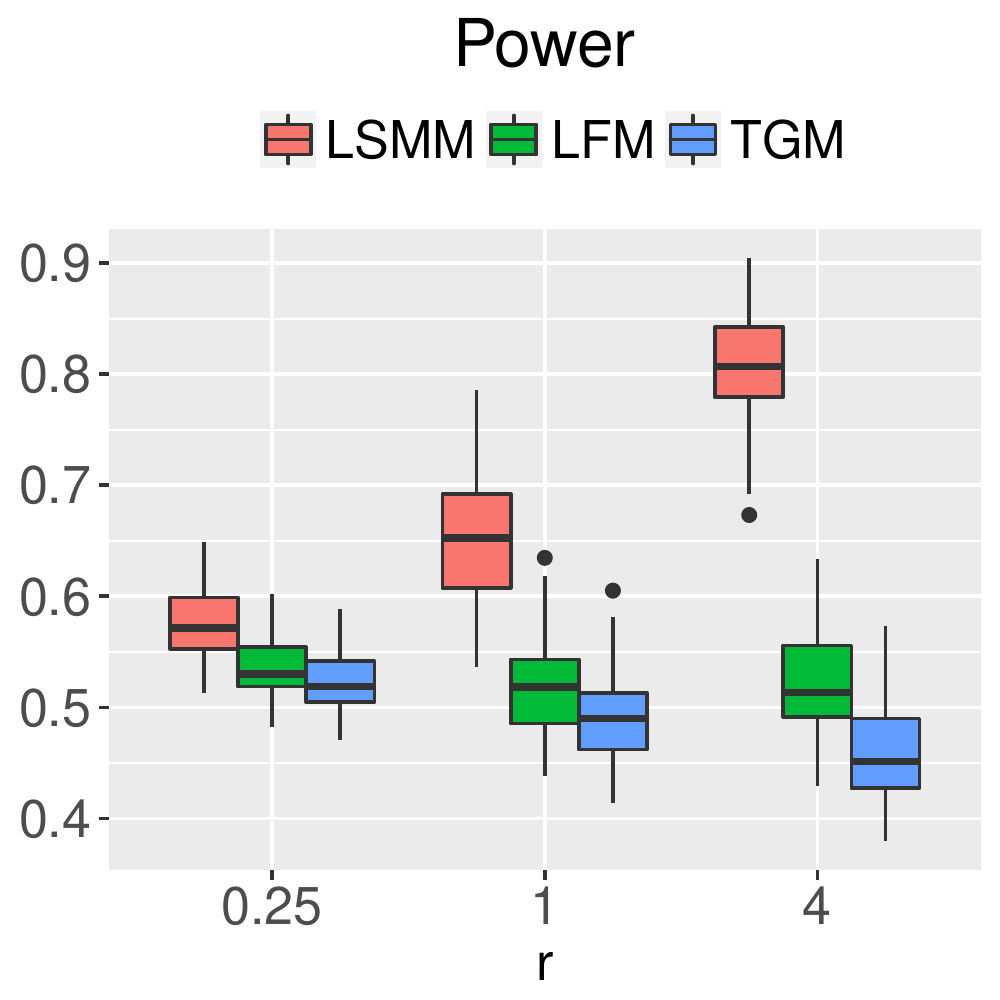}\includegraphics[scale=0.36]{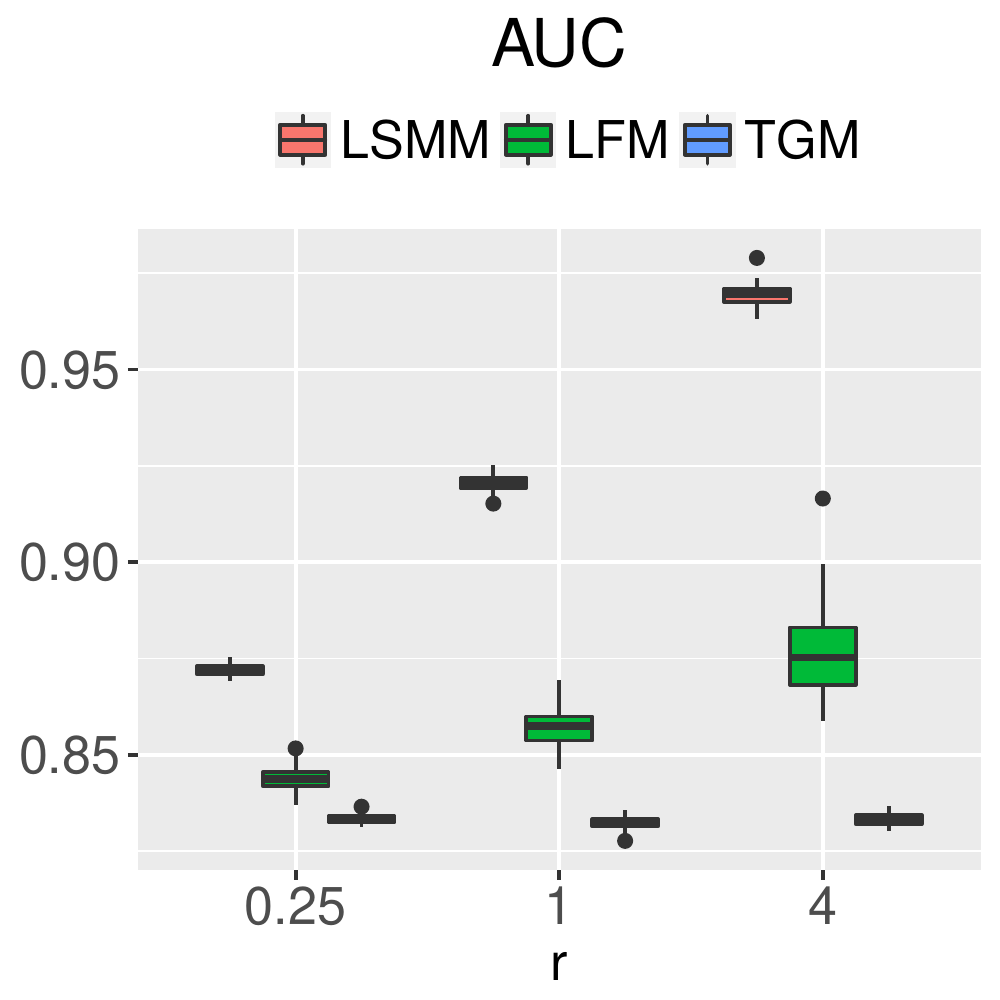}\includegraphics[scale=0.36]{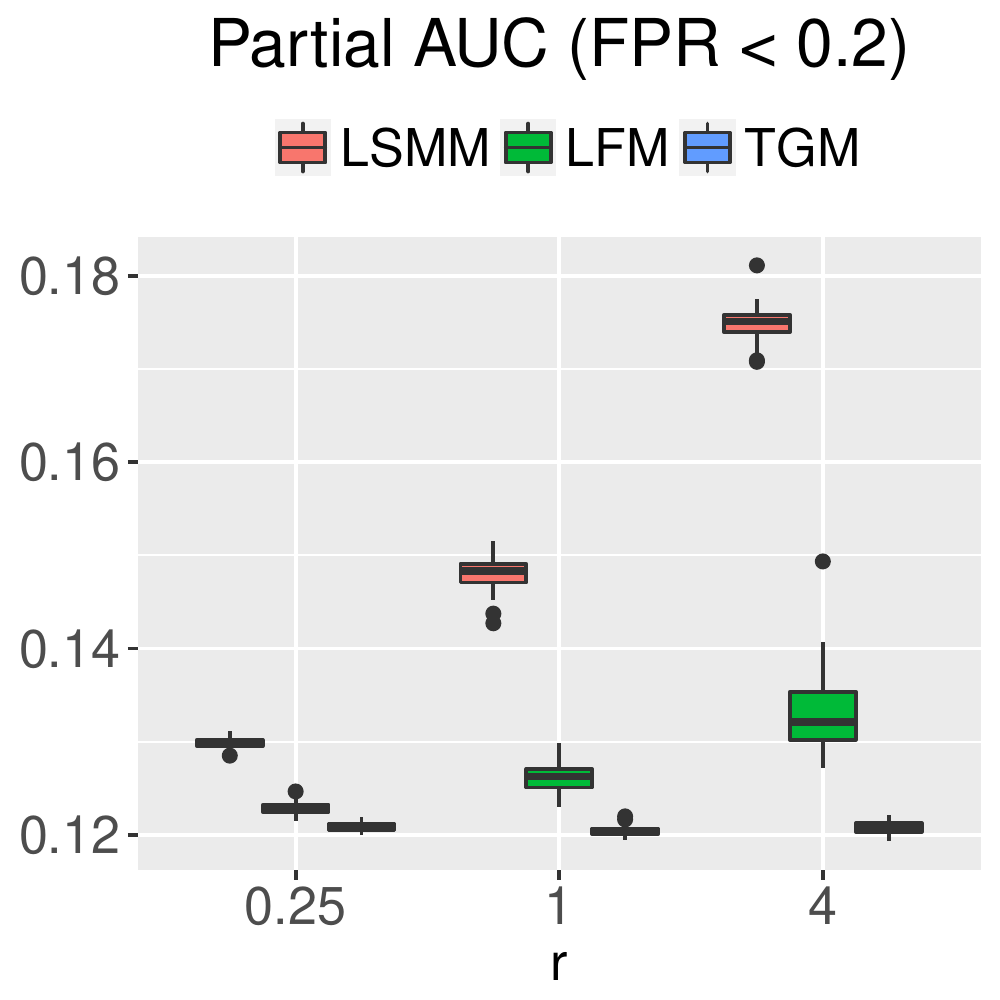}
		\par\end{centering}
	\caption{FDR, power, AUC and partial AUC of LSMM, LFM and TGM for identification
		of risk SNPs based on probit model with $K=100$. We controlled global
		FDR at 0.1 to evaluate empirical FDR and power. The results are summarized
		from 50 replications.}
\end{figure}

\begin{figure}[H]
	\begin{centering}
		\includegraphics[scale=0.36]{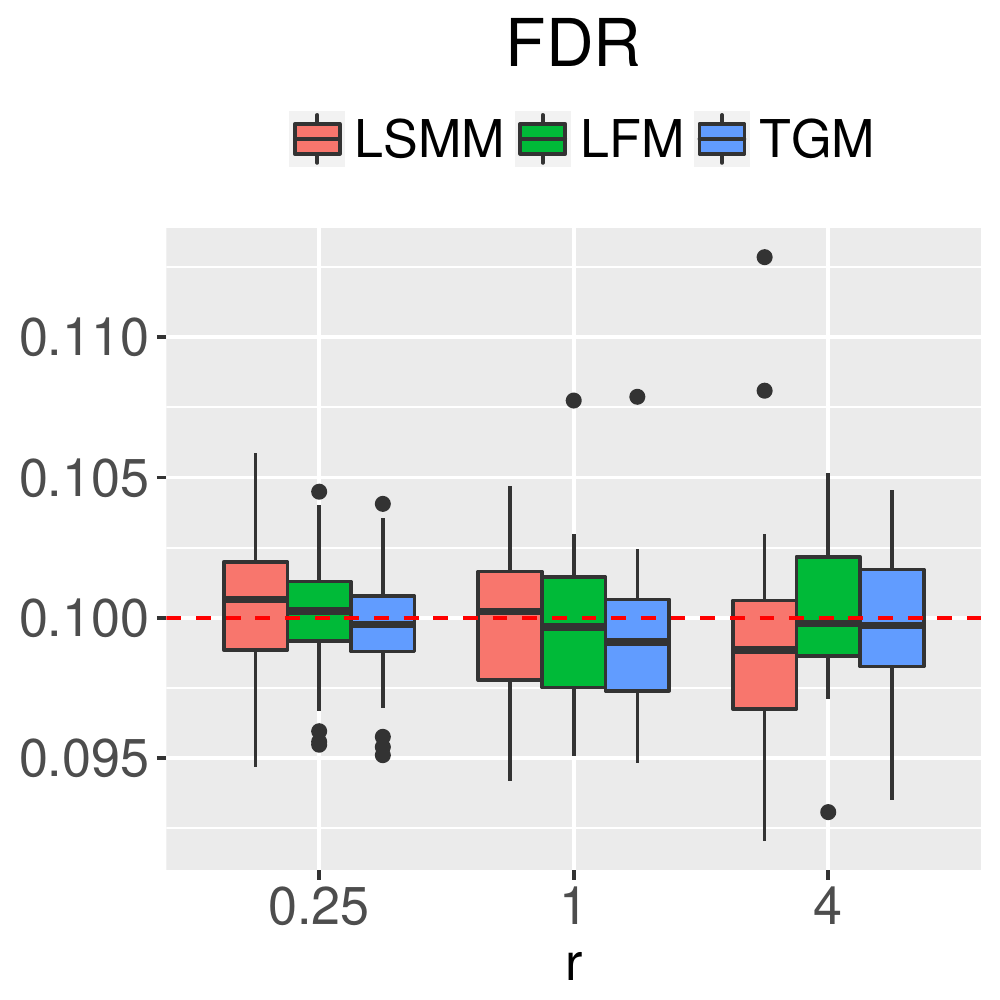}\includegraphics[scale=0.36]{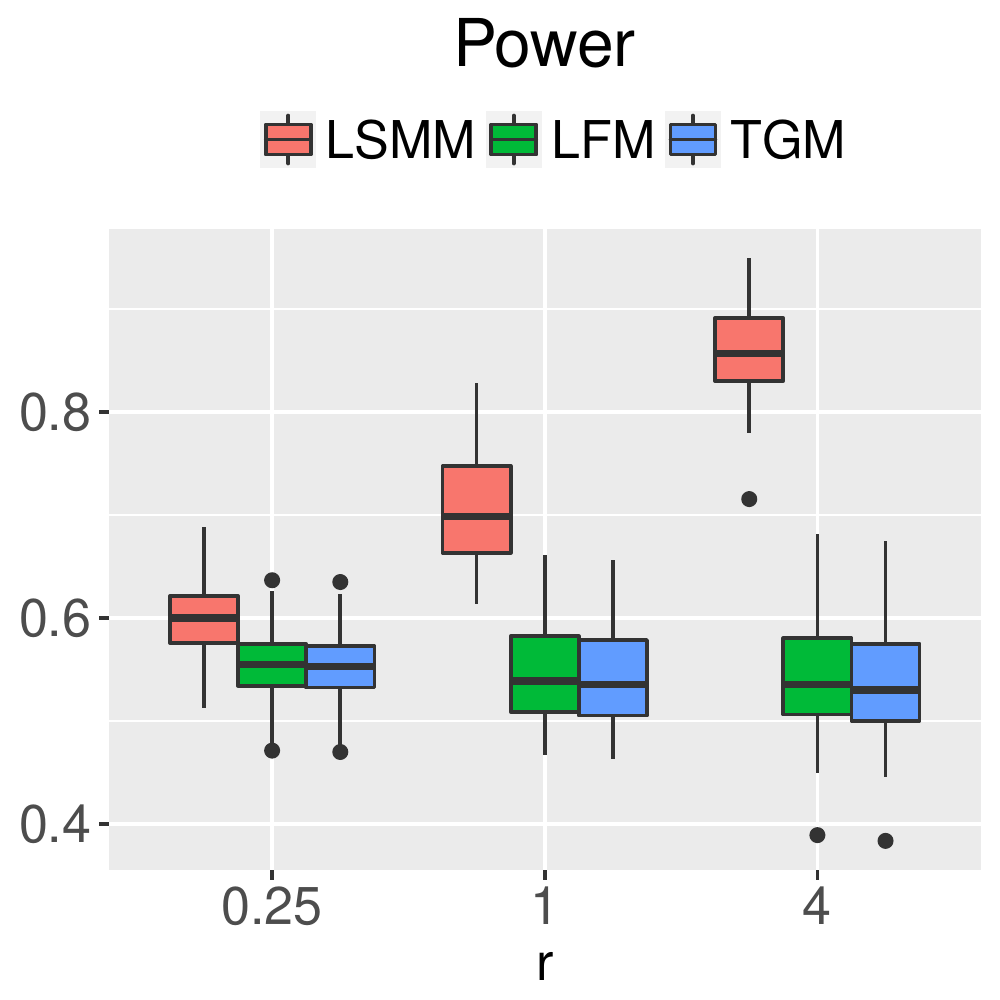}\includegraphics[scale=0.36]{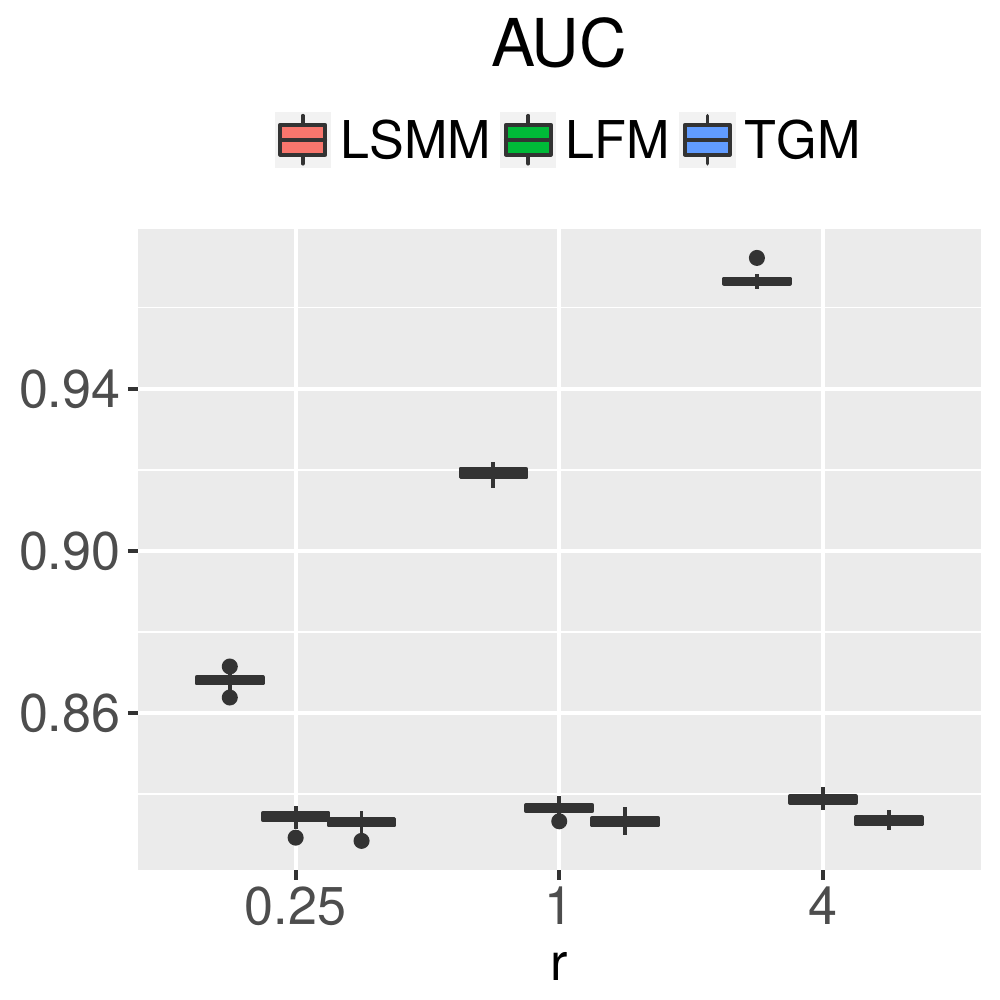}\includegraphics[scale=0.36]{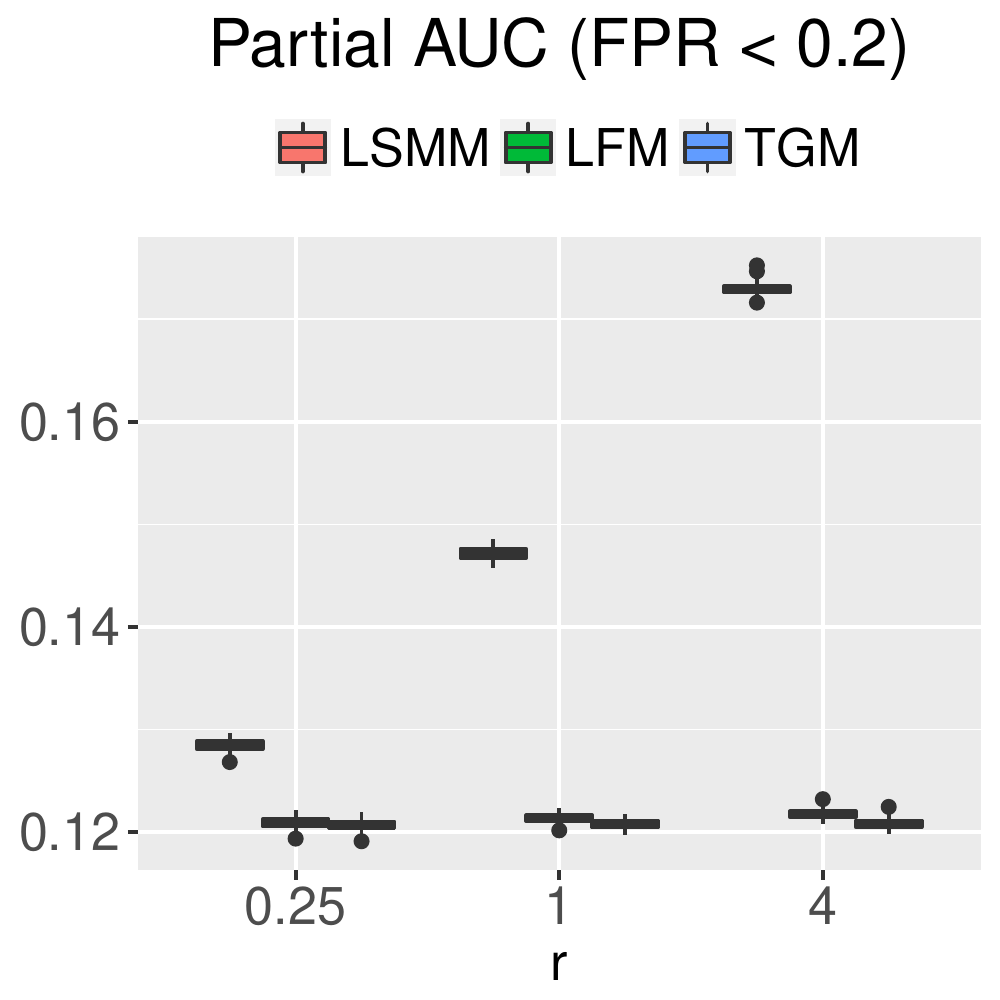}
		\par\end{centering}
	\caption{FDR, power, AUC and partial AUC of LSMM, LFM and TGM for identification
		of risk SNPs based on probit model with $K=1000$. We controlled global
		FDR at 0.1 to evaluate empirical FDR and power. The results are summarized
		from 50 replications.}
\end{figure}

\begin{figure}[H]
	\begin{centering}
		\includegraphics[scale=0.36]{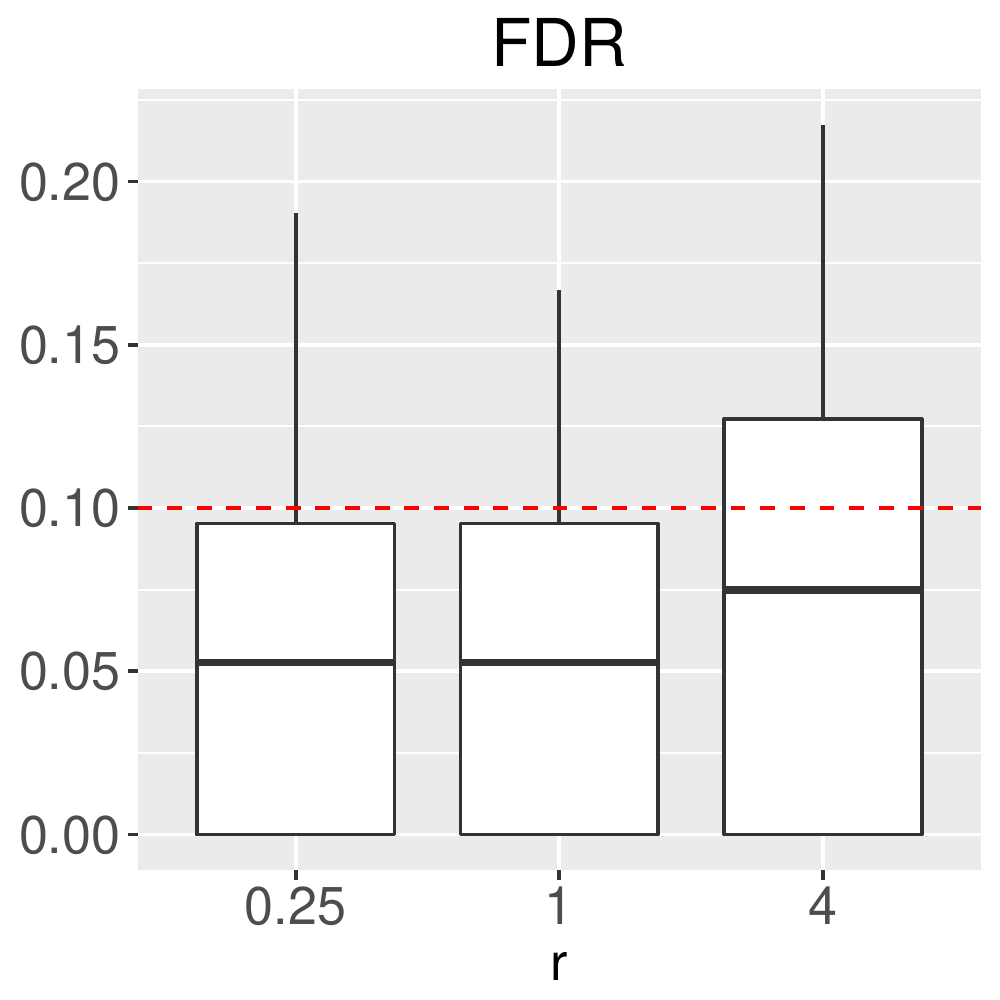}\includegraphics[scale=0.36]{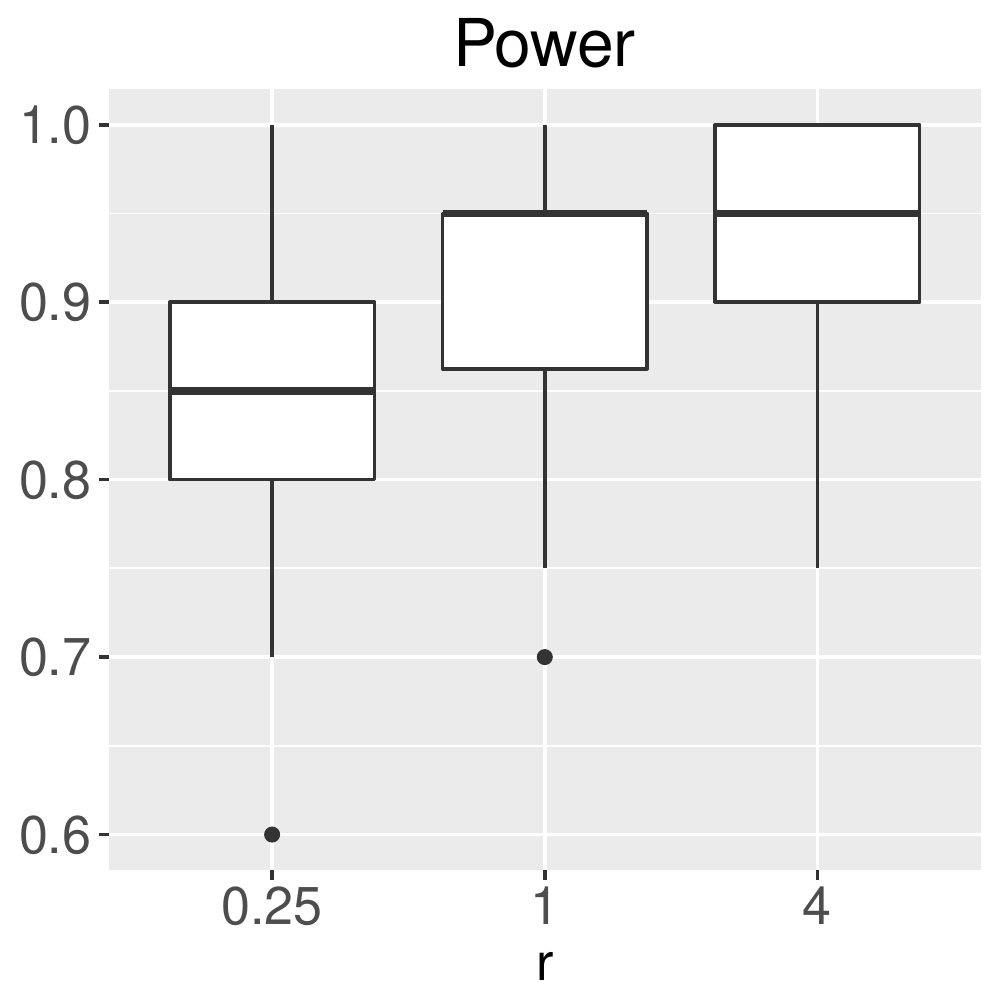}\includegraphics[scale=0.36]{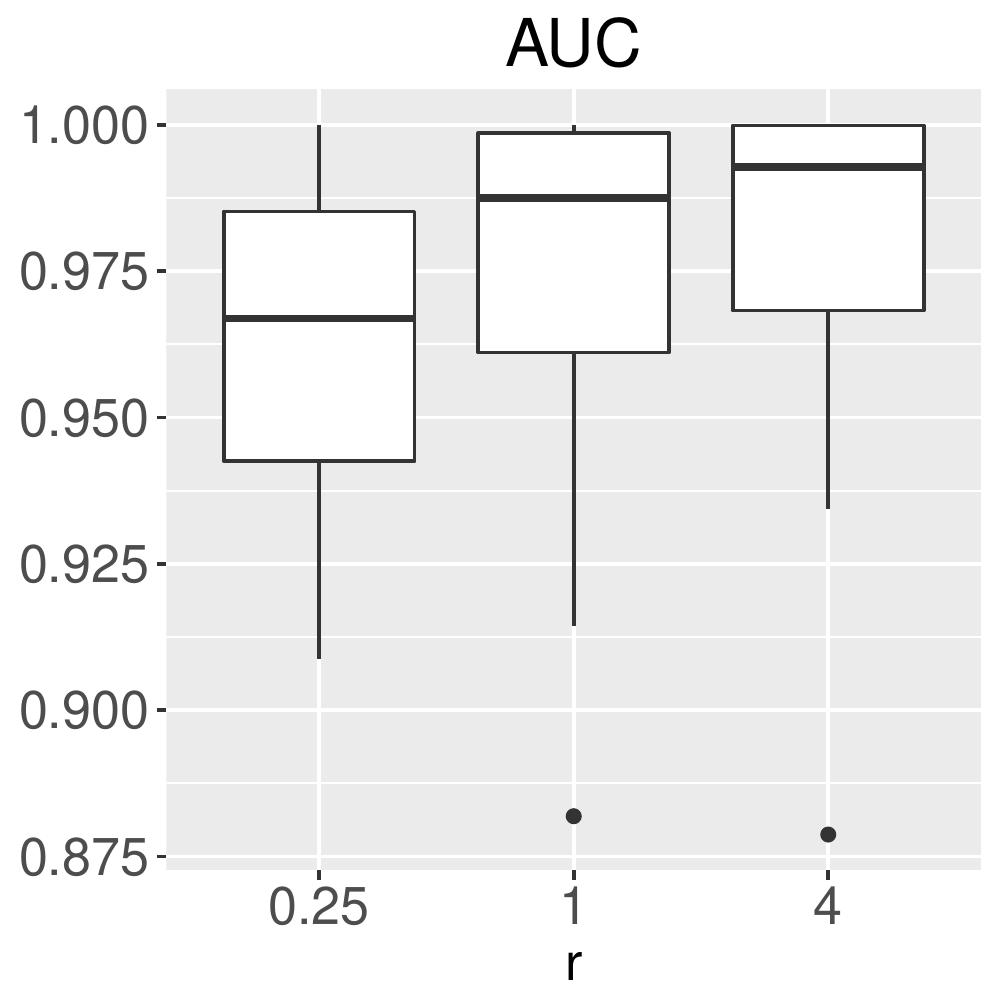}\includegraphics[scale=0.36]{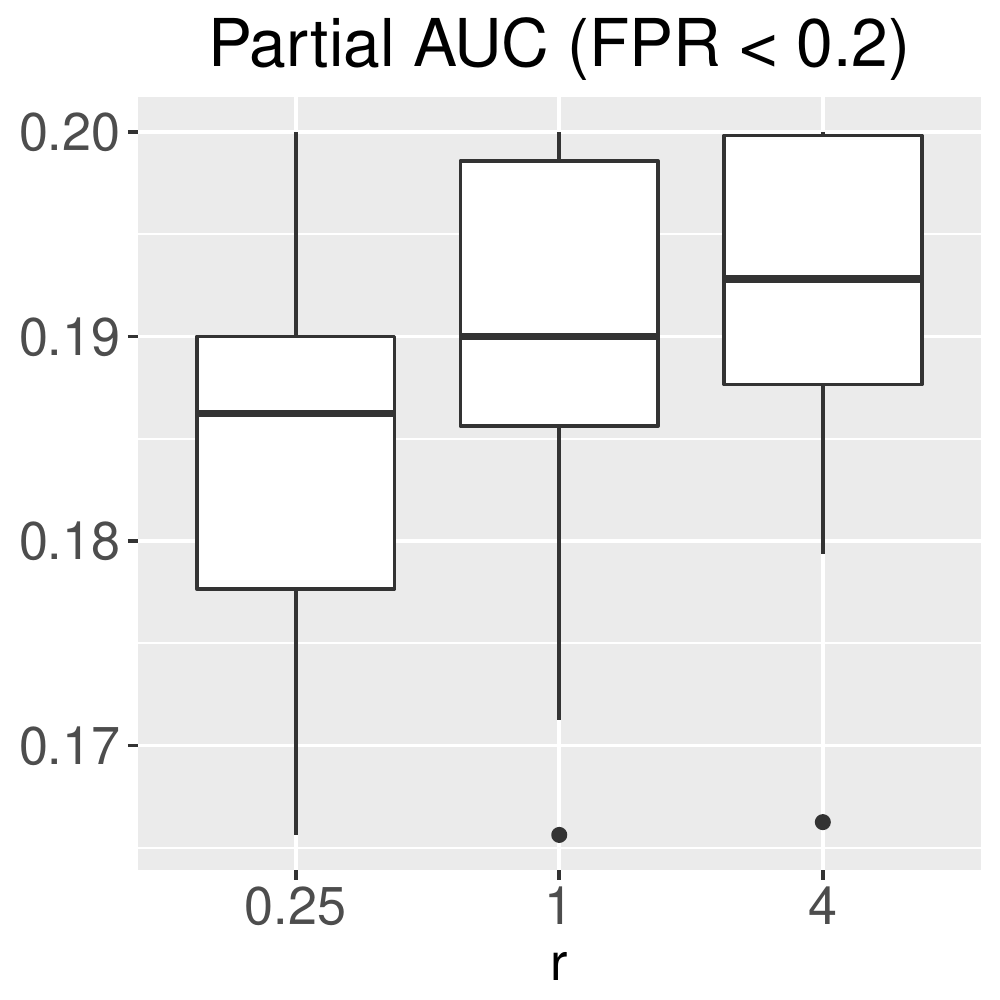}
		\par\end{centering}
	\caption{FDR, power, AUC and partial AUC of LSMM, LFM and TGM for detection
		of relevant annotations based on probit model with $K=100$. We controlled
		global FDR at 0.1 to evaluate empirical FDR and power. The results
		are summarized from 50 replications.}
\end{figure}

\begin{figure}[H]
	\begin{centering}
		\includegraphics[scale=0.36]{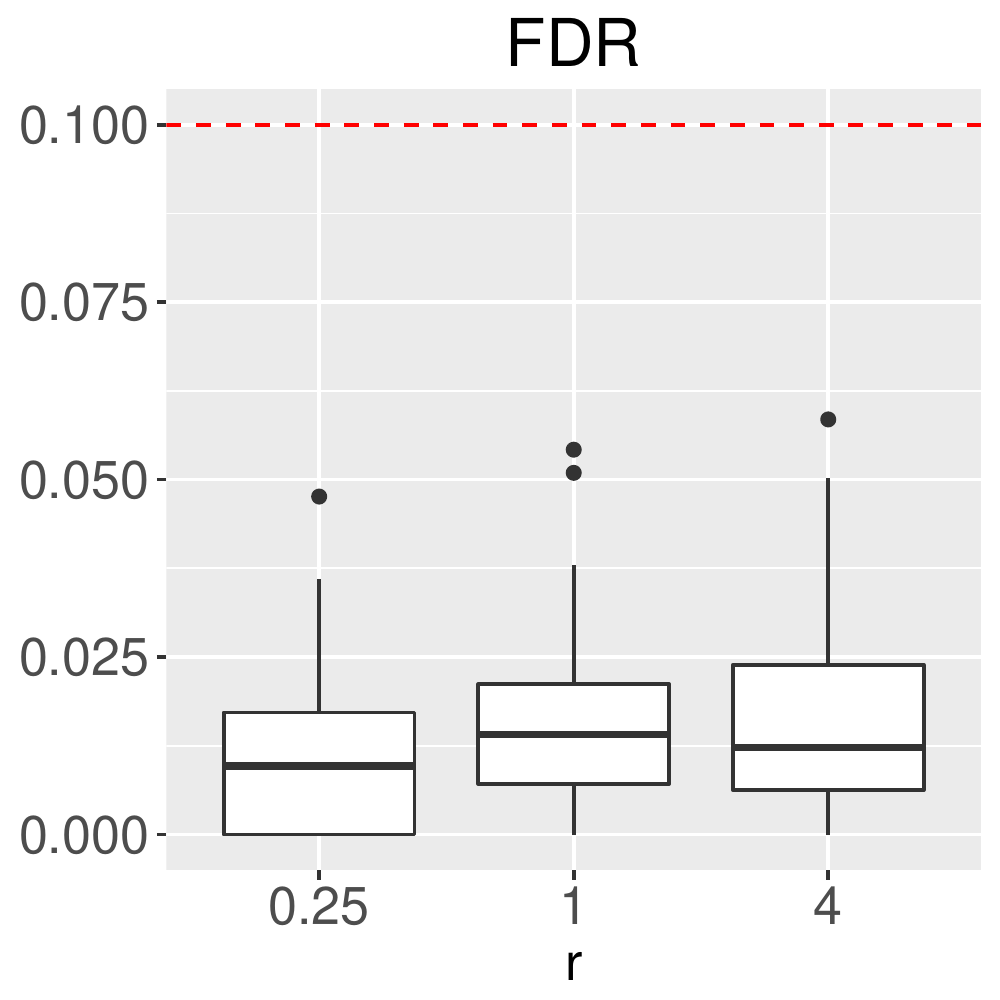}\includegraphics[scale=0.36]{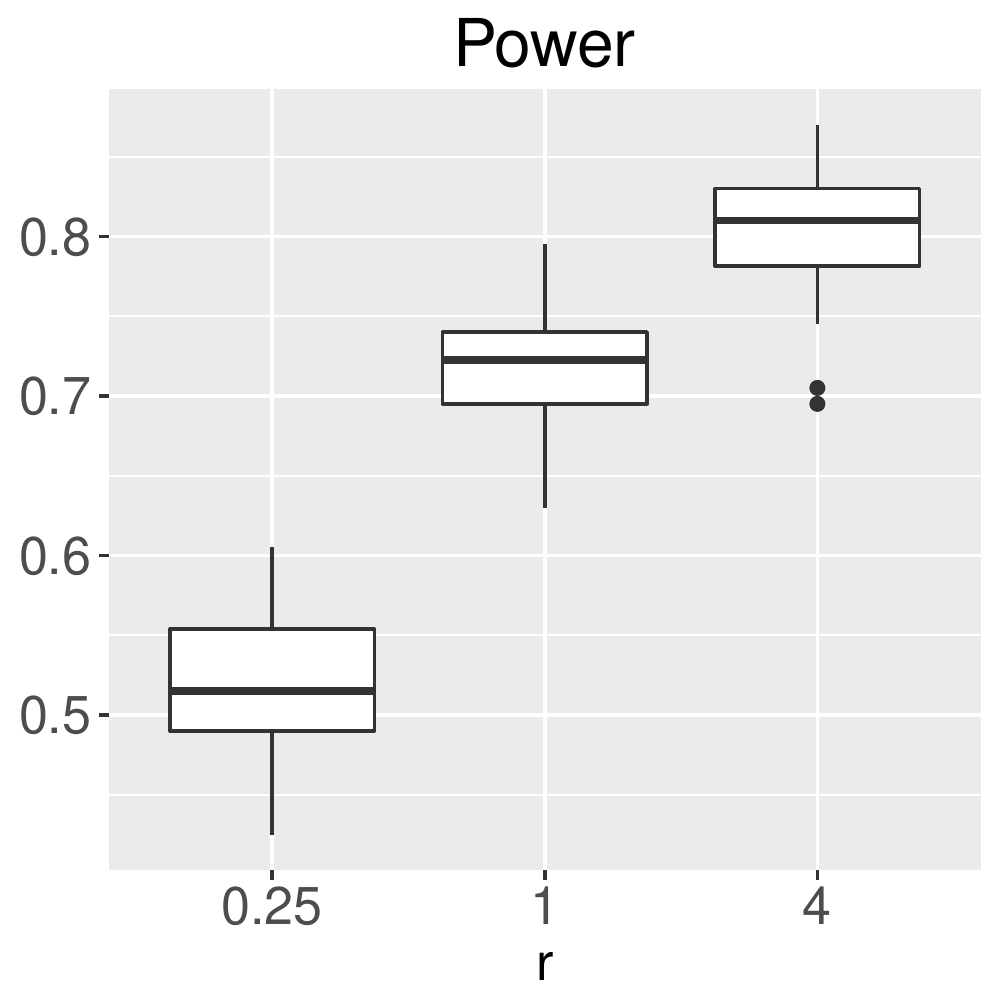}\includegraphics[scale=0.36]{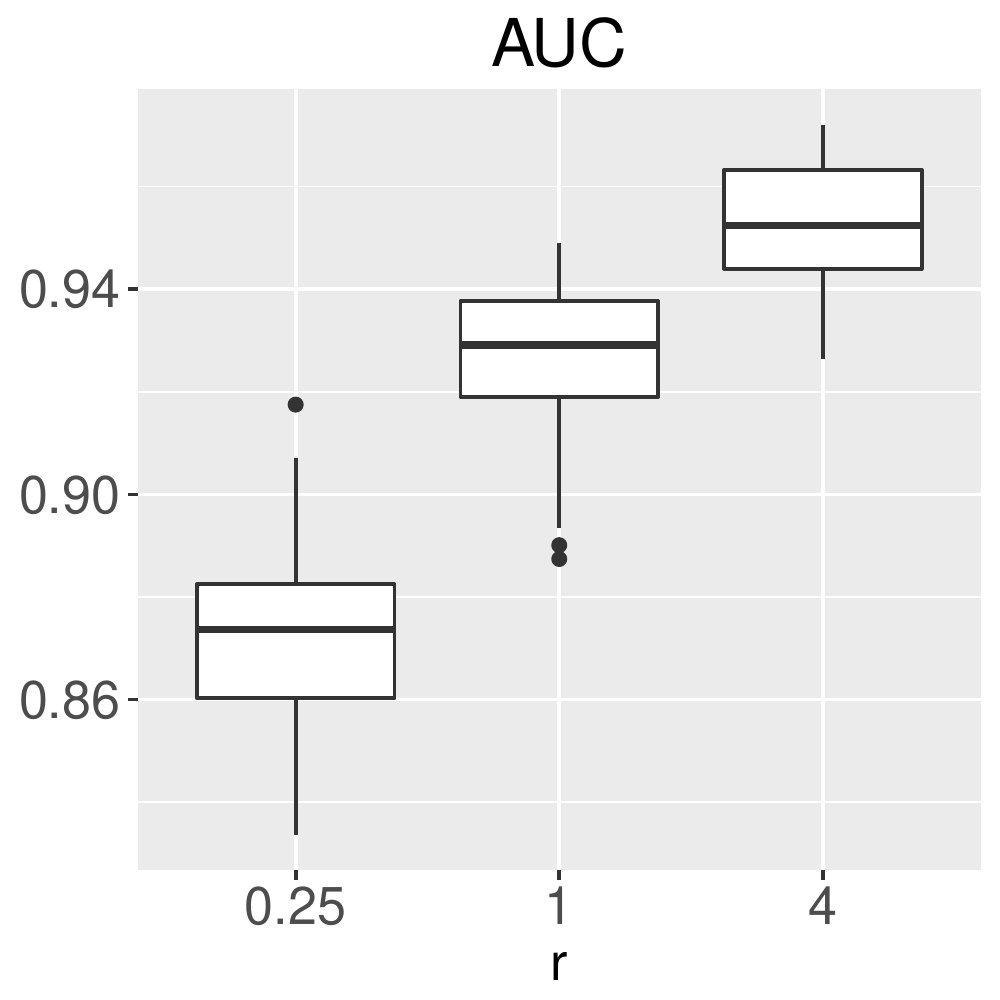}\includegraphics[scale=0.36]{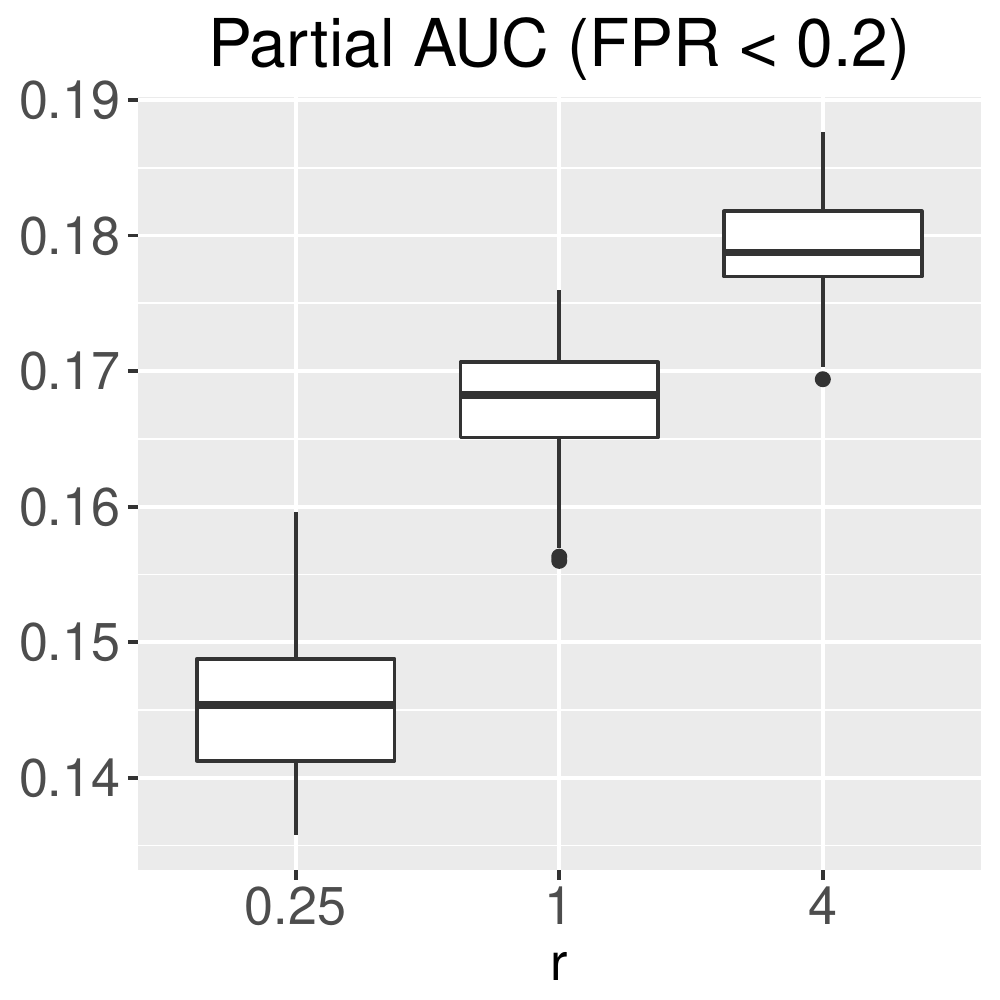}
		\par\end{centering}
	\caption{FDR, power, AUC and partial AUC of LSMM, LFM and TGM for detection
		of relevant annotations based on probit model with $K=1000$. We controlled
		global FDR at 0.1 to evaluate empirical FDR and power. The results
		are summarized from 50 replications.}
\end{figure}

\subsection{Simulations if $p$-values are not from beta distribution}

\begin{figure}[H]
	\begin{centering}
		\includegraphics[scale=0.36]{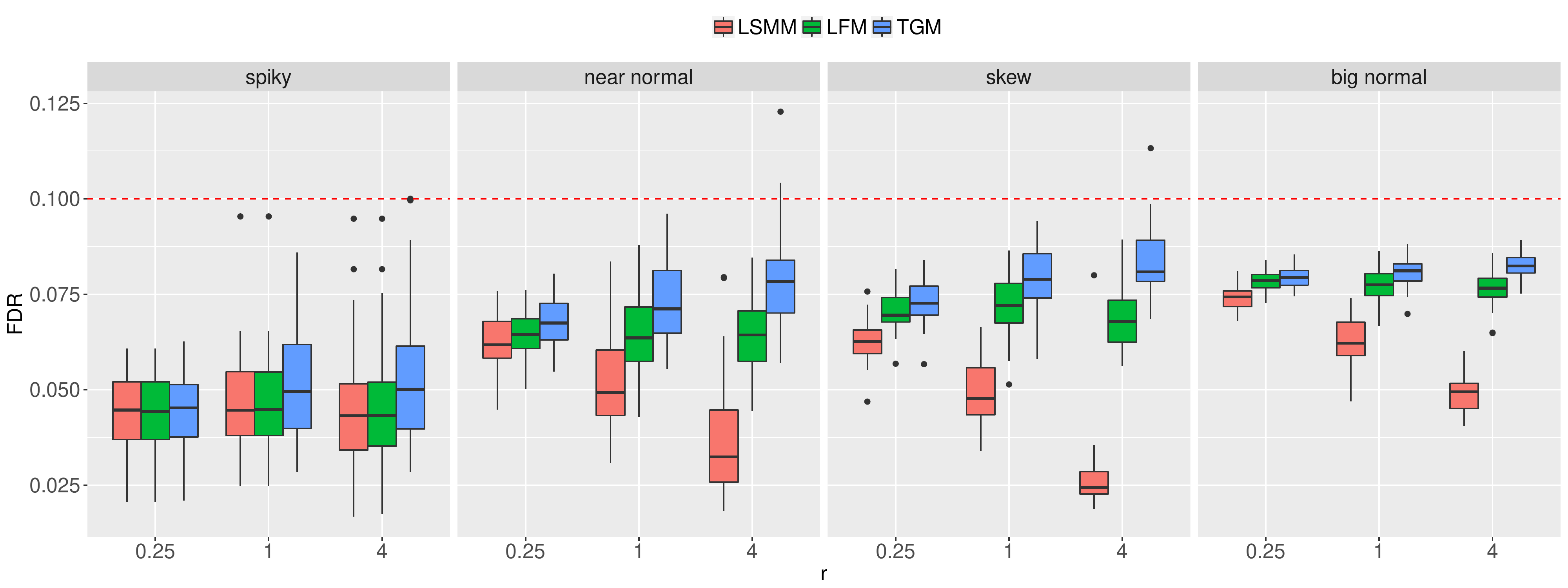}
		\par\end{centering}
	\caption{FDR of LSMM, LFM and TGM with $K=100$. We controlled global FDR at
		0.1 to evaluate empirical FDR. The results are summarized from 50
		replications.\label{fig:robust1}}
\end{figure}

\begin{figure}[H]
	\begin{centering}
		\includegraphics[scale=0.36]{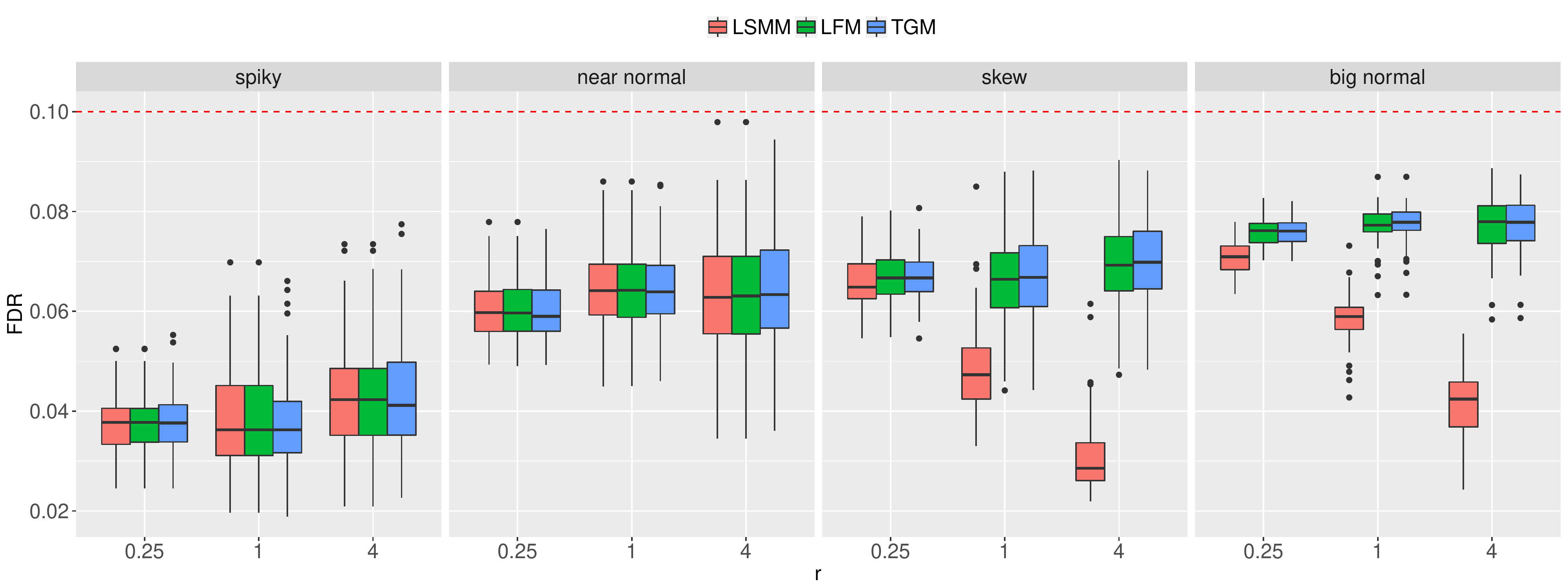}
		\par\end{centering}
	\caption{FDR of LSMM, LFM and TGM with $K=1000$. We controlled global FDR
		at 0.1 to evaluate empirical FDR. The results are summarized from
		50 replications.\label{fig:robust3}}
\end{figure}

In the model setting of the LSMM, we assume that $p$-values are from
the mixture of uniform and Beta distributions. To check the robustness
of our method, we conducted simulations as follows. We first generated
$z$-scores and then converted them to $p$-values. Here $z$-values
from the null group follow the standard normal distribution and $z$-values
from the non-null group follow the alternative distributions in Table
\ref{tab:distribution}. In these simulations, the $p$-values in
non-null group converted from $z$-scores will not from Beta distribution.
We evaluated the FDR, power and AUC. The results are shown in Figures
\ref{fig:robust1}-\ref{fig:robust3}. 

\begin{table}[H]
	\begin{centering}
		{\footnotesize{}}%
		\begin{tabular}{|c|c|}
			\hline 
			{\footnotesize{}Scenario} & {\footnotesize{}Distribution}\tabularnewline
			\hline 
			\hline 
			{\footnotesize{}spiky} & {\footnotesize{}$0.4N\left(0,0.25^{2}\right)+0.2N\left(0,0.5^{2}\right)+0.2N\left(0,1^{2}\right)+0.2N\left(0,2^{2}\right)$}\tabularnewline
			\hline 
			{\footnotesize{}near normal} & {\footnotesize{}$\frac{2}{3}N\left(0,1^{2}\right)+\frac{1}{3}N\left(0,2^{2}\right)$}\tabularnewline
			\hline 
			{\footnotesize{}skew} & {\footnotesize{}$\frac{1}{4}N\left(-2,2^{2}\right)+\frac{1}{4}N\left(-1,1.5^{2}\right)+\frac{1}{3}N\left(0,1^{2}\right)+\frac{1}{6}N\left(1,1^{2}\right)$}\tabularnewline
			\hline 
			{\footnotesize{}big-normal} & {\footnotesize{}$N\left(0,4^{2}\right)$}\tabularnewline
			\hline 
		\end{tabular}
		\par\end{centering}{\footnotesize \par}
	\caption{Alternative distributions for $z$-scores. \label{tab:distribution}}
\end{table}

\begin{figure}[H]
	\begin{centering}
		\includegraphics[scale=0.36]{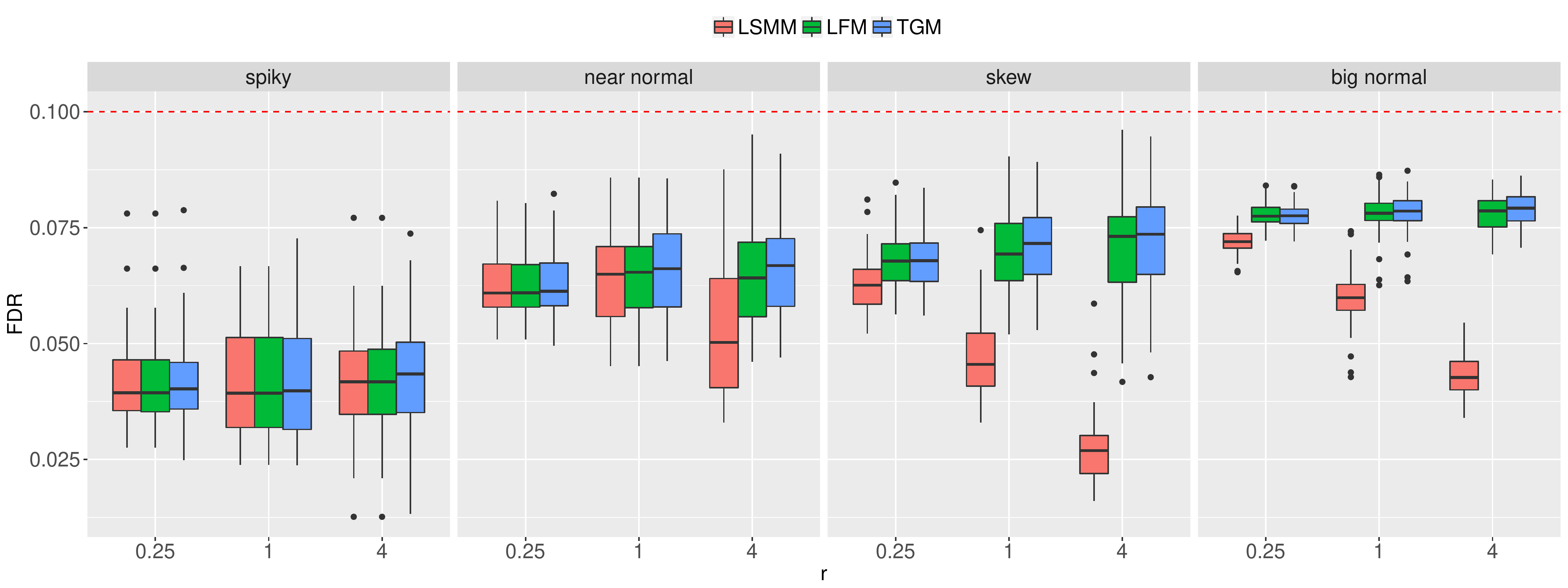}
		\par\end{centering}
	\caption{FDR of LSMM, LFM and TGM with $K=500$. We controlled global FDR at
		0.1 to evaluate empirical FDR. The results are summarized from 50
		replications.}
\end{figure}

\subsection{Comparison between LSMM and GPA}

\begin{figure}[H]
	\begin{centering}
		\includegraphics[scale=0.36]{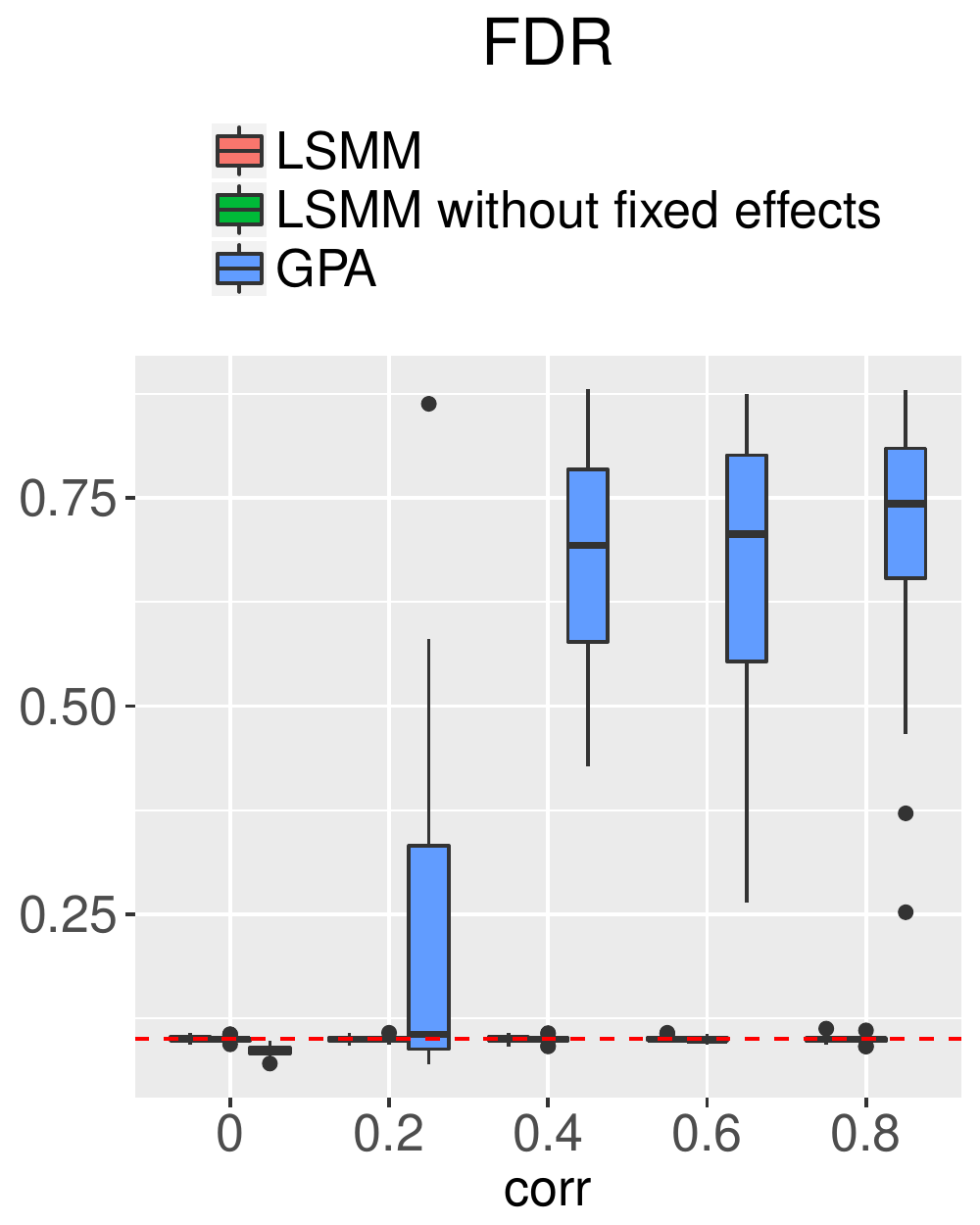}\includegraphics[scale=0.36]{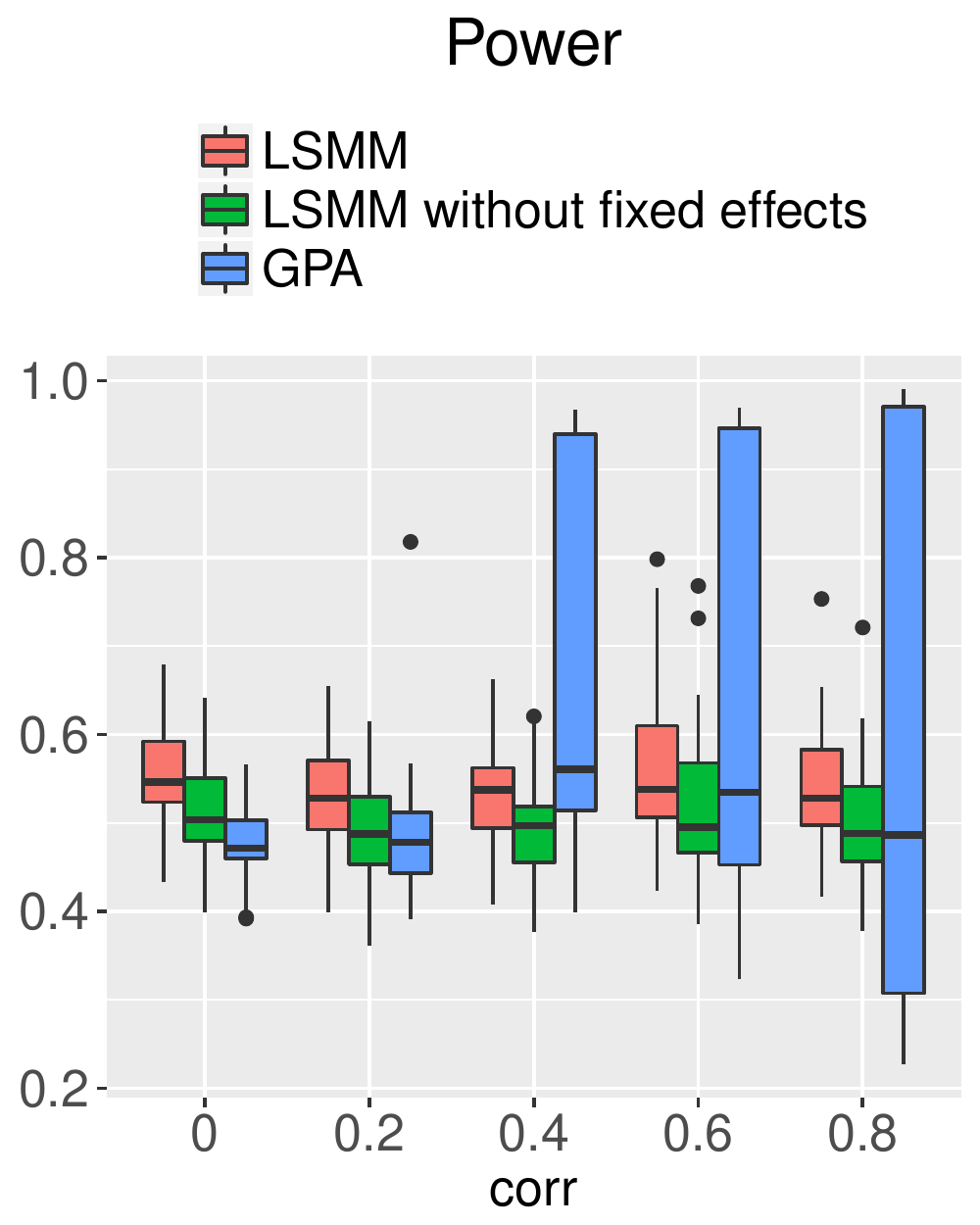}\includegraphics[scale=0.36]{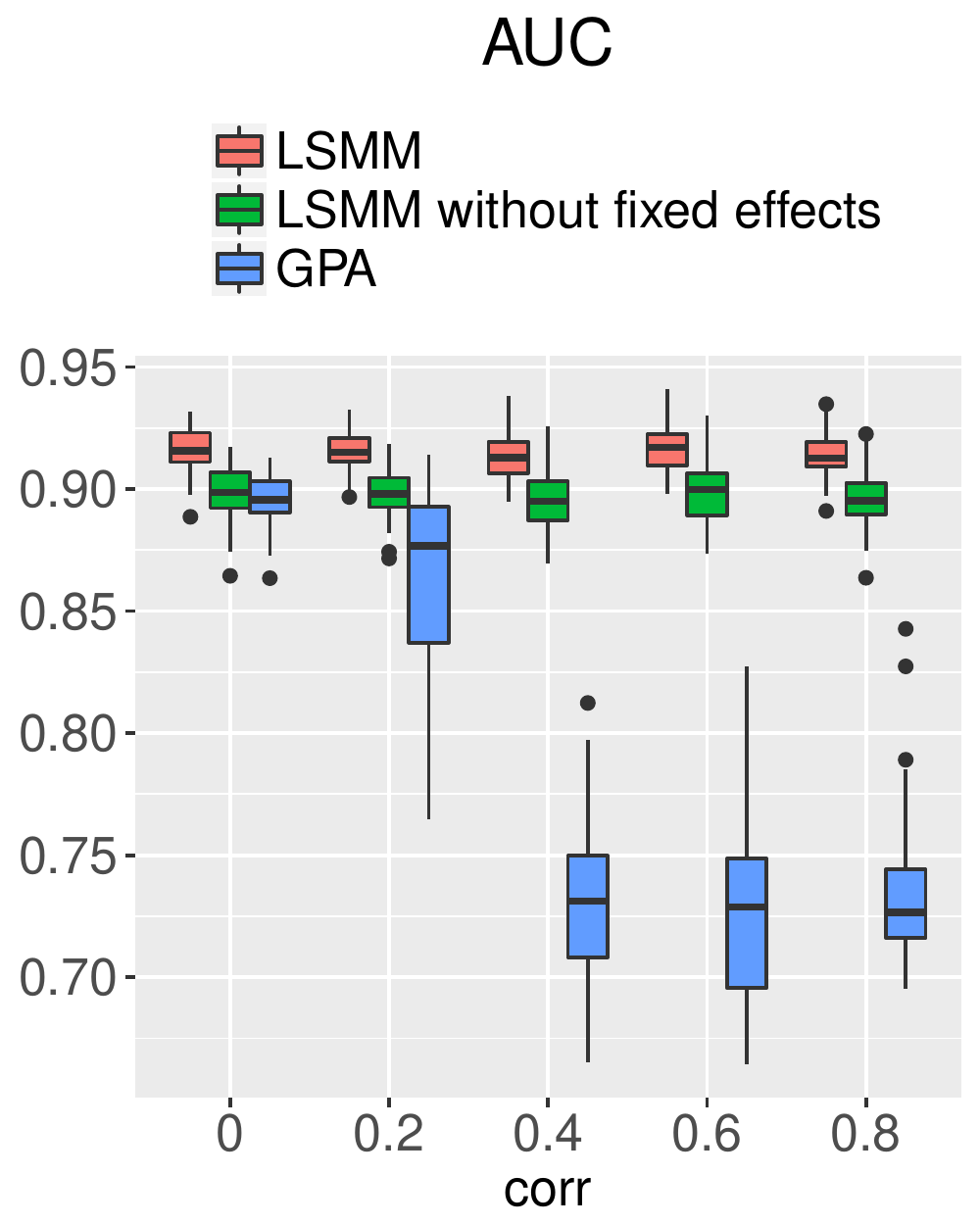}\includegraphics[scale=0.36]{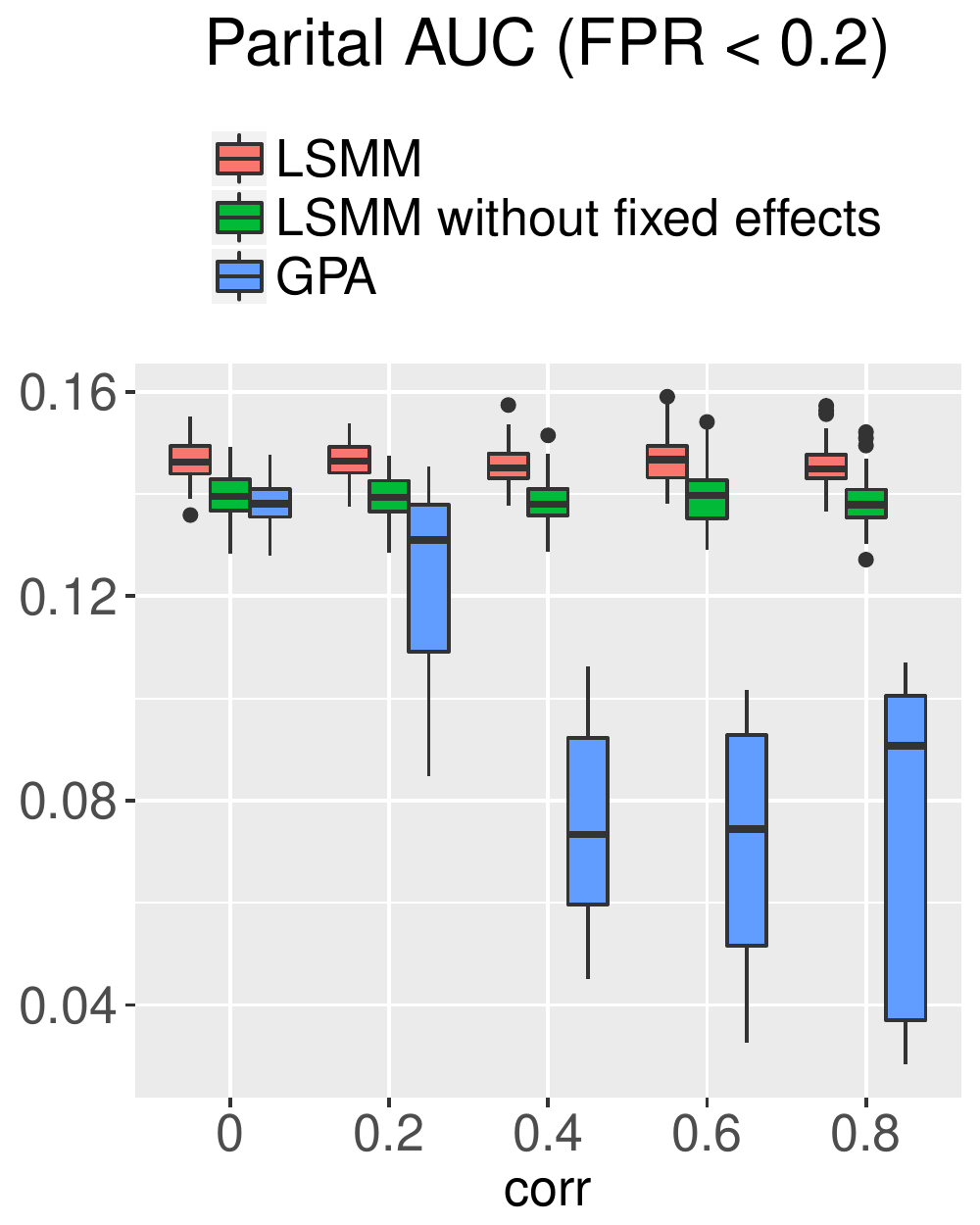}
		\par\end{centering}
	\caption{FDR, power, AUC and partial AUC LSMM, LSMM without fixed effects and
		GPA for identification of risk SNPs with $K=100$. We controlled global
		FDR at 0.1 to evaluate empirical FDR and power. The results are summarized
		from 50 replications. }
	
\end{figure}

\begin{figure}[H]
	\begin{centering}
		\includegraphics[scale=0.36]{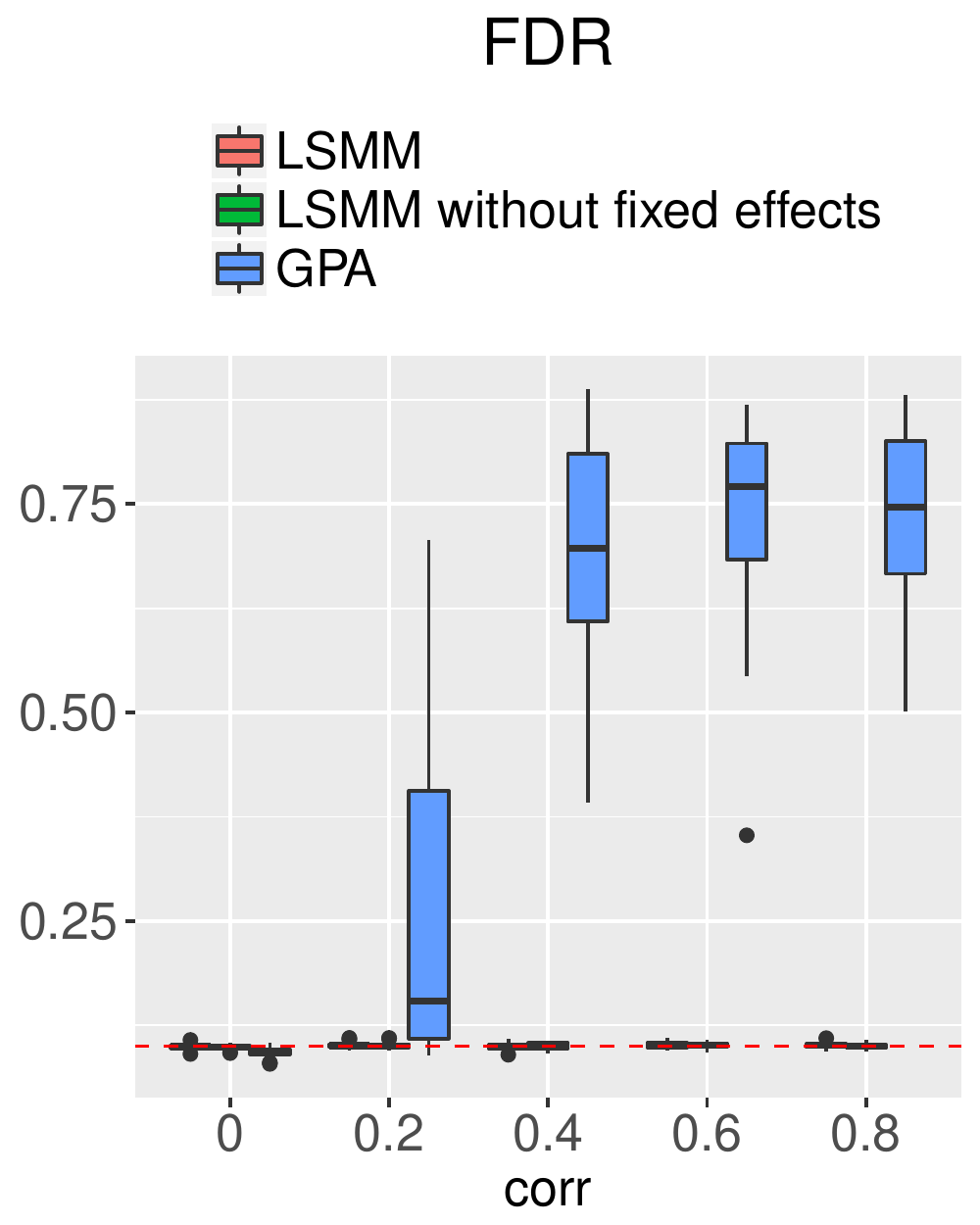}\includegraphics[scale=0.36]{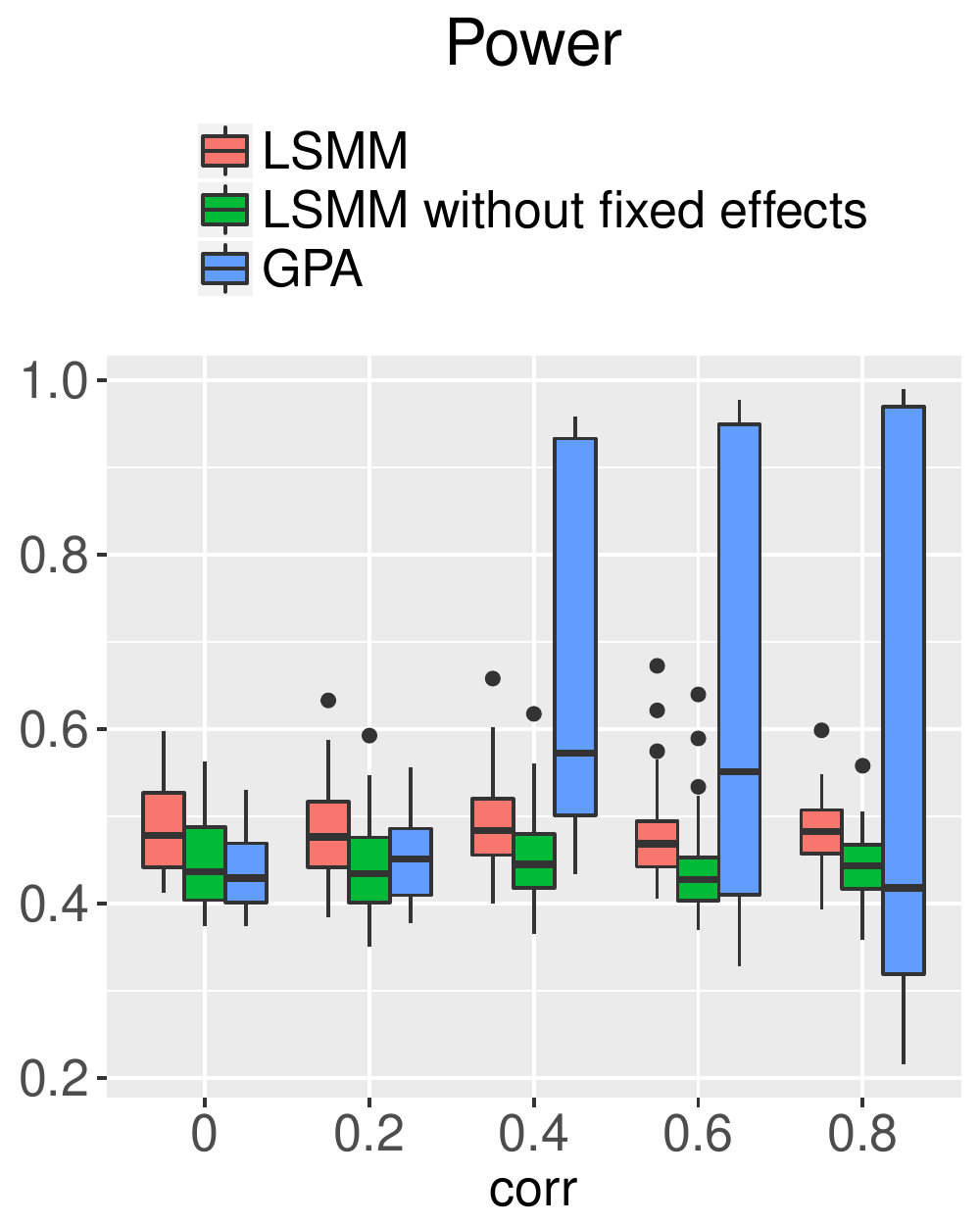}\includegraphics[scale=0.36]{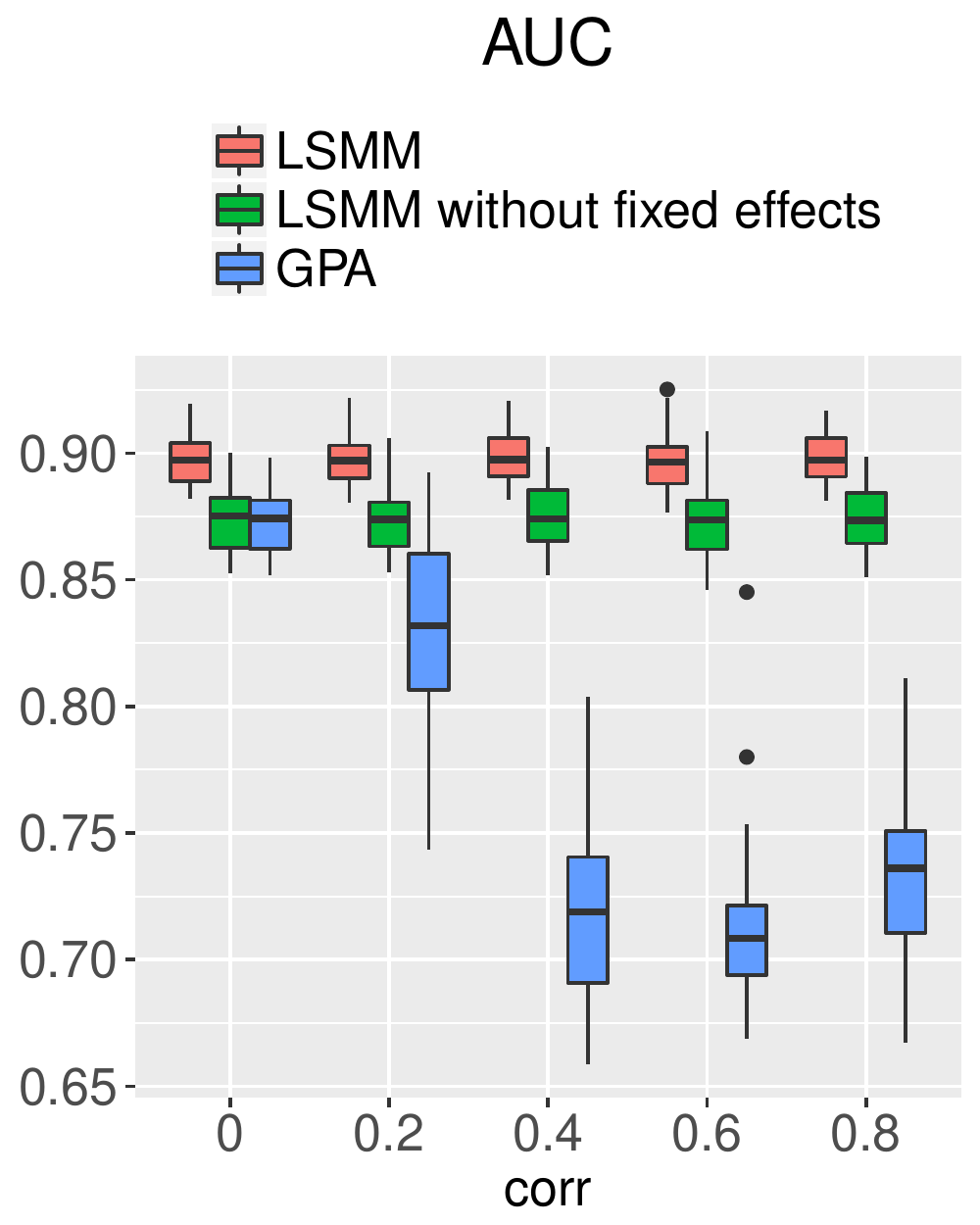}\includegraphics[scale=0.36]{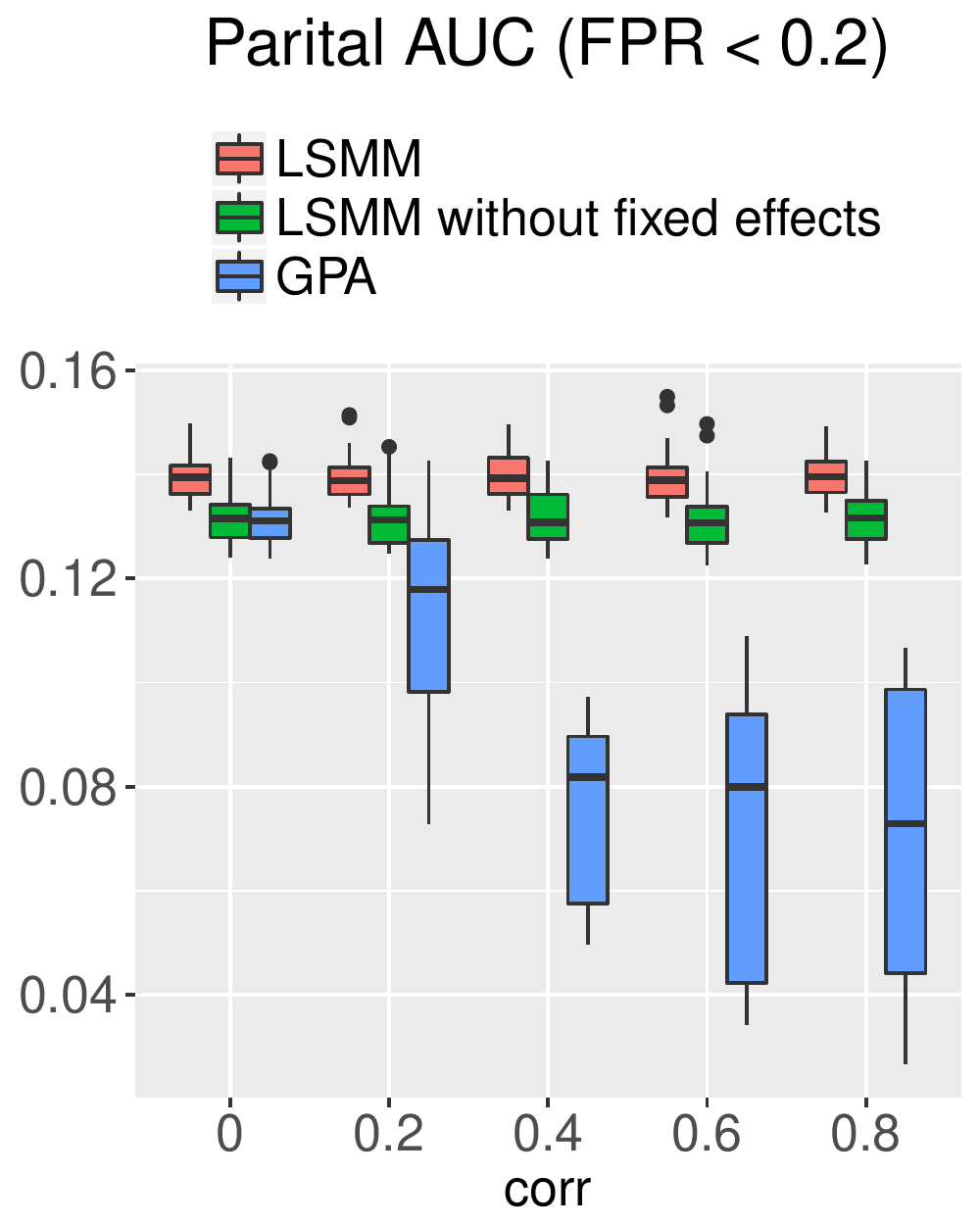}
		\par\end{centering}
	\caption{FDR, power, AUC and partial AUC LSMM, LSMM without fixed effects and
		GPA for identification of risk SNPsn with $K=50$. We controlled global
		FDR at 0.1 to evaluate empirical FDR and power. The results are summarized
		from 50 replications. }
\end{figure}

\begin{figure}[H]
	\begin{centering}
		\includegraphics[scale=0.36]{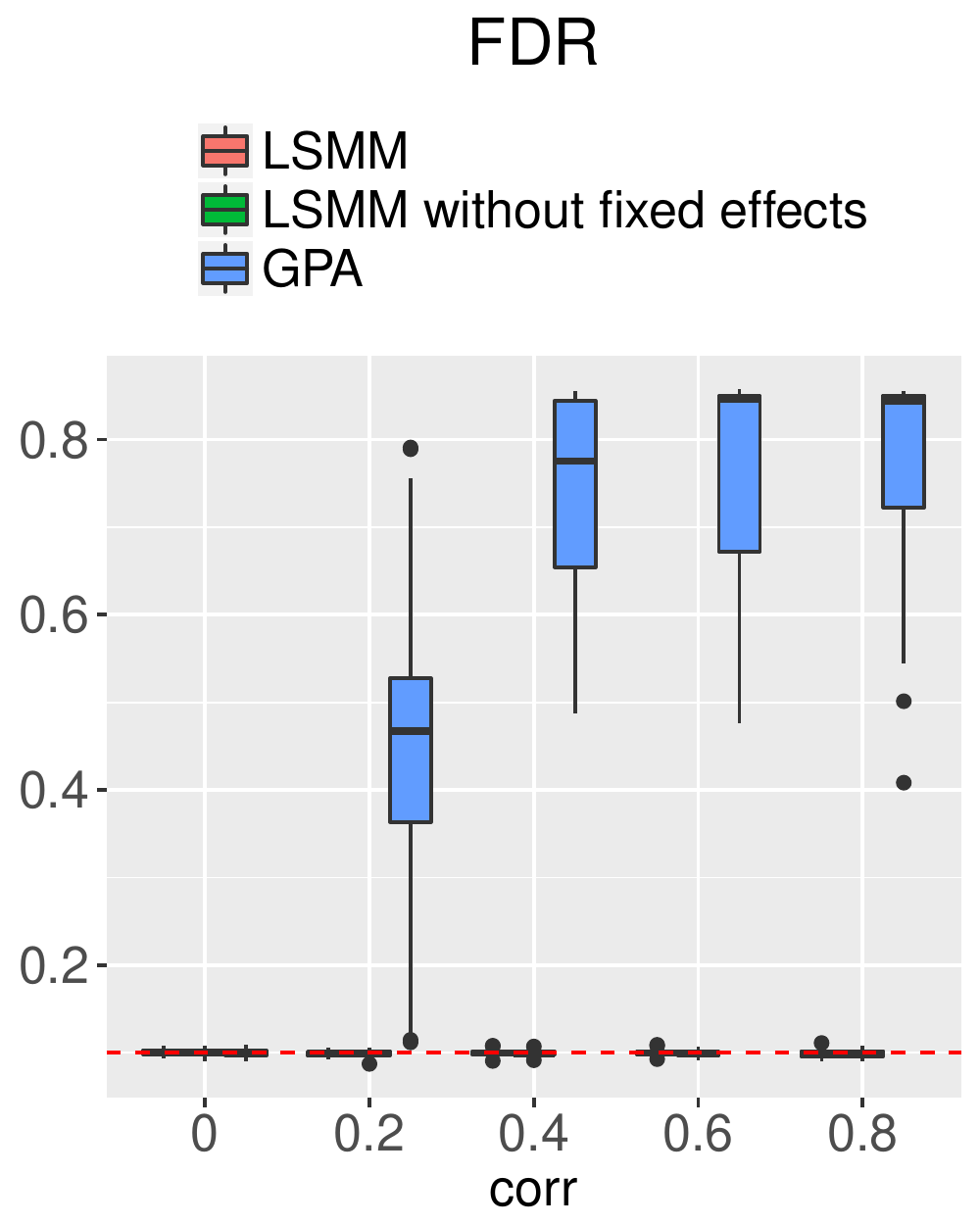}\includegraphics[scale=0.36]{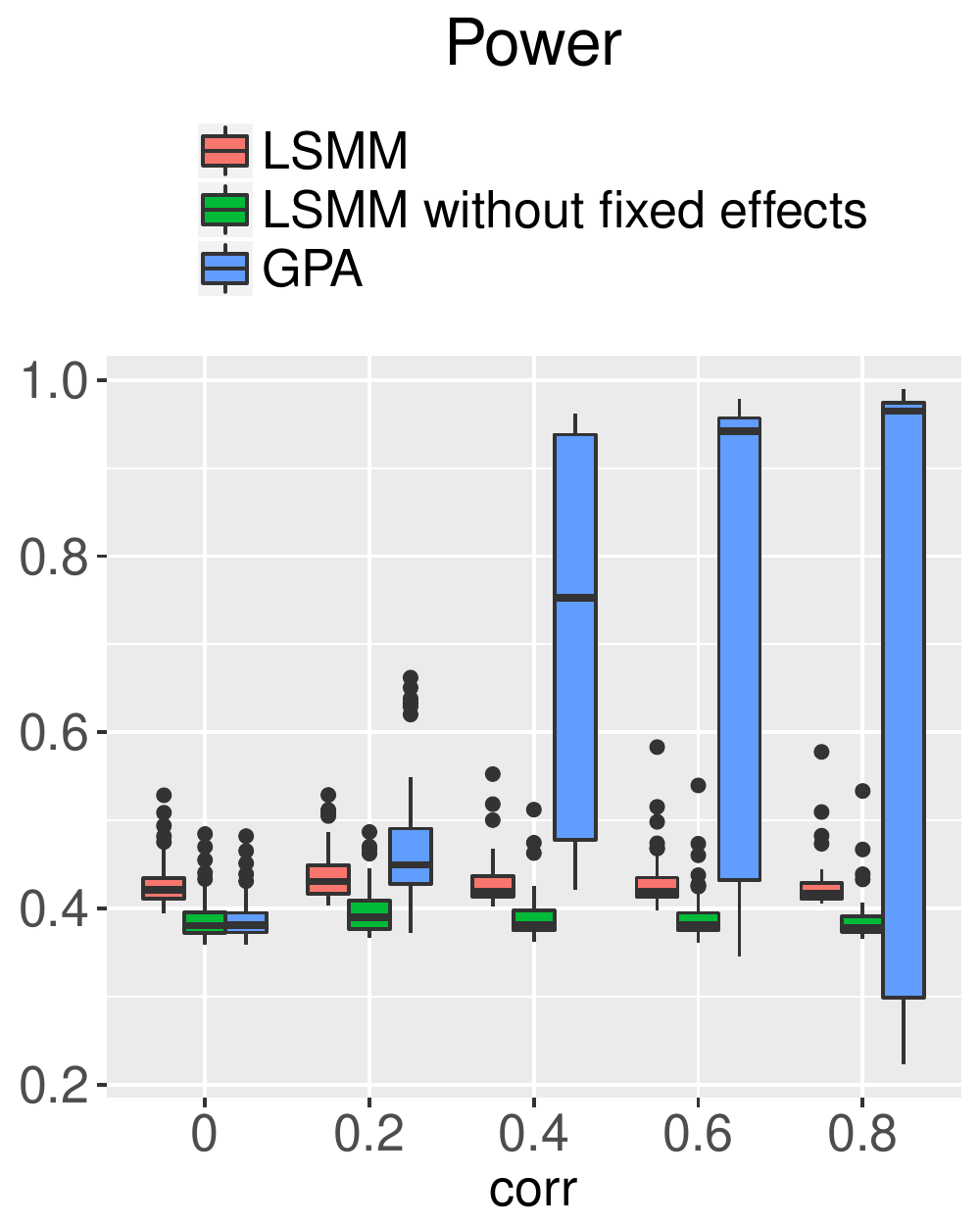}\includegraphics[scale=0.36]{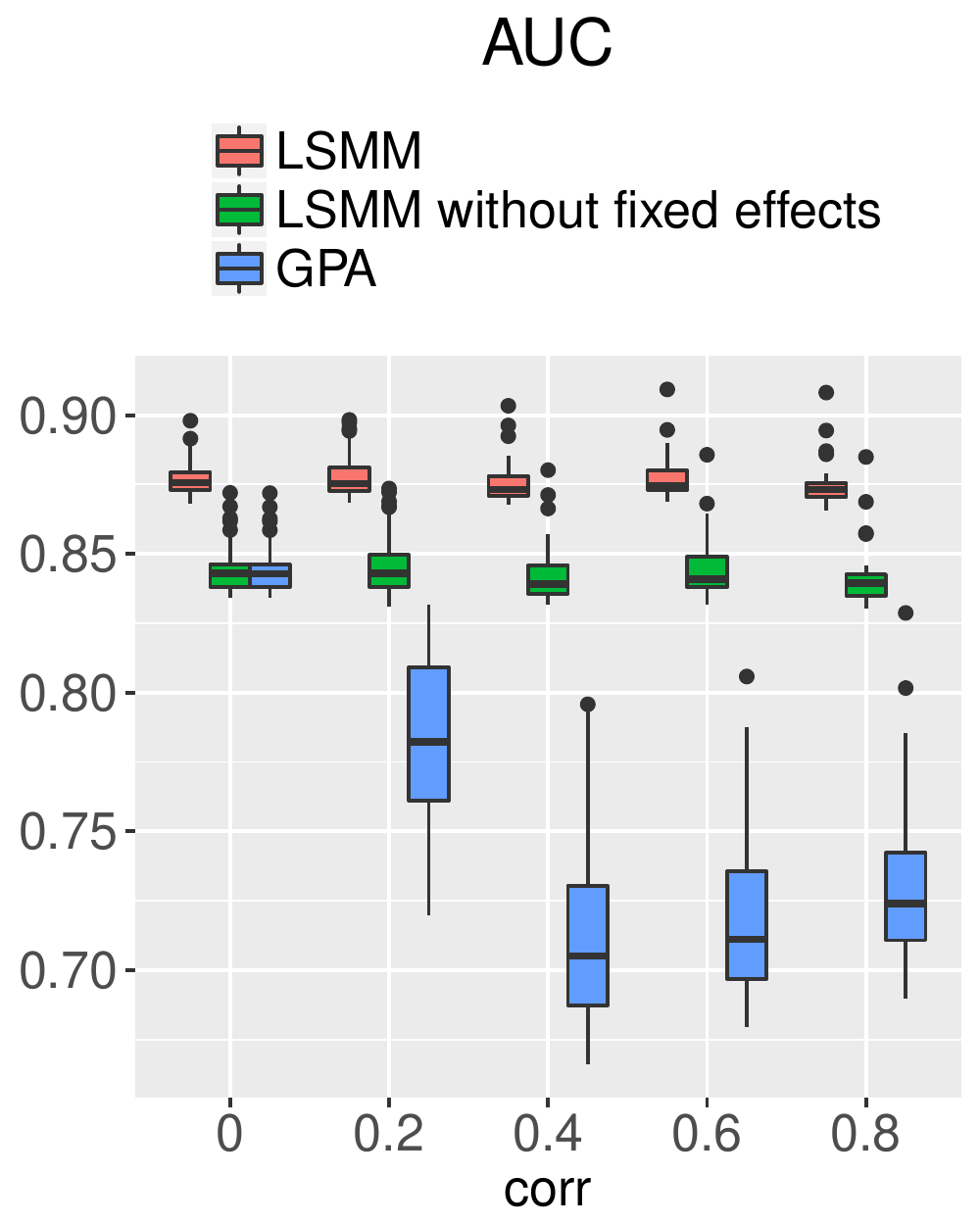}\includegraphics[scale=0.36]{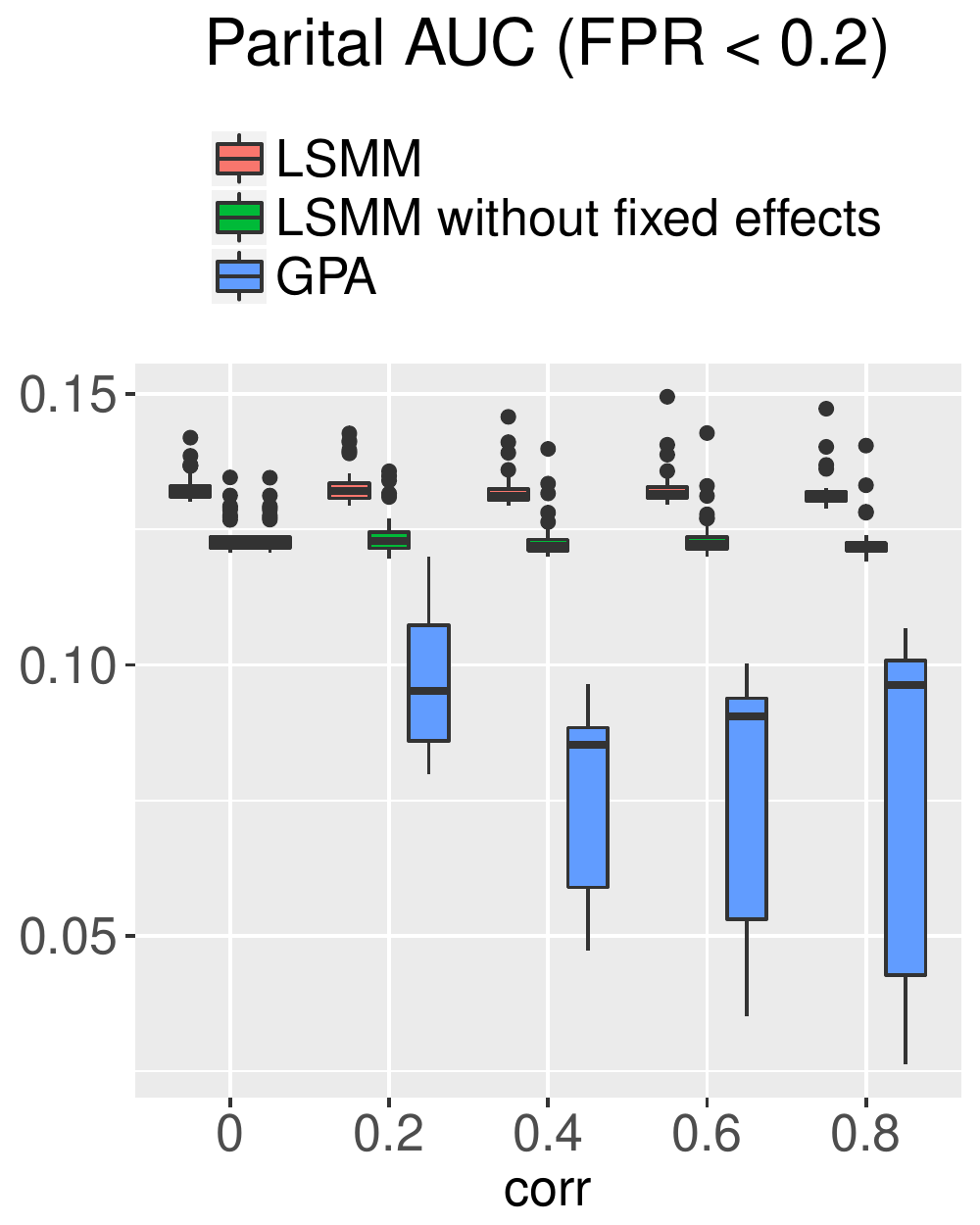}
		\par\end{centering}
	\caption{FDR, power, AUC and partial AUC LSMM, LSMM without fixed effects and
		GPA for identification of risk SNPs with $K=10$. We controlled global
		FDR at 0.1 to evaluate empirical FDR and power. The results are summarized
		from 50 replications. }
\end{figure}

\subsection{Comparison between LSMM and cmfdr}

\begin{figure}[H]
	
	\begin{centering}
		\includegraphics[scale=0.5]{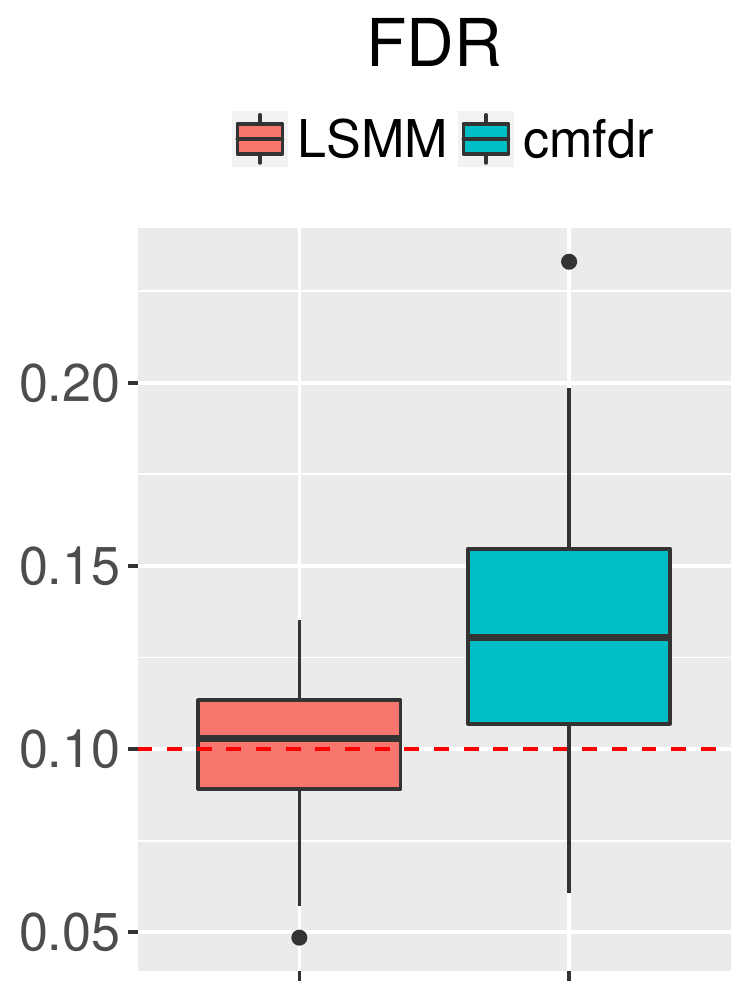}\includegraphics[scale=0.5]{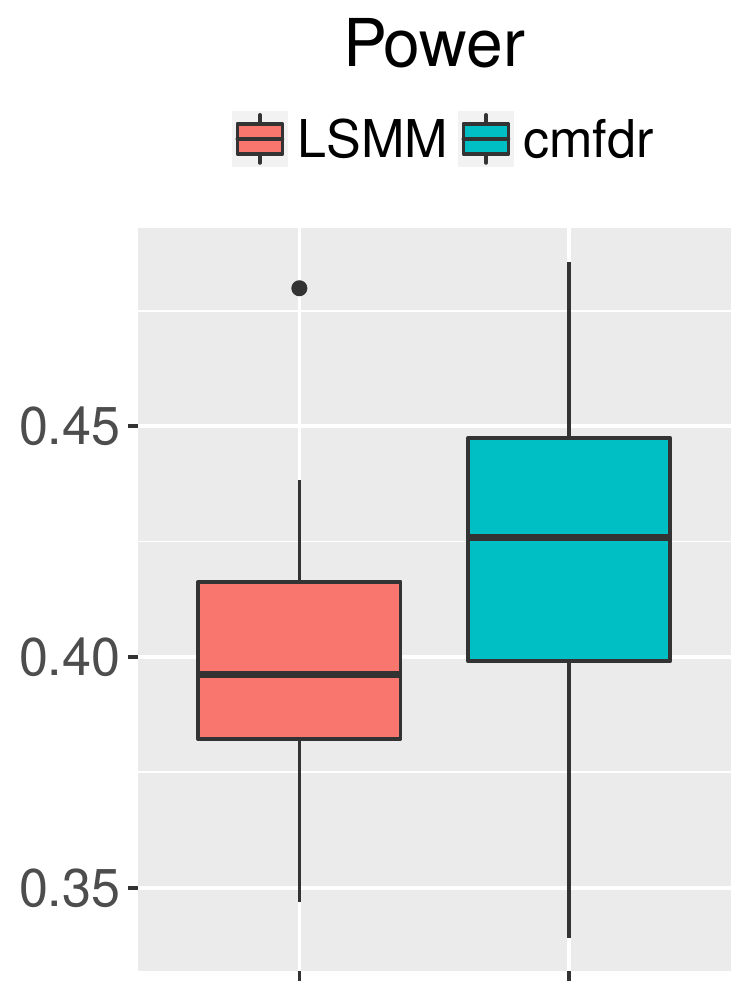}\includegraphics[scale=0.5]{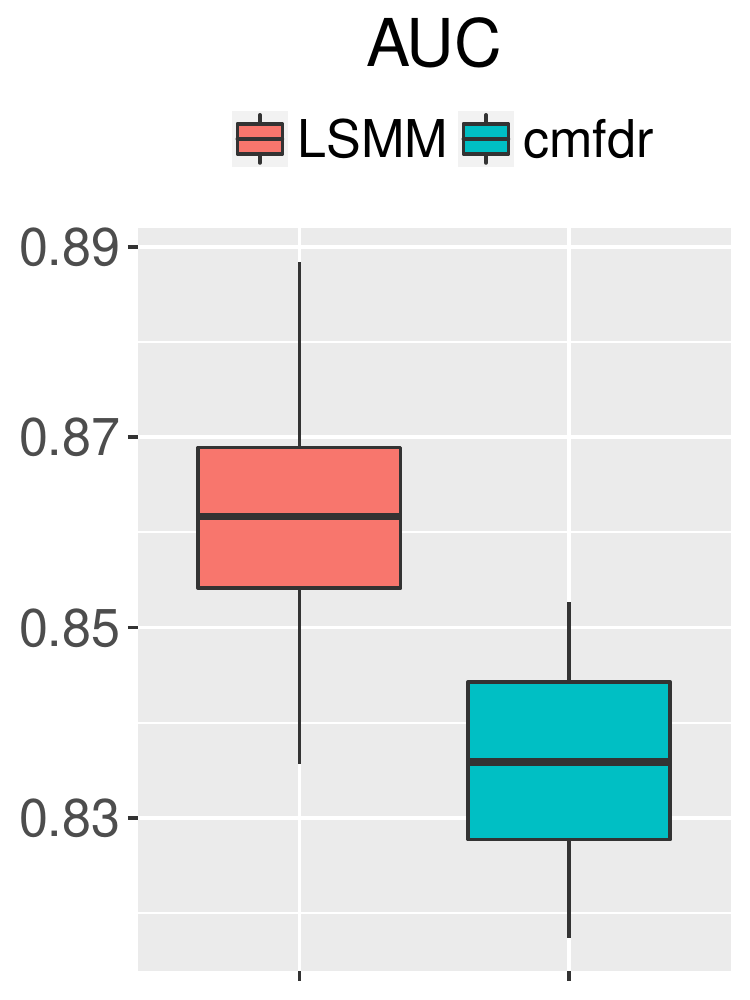}\includegraphics[scale=0.5]{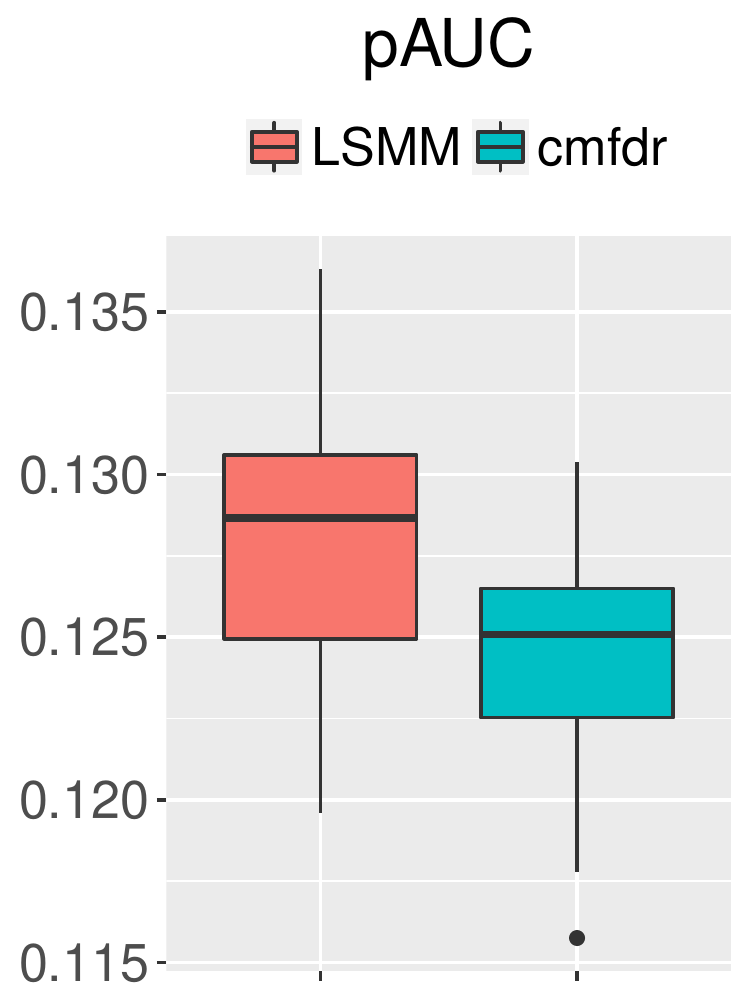}
		\par\end{centering}
	\caption{FDR, power, AUC and partial AUC of LSMM and cmfdr for identification
		of risk SNPs. We controlled global FDR at 0.1 to evaluate empirical
		FDR and power. The results are summarized from 50 replications.\label{fig:cmfdr}}
	
\end{figure}

We compared LSMM with cmfdr. As cmfdr is not able to handle a large
number of covariates and the MCMC sampling algorithm it derived is
time-consuming, we set $M=5000$, $L=5$, $K=5$ and run 2500 iterations
with 2000 retained draws for cmfdr. The comparison between LSMM and
cmfdr are shown in Figure \ref{fig:cmfdr}.

\subsection{Estimation of parameters}

\subsubsection{Estimation of $\alpha$}

\begin{figure}[H]
	\begin{centering}
		\includegraphics[scale=0.36]{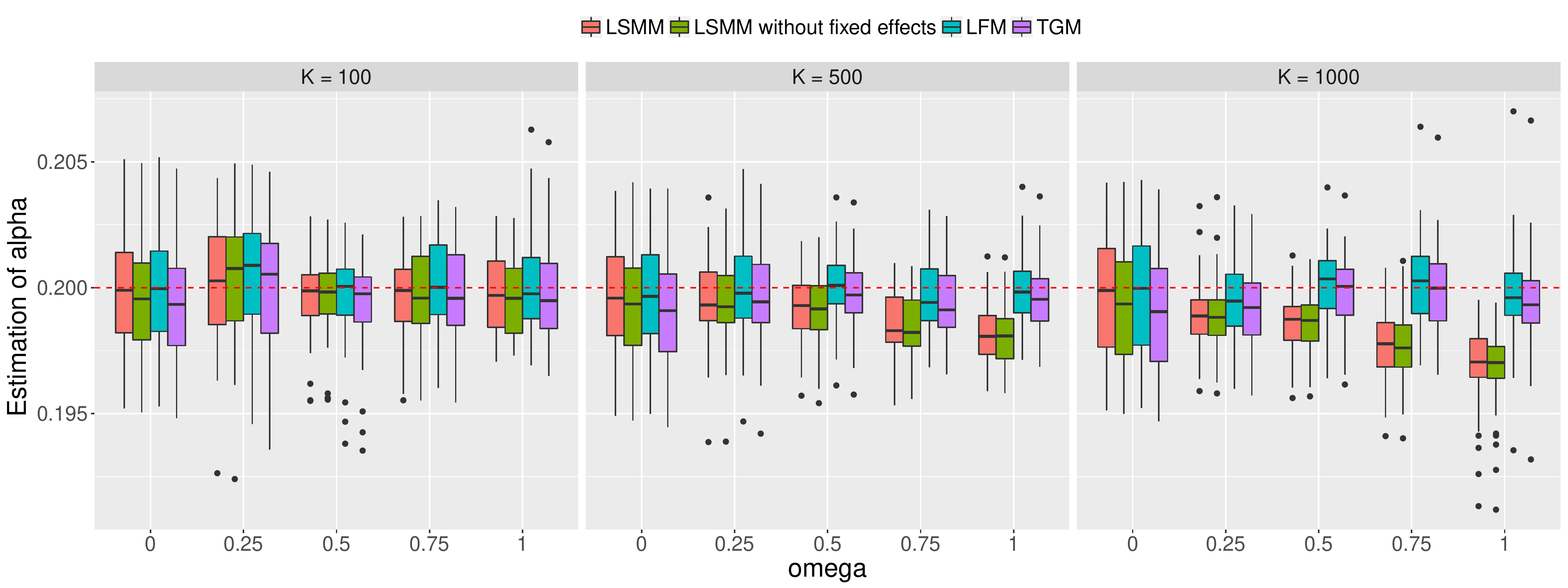}
		\par\end{centering}
	\caption{Perfermance in estimation of parameter $\alpha$ when the true $\alpha=0.2$.\label{fig:est alpha 0.2}}
\end{figure}

\begin{figure}[H]
	\begin{centering}
		\includegraphics[scale=0.36]{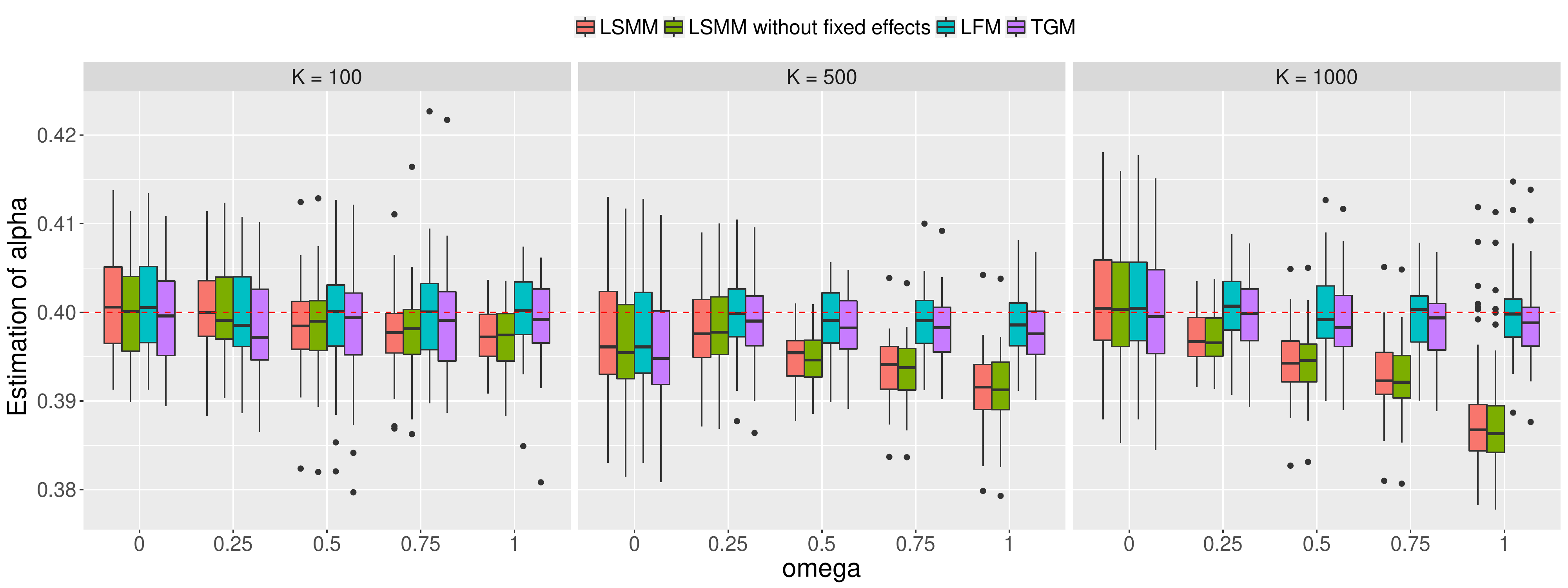}
		\par\end{centering}
	\caption{Perfermance in estimation of parameter $\alpha$ when the true $\alpha=0.4$.\label{fig:est alpha 0.4}}
\end{figure}

\begin{figure}[H]
	\begin{centering}
		\includegraphics[scale=0.36]{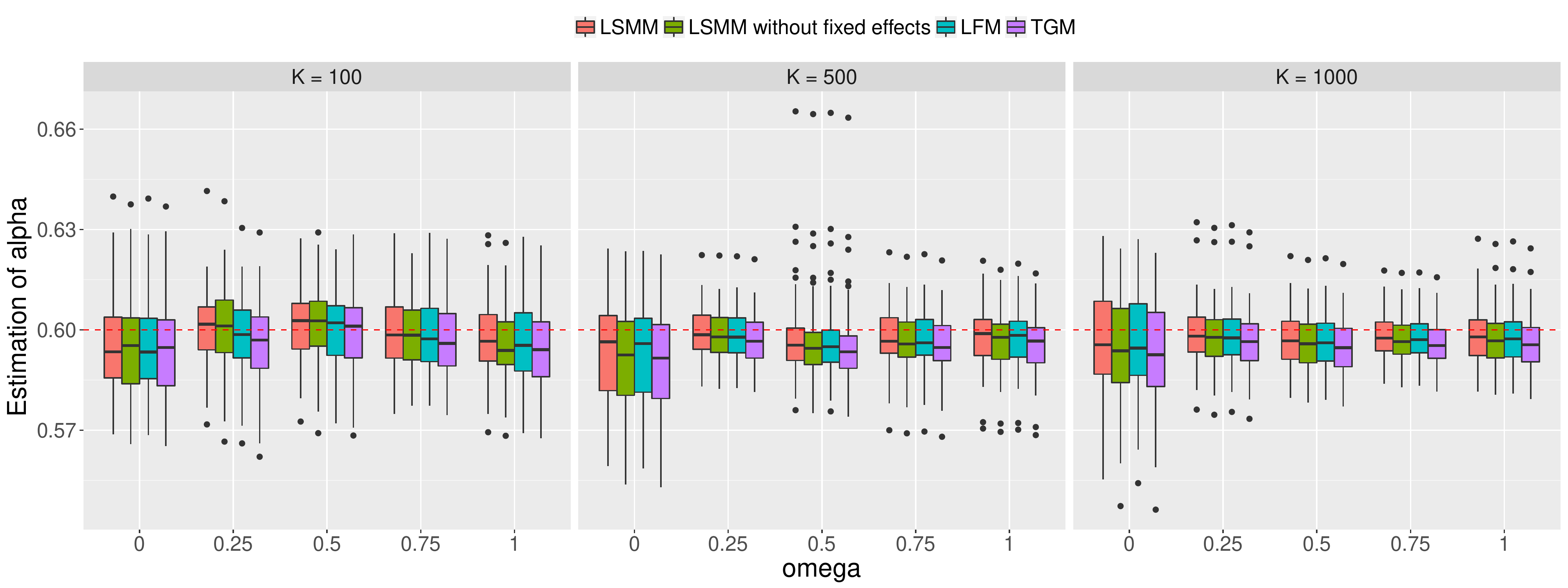}
		\par\end{centering}
	\caption{Perfermance in estimation of parameter $\alpha$ when the true $\alpha=0.6$.\label{fig:est alpha 0.6}}
\end{figure}

We evaluate the performance of LSMM in estimation of parameter $\alpha$
in the beta distribution. We compare LSMM with the other three methods,
TGM (without fixed effects and random effects), LFM (with only fixed
effects) and LSMM without fixed effects. We varied $\omega$ at $\left\{ 0,0.25,0.5,0.75,1\right\} $.
Figures \ref{fig:est alpha 0.2}-\ref{fig:est alpha 0.6} show the
comparision among these methods with $\alpha=0.2$, $0.4$ and $0.6$
respectively.

\subsubsection{Estimation of $\boldsymbol{b}$}

\begin{figure}[H]
	\begin{centering}
		\includegraphics[scale=0.36]{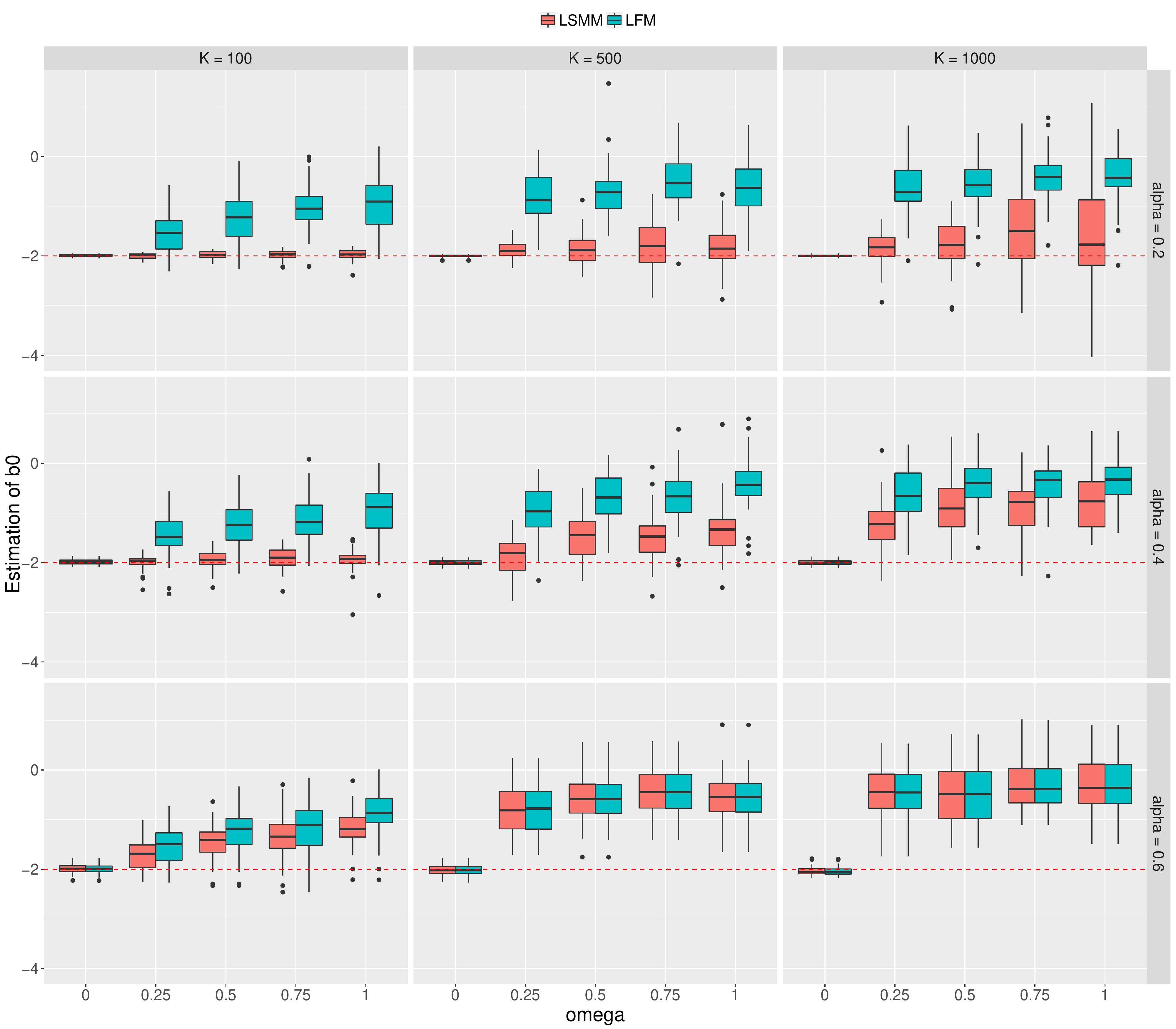}
		\par\end{centering}
	\caption{Perfermance in estimation of parameter $b_{0}$.\label{fig:est b0}}
\end{figure}

\begin{figure}[H]
	\begin{centering}
		\includegraphics[scale=0.36]{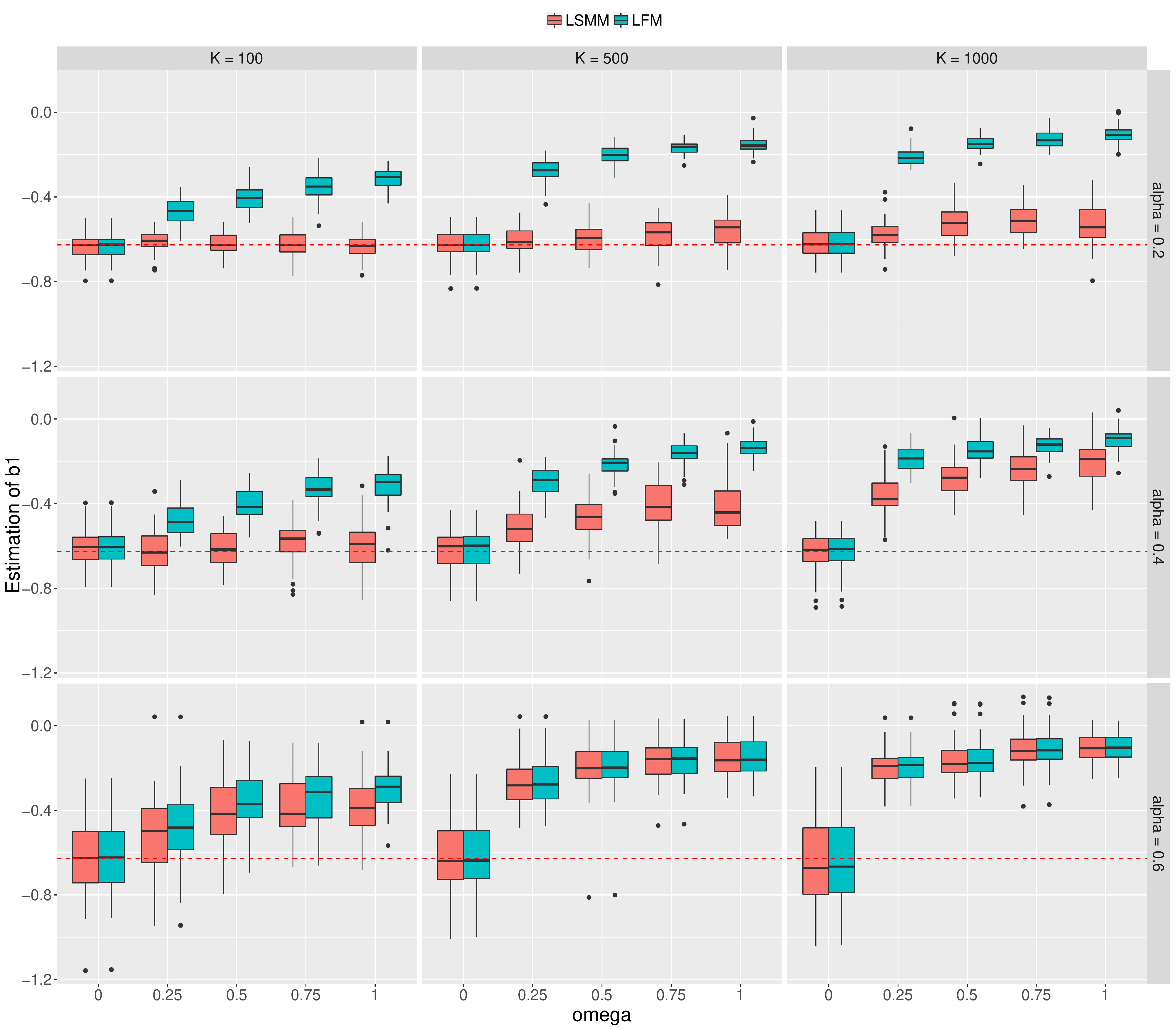}
		\par\end{centering}
	\caption{Perfermance in estimation of parameter $b_{1}$.\label{fig:est b1}}
	
\end{figure}

\begin{figure}[H]
	\begin{centering}
		\includegraphics[scale=0.36]{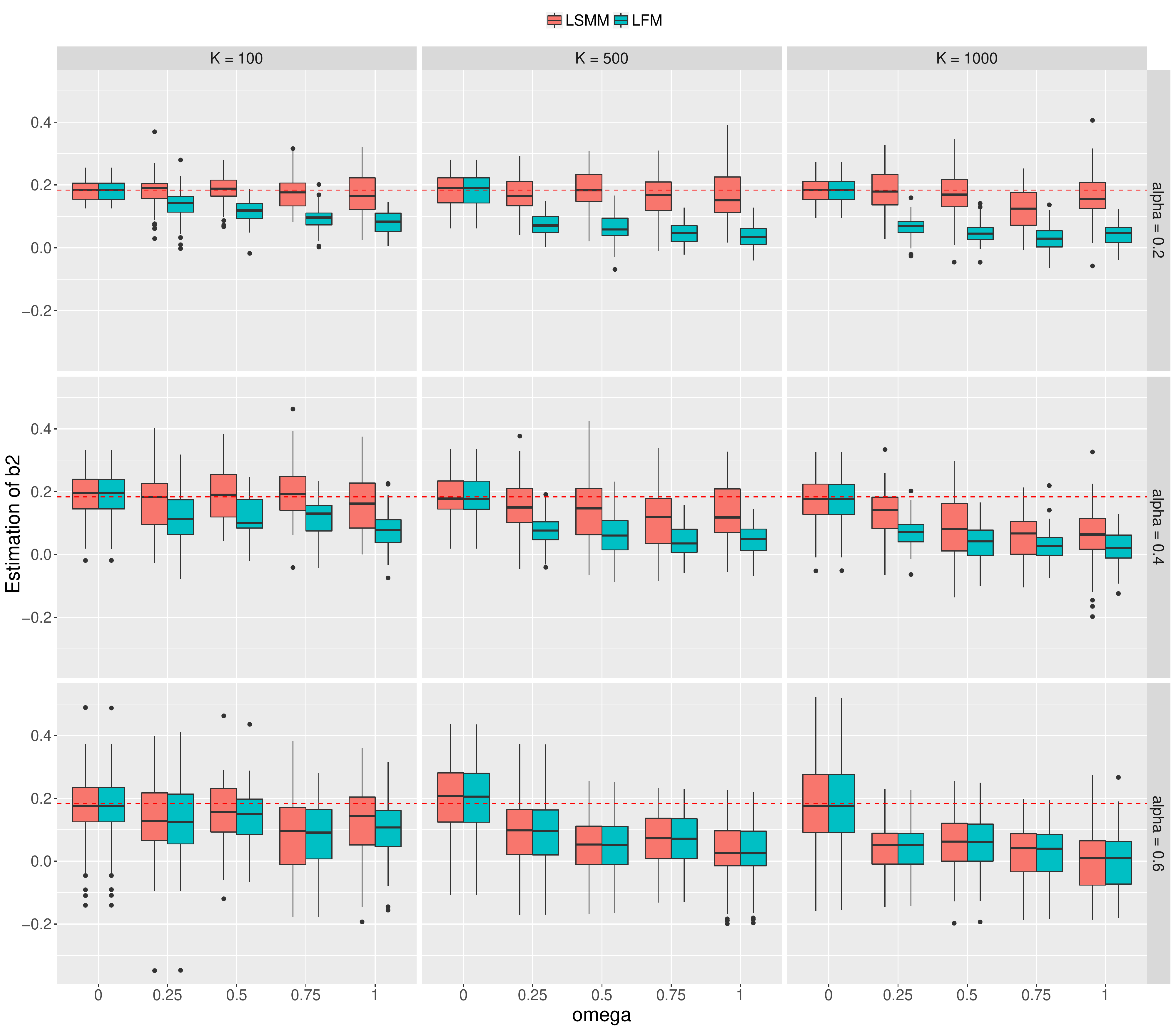}
		\par\end{centering}
	\caption{Perfermance in estimation of parameter $b_{2}$.\label{fig:est b2}}
\end{figure}

\begin{figure}[H]
	\begin{centering}
		\includegraphics[scale=0.36]{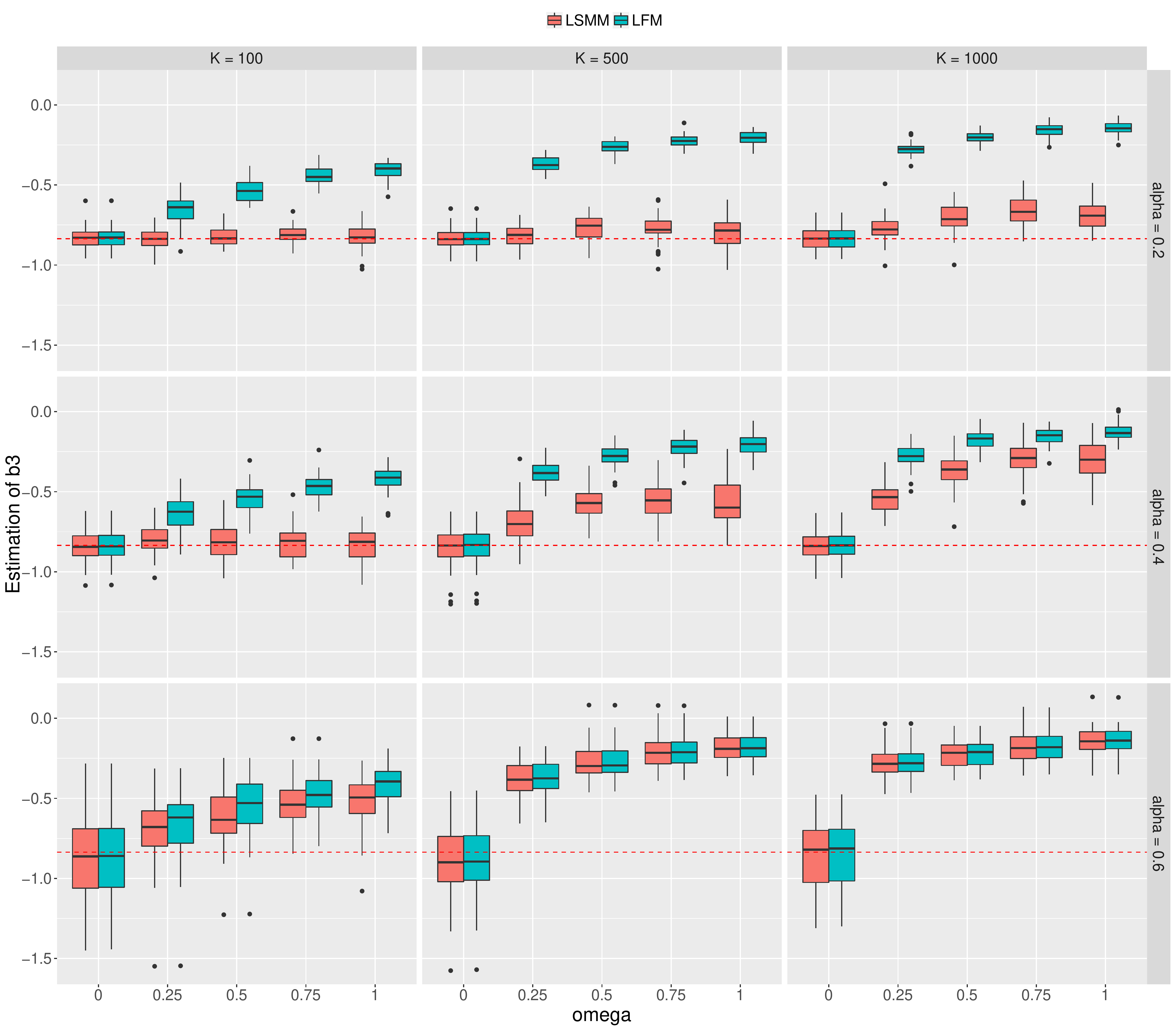}
		\par\end{centering}
	\caption{Perfermance in estimation of parameter $b_{3}$.\label{fig:est b3}}
\end{figure}

\begin{figure}[H]
	\begin{centering}
		\includegraphics[scale=0.36]{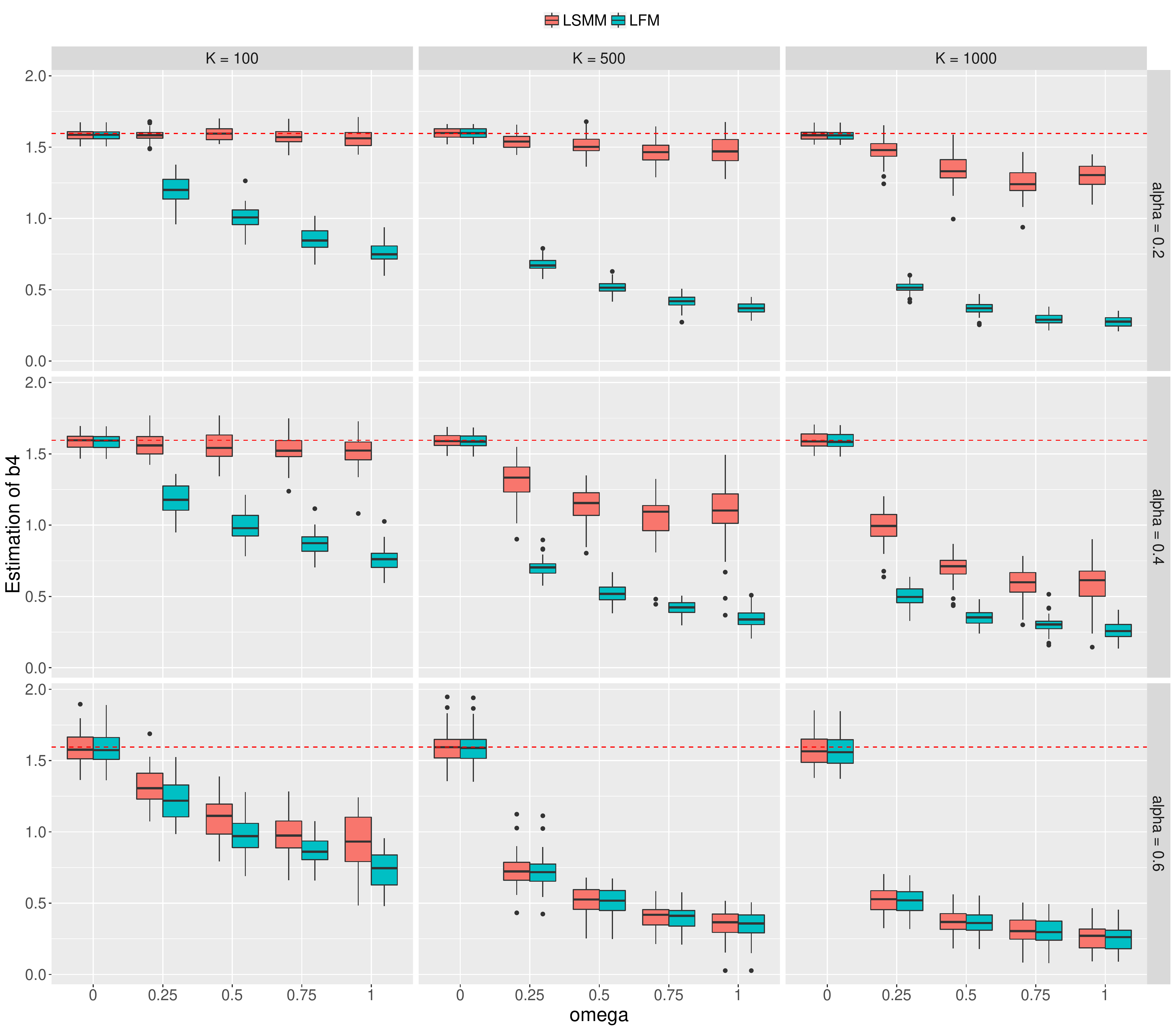}
		\par\end{centering}
	\caption{Perfermance in estimation of parameter $b_{4}$.\label{fig:est b4}}
\end{figure}

\begin{figure}[H]
	\begin{centering}
		\includegraphics[scale=0.36]{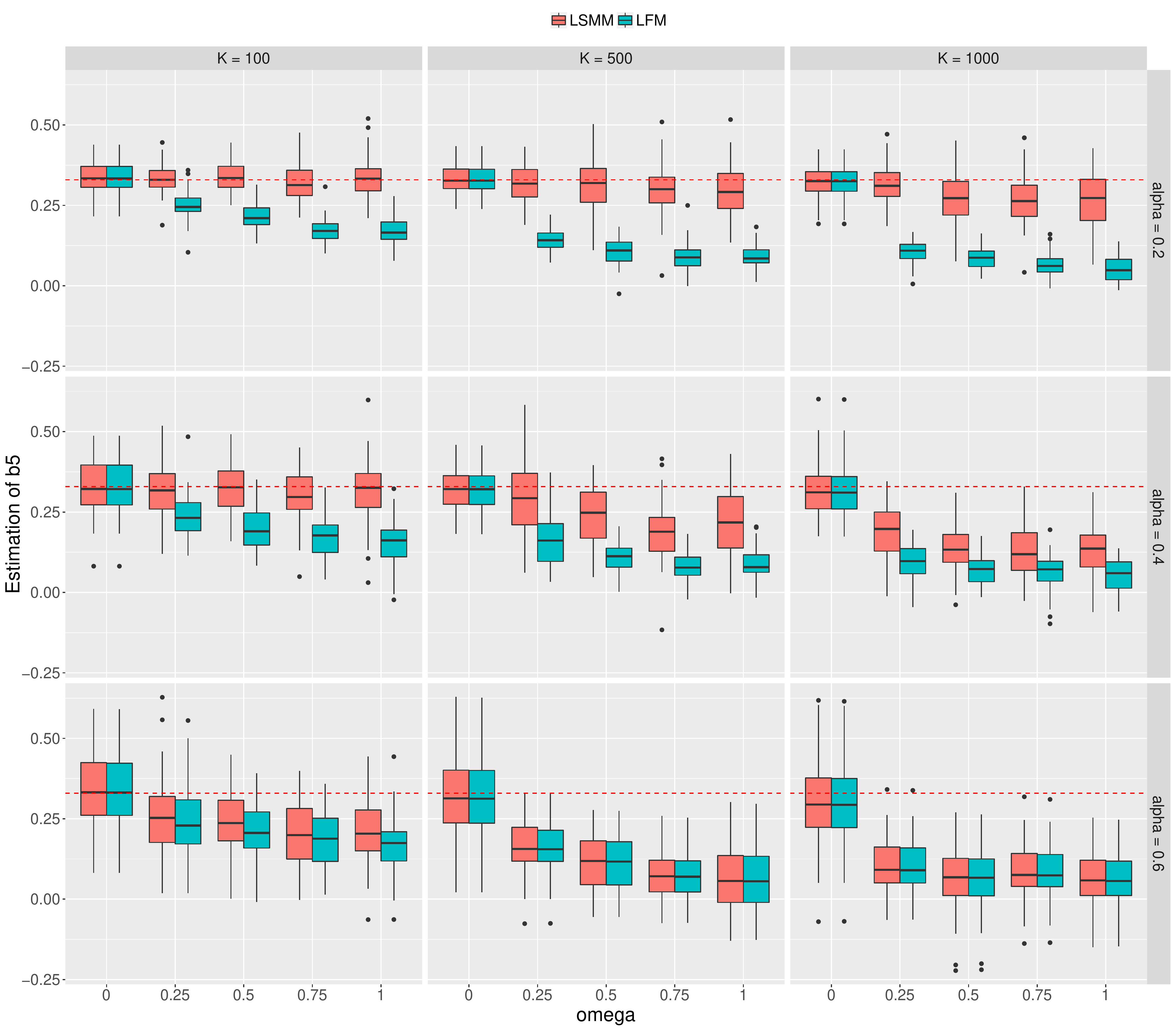}
		\par\end{centering}
	\caption{Perfermance in estimation of parameter $b_{5}$.\label{fig:est b5}}
\end{figure}

\begin{figure}[H]
	\begin{centering}
		\includegraphics[scale=0.36]{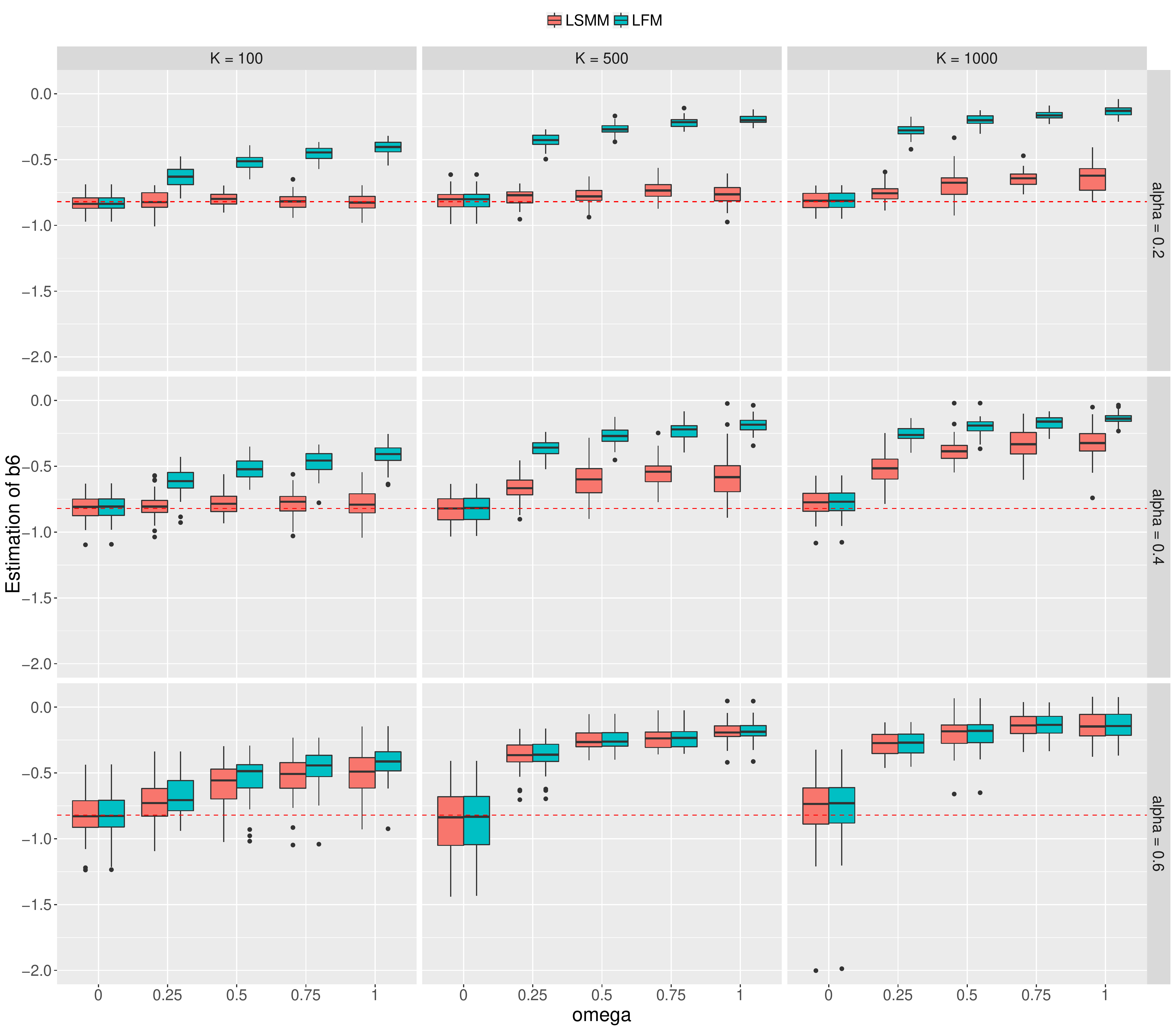}
		\par\end{centering}
	\caption{Perfermance in estimation of parameter $b_{6}$.\label{fig:est b6}}
\end{figure}

\begin{figure}[H]
	\begin{centering}
		\includegraphics[scale=0.36]{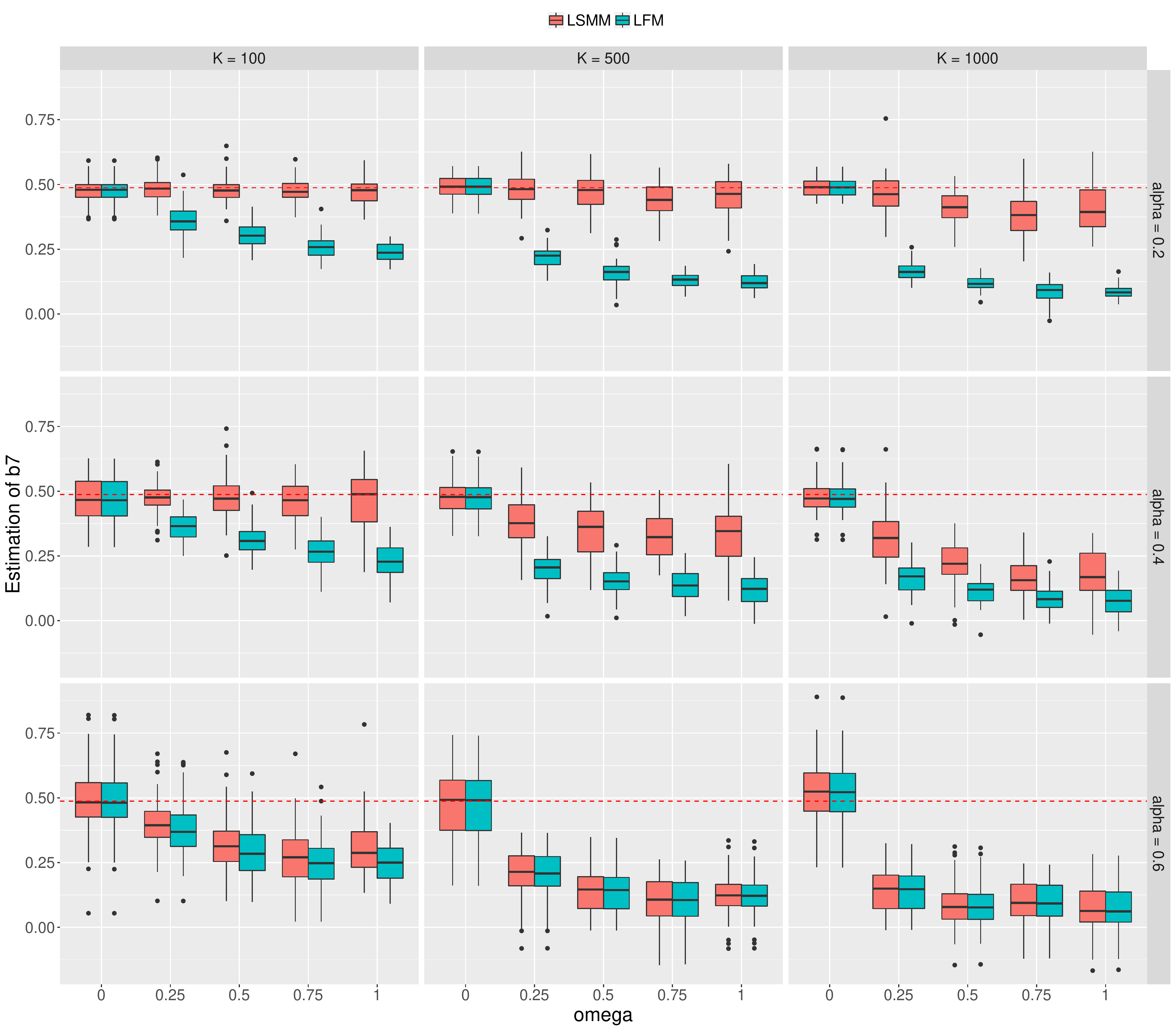}
		\par\end{centering}
	\caption{Perfermance in estimation of parameter $b_{7}$.\label{fig:est b7}}
\end{figure}

\begin{figure}[H]
	\begin{centering}
		\includegraphics[scale=0.36]{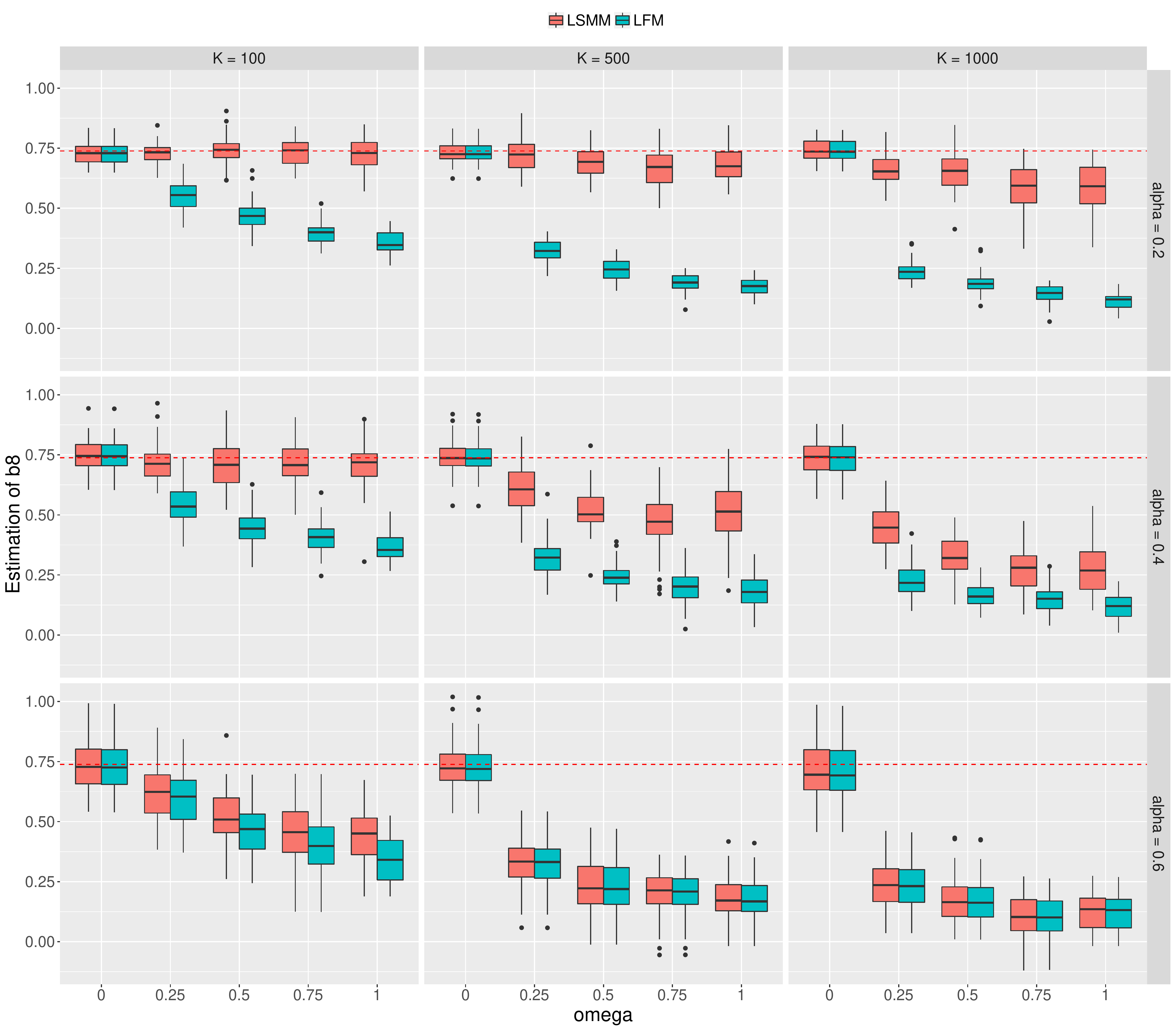}
		\par\end{centering}
	\caption{Perfermance in estimation of parameter $b_{8}$.\label{fig:est b8}}
\end{figure}

\begin{figure}[H]
	\begin{centering}
		\includegraphics[scale=0.36]{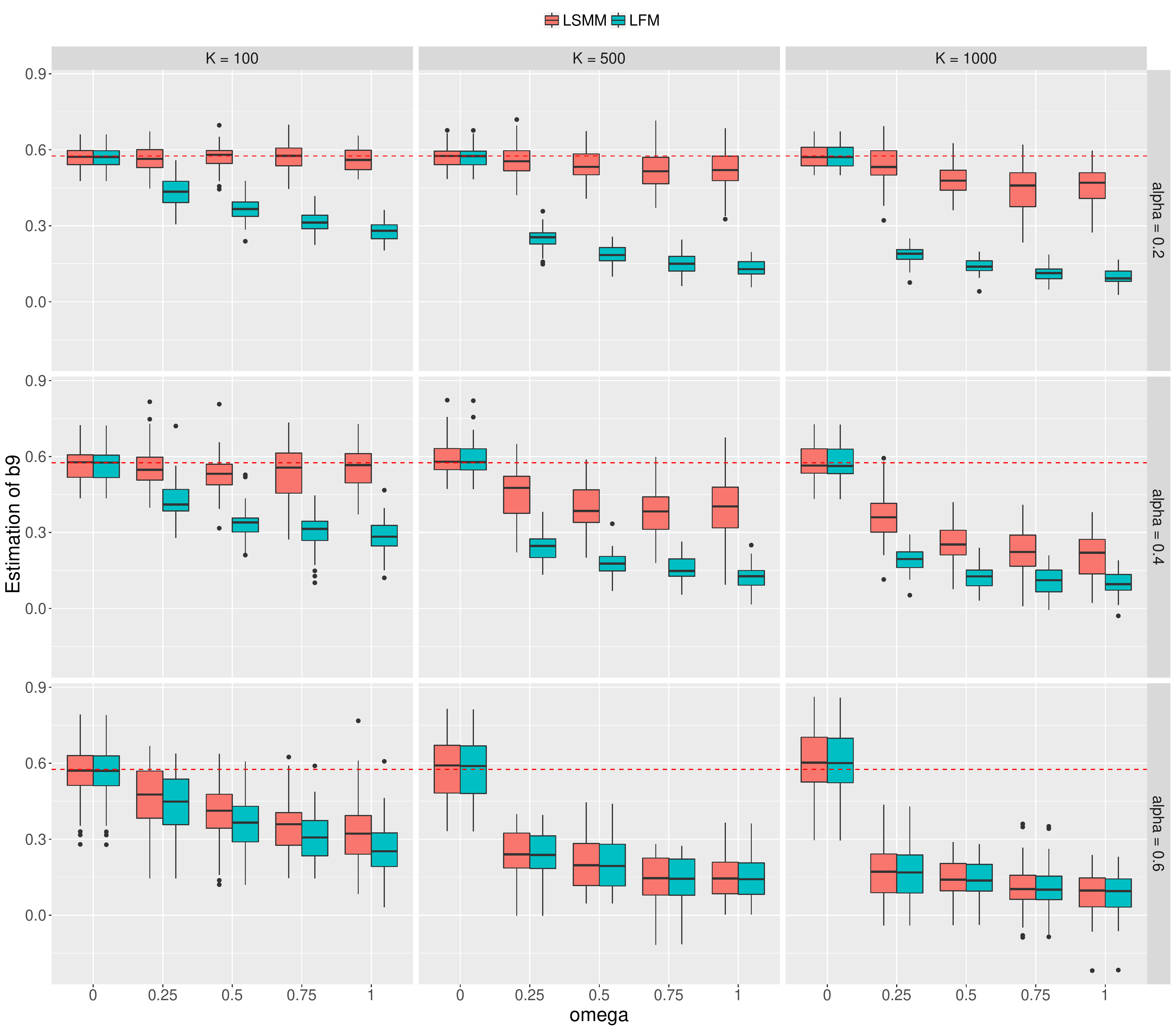}
		\par\end{centering}
	\caption{Perfermance in estimation of parameter $b_{9}$.\label{fig:est b9}}
\end{figure}

\begin{figure}[H]
	\begin{centering}
		\includegraphics[scale=0.36]{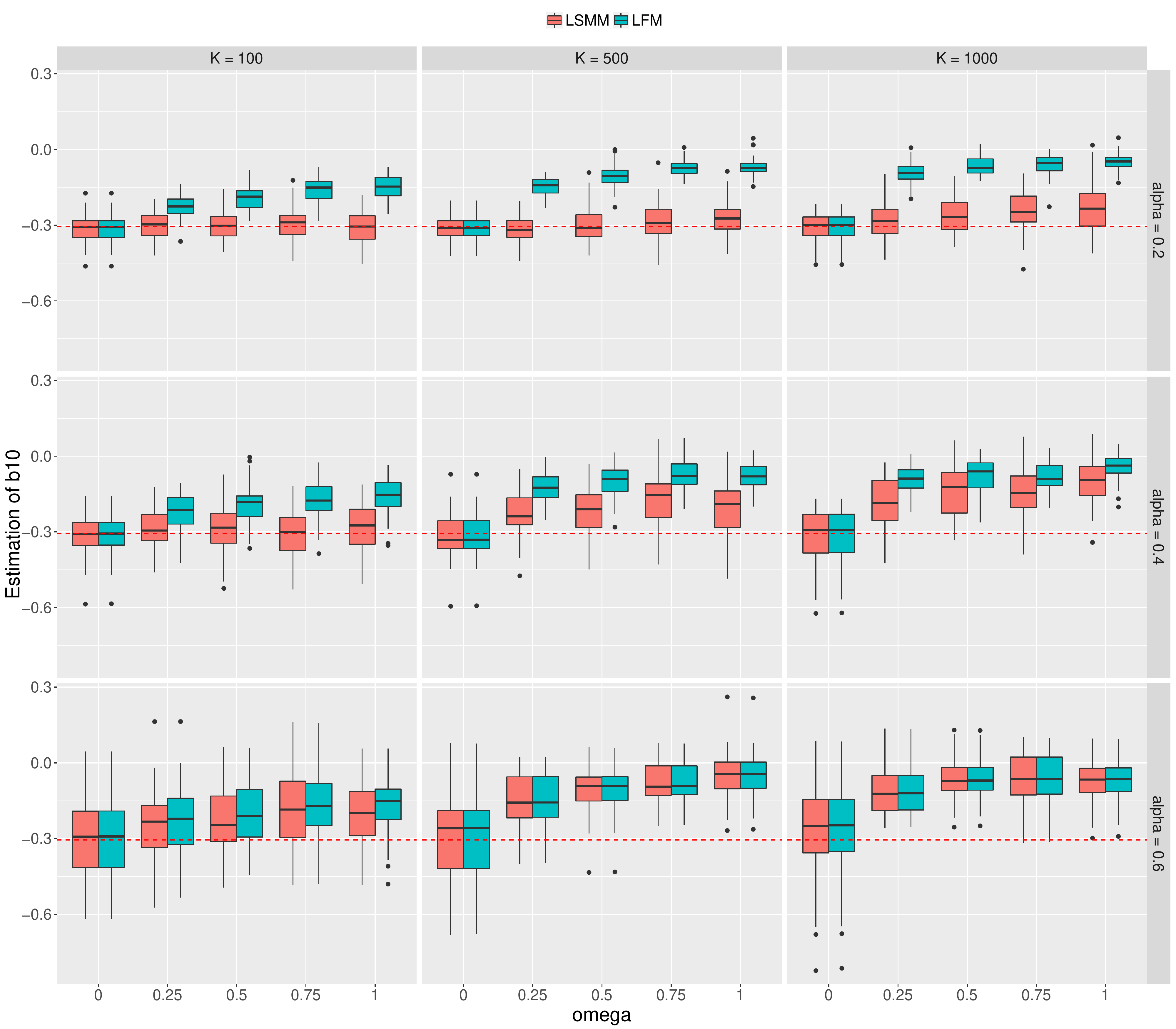}
		\par\end{centering}
	\caption{Perfermance in estimation of parameter $b_{10}$.\label{fig:est b10}}
\end{figure}

We evaluate the performance of LSMM in estimation of parameter $\beta_{0}$
and $b$. We varied $\omega$ at $\left\{ 0,0.25,0.5,0.75,1\right\} $.
Figures \ref{fig:est b0}-\ref{fig:est b10} show the comparision
between LSMM and LFM (with only fixed effects) with $\alpha=0.2$,
$0.4$ and $0.6$.

\subsubsection{Estimation of $\omega$}

\begin{figure}[H]
	\begin{centering}
		\includegraphics[scale=0.36]{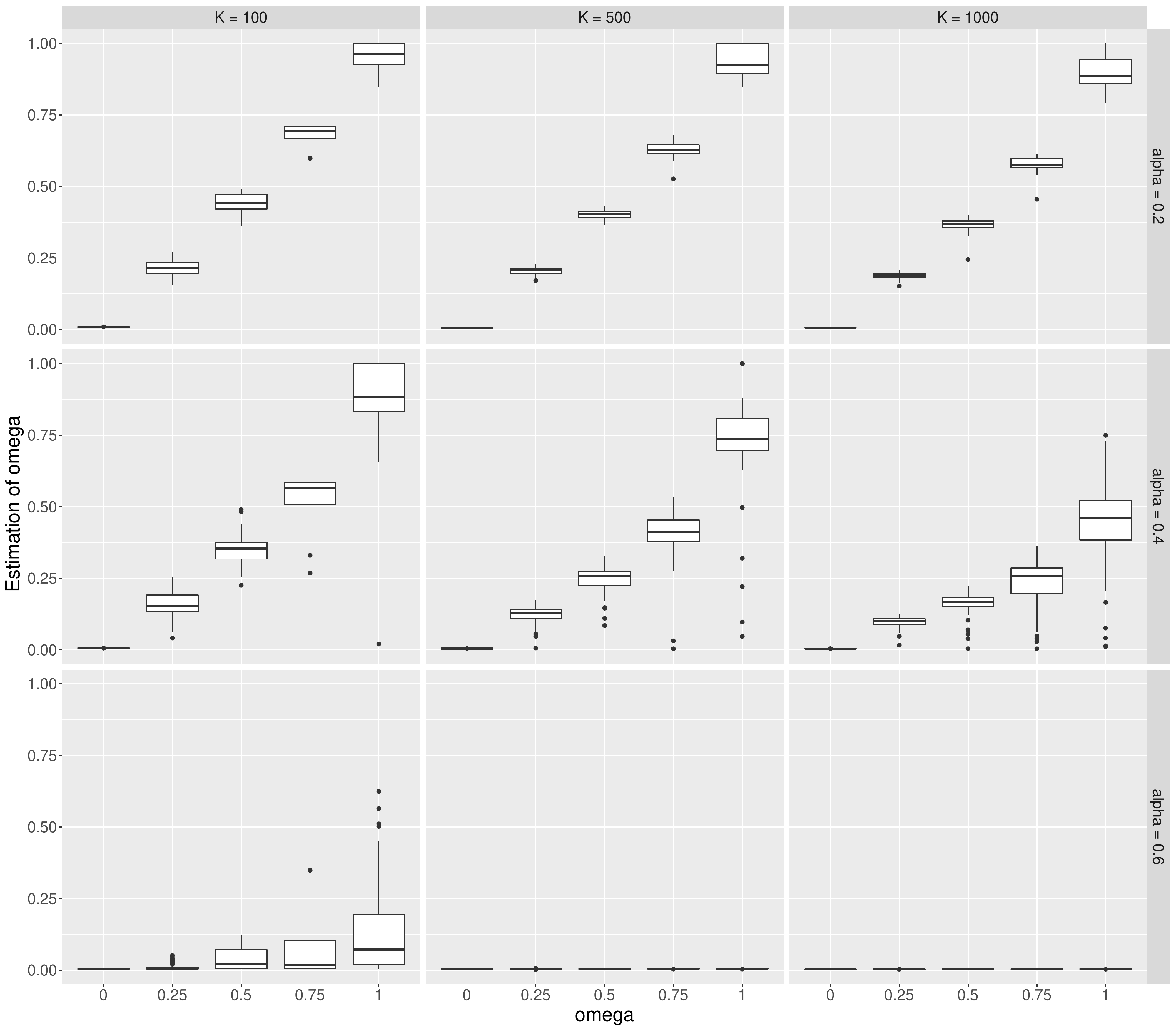}
		\par\end{centering}
	\caption{Perfermance in estimation of parameter $\omega$. \label{fig:est omega}}
\end{figure}

We evaluate the performance of LSMM in estimation of parameter $\omega$
which measures the proportion of relevant annotations. We varied $\omega$
at $\left\{ 0,0.25,0.5,0.75,1\right\} $. Figure \ref{fig:est omega}
shows the results with $\alpha=0.2$, $0.4$ and $0.6$.

\section{More about real data analysis }

\subsection{The source of the 30 GWAS}

\begin{table}[H]
	\begin{singlespace}
		\begin{centering}
			{\scriptsize{}}%
			\begin{tabular}{c|l}
				\hline 
				{\scriptsize{}Alzheimer} & {\scriptsize{}\citealp{lambert2013meta}, Nature Genetics. https://data.broadinstitute.org/alkesgroup/sumstats\_formatted/}\tabularnewline
				\hline 
				{\scriptsize{}BMI} & {\scriptsize{}\citealp{speliotes2010association}, Nature Genetics.
					https://data.broadinstitute.org/alkesgroup/sumstats\_formatted/}\tabularnewline
				\hline 
				{\scriptsize{}Bipolar Disorder} & {\scriptsize{}\citealp{psychiatric2011large}, Nature Genetics}\tabularnewline
				& {\scriptsize{}https://data.broadinstitute.org/alkesgroup/sumstats\_formatted/}\tabularnewline
				\hline 
				{\scriptsize{}Coronary Artery Disease} & {\scriptsize{}\citealp{schunkert2011large}, Nature Genetics. http://www.cardiogramplusc4d.org/data-downloads}\tabularnewline
				\hline 
				{\scriptsize{}Crohns Disease} & {\scriptsize{}\citealp{jostins2012host}, Nature. https://data.broadinstitute.org/alkesgroup/sumstats\_formatted/}\tabularnewline
				\hline 
				{\scriptsize{}Height} & {\scriptsize{}\citealp{wood2014defining}, Nature Genetics}\tabularnewline
				& {\scriptsize{}http://portals.broadinstitute.org/collaboration/giant/index.php/GIANT\_consortium\_data\_files}\tabularnewline
				\hline 
				{\scriptsize{}High-density Lipoprotein} & {\scriptsize{}\citealp{global2013discovery}, Nature Genetics}\tabularnewline
				& {\scriptsize{}http://csg.sph.umich.edu//abecasis/public/lipids2013/}\tabularnewline
				\hline 
				{\scriptsize{}HIV} & {\scriptsize{}\citealp{mclaren2013association}, PLoS Pathogens}\tabularnewline
				& {\scriptsize{}http://journals.plos.org/plospathogens/article?id=10.1371\%2Fjournal.ppat.1003515}\tabularnewline
				\hline 
				{\scriptsize{}Inflammatory Bowel Disease} & {\scriptsize{}\citealp{jostins2012host}, Nature. https://data.broadinstitute.org/alkesgroup/sumstats\_formatted/}\tabularnewline
				\hline 
				{\scriptsize{}Low-density Lipoprotein} & {\scriptsize{}\citealp{global2013discovery}, Nature Genetics}\tabularnewline
				& {\scriptsize{}http://csg.sph.umich.edu//abecasis/public/lipids2013/}\tabularnewline
				\hline 
				{\scriptsize{}Lupus} & {\scriptsize{}\citealp{bentham2015genetic}, Nature Genetics}\tabularnewline
				& {\scriptsize{}https://www.immunobase.org/downloads/protected\_data/GWAS\_Data/}\tabularnewline
				\hline 
				{\scriptsize{}Mean Cell Haemoglobin} & {\scriptsize{}\citealp{pickrell2014joint}, The American Journal of
					Human Genetics}\tabularnewline
				& {\scriptsize{}https://ega-archive.org/studies/EGAS00000000132}\tabularnewline
				\hline 
				{\scriptsize{}Mean Cell Volume} & {\scriptsize{}\citealp{pickrell2014joint}, The American Journal of
					Human Genetics}\tabularnewline
				& {\scriptsize{}https://ega-archive.org/studies/EGAS00000000132}\tabularnewline
				\hline 
				{\scriptsize{}Menopause} & {\scriptsize{}\citealp{day2015large}, Nature Genetics. http://www.reprogen.org/data\_download.html}\tabularnewline
				\hline 
				{\scriptsize{}Multiple Sclerosis} & {\scriptsize{}\citealp{sawcer2011genetic}, Nature. https://www.immunobase.org/downloads/protected\_data/GWAS\_Data/}\tabularnewline
				\hline 
				{\scriptsize{}Neuroticism} & {\scriptsize{}\citealp{okbay2016genetic}, Nature Genetics. http://ssgac.org/documents/Neuroticism\_Full.txt.gz}\tabularnewline
				\hline 
				{\scriptsize{}Primary Biliary Cirrhosis} & {\scriptsize{}\citealp{cordell2015international}, Nature Communications}\tabularnewline
				& {\scriptsize{}https://www.immunobase.org/downloads/protected\_data/GWAS\_Data/}\tabularnewline
				\hline 
				{\scriptsize{}Red Cell Count} & {\scriptsize{}\citealp{pickrell2014joint}, The American Journal of
					Human Genetics}\tabularnewline
				& {\scriptsize{}https://ega-archive.org/studies/EGAS00000000132}\tabularnewline
				\hline 
				{\scriptsize{}Rheumatoid Arthritis} & {\scriptsize{}\citealp{okada2014genetics}, Nature. https://data.broadinstitute.org/alkesgroup/sumstats\_formatted/}\tabularnewline
				\hline 
				{\scriptsize{}Schizophrenia1} & {\scriptsize{}\citealp{cross2013identification}, The Lancet.}\tabularnewline
				& {\scriptsize{}https://www.med.unc.edu/pgc/results-and-downloads (SCZ
					subset)}\tabularnewline
				\hline 
				{\scriptsize{}Schizophrenia2} & {\scriptsize{}\citealp{schizophrenia2011genome}, Nature
					Genetics.}\tabularnewline
				& {\scriptsize{}https://www.med.unc.edu/pgc/results-and-downloads (SCZ1)}\tabularnewline
				\hline 
				{\scriptsize{}Schizophrenia3} & {\scriptsize{}\citealp{ripke2013genome}, Nature Genetics. https://www.med.unc.edu/pgc/results-and-downloads
					(Sweden+SCZ1)}\tabularnewline
				\hline 
				{\scriptsize{}Schizophrenia4} & {\scriptsize{}\citealp{ripke2014biological}, Nature. https://www.med.unc.edu/pgc/results-and-downloads
					(SCZ2)}\tabularnewline
				\hline 
				{\scriptsize{}Total Cholesterol} & {\scriptsize{}\citealp{global2013discovery}, Nature Genetics}\tabularnewline
				& {\scriptsize{}http://csg.sph.umich.edu//abecasis/public/lipids2013/}\tabularnewline
				\hline 
				{\scriptsize{}Triglycerides} & {\scriptsize{}\citealp{global2013discovery}, Nature Genetics}\tabularnewline
				& {\scriptsize{}http://csg.sph.umich.edu//abecasis/public/lipids2013/}\tabularnewline
				\hline 
				{\scriptsize{}Type 1 Diabetes} & {\scriptsize{}\citealp{bradfield2011genome}, PLoS Genetics}\tabularnewline
				& {\scriptsize{}https://www.immunobase.org/downloads/protected\_data/GWAS\_Data/}\tabularnewline
				\hline 
				{\scriptsize{}Type 2 Diabetes} & {\scriptsize{}\citealp{morris2012large}, Nature Genetics. http://diagram-consortium.org/downloads.html}\tabularnewline
				\hline 
				{\scriptsize{}Ulcerative Colitis} & {\scriptsize{}\citealp{jostins2012host}, Nature. https://data.broadinstitute.org/alkesgroup/sumstats\_formatted/}\tabularnewline
				\hline 
				{\scriptsize{}Years of Education1} & {\scriptsize{}\citealp{rietveld2013gwas}, Science. https://data.broadinstitute.org/alkesgroup/sumstats\_formatted/}\tabularnewline
				\hline 
				{\scriptsize{}Years of Education2} & {\scriptsize{}\citealp{okbay2016genome}, Nature. http://ssgac.org/documents/EduYears\_Main.txt.gz}\tabularnewline
				\hline 
			\end{tabular}
			\par\end{centering}{\scriptsize \par}
	\end{singlespace}
	\caption{The source of the 30 GWAS.}
\end{table}

\subsection{Four Schizophrenia GWAS with different sample sizes}

\begin{table}[H]
	\caption{Summary of results for Schizophrenia. \label{tab:SCZ}}
	
	\begin{centering}
		\begin{tabular}{c|c|c|c|c|c}
			\hline 
			\multirow{2}{*}{} & \multirow{2}{*}{$\hat{\alpha}$} & \multicolumn{4}{c}{No. of risk SNPs}\tabularnewline
			\cline{3-6} 
			&  & Bonferroni correction & TGM & LFM & LSMM\tabularnewline
			\hline 
			Schizophrenia1 & 0.677 & 2 & 470 & 527 & 527\tabularnewline
			Schizophrenia2 & 0.633 & 7 & 2,107 & 2,404 & 2,405\tabularnewline
			Schizophrenia3 & 0.562 & 126 & 6,811 & 7,541 & 7,545\tabularnewline
			Schizophrenia4 & 0.413 & 1110 & 48,802 & 50,481 & 50,990\tabularnewline
			\hline 
		\end{tabular}
		\par\end{centering}
	a. The estimate $\hat{\alpha}$ is obtained using LSMM.
	
	b. The number of risk SNPs is reported based on global $FDR\le0.1$.
\end{table}

\begin{figure}[H]
	\begin{centering}
		\includegraphics[scale=0.2]{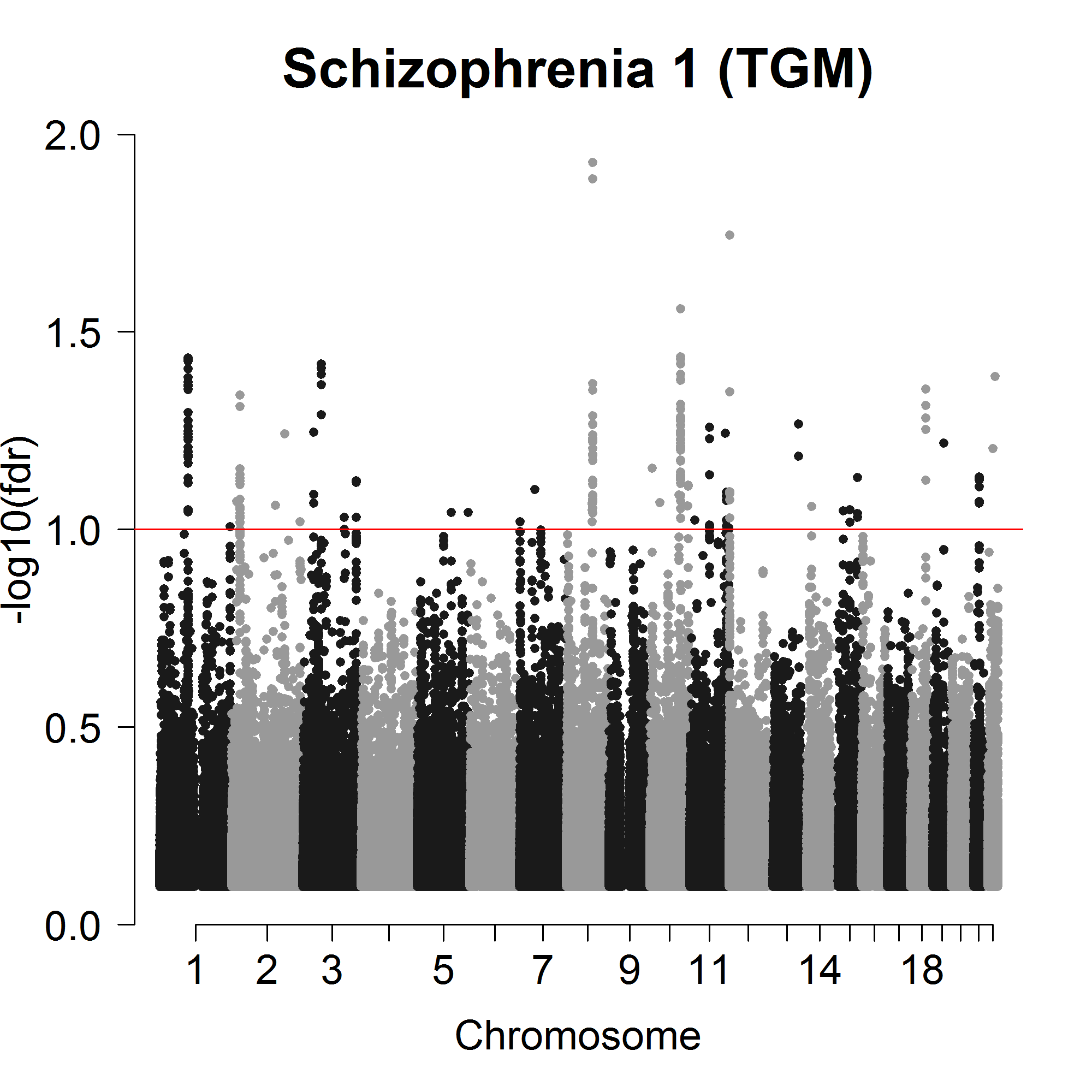}\includegraphics[scale=0.2]{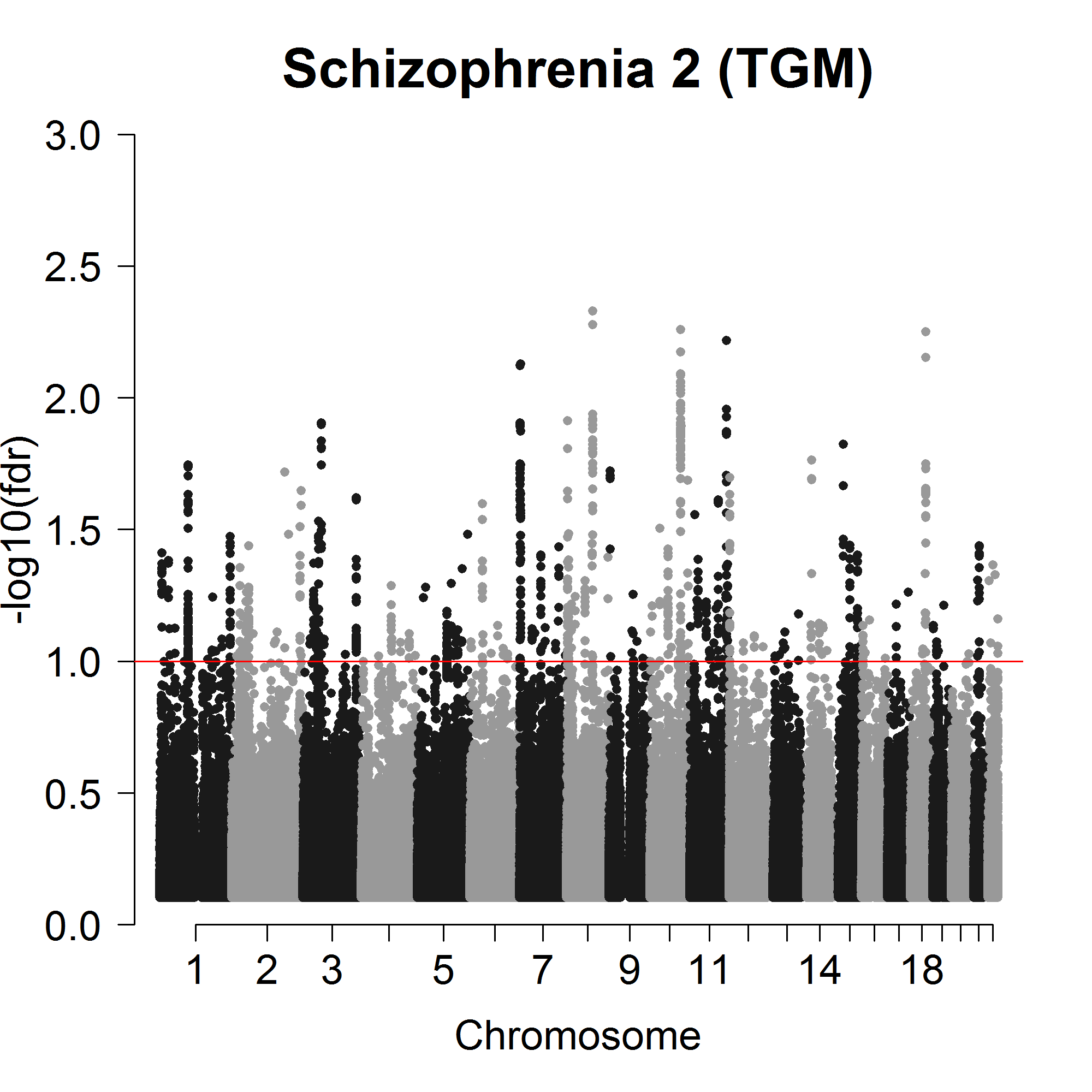}\includegraphics[scale=0.2]{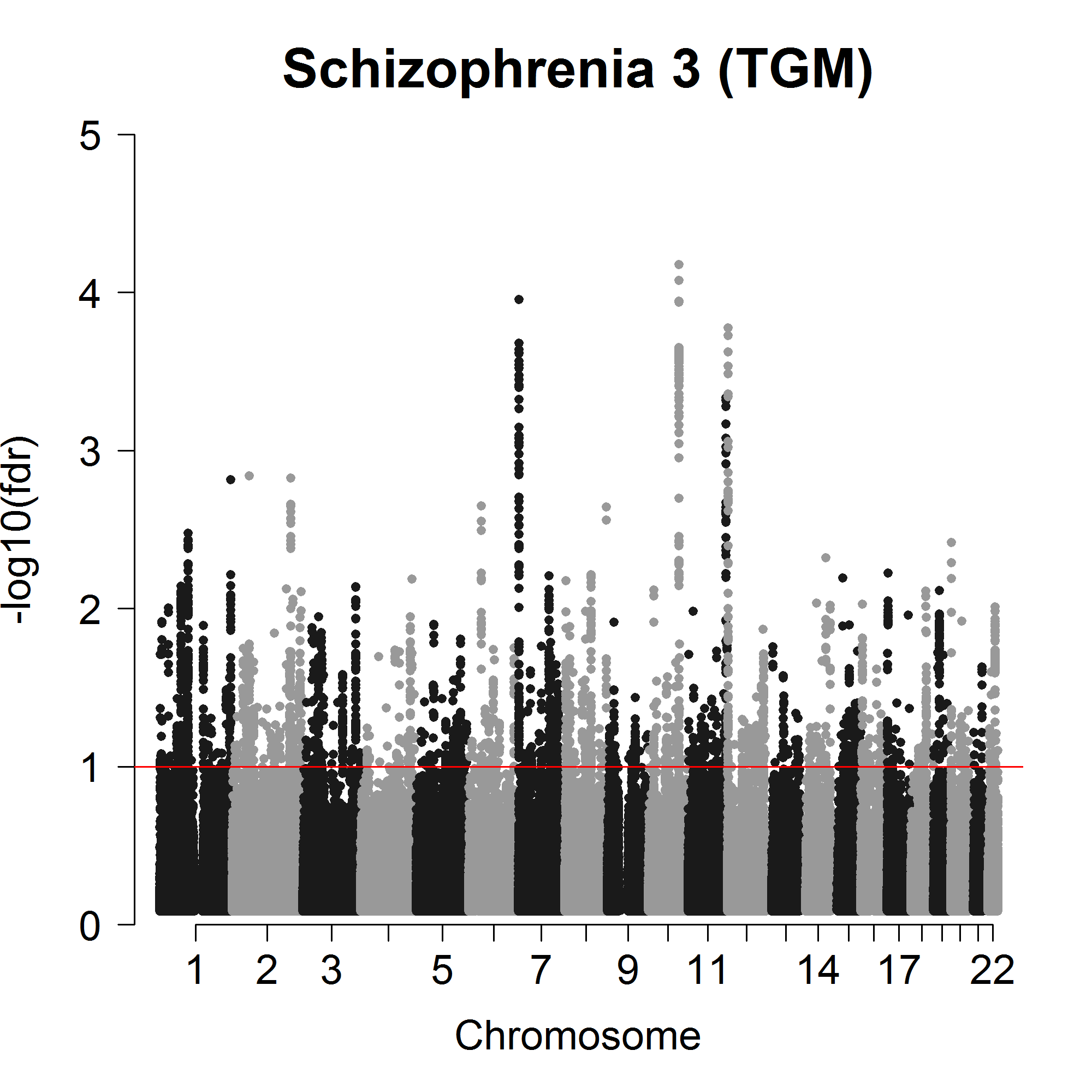}\includegraphics[scale=0.2]{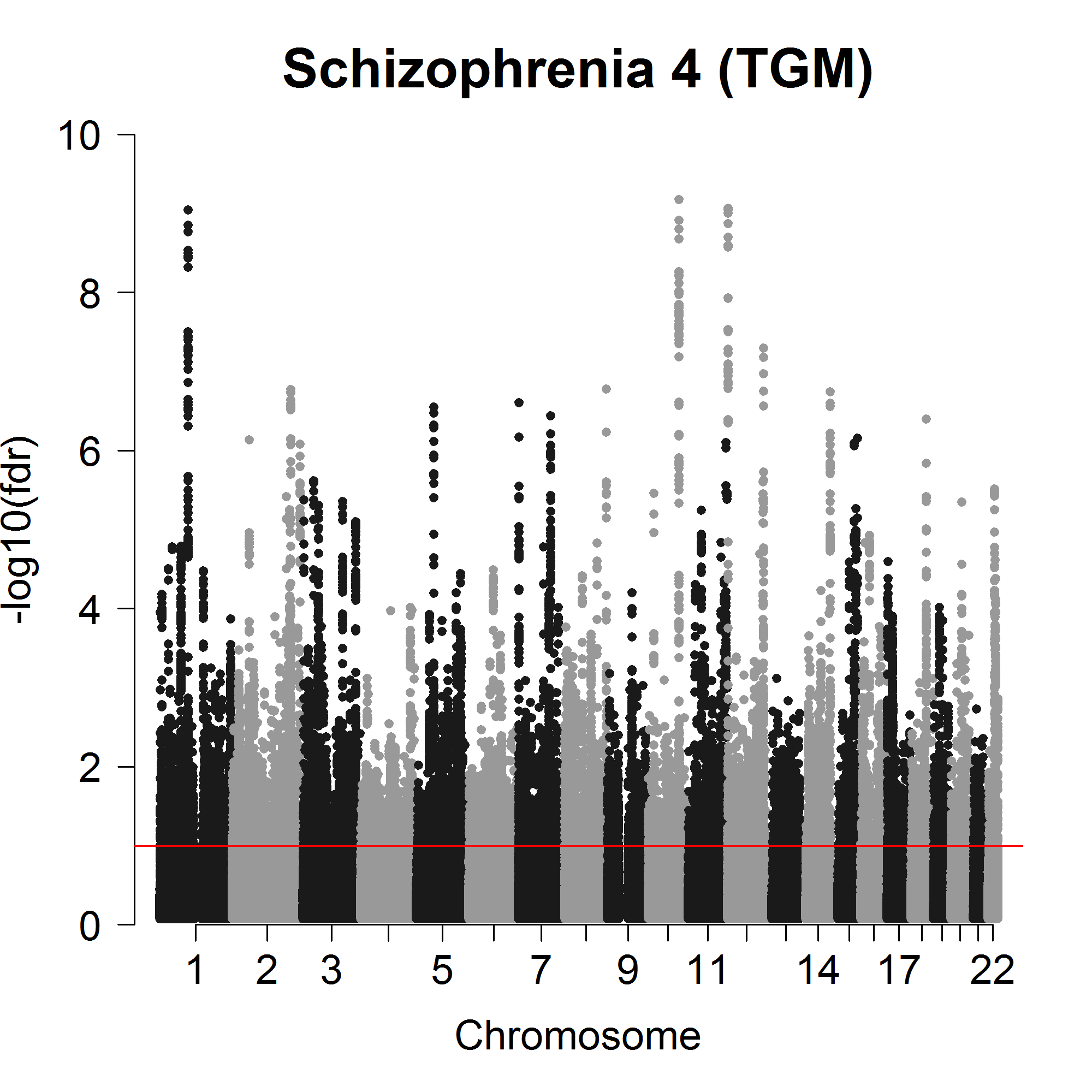}
		\par\end{centering}
	\begin{centering}
		\includegraphics[scale=0.2]{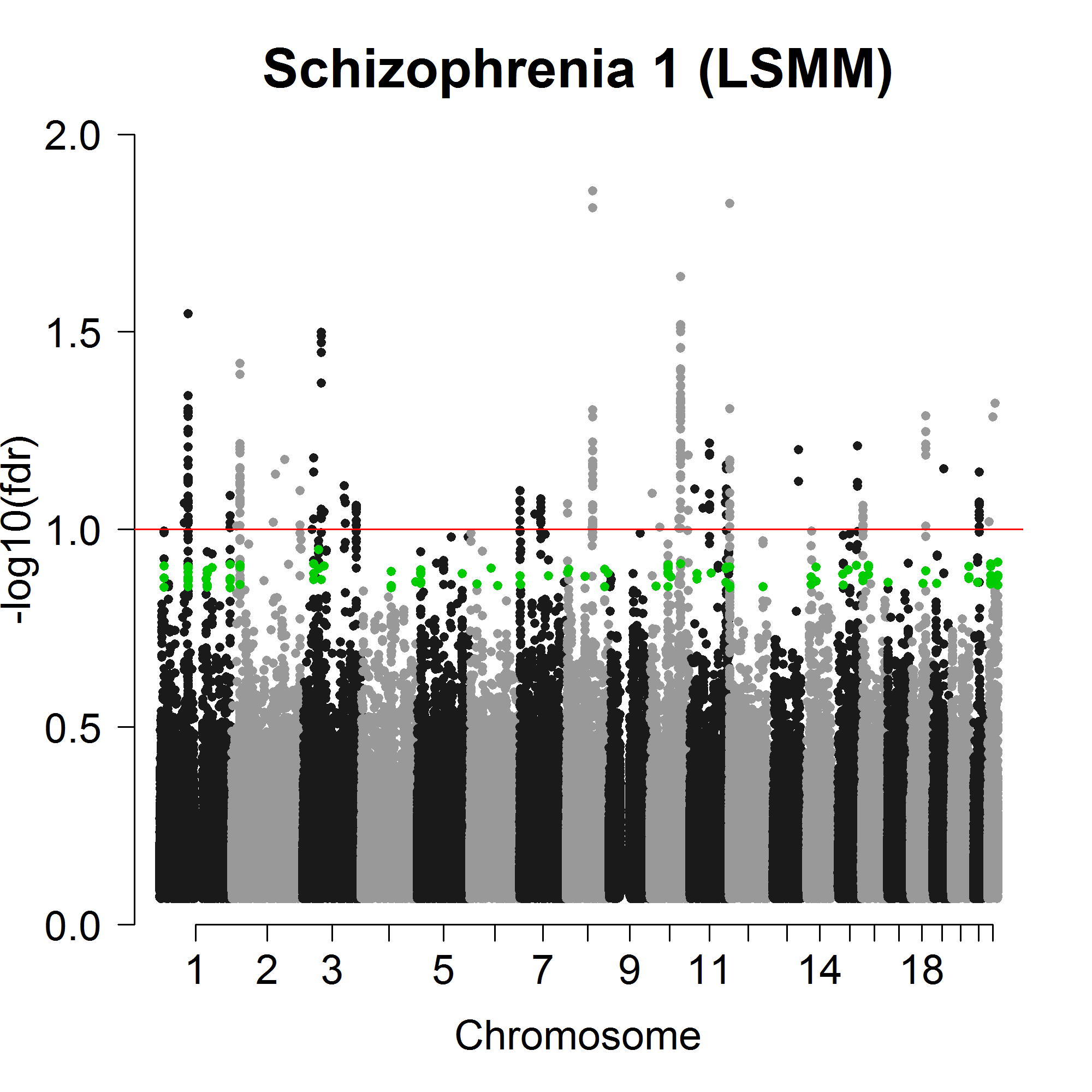}\includegraphics[scale=0.2]{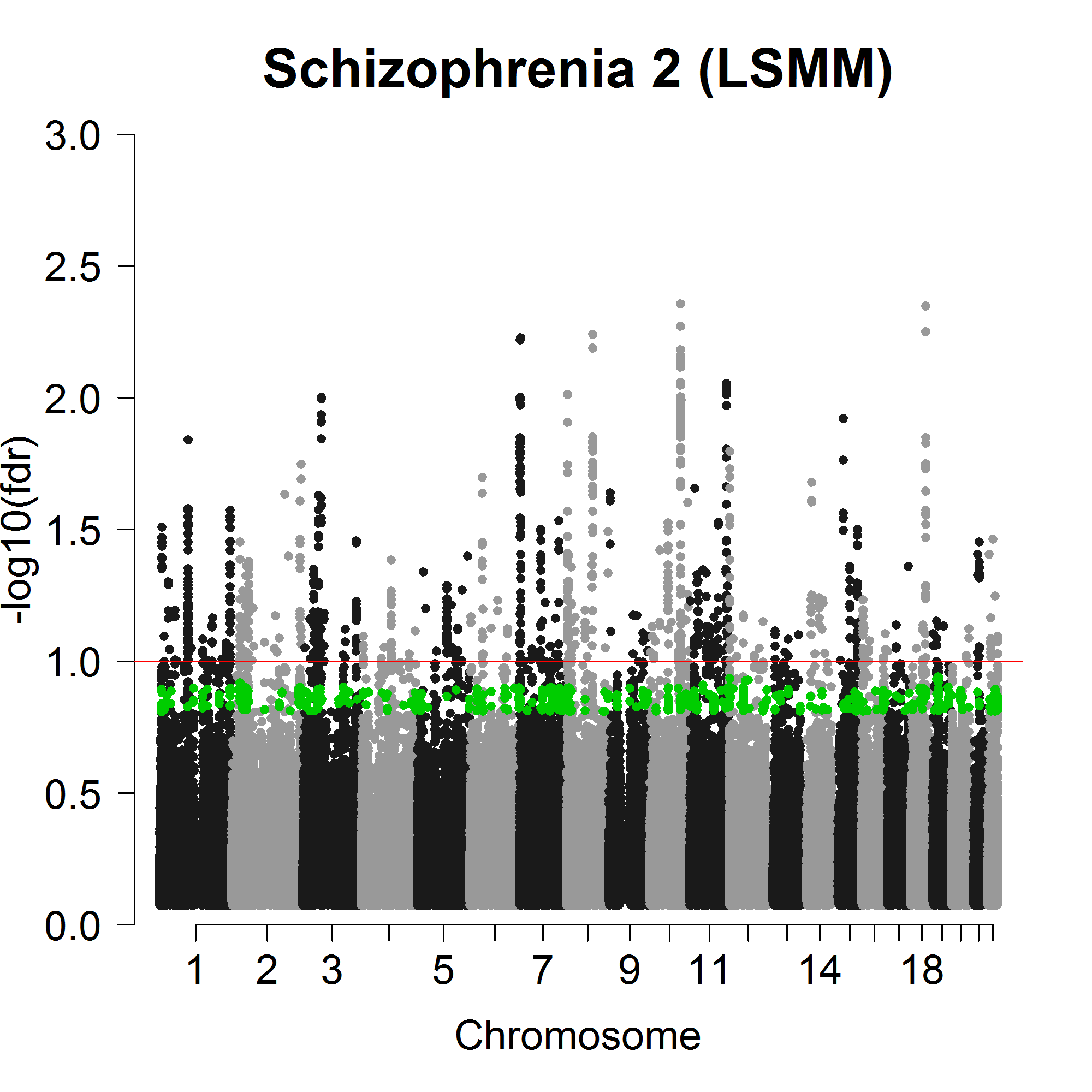}\includegraphics[scale=0.2]{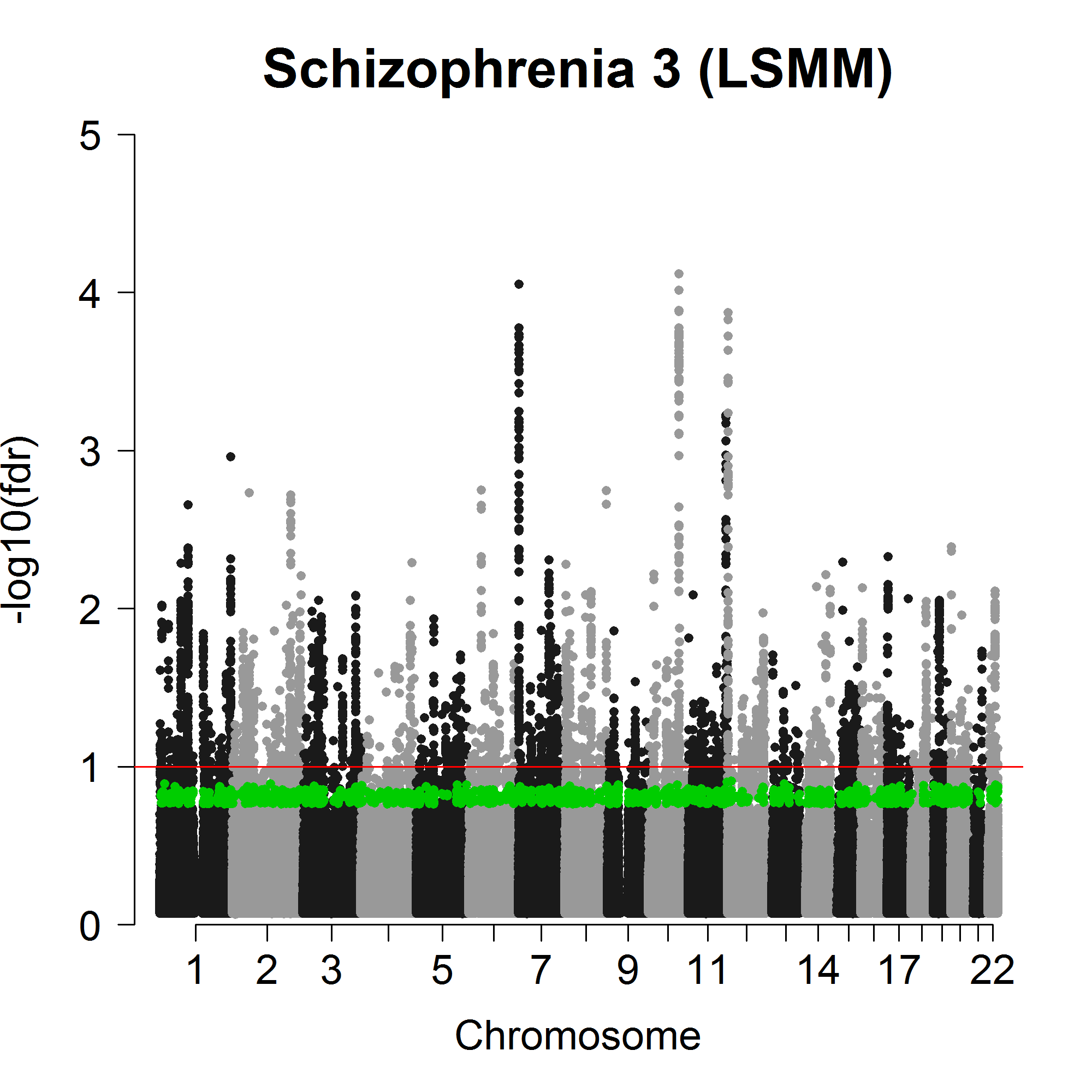}\includegraphics[scale=0.2]{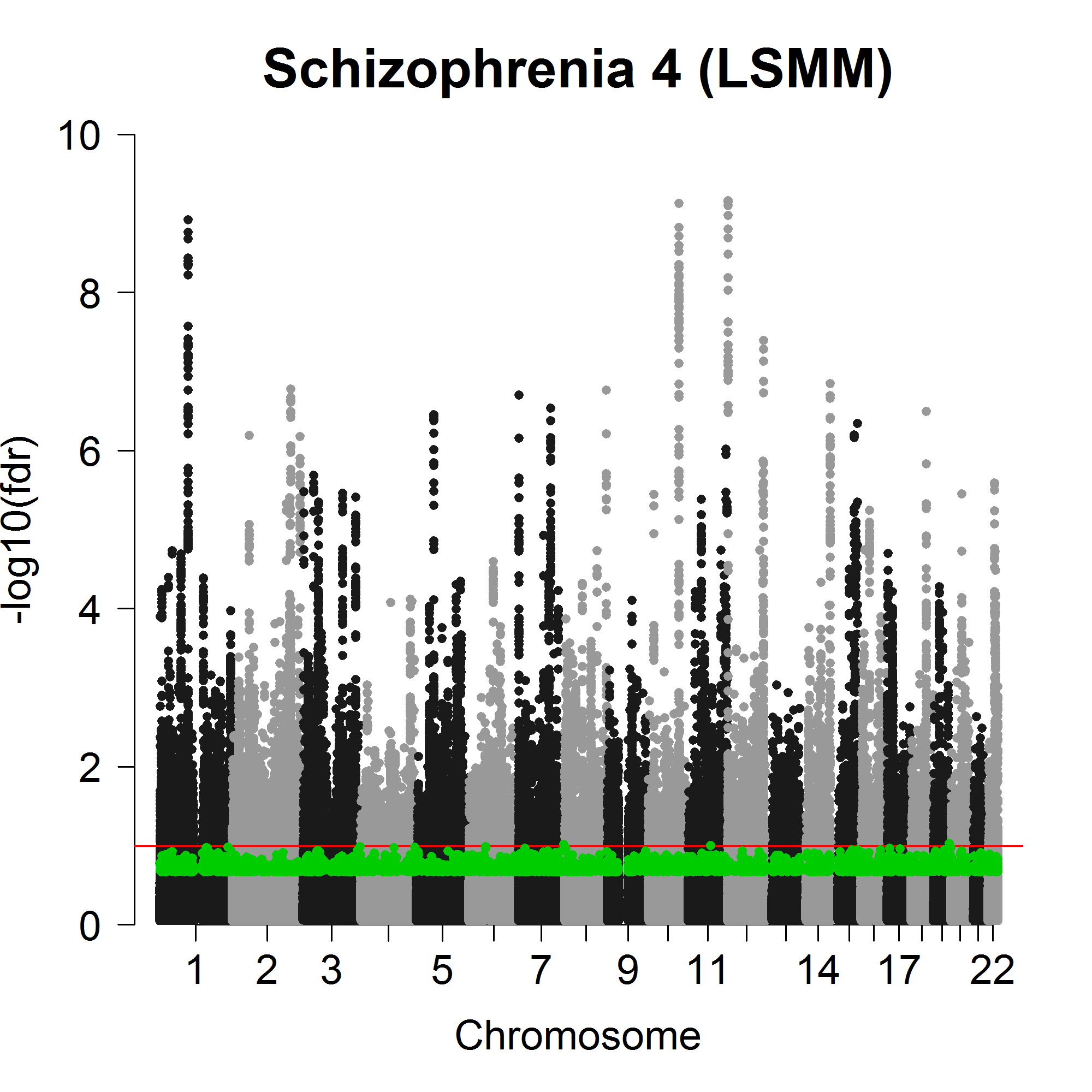}
		\par\end{centering}
	\caption{Manhattan plots of Schizophrenia1-4 using TGM and LSMM. The red lines
		indicate local $fdr=0.1$. The green points denote the additional
		SNPs LSMM identied with $FDR\le0.1$.\label{fig:man SCZ}}
\end{figure}

\subsection{Computational time for 30 GWAS}
\begin{figure}[H]
	\begin{centering}
		\includegraphics[scale=0.4]{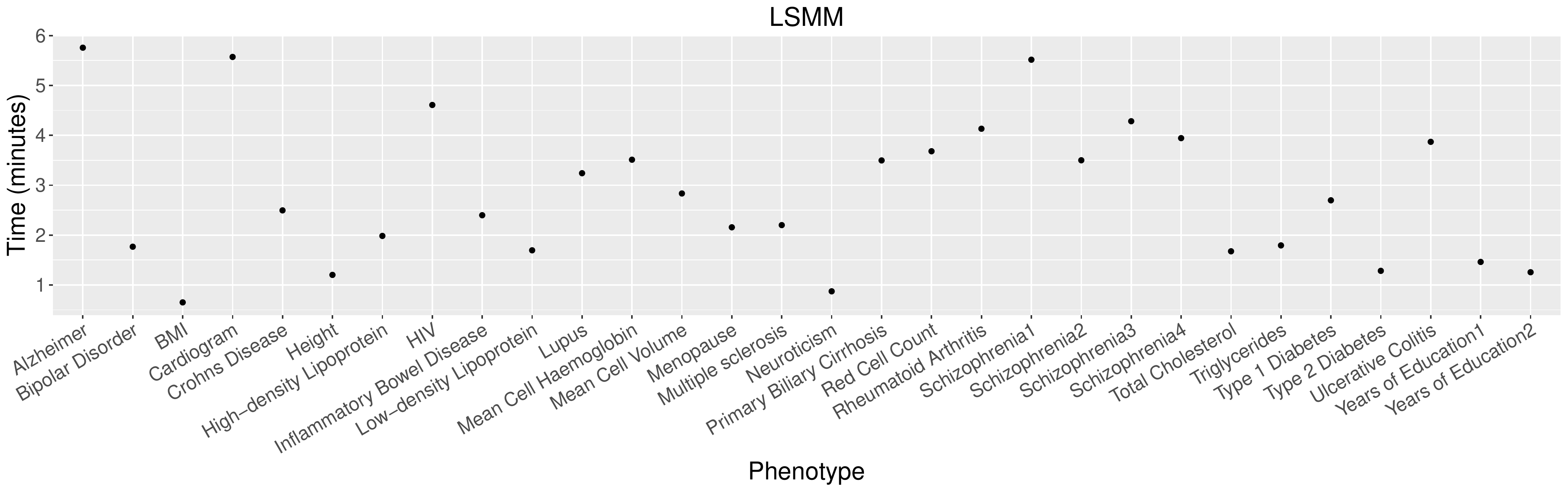}
		\par\end{centering}
	\begin{centering}
		\includegraphics[scale=0.4]{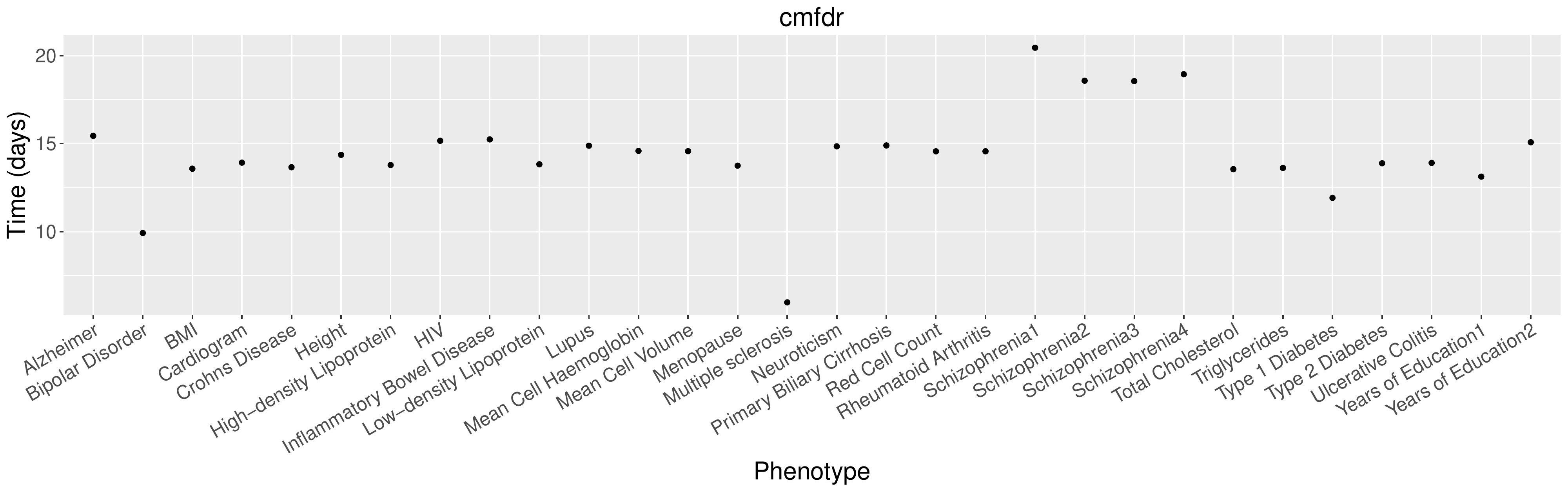}
		\par\end{centering}
	\caption{Computational time using LSMM and cmfdr for 30 GWAS.}
\end{figure}

\subsection{Relevant functional annotations for 30 GWAS without fixed effects}
\begin{figure}[H]
	\begin{centering}
		\includegraphics[scale=0.4]{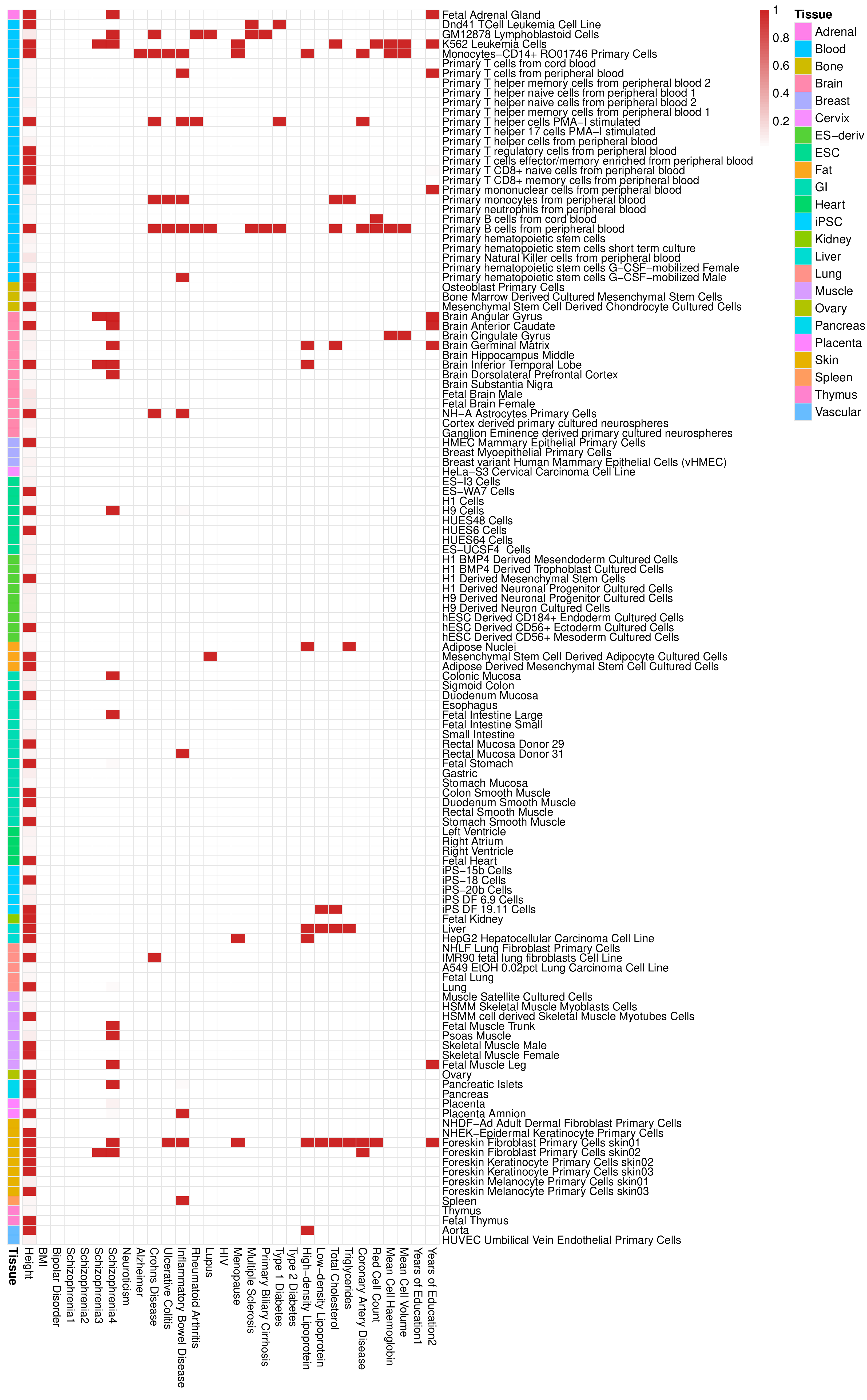}
		\par\end{centering}
	\caption{Relevant functional annotations for 30 GWAS without integrating genic category annotations.}
\end{figure}

\bibliographystyle{natbib}
\bibliographystyle{achemnat}
\bibliographystyle{plainnat}
\bibliographystyle{abbrv}
\bibliographystyle{bioinformatics}

\bibliographystyle{plain}

\bibliography{referenceLSMM}

\end{document}